\documentclass[a4paper,10pt,onesided]{anjbook}
\usepackage{times}
\usepackage{amsmath}
\usepackage{amssymb}
\usepackage{eucal}
\usepackage{multirow}
\usepackage{xspace}
\newif\ifpdf
\ifx\pdfoutput\undefined

\pdffalse % we are not running PDFLaTeX
\else
\pdfoutput=1 % we are running PDFLaTeX
\pdftrue
\fi

\ifpdf
\usepackage[pdftex]{graphicx}
\else
\usepackage{graphicx}
\fi

\usepackage{fullpage}
\usepackage[left=2.0cm,right=2.0cm,top=2.0cm,bottom=3.0cm]{geometry}
\ifpdf

\usepackage[pdftex,colorlinks=false, pdfstartview=FitV, linkcolor=blue,
           citecolor=blue, urlcolor=blue]{hyperref}
\DeclareGraphicsExtensions{.pdf, .jpg, .tif}
\else
\DeclareGraphicsExtensions{.eps, .ps,.jpg}
\fi

\newcommand{\dowidely}{\renewcommand{\baselinestretch}{1.5}\selectfont} % was 1.3
\newcommand{\donarrowly}{\renewcommand{\baselinestretch}{1.0}\selectfont}

\newcommand{\es}{extended sampling}
\newcommand{\pc}{phase-constrained distribution}
\newcommand{\ov}{\ensuremath{\tilde{O}}}
\newcommand{\pia}{{\ensuremath{\pi^c_A}}}
\newcommand{\pib}{{\ensuremath{\pi^c_B}}}

\newcommand{\qpia}[1]{{\ensuremath{\pi^{q}_{A,{#1}}}}}
\newcommand{\qpib}[1]{{\ensuremath{\pi^{q}_{B,{#1}}}}}

\newcommand{\pips}{{\ensuremath{\pi^c_{PS}}}}

\newcommand{\tpips}{{\ensuremath{{\tilde{\pi}}^c_{PS}}}}
\newcommand{\tpia}{{\ensuremath{\tilde{\pi}^c_A}}}
\newcommand{\tpib}{{\ensuremath{\tilde{\pi}^c_B}}}

\newcommand{\palalp}{{\ensuremath{\al\rightarrow\alp}}}
\newcommand{\palpal}{{\ensuremath{\alp\rightarrow\al}}}

\newcommand{\avec}[1]{{\ensuremath{\bf #1}}}

\newcommand{\stbox}{\vspace{0.2cm}\begin{tabular}{|c|}\hline
\begin{minipage}{0.9\textwidth}\vspace*{0.4cm}}
\newcommand{\equaldot}{{\ensuremath{\stackrel {\mathrm{.}}{=}}}}
\newcommand{\tpi}{\ensuremath{\tilde{\pi}}}
\newcommand{\deb}{\ensuremath{\tilde{D}}}

\newcommand{\hhat}{\ensuremath{\hat{H}}}
\newcommand{\that}{\ensuremath{\hat{T}}}

\newcommand{\ket}[1]{\ensuremath{|{#1}>}}
\newcommand{\bra}[1]{\ensuremath{<{#1}|}}
\newcommand{\braket}[2]{\ensuremath{<{#1}|{#2}>}}
\newcommand{\est}{\ensuremath{\stackrel {\mathrm{e.b.}}{=}}}
\newcommand{\finbox}{\vspace*{0.2cm}\end{minipage}\\ \hline \end{tabular}}

\newcommand{\mba}{\ensuremath{M_{BA}}}

\newcommand{\mal}{\ensuremath{M_{\alp\al}}}
\newcommand{\mbai}{\ensuremath{M_{BA,i}}}
\newcommand{\mbak}{\ensuremath{M_{BA,k}}}

\newcommand{\mab}{\ensuremath{M_{AB}}}

\newcommand{\ma}{{$M$}}

\newcounter{abc}

\newcommand{\discard}[1]{}

\newcommand{\rvec}{\avec{r}}
\newcommand{\arvec}{\avec{r}}
\newcommand{\aevec}{\avec{e}}
\newcommand{\ecal}{{\cal {E}}}

\newcommand{\uvec}{\ensuremath{\avec{u}}}
\newcommand{\auvec}{\avec{u}}

\newcommand{\vvec}{\ensuremath{\avec{v}}}
\newcommand{\Vvec}{\ensuremath{\avec{V}}}
\newcommand{\avvec}{{\avec{v}}}

\newcommand{\Rvec}{\ensuremath{\avec{R}}}
\newcommand{\Dvec}{\ensuremath{\avec{D}}}
\newcommand{\aRvec}{\avec{R}}

\newcommand{\rpf}{ratio of the partition functions}
\newcommand{\metrop}{Metropolis algorithm}
\newcommand{\cs}{configuration space}
\newcommand{\ecs}{(effective) configuration space}
\newcommand{\ce}{configurational energy}
\newcommand{\ces}{configurational energies}

\discard{

}

\newcommand{\al}{\ensuremath{{\gamma}}}
\newcommand{\alp}{\ensuremath{{\tilde{\gamma}}}}
\newcommand{\m}{\ensuremath{\mu}}
\newcommand{\ga}{\ensuremath{\gamma}}
\newcommand{\la}{\ensuremath{{\lambda}}}

\newcommand{\RBA}{\ensuremath{{{R}_{BA}}}}
\newcommand{\RBAcal}{\ensuremath{{\cal{{R}_{BA}}}}}

\newcommand{\Ralcal}{\ensuremath{{\cal{{R}_{\alp\al}}}}}
\newcommand{\RBAP}{\ensuremath{{\cal{R}_{BA,P}}}}

\newcommand{\RBAPq}{\ensuremath{{{\cal{R}^q}_{BA,P}}}}
\newcommand{\RBAq}{\ensuremath{{{\cal{R}^q}_{BA}}}}

\newcommand{\z}{\ensuremath{{\tilde{Z}}}}

\newcommand{\wal}{\ensuremath{{{W}_{\alp\al}}}}
\newcommand{\wba}{\ensuremath{{{W}_{BA}}}}
\newcommand{\wbai}{\ensuremath{{W_{BA,i}}}}

\newcommand{\hcal}[2]{\ensuremath{{\cal{H}}_{#1} ({#2})}}
\newcommand{\hcalone}[1]{\ensuremath{{\cal{H}}_{#1}}}

\newcommand{\delT}{\ensuremath{{\triangle T}}}
\newcommand{\delt}{\ensuremath{{\triangle t}}}

\newcommand{\delela}{\ensuremath{{\triangle\lambda}}}

\newcommand{\f}[2]{\ensuremath{{\frac{#1}{#2}}}}
\newcommand{\s}[2]{\ensuremath{{\sum_{#1}^{#2}}}}
\newcommand{\mf}[1]{\ensuremath{{\mathbf{#1}}}}

\newcommand{\p}[2]{\ensuremath{{\prod_{#1}^{#2}}}}

\newcommand{\set}[1]{\ensuremath{{\{{#1}_i\}}}}
\newcommand{\seta}[1]{\ensuremath{{\{{#1}\}}}}

\newcommand{\h}{\ensuremath{\f 1 2}}

\newcommand{\fig}{Figure}
\newcommand{\eq}{Equation}
\newfont{\sss}{cmsy10}
\newcommand{\exi}{\text{\sss\symbol{69}}} %excitation energy

\newcommand{\tshow}[1]{}
\newcommand{\hidefigure}[1]{#1}
\begin{document}
\setlength{\headsep}{1cm}

\mainmatter
\pagestyle{headings}
\dowidely

% How to number equations
\numberwithin{equation}{chapter}
% or \numberwith{equation}{section}

%==========================================================

\begin{titlepage}
\vspace*{0.5cm}
\begin{center}
{\bf\huge\sffamily Free energy differences: }
\end{center}

\begin{center}
{\bf\Large\sffamily Representations, estimators, and sampling strategies}
\end{center}

\vspace*{0.5cm}
{\Large
\begin{center}
Arjun R. Acharya
\end{center}
\vspace*{2.5cm}
\begin{center}
\includegraphics[scale=0.5]{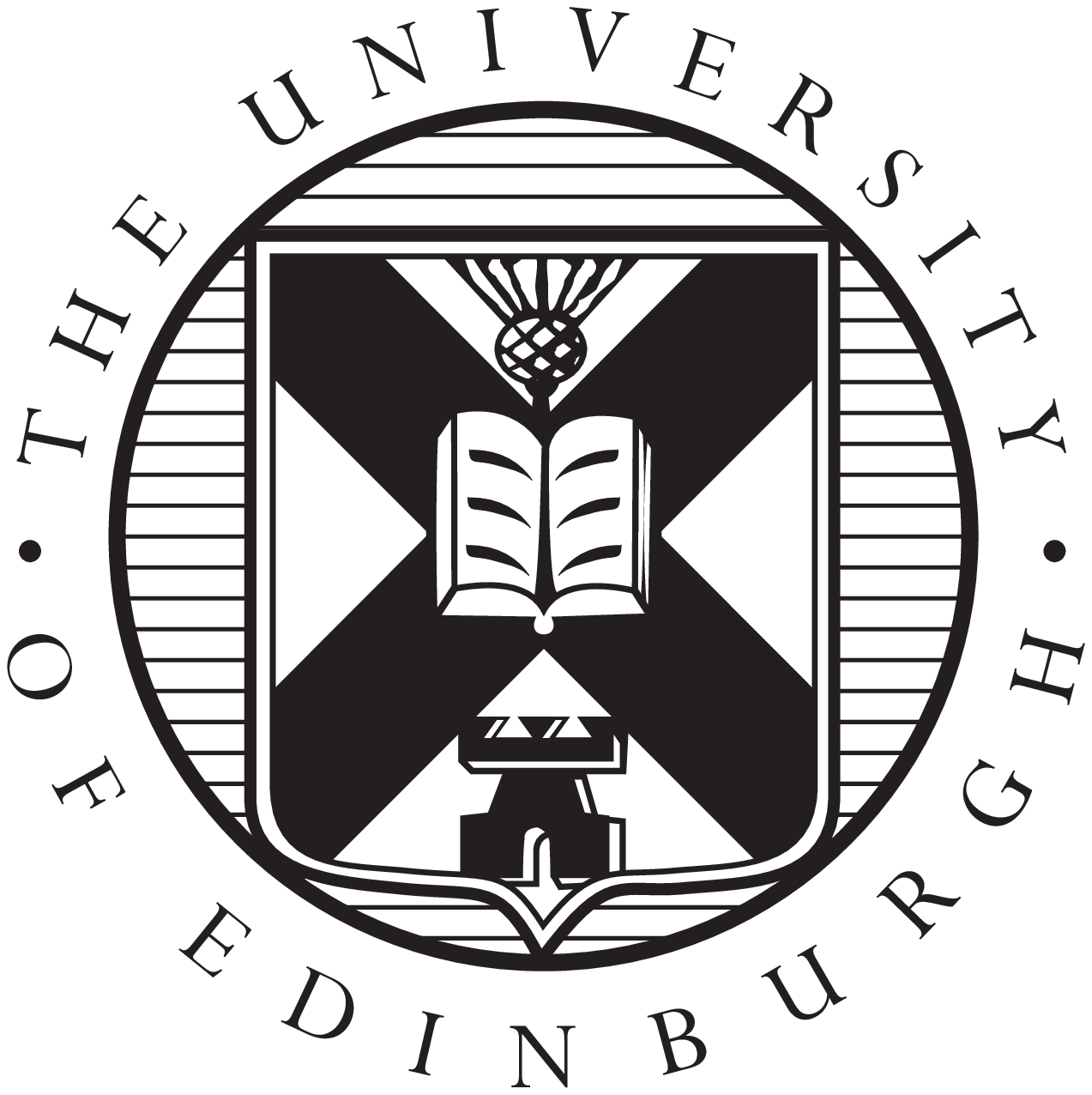}
\end{center}
\vspace*{3.5cm}
\begin{center}
Doctor of Philosophy
\end{center}
\begin{center}
The University of Edinburgh
\end{center}
\begin{center}
June 2004
\end{center}
}

\end{titlepage}

\chapter*{Abstract}

Within the framework of equilibrium statistical mechanics the free energy of a phase gives a measure of its associated probabilistic weight. In order to determine phase boundaries one must then determine the conditions under which the free energy difference (FED) between two phases is zero. The underlying complexity usually rules out any analytical approaches to the problem, and one must therefore adopt a  computational approach. The focus of this thesis is on (Monte Carlo) methodologies for FEDs. In order to determine FEDs via Monte Carlo, the simulation must (in principle) be able to visit the regions of configuration space associated with {\em both} phases in a single simulation. Generally however one finds that these regions are  significantly dissimilar, and are separated by an intermediate region of configuration space of intrinsically low probability, so that  a simulation initiated in either of the phases will tend to remain in that phase. This effect is generally referred to as the {\em overlap problem} and is the most significant obstacle that one faces in the task of estimating FEDs.

In chapter \ref{chap:review} we start by formulating the FED problem using the Phase Mapping (PM) technique of \cite{p:LSMC}. This technique allows one to circumvent the intermediate regions of configuration space altogether by mapping configurations of one phase directly onto those of the other. Despite the improvement that one gets when formulating the problem via the PM, the overlap problem persists, albeit to a lesser degree. In  chapter \ref{chap:review} we define precisely what we mean by overlap and then discuss a range of methods that are available to us for calculating the FEDs within the PM formalism. In the subsequent chapters we then focus on the three generic strategies that arise in addressing the overlap problem.

The first strategy that we focus on (chapter \ref{chap:tune}) is that of the {\em representation}. This corresponds to the choice of coordinate system with which one parameterises the degrees of freedom of the two phases. The PM works by {\em matching} these coordinates in the two systems, and therefore its efficiency in generating overlap is dependent on the choice of representation. We examine a particular representation in which the PM matches the {\em fourier} coordinates of the two phases in such a way so as to achieve perfect overlap in the harmonic limit (for structurally ordered phases). Previous formulations \cite{p:LSMC} have been limited to the real-space PM (RSM). By comparing the RSM to the fourier version (FSM), we show that for a range of temperatures the overlap associated with the FSM is considerably better than that of the RSM, thus allowing one to determine the FED efficiently under conditions in which the RSM would otherwise fail.

The second strategy that we study (in chapter \ref{chap:estsamp}) is that of the {\em estimator} which one uses to determine the free energy difference. The different estimators {\em pool} the data from the various regions of configurations space in different ways, and therefore have different (systematic and statistical) errors associated with them. As a consequence the  severity of the overlap problem is dependent on the estimator that one employs. We examine  conditions under which the different estimators are able to arrive at estimates which are free of systematic errors.

Generally however, the scope for refinement of the representation and the estimator is limited. In chapter \ref{chap:sampstrat} we deal with the third and final strategy which appears in the FED problem; the { refinement} of the {\em sampling strategy}. This involves the construction of a sampling distribution which explores regions of configurations space {\em outside} those typical of the two phases, and thereby {\em engineers} overlap. In particular we examine three strategies. The first is the Multicanonical strategy, and involves the introduction of corrections to the configurational energy appearing in the Metropolis acceptance probabilities, so as to force the simulation outside the regions associated with the corresponding phase. The second that we examine, and which has been developed here for the first time, is the Multihamiltonian strategy. This involves the independent  simulation of several systems, which overlap in the regions of configuration space that they explore, in a way which allows one to determine the FED. The advantage of this method is that it is  highly parallelizable in a way that is not possible for the other two methods. The third and final method that we study is the Fast Growth method, and involves performing non-equilibrium work on the system so as to force it from the regions of configuration space associated with one phase to those of the other. We demonstrate that all three methods are effective in overcoming the overlap problem.

The discussion until this point will be limited to the task of determining phase boundaries within the purely classical framework. At low temperatures, however,  quantum effects begin to become increasingly important; this is especially true for particles with light masses. Generally there are two types of quantum effects that arise. The first are quantum discretisation effects, which arise when the typical energy is of the order to the phonon excitation energies. The second effect is that of quantum exchange, which arises when the de Broglie wavelength becomes roughly of the order of the interatomic spacing. In the final part of this thesis (chapter \ref{chap:quantum}) we address the additional phenomena which  arise from the quantum discretisation effects by generalising the PM formalism so as to be applicable within the path integral formulation of statistical mechanics. The path integral approach allows one to obtain thermodynamic information of a quantum system by modelling a classical system in which the interacting particles are replaced by interacting polymers. The simulation of such a system expends considerably more computational effort than its classical counterpart, and as a result makes the calculation of the quantum FED considerably more difficult. We get around this problem by using the parallelizability of the  Multihamiltonian method to divide the computation of the quantum FED amongst several processors in a parallel cluster. This allows us to examine the importance of zero point motion on the quantum Lennard-Jones phase diagram.

\chapter*{Declaration}

This thesis has been written entirely by me and has not been submitted in any previous application for a degree. Except where stated, all the work detailed in this thesis was carried out by me.

\vspace*{1.5cm}

\noindent Arjun R. Acharya

\noindent June 2004

\chapter*{}

\vspace*{2cm}

\hspace*{10cm}{ To my parents and my teachers}

\chapter*{Acknowledgements}

\noindent I would like to thank the following people: first and foremost Alastair D. Bruce for his guidance and support during the PhD; Graeme J. Ackland for support and for allocating me time on the EPCC HPCx parallel systems; Andrew Jackson for making available his Lattice Switch code, upon which all the work contained in this thesis has been based. I am also very grateful to Rafael Ramirez for posting me information from Spain to get me started on Path Integral Monte Carlo.

\noindent I gratefully acknowledge the support of an EPSRC research studentship.
\setcounter{tocdepth}{5}
\tableofcontents
\newpage
\setcounter{secnumdepth}{0}
\section{List of Symbols}

\begin{itemize}

\item $\equaldot$ : equality up to a normalisation constant which is not known a-priori.

\item $<Q>_\pi$ : this denotes the expectation of a macrovariable Q with respect to the sampling distribution $\pi$ (see Eq. \ref{eq:ensembleav}).

\item $\gamma$ : this variable denotes the phase label. The two possible values for $\gamma$ are $\gamma=A$ and $\gamma=B$.

\item $\beta = \frac {1} {kT}$ where k is the Boltzmann constant

\item $\delela_i \equiv \la_{i+1} -\la_i$. This is the increment in the field parameter. For all the simulations employed in this thesis, the increments were the same, so that $\delela_i = \delela$ (see section \ref{sec:ms} and section \ref{sec:fg}).

\item \delT\ : this corresponds to the time (in the FG method) for which the system is equilibrated, before work is subsequently performed on it.

\item $\triangle_\gamma (\rvec)$ : variable which keeps track of the phase. It assumes the value unity if $\rvec$ corresponds to a configuration generated in phase $\gamma$ and zero if not. In order for this function to be able to work, one must (in the most general case) keep track of the phase label by appeal to the reference configuration (Eq. \ref{eq:rinu}, Eq. \ref{eq:rintermsofv}) about which the particles are displaced.

\item $\triangle_\gamma [M]$ : variable which assumes the value unity if $M \epsilon [M]_\gamma$ and zero otherwise. In the case where $[M]_A$ and $[M]_B$ do not overlap (i.e. when M is an order parameter for the two systems), then this function may be used to identify the phases.

\item $\la$ : the field parameter used to morph the configurational energy of one phase onto that of the other phase. See Eq. \ref{eq:linearhamiltonian} for a simple example of this.

\item $\la_q$ : See Eq. \ref{eq:debrog}.

\item $\pi^c(\arvec)$ : the Boltzmann sampling distribution, Eq. \ref{eq:boltzsamp}

\item $\pi^c_\gamma(\arvec)$ : the Boltzmann sampling distribution when constrained to phase $\gamma$ (see Eq. \ref{eq:sampconstrain}). Since a phase essentially corresponds to a local basin of attraction in the configurational energy $E(\rvec)$, this distribution may be realised by implementing a simulation initiated in phase $\al$. The local structure of the \cs\ will ensure that the simulation will {\em remain} in that phase.

\item $\tpi^c_\al(\Vvec)$ : the canonical MH sampling distribution given by Eq. \ref{eq:mhsamp} in terms of the collective configuration of the composite (multi-replica) system.

\item $\pips=\pips(\vvec,\al)$ : the canonical PS sampling distribution (Eq. \ref{eq:PSsamp}) in terms of the effective configuration \vvec\ of the system, where the phase label is a stochastic variable.

\item $\pi^m(\vvec)$ : the superscript m denotes a MUCA sampling distribution.

\item $\pi^q_\al(\seta {\uvec})$ : the quantum sampling distribution of phase \al\ (see Eq. \ref{eq:qsamp}).

\item $\tau$ : the effective temperature in the quantum system (Eq. \ref{eq:eftemp}).

\item A(x) : the metropolis acceptance function, Eq. \ref{eq:metro}

\item \Dvec\ : global translation vector appearing in the PM formulation. Figure \ref{phase_in_realspace_withswitch}, Eq. \ref{eq:DLS}. 

\item \deb\ : The De Boer parameter (Eq. \ref{eq:pdeb}) appears in quantum Lennard-Jones systems and fixes the temperatures at which the different quantum effects become important (see also appendix \ref{app:deboer}).

\item $E(\arvec)$ : the configurational energy of the system.

\item $E_\gamma (\vvec)$ : the configurational energy of phase $\gamma$ in the \vvec\ representation, Eq. \ref{eq:absoluteeinv}. 

\item $\triangle E^o_{BA}$ : the difference in energies between the reference configuration of phase B and that of phase A (see Eq. \ref{eq:gse}).

\item ${\mf e}^{ij}_\al$ : the i'th component of the j'th eigenvector of the dynamical matrix ${\mf K}_\al$ of phase \al .

\item $\ecal_\gamma (\vvec)$ : difference between the actual energy of the system and the energy of the reference configuration (Eq. \ref{eq:expandenergy}). In the case of the reference configuration being the ground state (as it is in the case of crystalline solids, where the reference configurations correspond to the lattice sites), $\ecal_\gamma (\vvec)$ corresponds to the excitation energy of the system. We will frequently refer to this simply as the 'configurational energy'.

\item $\ecal^h_\al (\vvec)$ : the harmonic contribution to the excitation energy of phase \al\ (see Eq. \ref{eq:energy_har} and Eq. \ref{eq:einv}).

\item $\ecal^a_\al (\vvec)$ : the anharmonic contribution to the excitation energy of phase \al .

\item $\ecal_\la$ : a configurational energy which is a function of the field parameter \la . One may use this function to construct a chain of configurational energies (see Eq. \ref{eq:hamilchain}) which links the \ce\ associated with phase A ($\ecal_A$) to that associated with phase B ($\ecal_B$). The most straightforward parameterisation, one which we will employ, is the linear one of Eq. \ref{eq:linearhamiltonian}.

\item $\ecal_q$ : the configurational energy of the quantum system.

\item $F_\gamma$ : the absolute free energy of phase $\gamma$, Eq. \ref{eq:fundamental}.

\item $\triangle F_{BA} = F_B - F_A$

\item $\triangle F^a_{BA}$ : the anharmonic contribution to the FED.

\item $H(M_{BA,i} | \pi)$ : the histogram recording the number of times a data output of the simulation falls in bin $M_{BA,i}$ under an experiment performed with the sampling distribution $\pi$.

\item $\hcalone {\al}$: In the MH method this corresponds to the configurational  energy 'associated' with the extended system (see Eq. \ref{eq:gen_hamil}).

\item $\hcalone {\al}$: In the quantum case this is used to refer to the configurational energy of the (classical) polymeric system representing phase \al\ (see Eq. \ref{eq:prim_pot}).

\item $\hat{H}$ : denotes the hamiltonian operator of the (quantum) system.

\item ${\mf K}_\al$ : the dynamical matrix of phase \al\ (see Eq. \ref{eq:dynamatrix}).

\item $k^i_\al$ : the eigenvalue corresponding to the eigenvector ${\mf e}^i_\al$.

\item $[M]_\gamma$ : the set of macrostates consistent with phase $\gamma$. Note, however,  that when one performs a simulation constrained to  phase $\gamma$ (via $\pi^c_\gamma (\rvec)$), the simulation will only visit a {\em subset} of $[M]_\gamma$.

\item $M_{\tilde{\gamma}\gamma}$ : the macrovariable in Eq. \ref{eq:op}. 

\item $M_{BA,i}$ : bin i in $M_{BA}$ space. (i=1,2,..,b where b = number of bins).

\item $\mba (i)$ : the i'th output of \mba\ during the course of  the simulation. (i=1,2,...,t), t being the final output of the simulation.

\item $\eta (M_{BA,i})$ : the MUCA weights.

\item $\eta_{ST}^{(i)}$ : the weight associated with sub-ensemble i for the simulated tempering method (see Eq. \ref{eq:partforexpanded}). Note that unlike the multicanonical weights $\eta(\mbai)$ these are not functions defined on \cs . 

\item N : number of particles

\item $\tilde{O}$: the overlap parameter, Eq. \ref{eq:overlap}. This variable assumes the value unity if there is perfect overlap between the two phase constrained distributions and 0 if there is no overlap.

\item P : the number of replicas in the polymeric system modelling the quantum phase (see Eq. \ref{eq:poly_part}).

\item ${\cal P}$ : the permutation operator.

\item $P(\rvec)$ : absolute canonical probability of observing a configuration $\rvec$ (Eq. \ref{eq:absoluteprob}).

\item $P(M)$ : absolute canonical probability of observing a macrostate M (Eq. \ref{eq:pmabsolute}).

\item $P(\rvec | \al)$ : absolute canonical probability of observing a configuration \rvec\ {\em conditional} on being in phase \al\ (Eq. \ref{eq:probgivenphase}).

\item $P(M | \al)$ : absolute canonical probability of observing a macrostate M conditional on being in phase \al\ (Eq. \ref{eq:pmphase}).

\item $P(M, \al)$ : the {\em joint} probability of observing a macrostate M and being in phase \al\ (Eq. \ref{eq:jointpdf}).

\item $P(\mba | T_i)$ : the distribution of \mba\ at timeslice i (when the configurational energy $\ecal_{\la}$ has been changed from  $\ecal_{\la_{i-1}}$ to  $\ecal_{\la_i}$, and {\em after} the system has been equilibrated for a time \delT ).

\item $\hat{P}(\wba| \pi^c_\al)$ : the {\em estimator} (see Eq. \ref{eq:probestfg}) of $P(\wba | \pi^c_\al)$ for the $\al \rightarrow \alp$ FG process. In terms of notation, this is equivalent to $\hat{P}(\wba| \zeta_{\al\rightarrow \alp})$.

\item ${\cal{P}}^c_{\al \rightarrow \alp} (\seta {\vvec})$ : this denotes the probability of obtaining a path \seta {\vvec} in the $\al\rightarrow\alp$ FG process, as described in section \ref{sec:fg}.

\item $\arvec$ : 3N dimensional column vector denoting the positions of all the particles.

\item $R_{BA}$ : the ratio of the absolute partition functions, Eq. \ref{eq:partrsestimator0}

\item $\RBAcal$ : the \rpf\ as given in  Eq. \ref{eq:RBA}

\item $\RBAPq$ : the \rpf\ of the P replica polymeric system representing the quantum phase (see Eq. \ref{quant_free}).

\item $\Rvec_\gamma$ : a reference configuration in phase $\gamma$. For crystalline structures, an appropriate reference  configuration  is the underlying lattice structure, corresponding to the (classical) ground state configuration.

\item $S_{BA}$ : the linear transformation used to map the displacements \uvec\ of phase A onto those of phase B, Eq. \ref{eq:genswitch}. This mapping ensures that the two phases share the same $\vvec$ coordinates.

\item $S_T$ : the total action of the classical polymeric system modelling the quantum phase (Eq. \ref{tot_action}).

\item $S_K$ : the terms in the total action $S_T$ which contain information relating to the kinetic properties of the quantum system (Eq. \ref{kinetic_action}).
\item $S_V$ : the terms in the total action $S_T$ which contain information relating to the configurational properties of the quantum system (Eq. \ref{potential_action}).

\item T : temperature of the heat bath

\item $T^*$ : effective temperature appearing in the Lennard-Jones system (Eq. \ref{eq:redtemp}).

\item ${\mf T}_\gamma$ : the linear transformation which relates the displacements \uvec\ to the effective configuration \vvec\ of phase $\gamma$, Eq. \ref{eq:vtoutran}.

\item $\hat{T}$ : denotes the kinetic energy operator (Eq. \ref{eq:keop}) of the (quantum) system.

\item $\uvec$ : the displacement of the particles about the reference configuration $\Rvec_\gamma$.

\item $\vvec$ : the effective configuration. These are generalised coordinates which may be used to parameterise the configuration space of the system. When the distinction between the \cs\ as described by the \rvec\ coordinates and that described by the \vvec\ coordinates is necessary, we will refer to the space spanned by the variables \rvec\ as the absolute configuration space, and those spanned by the variables \vvec\ as the effective configuration space.

\item V : volume of the system.

\item $w_n(\wba)$: This measures the contribution of the macrostate \wba\ to the {\em numerator} of the corresponding estimator. For the general DP estimator this given by Eq. \ref{eq:weinum0}, whereas for the EP estimator it is given by Eq. \ref{eq:epn}, and for the PS estimator it is given by Eq. \ref{eq:wpsn}.

\item $w_d(\wba)$: This measures the contribution of the macrostate \wba\ to the {\em denominator} of the corresponding estimator. For the general DP estimator this given by Eq. \ref{eq:weiden0}, whereas for the EP estimator it is given by Eq. \ref{eq:epd}, and for the PS estimator it is given by Eq. \ref{eq:wpsd}.
y

\item $\delta W_{BA,i}$ : the (temperature scaled) work incurred in incrementing the \ce\ from $\ecal_{\la_i}$ to $\ecal_{\la_{i+1}}$ whilst keeping the configuration $\vvec_i$ constant (see Eq. \ref{eq:totalwork}).

\item $W_{BA}$ : the net (temperature scaled) work (which we will simply refer to as work) appearing in the FG method, obtained on changing the \ce\ $\ecal_\la$ from $\ecal_A$ to $\ecal_B$ through a series of steps in which at each stage one increments \la\ and then equilibrates the system with the new \ce\ for a time \delT\ (see section \ref{sec:fg} for details).

\item $W_m$: the (reversible) work obtained in the limit of thermodynamic integration (see section \ref{sec:ms}, Eq. \ref{eq:MSTI}, Eq. \ref{eq:fgti}). This is also the point at which the two phase constrained distributions $P(\wba |\pia)$ and $P(\wba | \pib)$ intersect (see figure \ref{pic:overlapping}) so that $W_m = -\ln \RBAcal$.

\item $\omega_n$ : the n-th cumulant (see Eq. \ref{eq:cumulant}) of the probability distribution $P(M_{\alp\al} | \pi^c_\al)$.

\item $\zeta_{\al\rightarrow \alp}$ : this is used to denote the $\al\rightarrow \alp$ FG process, as described in section \ref{sec:fg}.

\item Z : the absolute partition function, Eq. \ref{eq:absolutepart}

\item $Z_\gamma$ : the absolute partition function of phase $\gamma$, Eq. \ref{eq:partofphase}

\item $\z_{\la_i}$ : the partition function associated with the configurational energy $\ecal_{\la_i}$ (see Eq. \ref{eq:partofstage}).

\end{itemize}
\newpage
\setcounter{secnumdepth}{0}
\section{List of Acronyms}

\begin{itemize}

\item AR : acceptance ratio method denotes the estimator of the FED in which one performs two independent simulations, one in each phase, and estimates the expectations of the acceptance probabilities (see Eq. \ref{eq:ar}, or more generally  Eq. \ref{eq:FGAR}).

\item DP : dual phase. This refers to the most general canonical perturbation formula (see Eq. \ref{eq:dual}).

\item EP : this refers to the exponential perturbation estimator (Eq. \ref{eq:estep}) of \RBAcal\ in which the FED is estimated from data extracted from a simulation constrained to a single phase.

\item FED : free energy difference, Eq. \ref{eq:rbafree}, Eq. \ref{eq:FED},  Eq. \ref{eq:FEDexpansion}. Since the problem of estimating \RBAcal\ is equivalent to that of estimating the FED of the two phases, we will frequently interchange the use of the terms \RBAcal\ and FED.

\item FF : this refers to the fermi function estimator corresponding to Eq. \ref{eq:fermiform}. The optimal C is obtained by recursively solving  Eq. \ref{eq:recurse1} and Eq. \ref{eq:recurse2}.

\item FG : the fast growth method process (see section \ref{sec:fg}, \ref{sec:asfg}).

\item FSM : the fourier space mapping is a particular realisation of the general phase mapping (Eq. \ref{eq:genswitch}) in which the fourier coordinates of one phase are mapped onto those of the other phase (see Eq. \ref{eq:fsstran}).

\item HOA : higher order approximation.

\item LJ: Lennard Jones (refers to the pairwise interactomic potential given in Eq. \ref{lennard}).

\item MH : the multi-hamiltonian strategy is an extended sampling strategy which involves the construction of several independent but overlapping distributions. These overlapping distribution then allow the construction of a path linking the typical macrostates of the two phases (see section \ref{sec:mh}).

\item MS : the multistage strategy is similar in principle to the MH methods (see section \ref{sec:ms}).

\item MH-PS : the PS method, as implemented within the framework of the MH extended sampling strategy (see section \ref{sec:mh}).

\item MUCA : multicanonical (see sections \ref{sec:umbrella}, \ref{sec:phaseswitch}, \ref{sec:mul}).

\item NVT : this refers to a system whose volume V and temperature T are maintained at a constant value and in which the number of particles within the system remains unchanged during the course of the simulation. Such a system is referred to as a canonical system and has a distribution given by the Boltzmann distribution (Eq. \ref{eq:absoluteprob}).

\item PA : primitive approximation

\item PM : 'phase mapping' refers to the scheme whereby the configurations of one phase are mapped onto those of the other phase. The employment of a phase mapping allows one to map the problem of estimating the FED between the two phases  onto that of estimating the FED between two systems with different \ces\ (see Eq. \ref{eq:RBA}).

\item PS : the 'phase switch' should not be confused with the PM. This corresponding to a simulation in which  attempts to switch phases are actually made, and whose corresponding estimator for \RBAcal\ essentially amounts to measuring the (unbiased) ratio of the times spent in the two phases (see Eq. \ref{eq:ratiooftimes}). See chapter \ref{chap:estsamp} for generalisations of this method.

\item Q-FSM : quantum fourier space mapping.

\item Q-RSM : quantum real space mapping.

\item REP : restricted exponential perturbation formula (see Eq. \ref{eq:dualphaserestrict}).

\item RDP : restricted dual phase perturbation formula (Eq. \ref{eq:dualphaserestrict}).

\item RSM : the real space mapping is a particular realisation of the PM (see Eq. \ref{eq:RSS}).

\item ST : simulated tempering (see section \ref{sec:st}).

\item WHAM : the weighted histogram analysis method (see section \ref{sec:wham}).

\end{itemize}
\newpage
%\chapter{Glossary}

\setcounter{secnumdepth}{0}
\section{Glossary}

\begin{itemize}

\item \noindent order parameter : macrovariable which assumes a different ranges of values in the different phases. These ranges are, by definition, non-overlapping

\item A : this is the phase label denoting the fcc structure.

\item B : this is the phase label denoting the hcp structure.

\item configuration space : this term is used both to refer to the space spanned by \rvec\ and that spanned by \vvec . When the distinction between the \cs\ as described by the \rvec\ coordinates and that described by the \vvec\ coordinates is necessary, we will refer to the space spanned by the variables \rvec\ as the absolute configuration space, and that spanned by the \vvec\ coordinates as effective configuration space.

\item canonical : in the context of sampling, this refers to sampling from a Boltzmann distribution (see Eq. \ref{eq:boltzsamp}).

\item conjugate phase : this corresponds to the phase which the simulation is currently not in and the phase {\em onto} which the configurations are being mapped. In the case of the phase switching method this changes during the course of the simulation. In the case of a phase constrained simulation this remains the same for the entire duration of the method.

\item dual phase : this refers to estimators of the form of Eq. \ref{eq:freebennett}, which explicitly involve the accumulation of data from two simulations, one constrained to each phase.

\item extended sampling strategy : this refers to the procedure whereby the sampling distribution is made to encompasses a wider (or extended) region of \cs\ than is typically associated with the canonical  distribution, which the expectations are performed with respect to. The desired expectations are recovered from Eq. \ref{eq:reweightav}.

\item macrostate : this corresponds to the collection of configurations which yield a particular value for a given macrovariable. (See also Eq. \ref{eq:pmabsolute} for the probability of observing a given macrostate).

\item parent phase : this refers to the phase which the simulation is currently in and {\em from} which the configurations are being mapped onto the other phase. In the case of the phase switching method this changes during the course of the simulation. In the case of a phase constrained simulation this remains the same for the entire duration of the method.

\item partial overlap : see note \cite{note:partialmean}

\item path : this refers to a sequence of (closely spaced) macrostates which are actually sampled during the course of a simulation and which connect the regions of \ecs\ associated with one phase to those associated with the other phase.

\item representation : this refers to the particular way  in which one expresses the degrees of freedom (\vvec) of the phase (see Eq. \ref{eq:absoluteeinv}). Since the PM matches the \vvec\ coordinates of the two phases, the representation directly affects the overlap obtained under the operation of the  PM.

\item system : We will frequently interchange this word with the word ''phase''

\item thermodynamic limit : limit of the system size tending to infinity. That is $N\rightarrow \infty$.

\end{itemize}

%\chapter{Glossary}

\setcounter{secnumdepth}{5}

\chapter{\label{chap:intro}Introduction}
\tshow{chap:intro}

\section{\label{sec:phases}Phases and their stability}
\tshow{sec:phases}

The material world around us comprises of matter and its interactions. Depending on the strengths and ranges of these interactions matter, on the macroscopic scale, displays a variety of collective properties. These collective properties result in the formation of different ''phases'' of matter such as gas, liquid, and solid. For these phases there are two levels of description, which are the microscopic and macroscopic approaches. The microscopic picture describes matter in terms of its constituent particles and their interactions, whereas the macroscopic description coarse grains the configurational and kinetic information of the constituent particles into a small set of so called macrovariables.

In the case of equilibrium \cite{p:vved}\nocite{b:reif}\nocite{b:honerkamp}-\cite{b:pathria} these macrovariables fluctuate in time about a mean which remains constant in time, and the corresponding theory that describes the interrelation of the means of these variables is thermodynamics. The fundamental parameters which enter into the theory are certain macrovariables (such as the configurational energy E), certain parameters called intensive variables (such as the temperature and the chemical potential) which do not explicitly depend on the system and instead describe the coupling of the system with the environment, and the concept of entropy, which is a measure of the amount of disorder present in the system.

Thermodynamics is useful in that it explains the interrelation amongst some of the most important macrovariables. However, the theory does not give one the power of being able to predict how these macrovariables (or more precisely the means of these macrovariables) vary as one changes the intensive parameters. In order to do this one must resort to the microscopic description of the phenomena. A full blown microscopic approach would be (in the classical case) to solve Newton's equations for the particles and then average the relevant macrovariables over sufficiently long times, or (in the quantum case), to solve the multi-particle Schrodinger equation and evaluate the time averages of the expectation of the relevant macrovariables. Such an approach is, however, analytically intractable and one must instead resort to a more approximate microscopic approach.

The relevant microscopic theory is that of statistical mechanics, a theory in which all temporal effects have been averaged out. The core ingredients of the theory are the set of spatial configurations which the system may assume, the set of intensive variables which describe the system-environment coupling, and the configurational energy of the system. Using these, one may then construct probabilities for finding the system in a given configuration at any given instant of time. Since the theory is independent of kinetics, a considerably reduced amount of effort is required in describing phenomena. For a more in-depth development of the points mentioned above, the reader is referred to some standard texts on statistical mechanics \cite{b:reif}\nocite{b:pathria}-\cite{b:feynmanstat}.

In order to describe phenomena directly via statistical mechanics we must first explain more precisely what exactly we mean by a phase. Within the framework of statistical mechanics a phase corresponds to the group of  microscopic configurations in which the constituent members of any given group exhibit some common property unique to that phase. For example, in the case of a crystalline solid phase, the associated group would correspond to all configurations in which the particles are displaced by ''small'' amounts about some lattice structure. This lattice structure, which is the common characteristic of all the configurations, is what identifies the group and different lattice structures yield different groups or different phases. By grouping the configurations in this way one may also calculate the probability associated with a phase simply by summing the probabilities of the constituent configurations. The result is proportional to a quantity called the partition function of the phase, which plays a central role in statistical mechanics. 

The properties of a phase can, within the framework of thermodynamics, be predicted through a central quantity called the free energy. On the other hand all such predictions will, within the scheme of statistical mechanics, stem from the probability distribution of the configurations associated with that given phase. 
Not surprisingly it turns out that the free energy of a phase is intimately related to the partition function of that phase, with the intensive variables being the common parameters in the two theories.
By finding the partition function of a phase,  one is able to predict its behaviour quantitatively in the macroscopic limit. 

Of all the predictions that statistical mechanics can make, we shall focus on one. Namely, given a set of candidate phases, which phase is the one that is actually going to be found in nature for a given set of constraints (of the environment on the system)? Within the framework of statistical mechanics this translates to the task of finding out which is the most probable phase, or correspondingly finding out which phase has the largest partition function. In the context of determining phase boundaries, where one is trying to determine the more probable out of  two candidate phases, one may reduce to number of calculations by merely focusing on the ratio of the partition functions (which entails a single calculation) as opposed to focusing ones efforts on the calculation of the individual partition functions (an approach which will require two separate calculations).

The analytic evaluation of the ratio of partition functions (or equivalently the free energy difference) is, however, no simple task.
Despite the fact that temporal effects have been averaged out in statistical mechanics so as to considerably simplify the theory, it turns out that, for most complex systems of interest, calculation of the desired properties via analytic techniques remains intractable. One instead has to resort to computation. 
However even within the framework of this approach, the task of determining the ratio of the partition functions still remains a difficult one. In the rest of this thesis we will primarily be concerned with the investigation and development of computational methods of determining the FED (or equivalently the \rpf) of two different phases.

In the next section we will introduce the necessary machinery which will enable us  to define exactly what we mean by the partition function of a phase. We will express the partition function of a phase as a multidimensional integral and illustrate how the \rpf\ can be thought of as the ratio of two multidimensional integrals in which the regions which contribute the most come from two non-overlapping regions of the space over which the integrals are defined. We will then briefly discuss the \metrop , which is a widely used computational technique to model equilibrium phase behaviour. This will be followed by a discussion of how the method may, in principle, be used to estimate the FED, and how in practice it fails.

\section{\label{sec:statmech}Statistical mechanics : The formulation of the problem}
\tshow{sec:statmech}

\subsection{Key concepts and definitions}

In this section we develop the necessary statistical mechanical theory \cite{p:brucereview}. Suppose that we have a system with a fixed number of particles N, at a fixed volume V, and at a fixed temperature T. Such a system is referred to as a canonical or NVT system. Let \arvec\ denote the 3N dimensional column vector containing  the positions of all the particles. Within the framework of equilibrium statistical mechanics it follows that, when the dynamics of the system have been averaged out over sufficiently long periods of time, the canonical probability of finding the system assuming a  configuration \arvec\ is given by the Boltzmann distribution:

\begin{equation}
P(\arvec) = \frac {1} {Z} e^{-\beta E(\arvec)}
\label{eq:absoluteprob}
\end{equation}\tshow{eq:absoluteprob}

\noindent where $E(\arvec)$ is the configurational energy of the system and Z is the absolute partition function:

\begin{equation}
Z = \int d \arvec e^{-\beta E(\arvec)}
\label{eq:absolutepart}
\end{equation}\tshow{eq:absolutepart}

\noindent and $\beta = 1 / kT$, where k is the Boltzmann constant. 
If we have a variable $M=M(\rvec)$ which is a function on configuration space, which we call a macrovariable, it follows that the canonical probability of \ma\ assuming a value $M^*$ (called a macrostate) is then given by:

\begin{equation}
P(M^*) = \int d\arvec \delta (M({\arvec})-M^*) P({\arvec})
\label{eq:pmabsolute}
\end{equation}\tshow{eq:pmabsolute}

\noindent A macrostate is essentially a collection of microscopic configurations \seta {\rvec}\ for  which a macrovariable assumes a particular value. The canonical probability of a macrostate (Eq. \ref{eq:pmabsolute}) accounts for  the fact that there may be a multiplicity of  microscopic configurations associated with a given macrostate.

Equipped with the armoury of the probabilities of microstates (Eq. \ref{eq:absoluteprob}) and of macrostates (Eq. \ref{eq:pmabsolute}), we may now proceed to define a phase. We first note that in thermodynamics one generally identifies a phase (which we label $\gamma$) through a macrovariable \ma\ , also called an order parameter, which spans  a set of values $[M]_\gamma$. For an order parameter the set of values (say $[M]_{\gamma}$ and $[M]_{\tilde{\gamma}}$) associated with the two different phases (\al\ and \alp ) do not overlap, allowing it to be used as an identifying variable for the phases in question. Carrying this idea over into statistical mechanics, one may construct a criterion for deciding whether a configuration \rvec\ belongs to a phase or not by virtue of the following function:

\begin{equation}
\triangle_\gamma[\rvec] \equiv \left\{\begin{array}
{r@{\quad:\quad}l}
1 & \mbox{if $ M(\arvec)\epsilon [M]_\gamma$} \\
0 & \mbox{otherwise}
\end{array}\right.
\label{eq:triangle}
\end{equation}\tshow{eq:triangle}

\noindent If $\triangle_\gamma[\rvec]$ is 1, then the configuration \rvec\ belongs to phase $\gamma$, otherwise it does not. This function essentially uses the property of \ma\ being an order parameter (that is assuming a unique set of values in the different phases) in order to determine whether a configuration belongs to a phase or not.

Using Eq. \ref{eq:triangle} one may immediately write down the partition function and the (canonical) conditional probabilities of finding a configuration $\arvec$ and that of finding a  macrostate $M^*$, in phase $\gamma$:

\begin{equation}
P({\rvec}|\gamma)  =  \frac {1} {Z_\gamma} e^{-\beta E(\rvec)} \triangle_\gamma [\rvec]
\label{eq:probgivenphase}
\end{equation}\tshow{eq:probgivenphase}

\noindent and

\begin{eqnarray}
P(M^*|\gamma)& = & \int  d\arvec \delta (M({\arvec})-M^*) P({\arvec}|\gamma)\nonumber\\
& = & \frac {1} {Z_\al}  \int  d\arvec \delta (M({\arvec})-M^*)  e^{-\beta E(\rvec)} \triangle_\gamma [\rvec]
\label{eq:pmphase}
\end{eqnarray}\tshow{eq:pmphase}

\noindent where $Z_\gamma$ denotes the partition function, or probabilistic weight, associated with phase $\gamma$:

\begin{equation}
Z_\gamma \equiv \int d \arvec e^{-\beta E(\arvec)} \triangle_{\gamma} [\rvec]  
\label{eq:partofphase}
\end{equation}\tshow{eq:partofphase}

\noindent Having now defined the concept of a phase and its corresponding weight (the partition function, Eq. \ref{eq:partofphase}) we may now write the \rpf\ of the two different phases as the ratio of two multidimensional integrals:

\begin{eqnarray}
\RBA \equiv\ \frac {Z_B} {Z_A}& =& \frac {\int  d \rvec e^{-\beta E({\rvec})} \triangle_{B} [{\arvec}]  }{\int d \rvec e^{-\beta E(\rvec)} \triangle_{A} [{\arvec}] }\nonumber\\
& = &  \frac {< \triangle_B [\rvec] >} {< \triangle_A [\rvec] >}
\label{eq:partrsestimator0}
\end{eqnarray}\tshow{eq:partrsestimator0}

\noindent where the angular brackets $<>$ denote an expectation with respect to the distribution $P(\rvec)$ (defined more explicitly in  Eq. \ref{eq:ensembleav} below).
Alternatively, by using the order parameter \ma\,   one may write the \rpf\ as the ratio of two one dimensional integrals:

\begin{eqnarray}
\RBA & = & \frac {\int dM \triangle_B [M] P(M)} {\int dM \triangle_A [M] P(M)}\nonumber\\
& = & \frac {< \triangle_B [M] >} {< \triangle_A [M] >}
\label{eq:partrsestimator}
\end{eqnarray}\tshow{eq:partrsestimator}

\noindent where 

\begin{equation}
\triangle_\gamma[M] \equiv \left\{\begin{array}
{r@{\quad:\quad}l}
1 & \mbox{if $ M\epsilon [M]_\gamma$} \\
0 & \mbox{otherwise}
\label{eq:triangleM}
\end{array}\right.
\end{equation}\tshow{eq:triangleM}

\noindent The strategy of re-writing \RBA\ as has been done in Eq. \ref{eq:partrsestimator} is a highly advantageous one since it reduces the multidimensional problem in Eq. \ref{eq:partrsestimator0} to the one dimensional problem of Eq. \ref{eq:partrsestimator}. It crucially depends on ones ability to find a suitable order parameter M, which may not be possible for smaller systems. In such situations one may instead have to be content with a macrovariable M  which spans an overlapping range of values ($[M]_{\gamma}$ and $[M]_{\tilde{\gamma}}$) in the two phases. In this case one must distinguish the two phases on a microscopic level. For example, in the case of crystalline phases, one may do this by keeping track of the lattice vectors about which the particles of the system are displaced. In this more general case Eq. \ref{eq:probgivenphase}, Eq. \ref{eq:partofphase}, and Eq. \ref{eq:partrsestimator0} will continue to hold provided $\triangle_\gamma$ is more broadly defined as:

\begin{equation}
\triangle_\gamma[\rvec] \equiv \left\{\begin{array}
{r@{\quad:\quad}l}
1 & \mbox{if $ \rvec \epsilon \{\rvec\}_\gamma$} \\
0 & \mbox{otherwise}
\label{eq:triangle2}
\end{array}\right.
\end{equation}\tshow{eq:triangle2}

\noindent where $\{\rvec\}_\gamma$ denotes the set of configurations which one would typically associate with phase $\gamma$. In the case where $[M]_A$ and $[M]_B$ partially overlap, the expression in Eq. \ref{eq:partrsestimator} for  \RBA\ no longer holds, and must instead be expressed in terms of the joint probability distribution of \ma\ and  $\gamma$:

\begin{eqnarray}
P(M^*, \gamma) & = & P(M^* | \al).P(\al)\nonumber\\
& = & \int  d\arvec \delta (M({\arvec})-M^*)  e^{-\beta E(\rvec)} \triangle_\gamma [\rvec]
\label{eq:jointpdf}
\end{eqnarray}\tshow{eq:jointpdf}

\noindent where $\triangle_\gamma[\rvec]$ is now given by Eq. \ref{eq:triangle2}. It then follows that the \rpf\ may now be expressed more generally as:

\begin{eqnarray}
\RBA & = & \frac {\int dM \triangle_B [M] P(M,B)} {\int dM \triangle_A [M] P(M,A)}\nonumber\\
& = & \frac {\int dM P(M,B)} {\int dM P(M,A)}
\label{eq:partrsestimator2}
\end{eqnarray}\tshow{eq:partrsestimator2}

\noindent It is clear that in Eq. \ref{eq:partrsestimator2} the macrovariable \ma\ is in fact a redundant variable. Its utility, however,  lies in the estimation of Eq. \ref{eq:partrsestimator2} via simulations, where the macrovariable M is used to guide the simulation to certain regions of configuration space . We will have more to say about this in section \ref{sec:review} (in particular sections \ref{sec:umbrella} and \ref{sec:phaseswitch}).

\subsection{The link to thermodynamics}

In order to establish the connection between statistical mechanics and thermodynamics we first note that, in statistical mechanics, questions as to the relative stability of phases may be entirely addressed through the quantity \RBA, (Eq. \ref{eq:partrsestimator0}, Eq. \ref{eq:partrsestimator}, and Eq. \ref{eq:partrsestimator2}, ). If this quantity is greater than unity, phase B is the more stable. Otherwise phase A is the more stable of the two.
 Thermodynamics, on the other hand, extracts the corresponding information through the free energy ($F_\gamma$) of the phase. The phase which has the lower free energy is the more stable of the two.
The identity which bridges the two theories is the following:

\begin{equation}
F_\gamma \equiv - \beta^{-1} \ln\ Z_\gamma
\label{eq:fundamental}
\end{equation}\tshow{eq:fundamental}

\noindent It follows from Eq. \ref{eq:fundamental} that the \rpf\ is intimately related to the FED of the two phases:

\begin{equation}
\RBA =  e^{-\beta \triangle F_{BA}}
\label{eq:rbafree}
\end{equation}\tshow{eq:rbafree}

\noindent where $F_{BA}$ is the free energy difference:

\begin{equation}
F_{BA} = F_B - F_A
\label{eq:FED}
\end{equation}\tshow{eq:FED}

\noindent From thermodynamics we know that the equilibrium phase (that is the one which is found in nature, subject to the necessary constraints) is the one with the minimum free energy. This is consistent with the statistical mechanical formulation since from Eq. \ref{eq:fundamental} this merely corresponds to the phase with the maximum probabilistic weight $Z_\gamma$. Furthermore since the free energy, and hence the free energy difference, is an extensive quantity (that is $\triangle F_{BA} \propto N$), it follows  from Eq. \ref{eq:rbafree}  that in the limit of $N\rightarrow \infty$ (called the thermodynamic limit) the difference in partition functions of the two phases will magnify so as to make one of the phases overwhelmingly more probable than the other. This is in line with the thermodynamic observation of there being only one phase that is consistent with the constraints imposed on the system \cite{note:thermominfree}.

\subsection{Summary}

Summarising, if one has two candidate phases, and one wants to find out which will appear in nature, one can construct a finite system and  calculate the \rpf\  \RBA\ via Eq. \ref{eq:partrsestimator0}, Eq. \ref{eq:partrsestimator2}, Eq. \ref{eq:partrsestimator}. This allows one to determine the more stable (or more probable) of the two phases. It then follows that in the thermodynamic limit this phase then becomes overwhelmingly more probable that the other and as a result will be the one found in nature \cite{note:wayscale}.

For most interesting systems, even for a finite system the underlying complexity rules out any analytic approach. One must instead resort to computational techniques. The Monte Carlo method is a computational approach which is particularly suited for the simulation of equilibrium systems in which one is not concerned with the dynamics but merely static, time averaged quantities.
 In the next section we will briefly introduce a particular type of Monte Carlo method, called the Metropolis algorithm, and then discuss in section \ref{sec:samplingstrat} how this method may, in principle, be used to tackle the problem of estimating the \rpf\ \RBA.

\section{Simulation tools}

\subsection{\label{sec:metrop}The Metropolis algorithm}
\tshow{sec:metrop}

\subsubsection{Constructing the method}

There are two main simulations techniques which are employed to sample configuration space distributed according to Eq. \ref{eq:absoluteprob}, \cite{note:propp},  \cite{note:phasereview}. The first is molecular dynamics, a method which we do not employ in this thesis. For further information refer to \cite{b:frenkel}. The second, and more natural (in the context of equilibrium statistical mechanics) is the \metrop\ \cite{p:metropolis}, \cite{p:hastings}. Unlike molecular dynamics, in which the dynamics is purely deterministic, the \metrop\ is a purely probabilistic method. We will now describe the method in some detail.

The \metrop\ works by generating a sequence of configurations \seta {{\arvec}(1),{\arvec}(2),....,{\arvec}(t) } in which the probability of generating a configuration ${\arvec}({t+1})$ is only dependent on the current configuration ${\arvec}({t})$. This algorithm may be constructed in such a way so as to ensure that in the infinite time limit ($t\rightarrow \infty$), the relative probabilities of configurations appearing in the chain satisfy any arbitrary sampling distribution $\pi ({\arvec} )$. To see how this is done consider the rate equation for $\pi ({\arvec})$:

\begin{equation}
\frac {\partial \pi ({\arvec})} {\partial t} = \int d\arvec ' P_S ({\arvec '}\rightarrow {\arvec}). \pi ({\arvec '}) - \pi ({\arvec }). \int d \arvec ' P_S({\arvec}\rightarrow {\arvec '})
\label{eq:rateeq}
\end{equation}\tshow{eq:rateeq}

\noindent where $P_S({\arvec '}\rightarrow {\arvec})$ denotes the transition (or sampling)  probability of the algorithm from a configuration ${{\arvec '}}$ to a configuration ${{\arvec}}$. If the transition probability of the algorithm is to yield a process with a stationary distribution (that is a distribution $\pi ({\arvec})$ which does not change in time) one must have:

\begin{equation}
\frac {\partial \pi ({{\arvec}})} {\partial t} = 0
\label{eq:staticprobdistrib}
\end{equation}\tshow{eq:staticprobdistrib}

\noindent Clearly one way, but by no means the only way, in which Eq. \ref{eq:staticprobdistrib} may be satisfied is by assuming that:

\begin{equation}
P_S({\arvec '}\rightarrow {\arvec}) \pi ({\arvec '}) = P_S({\arvec}\rightarrow {\arvec '}) \pi ({\arvec })
\label{eq:detailedbalance}
\end{equation}\tshow{eq:detailedbalance}

\noindent The constraint on $P_S({\arvec '}\rightarrow {\arvec})$ in Eq. \ref{eq:detailedbalance} is called the condition of detailed balance and is used by the \metrop\ in order to produce a chain of configurations in which different configurations appear with relative frequencies which are consistent with $\pi(\rvec)$ \cite{note:relativefreq}.

In the \metrop\ the procedure of sampling is divided into two stages. The first stage involves generating a new configuration ${\arvec '}$ given a current configuration ${\arvec}$. The second stage is that of accepting or rejecting the proposed moves. Let $P_G({\arvec '}|{\arvec})$  denote the probability of generating $\rvec '$ given \rvec, and let $P_a({\arvec }\rightarrow {\arvec '})$ denote the corresponding acceptance probability. It follows that the sampling probability may be written as:

\begin{equation}
P_S({\arvec }\rightarrow {\arvec '}) = P_G({\arvec '}|{\arvec })  P_a({\arvec }\rightarrow {\arvec '})
\label{eq:chonetemp}
\end{equation}\tshow{eq:chonetemp}

\noindent Using Eq. \ref{eq:detailedbalance} and \ref{eq:chonetemp} it is easy to show that:

\begin{equation}
\frac {P_a(\rvec\rightarrow \rvec ')} {P_a(\rvec ' \rightarrow \rvec)} = \frac { \pi (\rvec ')  P_G(\rvec | \rvec ')} {\pi(\rvec)  P_G(\rvec ' | \rvec) }
\label{eq:critter}
\end{equation}\tshow{eq:critter}

\noindent Using this one may easily verify that a suitable $P_a({\arvec }\rightarrow {\arvec '})$ is of the form \cite{p:metropolis}, \cite{p:hastings}, \cite{note:sampling}:

\begin{equation}
P_a({\arvec }\rightarrow {\arvec '}) = \mbox{Min} \seta {1, \frac {\pi ({\arvec '}) P_G({\arvec }|{\arvec '})} {\pi ({\arvec }) P_G({\arvec '}|{\arvec })} }
\label{eq:acceptancerate}
\end{equation}\tshow{eq:acceptancerate}

\noindent Eq. \ref{eq:acceptancerate} is called the Metropolis acceptance criterion. An alternative, which also satisfies Eq. \ref{eq:critter}, is given by:

\begin{equation}
P_a (\rvec \rightarrow \rvec ') = \frac {1} {1+ \frac {\pi(\rvec) P_G(\rvec ' | \rvec)}  {\pi(\rvec ') P_G(\rvec | \rvec ')}}
\label{eq:fermiacceptance}
\end{equation}\tshow{eq:fermiacceptance}

\noindent In the simulation of statistical mechanical systems a particular case of Eq. \ref{eq:acceptancerate} is generally adopted. Consider a simulation performed via the \metrop\ in which the generation of a trial configuration involves perturbing a randomly chosen particle to a random position chosen to lie within a specified volume \cite{note:mildet} about the particle's initial point \cite{b:frenkel}. For such an algorithm, the probability of generating a new configuration ${\arvec '}$, given a current configuration {\arvec}, is symmetrical in the following way:

\begin{equation}
P_G({\arvec '}|{\arvec}) = P_G({\arvec  }|{\arvec '})
\label{eq:symgen}
\end{equation}\tshow{eq:symgen}

\noindent As a result Eq. \ref{eq:acceptancerate} simplifies to:

\begin{equation}
P_a({\arvec }\rightarrow {\arvec '}) = \mbox{Min} \seta {1, \frac {\pi ({\arvec '})} {\pi ({\arvec })} }
\label{eq:acceptancerate2}
\end{equation}\tshow{eq:acceptancerate2}

\noindent For the particular case where the sampling distribution is the Boltzmann distribution (Eq. \ref{eq:absoluteprob}):

\begin{equation}
\pi = \pi^c \equaldot e^{-\beta E(\arvec)}
\label{eq:boltzsamp}
\end{equation}\tshow{eq:boltzsamp}

\noindent where $\equaldot$ denotes an equality up to a normalisation constant which is not known \cite{note:sampling}, Eq. \ref{eq:acceptancerate2} may be written as:

\begin{equation}
P_a({\arvec  }\rightarrow {\arvec '}) = A(\beta \triangle E)
\label{eq:sampbol}
\end{equation}\tshow{eq:sampbol}

\noindent where

\begin{equation}
\triangle E = E({\arvec '}) - E({\arvec  })
\end{equation}

\noindent and 

\begin{equation}
A(x) = \mbox{Min} \{1, e^{-x}\}
\label{eq:metro}
\end{equation}\tshow{eq:metro}

\noindent is the Metropolis acceptance function. The procedure of employing a $P_G(\rvec ' | \rvec)$ with the property given in Eq. \ref{eq:symgen} and the acceptance probability $P_a(\rvec\rightarrow \rvec ')$ given in Eq. \ref{eq:sampbol} forms the cornerstone of the original Metropolis method, and will be the one that is used in the canonical simulations (Eq. \ref{eq:absoluteprob}) performed in this thesis.

Summarising, if one performs a simulation in which one stochastically generates configurations and accepts via the acceptance probabilities of Eq. \ref{eq:acceptancerate}, one generates a chain of configurations in which the frequencies of the appearance of different configurations are proportional to their probabilities $\pi(\rvec)$. We will now show how this property  may be used to estimate the expectation of macrovariables.

\subsubsection{Estimating the expectation of macrovariables}

Suppose now that one wants to evaluate the expectation of some function Q of a macrovariable M with respect to the sampling distribution $\pi$:

\begin{equation}
<Q>_\pi \equiv \frac {\int  d\arvec \pi ({\arvec}) Q(M({\arvec}))} {\int  d\arvec \pi ({\arvec})}
\label{eq:ensembleav}
\end{equation}\tshow{eq:ensembleav}

\noindent By using the \metrop\ (that is the stochastic algorithm in which proposed moves are accepted via eq. \ref{eq:acceptancerate}), one may estimate the expectation $<Q>_\pi$ via the following scheme:

\begin{eqnarray}
<Q>_\pi & \est & \frac {\s {i=1} {b} Q(M_i) H(M_i | \pi)} {\s {i=1} {b} H(M_i | \pi)}\nonumber\\
& = & \frac {1} {t} \s {i=1} {t} Q(M{(i)})
\label{eq:estimateexpectation}
\end{eqnarray}\tshow{eq:estimateexpectation}

\noindent where $M(i)$ denotes the i-th output of the macrostate $M$ by the simulation and  where $H(M_i | \pi)$ denotes the histogram count for bin $M_i$ under a sampling experiment performed via the sampling distribution $\pi$. It is important to keep in mind that it is the lack of knowledge of the normalising constant of $\pi$ which necessitates the inclusion of the integral $\int  d\arvec \pi ({\arvec})$ in the denominator of Eq. \ref{eq:ensembleav} \cite{note:denominator}. This will have important consequences for the task of estimating the FEDs (see section \ref{sec:umbrella}).

 A finite sample estimate given in Eq. \ref{eq:estimateexpectation} will generally have a {\em statistical} error associated with it. This arises from the fact that one is trying to reconstruct the relevant probability distribution from a finite number of samples, or equivalently from a finite time simulation \cite{note:error}. In the case of FED calculations, one has, in addition to this, {\em systematic} errors. These arise (in the context of FED calculations) from not sampling the regions of \cs\ which contribute the most significantly to the relevant estimator. Once again this arises from the fact that one is running the simulation for a finite amount of time. The differences in the two types of errors  lie in the time scales needed to reduce the error to an acceptable level, and therefore in some circumstances the distinction can become blurred. One may generally think of statistical errors as those which may be decreased to a desired level merely by running the simulation for sufficiently long times, where the lengths of time in question are generally those for which one would be prepared to run a simulation. In the case of systematic errors, the times needed to reduce them to an acceptable level are generally considerably  greater (by at least several orders) than one would be prepared to wait. The methods which are successful in estimating the FEDs are those which overcome such systematic errors. We will have more to say about the way in which they do this in chapter \ref{chap:review} and chapter \ref{chap:sampstrat}.

Eq. \ref{eq:estimateexpectation} tells us how we may estimate the expectation of a macrovariable (with respect to a sampling distribution $\pi$) based on an experiment performed with the {\em same} sampling distribution . More generally one may need to estimate the expectation of Q with respect to a distribution (say  $\tpi$) which is {\em different} from the sampling distribution $\pi$ used to obtain the data. To do this we simply  re-write the expectation $<Q>_{\tpi}$ as an expectation with respect to the sampling distribution $\pi$:

\begin{eqnarray}
<Q>_{\tpi} & \equiv & \frac{\int d\arvec \tpi ({\arvec}) Q(M({\arvec}))}{\int  d\arvec \tpi ({\arvec})}\nonumber \\
& = & \frac {\int d\arvec \pi ({\arvec}) Q(M({\arvec})) \frac {\tpi({\arvec})} {\pi({\arvec})}} {\int d\arvec \pi ({\arvec}) \frac {\tpi({\arvec})} {\pi ({\arvec})}}\nonumber \\
& = & \frac {<Q(M({\arvec})) \frac {\tpi({\arvec})} {\pi({\arvec})}>_\pi} {<\frac {{\tpi}({\arvec})} {\pi({\arvec})}>_\pi}
\label{eq:reweightav}
\end{eqnarray}\tshow{eq:reweightav}

\noindent where, as is the case in Eq. \ref{eq:ensembleav}, the need to evaluate the denominator of Eq. \ref{eq:reweightav} essentially arises from the lack of knowledge of the {\em relative} normalisation constants of $\pi$ and $\tpi$. 
Provided $\frac {{\tpi}({\arvec})} {\pi({\arvec})}$ is well defined Eq. \ref{eq:reweightav} may be estimated by:

\begin{equation}
<Q>_{\tpi}  \est  \frac {\s {i=1} {t} Q(M(i)) . \frac {\tpi ({\arvec}(i))} {\pi ({\arvec}(i))}} {\s {i=1} {t} \frac {\tpi ({\arvec}(i))} {\pi ({\arvec}(i))}}
\label{eq:reweightest}
\end{equation}\tshow{eq:reweightest}

\noindent  where \seta {\rvec{(1)}, \rvec{(2)}, ..., \rvec{(t)}}\ denotes the sequence of configurations generated by the simulation and where $M(i)=M(\rvec(i))$. In the special case where  $\frac {\tpi ({\arvec})} {\pi ({\arvec})}$ is a function  $f(M(\rvec))$ of $M(\rvec)$, that is:

\begin{equation}
\frac {\tpi ({\arvec})} {\pi ({\arvec})} \equaldot f(M(\rvec))
\label{eq:probrelation}
\end{equation}\tshow{eq:probrelation}

\noindent we may write Eq. \ref{eq:reweightav} as:

\begin{equation}
<Q>_{\tpi}  = \frac {<Q(M)f(M)>_\pi} {<f(M)>_\pi}
\label{eq:reweightav2}
\end{equation}\tshow{eq:reweightav2}

\noindent and we may re-write the corresponding estimator (Eq. \ref{eq:reweightest}) as:

\begin{equation}
<Q>_{\tpi} \est  \frac {\s {i=1} {b} Q(M_i) f(M_i) H(M_i |\pi)} {\s {i=1} {b} f(M_i) H(M_i |\pi)}
\label{eq:reweightest2}
\end{equation}\tshow{eq:reweightest2}

\noindent Eq. \ref{eq:reweightest} and Eq. \ref{eq:reweightest2} play a central role in the task of estimating FEDs via computational techniques. We will see in sections \ref{sec:review}, and more clearly in chapter \ref{chap:sampstrat}, that at the heart of all the  methods designed to tackle the problem of estimating FEDs is the construction of a sampling distribution $\pi$ which differs from the distribution \tpi\ with respect to which the expectations are performed. We will refer to this as the {\em extended sampling (ES) strategy} \cite{p:iba}. We will have more to say about these extended sampling strategies in section \ref{sec:umbrella} and chapter \ref{chap:sampstrat}. Before doing this we will describe three general techniques which may be used to estimate Eq. \ref{eq:partrsestimator0} via simulation and will then proceed to focus on one of these, namely the phase mapping method. In the next chapter we will then proceed to review the various methods that are available for estimating FEDs within the framework of this method.

\subsection{\label{sec:samplingstrat} Sampling strategies for estimating \RBA}
\tshow{sec:samplingstrat}

\noindent Broadly speaking there are (for NVT systems) three generic strategies which one may pursue in order to estimate \RBA\ \cite{note:otherphases}. They are the reference state technique, the continuous phase technique, and  the phase mapping (PM) technique. At the heart of all the techniques is the concept of a path, which we define to be a series of overlapping macrostates (obtained during the course of a simulation) connecting  the regions of \cs\ associated with one phase to those associated with the other.

In the reference state technique, a path is constructed which connects each phase to a reference system for which the partition function is known exactly. In this way one is able to estimate the absolute partition function of each phase.  In the continuous path technique, a continuous path is constructed from one phase to the other, thereby allowing one to estimate the \rpf . 
In the phase mapping (PM) technique, a path linking the two phases is constructed in which one ''leaps'' directly from one phase to the other, omitting all the regions of configuration space lying in between the two phases. We will now review these methods in greater detail.

\subsubsection{Reference State Technique}

\begin{figure}[tbp]
\begin{center}
\rotatebox{270}{
\hidefigure{\includegraphics[scale=0.5]{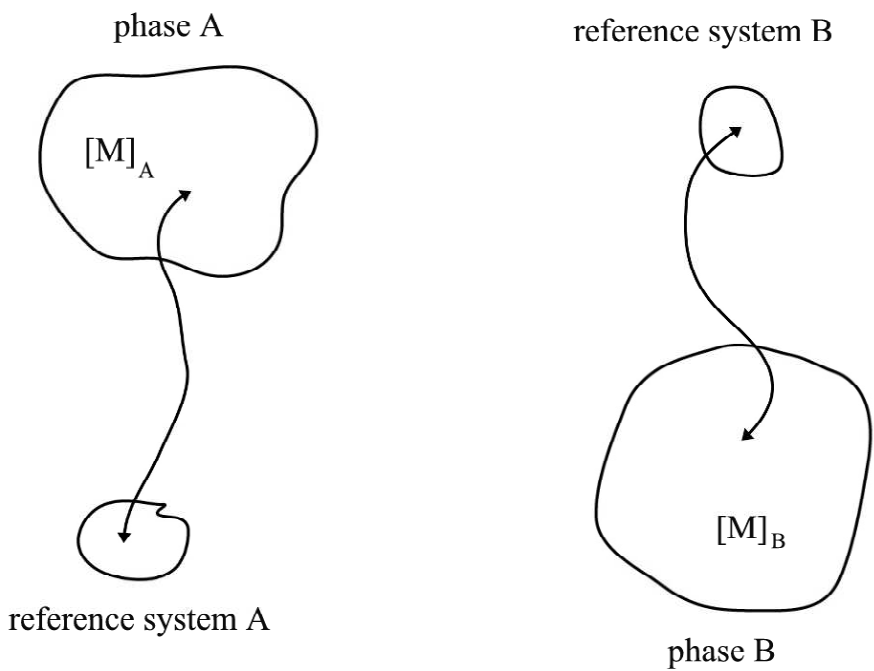}}
}
\end{center}
\caption{
Reference State Technique
}
In the reference state technique one performs a separate simulation for each phase, in which the simulation connects the phase to a reference system for which the partition function is known exactly. Knowledge of the absolute value of the partition function of the reference system and the \rpf\ of the phase and the reference system (which is estimated from the simulations) allows one to estimate the absolute value of the partition function of each phase. One can then determine the more probable of the two phases by noting which of the two has the larger partition function (or smaller free energy).\tshow{referencestate}

\begin{center}
{\bf{------------------------------------------}}
\end{center}

\label{referencestate}
\end{figure}

In the reference state technique (also called thermodynamic integration) \cite{p:frenkelladd}, \cite{p:fleischman}, \cite{p:straatsma}, the basic idea is to construct a path which connects the desired phase to a reference system for which the partition function is known exactly. This allows one to compute the FED between the given phase and the reference system. By performing two such simulations, connecting each phase to an appropriate reference system \cite{note:waapp}, one may infer the FEDs between the phases and their respective reference systems. Since the partition functions of these reference systems are known a-priori, one may use these results to determine the absolute values of the partition functions of each phase \cite{note:referencestate}. One may then proceed to determine which is the more stable phase of the two. A schematic is shown in figure \ref{referencestate}.

Technically, the way in which one links the desired system to the reference phase is as follows. One constructs a configurational energy $E_\la (\rvec)$ in which the field parameter \la\ assumes any value between 0 and 1. Furthermore suppose that at the extremities of $\la=0$ and $\la=1$ this \ce\ assumes the form of the \ces\ of the reference and desired phases respectively (i.e. $E_{\la=0} = E_r$ and $E_{\la=1} = E_\al$, where $E_r$ and $E_\al$ are the \ces\ of the reference and desired phases respectively). Then the fundamental relation upon which the method is based is:

\begin{eqnarray}
F_{\la=1} - F_{\la=0} & = & \int^{\la=1}_{\la=0} \frac {\partial F_\la} {\partial \la} d\la\nonumber\\
& = & \int^{\la=1}_{\la=0} <\frac {\partial E_\la} {\partial \la}>_{\pi_\la} d\la
\label{eq:thermody}
\end{eqnarray}\tshow{eq:thermody}

\noindent where $\pi_\la$ is given by:

\begin{equation}
\pi_\la(\rvec) \equaldot e^{-\beta E_\la(\rvec)}
\end{equation}

\noindent Eq. \ref{eq:thermody} must be estimated numerically by first dividing up the interval \seta {\la} into a discrete set

\noindent $\seta {\la_1=0, \la_2, ...., \la_n = 1}$, and then proceeding to estimate Eq. \ref{eq:thermody} via:

\begin{equation}
F_{\la=1} - F_{\la=0} \est \s {i=1} {n-1}  \widehat{< {\partial E_\la} / {\partial \la}>_{\pi_{\la_i}}} (\la_{i+1}-\la_i)
\label{eq:thermodyest}
\end{equation}\tshow{eq:thermodyest}

\noindent where $\widehat{< {\partial E_\la}/ {\partial \la}>_{\pi_{\la_i}}}$ denotes an estimate of the expectation ${< {\partial E_\la} / {\partial \la}>_{\pi_{\la_i}}}$. 

Generally there will be two sources of error in the estimator of Eq. \ref{eq:thermodyest}. The first will be statistical errors in estimating ${< {\partial E_\la} / {\partial \la}>_{\pi_{\la_i}}}$ and the second will be systematic errors arising from the discretisation of the interval \seta{\la} \cite{note:newsys}.
Whereas the statistical errors are made smaller simply by increasing the duration of the simulations, the systematic errors can be made smaller by reducing the size of the increments $\la_{i+1}-\la_i$. By ensuring that the increments  $\la_{i+1}-\la_i$ are sufficiently small, one may ensure that the systematic errors are smaller than the corresponding statistical errors, though a-priori it is not clear how small these increments have to be in order to ensure that this is indeed the case.

The presence of systematic errors is one point which makes the reference state technique, as formulated here, a slightly unattractive one. Furthermore the reference state technique requires the estimate of  two separate quantities, that is the partition functions of the individual phases, when one is in fact only interested in the single quantity corresponding to the {\em ratio} of these quantities. Clearly a single calculation which directly estimates this ratio would be preferable \cite{note:absolutevsratio}.

\subsubsection{Continuous Path Technique}

In the continuous path technique \cite{p:errington}, one performs a simulation which travels from one phase to the other via a continuous path. In order to estimate the \rpf\ (or equivalently the FEDs) of the two phases via Eq. \ref{eq:partrsestimator0}, Eq. \ref{eq:partrsestimator}, one performs a simulation in which one keeps track of the order parameter \ma. Using Eq. \ref{eq:estimateexpectation} one may then estimate \RBA\ from Eq. \ref{eq:partrsestimator} via the identity:

\begin{equation}
\RBA =\frac {<\triangle_B[M]>} {<\triangle_A[M] >} \est \frac {\s {i=1} {t} \triangle_B[M{(i)}]} {\s {i=1} {t} \triangle_A[M{(i)}]}
\label{eq:estrba1}
\end{equation}\tshow{eq:estrba1}

\noindent where $\triangle_\gamma [M]$ is given by Eq. \ref{eq:triangleM}.

\begin{figure}[tbp]
\begin{center}
\rotatebox{270}{
\hidefigure{\includegraphics[scale=0.5]{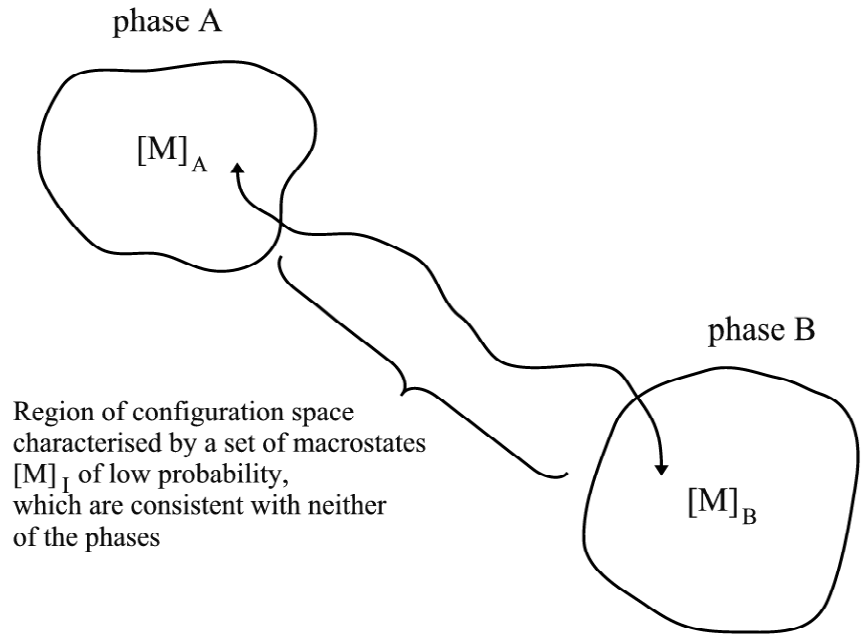}}
}
\end{center}
\caption{
Continuous Path Technique
}
In this method the simulation traverses between the two phases via a continuous path which involves crossing through regions of configuration space which are characterised by mixed phase-configurations. These configurations do not contribute to the relevant \RBA\, and are characterised by defect-rich structures.\tshow{phase-in-realspace}
\begin{center}
{\bf{------------------------------------------}}
\end{center}

\label{phase_in_realspace}
\end{figure}

In the case where the set of values $[M]_A$ and $[M]_B$ overlap (that is when \ma\ is no longer strictly an order parameter), one must instead use Eq. \ref{eq:partrsestimator2}. In this case the macrovariable \ma\ becomes redundant, and one instead estimates \RBA\ via the identity:

\begin{equation}
\RBA \est \frac {\s {i=1} {t} \triangle_B[\rvec(i)]} {\s {i=1} {t} \triangle_A[\rvec(i)]}
\label{eq:estrba2}
\end{equation}\tshow{eq:estrba2}

\noindent where $\triangle_\gamma [\rvec(i)]$ is given by Eq. \ref{eq:triangle2}. A schematic is shown in figure \ref{phase_in_realspace}

There are three problems with the method as it stands. The first is the fact that the estimators in Eq. \ref{eq:estrba1} and Eq. \ref{eq:estrba2} will generally fail. The reason for this lies in the fact that for straightforward Boltzmann sampling, in which one samples according Eq. \ref{eq:boltzsamp}, one generates configurations whose frequencies of appearance are in accordance with their canonical probabilities. Since the two phases are separated by a region of configuration space characterised by macrostates $[M]_I$ of extremely low probability (see figure \ref{phase_in_realspace}) the probability of the simulation generating a sequence of configurations which traverses this region will be vanishingly small. As a result the simulation will remain stuck in one of the phases, making it impossible to estimate \RBA\ from equation \ref{eq:estrba1} or \ref{eq:estrba2}, since either the numerator or the denominator of these estimators will be zero. This problem, which is called the {\em overlap problem} and is the origin of the systematic errors we were alluding to in section \ref{sec:metrop}, may be overcome with the adoption of appropriate extended sampling strategies \cite{p:errington}. We will have more to say about this in section \ref{sec:review} \cite{note:referenceprobs}.

The second problem, which is a problem afflicting the case of (structurally) ordered phases, arises from the fact that  in the process of going from one phase to the other (via $[M]_I$ in figure \ref{phase_in_realspace}) the simulation will in general have to traverse through regions of configuration space which are characterised by mixed-phase or disordered configurations. That is  the transition from one phase to the other will involve the disassembling of a phase, followed by the organisational restructuring resulting in the assembling of the other phase. This will result in the formation of a defect-rich final structure in the case where one of the phases is a crystalline solid. As a result one will not obtain a correct estimate for \RBA\ \cite{note:magnetic}. Note that even though the first problem, that is the problem of interphase traverse, may be overcome by the use of extended sampling (see for example \cite{p:errington}), this second problem will continue to persist in the case of ordered phases \cite{note:referenceprob2}.

The third problem, though not as serious as the previous two, arises from the fact that the regions of configurations space $[M]_I$ which one needs to traverse in going from one phase to the other do not actually contribute to \RBA . As a result, the simulation will waste large amounts of times in regions of configuration space which do not actually contribute to the final estimate of \RBA .

\subsubsection{\label{sec:pmtt}Phase Mapping Technique}
\tshow{sec:pmtt}

\begin{figure}[tbp]
\begin{center}
\rotatebox{270}{
\hidefigure{\includegraphics[scale=0.5]{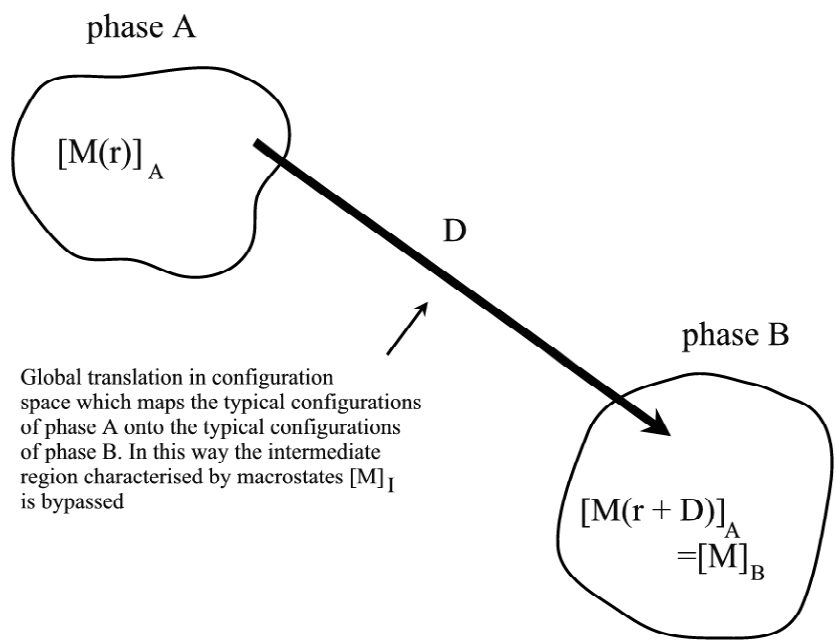}}
}
\end{center}
\caption{
PM technique for a simulation initiated in phase A
}

By employing a global configuration space displacement vector \Dvec, one may map the configurations of phase A directly onto those of phase B. The benefit of this is that it allows one to sit in the parent phase (A) and calculate \RBA\ simply by mapping the configurations onto those of the conjugate phase (B). This operation {\em does not} require one to traverse the intermediate region of configuration space characterised by the macrostates $[M]_I$. However one finds that for most systems of interest, the configurations of typical (or probable) macrostates of phase A are mapped onto the configurations of a-typical macrostates (macrostates of low probability) of phase B. This is again another manifestation of the overlap problem (see section \ref{sec:overlap}), and results in a poor estimate of \RBA\ via Eq. \ref{eq:abswitchest}.\tshow{phase-in-realspace-withswitch}

\begin{center}
{\bf{------------------------------------------}}
\end{center}

\label{phase_in_realspace_withswitch}
\end{figure}

A method which overcomes the last two problems of the continuous path technique, namely that of sampling the regions of configuration space $[M]_I$ characterised by mixed phase-configurations and that of sampling regions of configuration space which do not contribute directly to \RBA , is the phase mapping technique (PM) \cite{p:LSMC}, \cite{p:LSMCprehard}\nocite{p:LSMCpresoft}\nocite{p:solidliquid}-\cite{p:LSMCerrington} (see also \cite{p:voter}, \cite{p:kuchta}). This method  avoids the intermediate regions (characterised by the set of macrostates $[M]_I$ in figure \ref{phase_in_realspace})  altogether by using a global configuration space shift \cite{p:voter} to map configurations of one phase onto those of the other phase. Suppose that one is in phase A, with a configuration {\arvec}. The basic idea is to find a constant displacement \Dvec\  such that {\arvec + \Dvec} denotes a characteristic configuration of phase B (see figure \ref{phase_in_realspace_withswitch}). The result is a mapping of the configurations of one phase onto those of the other via the operation:

\begin{equation}
{\arvec} \rightarrow {\arvec + \Dvec}
\label{eq:voter}
\end{equation}\tshow{eq:voter}

\noindent In order to make use of this mapping one re-writes Eq. \ref{eq:partrsestimator0} as:

\begin{equation}
\RBA\ =  \frac {\int  d\arvec e^{-\beta E({\arvec + {\Dvec}})} \triangle_B [M({\arvec + {\Dvec}} )]  } {\int  d\arvec e^{-\beta E({\arvec })} \triangle_A [M({\arvec } )] }
\label{eq:switchtemp}
\end{equation}\tshow{eq:switchtemp}

\noindent By writing this as:

\begin{equation}
\RBA = \frac {\int d\rvec \frac {e^{-\beta E(\rvec + \Dvec)} \triangle_B[M(\rvec + \Dvec)]} {e^{-\beta E(\rvec)} \triangle_A[M(\rvec)]} e^{-\beta E(\rvec)} \triangle_A [M(\rvec)]} {\int d\rvec e^{-\beta E(\rvec)} \triangle_A[M(\rvec)]}
\end{equation}

\noindent Eq. \ref{eq:switchtemp} may be written as:

\begin{eqnarray}
\RBA\ &  = &  <e^{-\beta [E({\arvec + \Dvec}) - E({\arvec })]} \frac {\triangle_B [M({\arvec + \Dvec} )]}{\triangle_A [M({\arvec} )]}>_{\pi^c_A}\nonumber\\
& = & <e^{-\beta [E({\arvec + \Dvec}) - E({\arvec })]}  {\triangle_B [M({\arvec + \Dvec} )]}>_{\pi^c_A}
\label{eq:absoluteswitch}
\end{eqnarray}\tshow{eq:absoluteswitch}

\noindent where $\pi^c_A$ denotes that the expectation is performed with respect to configurations constrained to phase A \cite{note:singlepdf} :

\begin{equation}
\pi^c_\gamma \stackrel {\mathrm{.}}{=} P(\rvec | \gamma)
\label{eq:sampconstrain}
\end{equation}\tshow{eq:sampconstrain}

\noindent From Eq. \ref{eq:estimateexpectation} it follows that \RBA\ may then be estimated via:

\begin{eqnarray}
\RBA & \est & \frac {\s {i=1} {t} e^{-\beta [E(\arvec(i) + \Dvec) - E({\arvec}(i))]} \triangle_B [M(\arvec(i) + \Dvec)] } { \s {i=1} {t}  1 }\nonumber\\
 & = & \frac {1} {t} {\s {i=1} {t} e^{-\beta [E(\arvec(i) + \Dvec) - E(\arvec(i))]} \triangle_B [M(\arvec(i) + \Dvec)] }
\label{eq:abswitchest}
\end{eqnarray}\tshow{eq:abswitchest}

\noindent where the data is obtained from a simulation constrained to phase A. Since Eq. \ref{eq:abswitchest} is essentially an estimator involving a sampling distribution constrained to a given phase, we will refer to Eq. \ref{eq:abswitchest} as a {\em phase-constrained} estimator. For future use, we will refer to the phase which the simulation is actually in as the parent phase (phase A, in the case of Eq. \ref{eq:abswitchest}), and the phase {\em onto} which the configurations are being mapped as the conjugate phase (which in this case is phase B). Eq. \ref{eq:abswitchest} is just one example of a phase-constrained estimator. Looking forward, we note that these estimators may be most generally written of the form of  Eq. \ref{eq:dual}. 

The expression in Eq. \ref{eq:absoluteswitch} is essentially a way of estimating FEDs by performing a simulation in a single phase. 
By employing a global configuration space shift \Dvec, one is able to bypass the intermediate regions of configuration space region characterised by the set of macrostates $[M]_I$ in figure  \ref{phase_in_realspace}. The core idea behind the method is to find a global configuration displacement \Dvec\ which will allow one to  sample the configurations important to $Z_B$ {\em as well as} those important to $Z_A$ in a single simulation performed in phase A.

The problem with the method as it stands is that it is not sufficient that one finds a \Dvec\ such that the macrostates $[M]_A$ of the parent phase (A) are  mapped onto those, $[M]_B$, of the conjugate phase (B). In order for Eq. \ref{eq:abswitchest} to serve as an efficient estimator of \RBA, one must ensure that the configurations of {\em probable} macrostates in $[M]_A$ are mapped onto the configurations of {\em probable} macrostates in $[M]_B$. Generally one will not  be able find a suitable $\Dvec$ which ensures that this is the case. One will instead find that configurations of probable macrostates belonging to $[M]_A$ are mapped onto configurations of  improbable macrostates of $[M]_B$. 
This is another form of the overlap problem and results, in a way that will be explained in much greater detail in section \ref{sec:overlap}, in the failure of Eq. \ref{eq:abswitchest} to serve as an efficient estimator for \RBA .
However since we avoid the region of configuration space characterised by the macrostates $[M]_I$, the magnitude of the overlap problem that we face in dealing with Eq. \ref{eq:abswitchest} is considerably less than that of estimating \RBA\ via the continuous path technique (Eq. \ref{eq:estrba2}).

\section{Summary}

In seeking to model the equilibrium behaviour of bulk material in terms of its constituent components a useful microscopic theory is that of statistical mechanics. This theory works by associating a probability with each configuration that the system may assume. Accordingly one may associate a probability with a phase simply by summing up all the probabilities of the configurations consistent with that phase. The net result is proportional to a quantity called the partition function $Z_\gamma$, given by Eq. \ref{eq:partofphase}.

For a given set of external constraints, one finds that for many non-trivial (finite) systems several candidate structures may be stable. Each of these structures, or phases, will have an associated probability (proportional to their respective partition functions $Z_\gamma$). As the size of the system increases one of the phases becomes overwhelmingly more probable than the others, resulting in that phase being the one that is found in nature.

In the case of finding phase boundaries, one is merely interested in determining the more probable out of two candidate structures \cite{note:freecomplex}. Therefore it suffices to concentrate ones efforts on the determination of \RBA\ rather than the individual partition functions themselves. By using the \metrop\  one may, in principle, estimate \RBA\ by taking the ratio of the times spent in the two phases (Eq. \ref{eq:partrsestimator0}).  In practice  however, a  \metrop\ which generates (via  the sampling distribution $\pi^c_\gamma$, Eq. \ref{eq:sampconstrain}) macrostates according to their canonical probabilities  (Eq. \ref{eq:pmphase})  will remain trapped in the phase in which it is initiated, resulting in the systematic errors alluded to in section \ref{sec:metrop}. Though this problem may be overcome by appealing to an appropriate extended sampling strategy, the transition of the simulation through the intermediate regions $[M]_I$ results (in the case of transitions to an ordered phase) in the formation of defect-rich structures. One way around this to use the reference state technique. An alternative is the PM technique, in which one maps configurations of one phase directly onto those of the other phase (see figure \ref{phase_in_realspace_withswitch}). This in principle allows one to estimate \RBA\ from a simulation performed in a single phase via Eq. \ref{eq:abswitchest}. In practice however even this method fails (in a way that will be described in greater detail in section \ref{sec:overlap}) due to a milder version of the original overlap problem, which, in the context of the PM method, translates to  configurations of probable macrostates of $[M]_A$ being mapped onto configurations of  improbable macrostates of $[M]_B$ under the operation of the PM. As is the case with the continuous path technique this overlap problem may be overcome with the aid of a suitably refined extended sampling strategy (see section \ref{sec:umbrella}, \ref{sec:phaseswitch} and chapter \ref{chap:sampstrat}). 

The focus of this thesis will be on the development of methodologies of tackling the problem of estimating FEDs via the PM formalism, and will ultimately lead to their application to the calculation of quantum FEDs. In the next section we will start off by mapping (by using the framework of the PM method) the problem of determining the \rpf\ given in Eq. \ref{eq:partrsestimator0} onto that of evaluating the \rpf\ associated with two systems with {\em different} configurational energies. This will allow us to present the overlap problem in a quantitative fashion. We will then review the array of methods that have been engineered over the last 30 years in order to address this sort of computational problem, before proceeding to discuss our own investigations on the problem.

\chapter{\label{chap:review}Review}
\tshow{chap:review}

\section{Introduction}

In this section we will formulate the task of estimating the FED within the PM formalism, thus mapping the problem of evaluating the ratio of two integrals involving a  {\em single} \ce\ (see Eq. \ref{eq:partrsestimator0}) onto that of determining the ratio of integrals involving two  {\em different} \ces . We will proceed to define what we mean by overlap and then discuss the overlap problem in the context of the PM formalism. Following this we will review an array of methods which have been designed to tackle the problem of evaluating the ratio of integrals involving two different \ces . Some of what is presented here is new (in particular the (unifying) formulation of the estimators in terms of the macrovariable \mba\ and the process switching generalisation of the fast growth method, section \ref{sec:pfg}). We include it  here for the sake of providing a coherent presentation.

\section{\label{sec:phaseswitchscene} Formulation of the problem}
\tshow{sec:phaseswitchscene}

Suppose that $\avec {R}_\gamma$ denotes a reference configuration that is consistent with phase $\gamma$. One may then express the position vector $\arvec$  in terms of the displacements \auvec\ about $\aRvec_\gamma$:

\begin{equation}
\arvec = \aRvec_\gamma + \auvec
\label{eq:rinu}
\end{equation}\tshow{eq:rinu}

\noindent The degrees of freedom may now be parameterised through the variable \uvec\ (instead of \rvec) so that Eq. \ref{eq:partrsestimator0} may be written as:

\begin{equation}
\RBA\ = \frac {\int e^{-\beta E(\aRvec_B + \auvec)} d\auvec} {\int e^{-\beta E(\aRvec_A + \auvec)} d\auvec}
\label{eq:RBA_in_uspace}
\end{equation}\tshow{eq:RBA_in_uspace}

\noindent where the Jacobian of the transformation from the variable \rvec\ to \uvec\ is unity. By expressing the degrees of freedom in Eq. \ref{eq:RBA_in_uspace} in terms of the variables \uvec , we have effectively switched from the \rvec\ representation (see  Eq. \ref{eq:partrsestimator0}) to the \uvec\ representation.

 By comparing Eq. \ref{eq:RBA_in_uspace} to Eq. \ref{eq:switchtemp}, we immediately see that the expression for \RBA\ in Eq. \ref{eq:RBA_in_uspace} involves a PM in which the configuration $\rvec_A$ of phase A is mapped onto a configuration $\rvec_B$ of phase B via the global configuration space displacement \cite{note:chooseD} :

\begin{equation}
\avec {D} = \aRvec_B -\aRvec_A
\label{eq:DLS}
\end{equation}\tshow{eq:DLS}

\noindent We will refer to this mapping, whereby configurations of one phase are mapped onto those of the other via the global configuration space displacement of Eq. \ref{eq:DLS} and in which the displacements of the particles from their lattice sites are matched in the two phases, as the real-space mapping (RSM)  \cite{p:LSMC}, \cite{p:LSMCprehard}\nocite{p:LSMCpresoft}\nocite{p:solidliquid}-\cite{p:LSMCerrington}. It is realised via the following operation:

\begin{equation}
\aRvec_\gamma , \auvec \rightarrow \aRvec_{\tilde{\gamma}} , \auvec
\label{eq:RSS}
\end{equation}\tshow{eq:RSS}

\noindent More generally, one may formulate the problem in an arbitrary representation. Consider writing the displacement \uvec\ as a linear transformation of some generalised coordinates \avvec\ (which we will call the effective configuration of the system):

\begin{equation}
\auvec = \mf {T}_\gamma \avvec
\label{eq:vtoutran}
\end{equation}\tshow{eq:vtoutran}

\noindent so that:

\begin{equation}
\arvec = \aRvec_\gamma + \avec{T}_\gamma \avvec
\label{eq:rintermsofv}
\end{equation}\tshow{eq:rintermsofv}

\noindent We may now use the effective configuration \vvec\ to parameterise the degrees of freedom of the phase.
Substituting Eq. \ref{eq:rintermsofv} into Eq. \ref{eq:partofphase} we find that:

\begin{equation}
Z_\gamma = \det {\avec{T}_\gamma} \int  d \avvec e^{-\beta E_\gamma (\avvec)}
\label{eq:partofphasev}
\end{equation}\tshow{eq:partofphasev}

\noindent where 

\begin{equation}
E_\gamma (\avvec) = E(\aRvec_\gamma + \avec {T}_\gamma \avvec)
\label{eq:absoluteeinv}
\end{equation}\tshow{eq:absoluteeinv}

\noindent If we express  the \ce\  about that of  the reference configuration:

\begin{equation}
E_\gamma(\avvec) = E (\aRvec_\gamma) + \ecal_\gamma (\avvec)
\label{eq:expandenergy}
\end{equation}\tshow{eq:expandenergy}

\noindent then the free energy difference between the two phases may be written as:

\begin{equation}
\triangle F_{BA} = \triangle E^o_{BA} - \beta^{-1} \ln \det S_{BA} - \beta^{-1} \ln \RBAcal
\label{eq:FEDexpansion}
\end{equation}\tshow{eq:FEDexpansion}

\noindent where

\begin{equation}
\triangle E^o_{BA} = E(\aRvec_B) - E(\aRvec_A)
\label{eq:gse}
\end{equation}\tshow{eq:gse}

\noindent  is the contribution of the reference state configurations to the FED, while 

\begin{equation}
S_{BA} = \avec {T}_B . \avec {T}^{-1}_A
\label{eq:stran}
\end{equation}\tshow{eq:stran}

\noindent with

\begin{equation}
\RBAcal = \frac {\z_B} {\z_A} = \frac {\int d\vvec e^{-\beta \ecal_B (\avvec)}} {\int d\vvec e^{-\beta \ecal_A (\avvec)}} 
\label{eq:RBA}
\end{equation}\tshow{eq:RBA}

\begin{figure}[tbp]
\begin{center}
\rotatebox{270}{
\hidefigure{\includegraphics[scale=0.6]{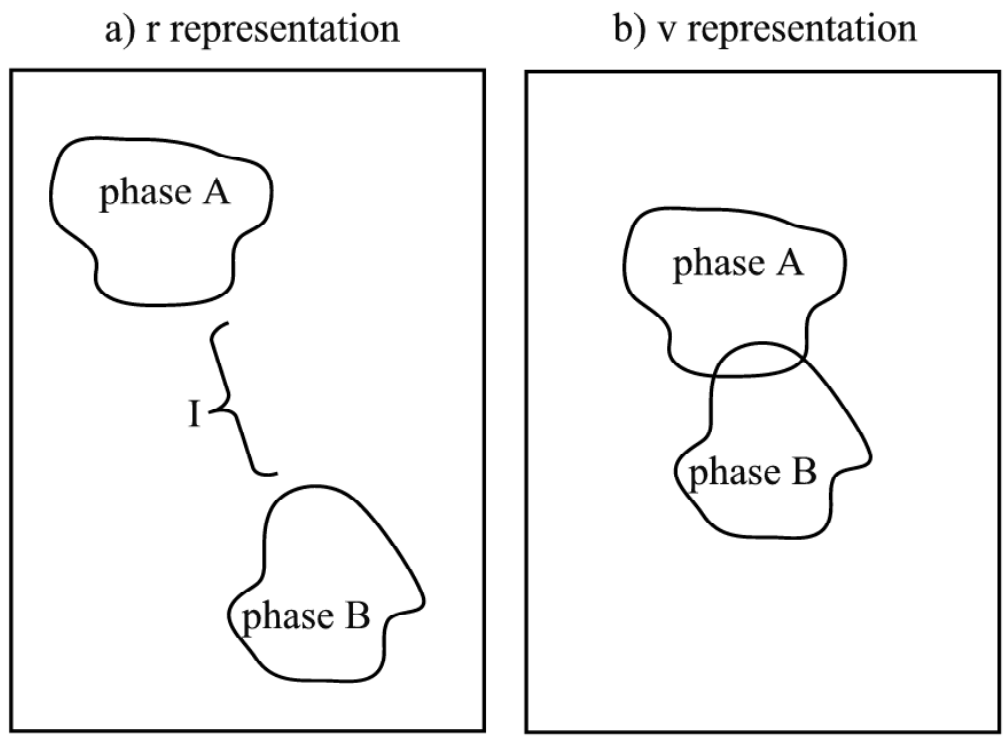}}
}
\end{center}
\caption{
Schematic of phases in different representations
}
In the original \rvec\ representation, the two phases correspond to two non-overlapping regions of (absolute) configuration space. Schematically this is shown in (a) by two widely separated regions of (absolute) \cs\ with the intermediate regions denoted by I. In the \vvec\ representation (of the PM formulation) the degrees of freedom of the phases are parameterised through the effective configuration \vvec\ (see Eq. \ref{eq:rintermsofv}). If a good choice of representation (i.e. \Rvec\ and \vvec ) is made then the distinction between the two phases becomes considerably less in the 'effective' configuration space (see (b)) than it is in the original \rvec\ representation (see (a)).
\begin{center}
{\bf{------------------------------------------}}
\end{center}

\tshow{fig:representations}
\label{fig:representations}
\end{figure}

\noindent By comparing Eq. \ref{eq:RBA} to Eq. \ref{eq:RBA_in_uspace}, we see that the expression for \RBAcal\ in Eq. \ref{eq:RBA} now involves a PM in which the effective configuration \vvec\ is preserved on the transition from one phase to the other; in other words the PM {\em matches} the \vvec\ coordinates of the two phases. This mapping, in which the  coordinates $\rvec_\al$ of phase \al\ are mapped  onto the coordinates $\rvec_\alp$ of phase \alp\ (such that they  share the same effective configuration \vvec ) may be realised in $\auvec$ space via the operation:

\begin{equation}
\aRvec_\gamma , \auvec \rightarrow \aRvec_{\tilde{\gamma}} , {\mf S}_{\tilde{\gamma} \gamma} \auvec
\label{eq:genswitch}
\end{equation}\tshow{eq:genswitch}

\noindent where ${\mf S}_{\tilde{\al}\al}$ is given by Eq. \ref{eq:stran}. The quantity $\RBAcal$ in Eq. \ref{eq:RBA} is  ubiquitous in a variety of fields. It is also the starting point for a string of literature concerned with the task of estimating the  FEDs \cite{p:LSMC}, \cite{p:voter}, 
\cite{p:barker}\nocite{p:zwanzig}\nocite{p:valleaucard}\nocite{p:bennett}\nocite{p:rahmanjacucci}\nocite{p:moodyray}\nocite{p:nezbedakolafa}\nocite{p:lyubartsev}\nocite{p:lyubartsevgeneral}\nocite{p:simulatedtempering}\nocite{p:kofkecummings}\nocite{p:radmer}-\cite{p:jarprl} 
(for reviews see \cite{p:brucereview}, \cite{b:frenkel}, \cite{note:otherphases}, \cite{p:kofkecummings}, \cite{p:chipotreview}\nocite{p:mezeireview}\nocite{b:tildesley}\nocite{p:acklandreview}\nocite{p:valleaureview}-\cite{p:gelmanreview}). We note that whereas Eq. \ref{eq:partrsestimator0} is the ratio of two integrals involving a {\em single} \ce , Eq. \ref{eq:RBA} is the ratio of two integrals  with {\em different} configurational energies. In both cases the regions which contribute most significantly to the two integrals come from different regions of configuration space. However in the case of Eq. \ref{eq:RBA} the disparity that exists between these two regions of (effective) configuration space may not be as great as it is for the two regions of (absolute) configuration space within the original \rvec\ formulation (see Eq. \ref{eq:partrsestimator0}).   We illustrate this idea schematically in figure \ref{fig:representations}.

Despite this scope for improvement, the overlap problem, albeit a milder version, generally persists. In figure \ref{fig:representations} (b) this corresponds to the fact that the most typical regions associated with phase A do not overlap with the most typical regions associated with phase B.  In the following section we will explain exactly why this poses a problem for the task of estimating the FED. In section  \ref{sec:review} we will then  proceed to review the methods that have been developed in order to estimate quantities of the form of Eq. \ref{eq:RBA}.

\section{\label{sec:overlap}The overlap problem}
\tshow{sec:overlap}

In order to discuss the overlap problem in a semi-quantitative way, let us first define a quantity which measures the energy cost in switching from a configuration in phase \al\ to a configuration in phase \alp\ via the mapping of Eq. \ref{eq:genswitch}:

\begin{equation}
M_{\alp\al} (\avvec) = \beta \{\ecal_\alp (\avvec) - \ecal_\al (\avvec)\}
\label{eq:op}
\end{equation}\tshow{eq:op}

\noindent Following the earlier definition of the conditional probability of a macrovariable (Eq. \ref{eq:pmphase}), one may define  the probability of observing a macrostate $\mba^*$ conditional on the  sampling distribution of phase \al, $\pi^c_\al$,  as:

\begin{equation}
P(M_{BA}^* | \pi^c_\al) \equiv \int \delta (\mba(\vvec) - \mba^*) \pi^c_\al(\vvec) d\vvec
\label{eq:condprobofmba}
\end{equation}\tshow{eq:condprobofmba}

\noindent We will refer to this as the phase constrained distribution of phase \al .

\begin{figure}[tbp]
\begin{center}
\rotatebox{270}{
\hidefigure{\includegraphics[scale=0.6]{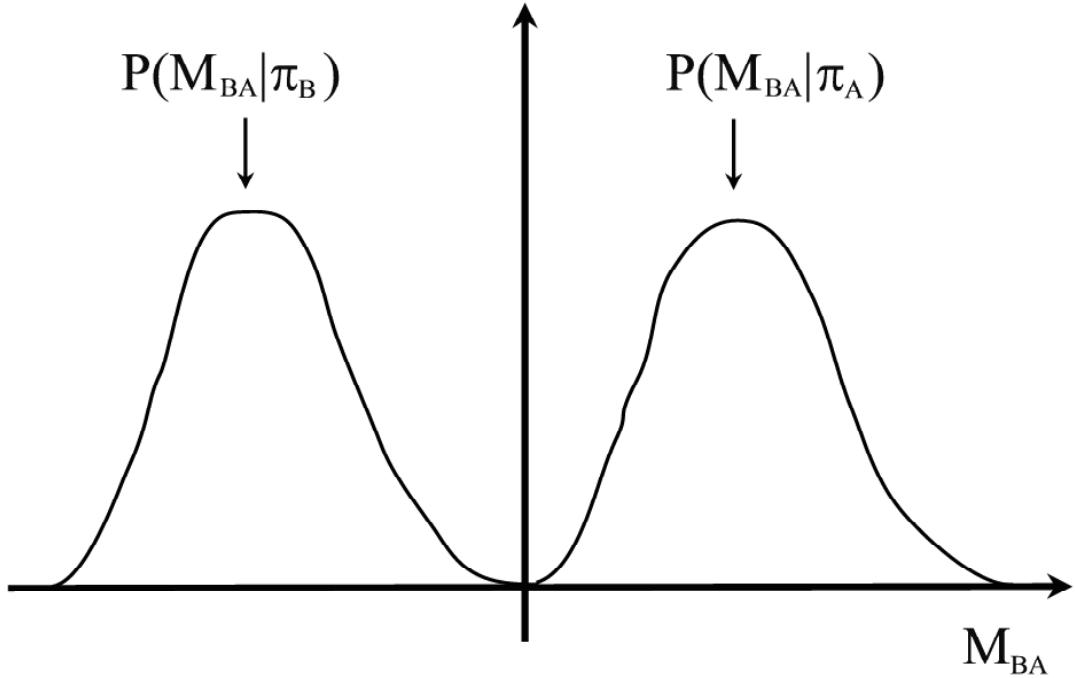}}
}
\end{center}
\caption{Schematic for ${P}(M_{BA}|\pi^c_A)$ and ${P}(M_{BA}|\pi^c_B)$}

This figure is a schematic for what one would typically expect for the distributions of \mba\ for the two phases. For a simulation in phase A, the effective configurations \avvec\ that will be sampled will be typical of $\pia$ and atypical of $\pib$. As a consequence one will find that for most configurations $\ecal_B(\avvec)> \ecal_A(\avvec)$ (meaning that \mba\ is  positive), which will yield the right-hand side peak shown in the figure above. For a simulation initiated in phase B the opposite will be true and this time the value of \mba\ will be, on average, negative, resulting in  the left hand peak. In this way \mba\ acts, in some sense, like the order parameter M described in section \ref{sec:pmtt}.

The overlap problem essentially amounts to the lack of overlap of the two peaks $P(\mba|\pi^c_A)$ and $P(\mba|\pi^c_B)$, and prevents one from obtaining an estimate of \RBAcal\ (via Eq. \ref{eq:overlapest}) which is free of systematic errors. Since the energy is an extensive quantity (so as to make \mba\ extensive) , the distance of the peaks from the origin will scale with N  and the spread will scale as $\sqrt{N}$. As a consequence the overlap will decrease (i.e. the overlap problem will get worse) as the system size increases.

\tshow{bimodaldistribution}

\begin{center}
{\bf{------------------------------------------}}
\end{center}

\label{bimodaldistribution}
\end{figure}

 In order to understand the behaviour of the phase-constrained distribution $P(\mba | \pi^c_\al)$ let us analyse the behaviour of the macrovariable $M_{\alp\al}$. This macrovariable is under certain circumstances an order parameter for the system. 
 To see this let us first consider a hard sphere system. For this type of system a mapping of the form of  Eq. \ref{eq:genswitch}\ will map  a configuration of phase $\gamma$ (for which the hard spheres do not overlap) onto a configuration of phase $\tilde{\gamma}$ in which the hard spheres typically penetrate each other's cores. As a consequence $M_{\alp\al}$ will be positive and infinite. A similar thing will happen for a simulation initiated in phase $\alp$ in which the mapping of Eq. \ref{eq:genswitch}\ is performed in the opposite direction. In this case $M_{\alp\al}$ will be negative and infinite. In this sense $M_{\alp\al}$ acts as an order parameter for the system. This idea carries over into systems with continuous configurational energies. In this case for typical effective configurations $\avvec$ of phase \al, the mapping in Eq. \ref{eq:genswitch} induces a configuration $\arvec_\alp$ (given by Eq. \ref{eq:rintermsofv}) of phase \alp\ of generally higher energy than the configuration $\arvec_\al$. In other words one will find that on average $M_{\alp\al} > 0$ for typical configurations of phase \al. Likewise the opposite will be true (that is that $M_{\alp\al} < 0$) for the typical configurations of phase \alp. The resulting distributions are shown in figure \ref{bimodaldistribution}.

We are now in a position to define (in a way which is free of ambiguity) what exactly we mean by overlap. Suppose that $H(M_{BA,i} | \pi^c_\al)$ denotes the histogram count for bin $M_{BA,i}$ for a simulation performed via $\pi^c_\al$. The estimator for $P(\mba | \pi^c_\al)$ is given by:

\begin{equation}
\hat{P}(M_{BA,i} | \pi_\al^c) = \frac {H(M_{BA,i} | \pi_\al^c)} {\s {i=1} {b} H(M_{BA,i} | \pi_\al^c)}
\label{eq:estcanp}
\end{equation}\tshow{eq:estcanp}

\noindent We then define the concept of overlap as follows:

\begin{center}
\framebox[6in]{
\begin{minipage}[t]{5.9in}
{\em The region of overlap is defined to be the set } \seta {M_{BA,i}\ }{\em \  over which }$H(M_{BA,i} | \pi^c_A)$ {\em and} $H(M_{BA,i} | \pi^c_B)$ ({\em or equivalently } $\hat{P}(M_{BA,i} | \pi^c_A)$ {\em and} $\hat{P}(M_{BA,i} | \pi^c_B)$ {\em })  {\em are  simultaneously non-zero}.
\end{minipage}
}
\end{center}

\noindent The overlap between $P(M_{BA} | \pia)$ and $P(M_{BA} | \pib)$  may be quantified \cite{p:bennett} by introducing the {\em overlap parameter} \ov :

\begin{equation}
\tilde{O} \equiv \int d\mba \frac {2 P(\mba | \pi^c_A) P(\mba | \pi^c_B)} {P(\mba | \pi^c_A) + P(\mba | \pi^c_B)}
\label{eq:overlap}
\end{equation}\tshow{eq:overlap}

\noindent \ov\ may be estimated by $\hat{\ov}$:

\begin{equation}
\hat{\ov} = \int d\mba \frac {2 \hat{P}(\mba | \pi^c_A) \hat{P}(\mba | \pi^c_B)} {\hat{P}(\mba | \pi^c_A) + \hat{P}(\mba | \pi^c_B)}
\label{eq:estov}
\end{equation}\tshow{eq:estov}

\noindent In the case where the estimators $\hat{P}(M_{BA} | \pi^c_A)$ and $\hat{P}(M_{BA} | \pi^c_B)$ overlap considerably, both \ov\ and $\hat{\ov}$ will assume a value close to unity. In the case where  $\hat{P}(M_{BA} | \pi^c_A)$ and $\hat{P}(M_{BA} | \pi^c_B)$ do not overlap at all \ov\ will assume a small (but non-zero) value, whilst $\hat{\ov}=0$.

 We now proceed to derive the overlap identity \cite{p:torrievalleau}, which relates the probabilities  of obtaining a given value of \mba\ in the two phases. From Eq. \ref{eq:condprobofmba} we see that:

\begin{eqnarray}
P(\mba^* | \pi_B^c)& = & \frac {1} {\z_B} \int \delta (\mba(\avvec) - \mba^*) e^{-\beta \ecal_B(\avvec)}d\avvec \nonumber\\
& = & \frac {\z_A} {\z_B} . \frac {1} {\z_A} \int \delta (\mba(\avvec) - \mba^*) e^{-\beta \ecal_A(\avvec)} e^{-\mba(\avvec)}d\avvec\nonumber\\
& = & \frac {\z_A}{\z_B} e^{-\mba^*} P(\mba^* | \pi_A^c) 
\label{eq:prelimOLI}
\end{eqnarray}\tshow{eq:prelimOLI}

\noindent Rearranging this equation one obtains the overlap identity:

\begin{equation}
\RBAcal = \frac {e^{-\mba} P(\mba | \pi_A^c)}{P(\mba | \pi_B^c) } 
\label{eq:overlapidentity}
\end{equation}\tshow{eq:overlapidentity}

\noindent This identity may be used to estimate \RBAcal\ via:

\begin{equation}
\RBAcal  \est  \frac {e^{-M_{BA,i}} \hat{P}(M_{BA,i} | \pi_A^c)} {\hat{P}(M_{BA,i} | \pi_B^c)}
\label{eq:overlapest}
\end{equation}\tshow{eq:overlapest}

\noindent where $M_{BA,i}$ is any bin for which one has a non-zero count for {\em both} simulations.

The overlap identity imposes several constraints on the arbitrariness of the phase constrained distributions $P(\mba | \pi^c_A)$ and $P(\mba | \pi^c_B)$. One such constraint is the value of \mba , which we label as $M_m$, at which point these two distributions intersect. Substituting $P(M_m | \pi^c_A)=P(M_m | \pi^c_B)$ into Eq. \ref{eq:overlapidentity} one obtains:

\begin{equation}
\RBAcal = e^{-M_m}
\end{equation} 

\noindent or:

\begin{eqnarray}
M_m & = & - \ln \RBAcal\nonumber\\
& = & \beta \triangle F_{BA}
\end{eqnarray}

\noindent Therefore the FED of the phases manifests itself as an {\em asymmetry} of the  point at which the two phase-constrained distributions intersect \cite{note:throw}.

Eq. \ref{eq:overlapest} is very important because it links the idea of the overlapping of  $P(\mba | \pi^c_A)$ and  $P(\mba | \pi^c_B)$ to ones ability to estimate \RBAcal .
 It is immediately clear that in order for Eq. \ref{eq:overlapest} to serve as an efficient estimator for \RBAcal\ there must be {\em some} regions of $\mba$ space over which  $P(\mba | \pi^c_A)$ and  $P(\mba | \pi^c_B)$ overlap. If this is not the case (as is the situation in figure \ref{bimodaldistribution}) then either the numerator or the denominator of  Eq. \ref{eq:overlapest} will be zero for any bin \mbai , yielding an incorrect estimate of \RBAcal.  It is in this way that the overlap problem  prevents one from arriving at an estimate of  \RBAcal\ which is free of systematic errors. Furthermore as the system size increases the extensivity of $M_{\alp\al}$ results in  the means and variances associated with the distributions $P(\mba |\pi^c_\al)$  scaling in such a way so as to reduce the overlap of the two distributions (see figure \ref{bimodaldistribution} for an explanation). As a consequence the overlap problem worsens as the system size increases.

In order to tackle the overlap problem, one needs to understand the factors which affect it. From our preceding discussion it is clear that the overlap is dependent on two factors:

\begin{itemize}
\item The choice of the translation vector $\avec {D}$ which maps configurations of one phase onto those of the other
\item The choice of representation \vvec\  \cite{note:bothimpo}, \cite{note:choosingperfect}.
\end{itemize}

\noindent The first point has been addressed before \cite{p:LSMCprehard} and is briefly discussed in section \ref{sec:pm}. The second point has not been investigated before and in chapter \ref{chap:tune}, we will show how one may choose a representation, based on the normal modes of the phase, which does, for structurally ordered phases,  cure the overlap problem as the harmonic limit is approached. Before doing this we will (in the next section) present an array of techniques that have been developed in order to estimate quantities of the form of Eq. \ref{eq:RBA}.

\section{\label{sec:review}Review of methods}
\tshow{sec:review}

\subsection{Introduction}

Over the years a spectrum of methods have been developed in order to address the evaluation of Eq. \ref{eq:RBA}. These methods include thermodynamic integration methods \cite{p:frenkelladd}\nocite{p:fleischman}-\cite{p:straatsma}, the canonical perturbation  methods \cite{p:barker}, \cite{p:zwanzig}, \cite{p:bennett},\cite{p:kofkecummings}, the simulated tempering methods \cite{p:nezbedakolafa}\nocite{p:lyubartsev}\nocite{p:lyubartsevgeneral}-\cite{p:simulatedtempering},  the slow growth methods \cite{p:slowgrowth}\nocite{p:hunteretal}-\cite{p:millerreinhardt}, and the umbrella sampling and the multistage methods \cite{p:valleaucard}, \cite{p:torrievalleau}, \cite{p:tvb}\nocite{p:tvb2}-\cite{p:torrievalleaucpl}. 
More recent developments include the fast growth methods \cite{p:jarprl}, \cite{p:crooks}\nocite{p:crookspre}\nocite{p:crookspre2}-\cite{p:hummerszabo}, and the phase switching method of \cite{p:LSMC}. We will now review a selection of the methods available for estimating FEDs, limiting  ourselves to NVT systems \cite{note:otherphases}. Though these approaches do, at first sight, appear to be quite different, there are common themes that run through all the methods; we shall try to make them clear. The interrelations between the methods can be most easily seen by observing the way in which the sampling distributions $P(\mba | \pi)$ are constructed and by expressing the estimators in terms of the macrostates \mba . We point out that some of the insights offered in this chapter were also part of the work of this thesis; in particular the \mba\ formulation of the simulated tempering (section \ref{sec:st}), the \mba\ formulation of the phase switch method (section \ref{sec:phaseswitch}), and the path sampling formulation of the fast growth method (section \ref{sec:pfg}). However for the sake of a coherent presentation of the ideas, we have put them in this section.

\subsection{\label{sec:perturbation}Canonical Perturbation Methods}
\tshow{sec:perturbation}

The perturbation methods are probably the simplest and earliest methods developed to tackle the problem of determining FEDs \cite{p:barker}\nocite{p:zwanzig}-\cite{p:bennett}. We will now describe the two most well known examples.

The first method may be most simply derived by integrating the overlap identity (Eq. \ref{eq:prelimOLI}) over all values of \mba, yielding:

\begin{equation}
1 = \int P(\mba | \pi^c_B) d\mba = \frac {\z_A} {\z_B} \int e^{-\mba} P(\mba | \pi^c_A) d\mba
\label{eq:tempperturb}
\end{equation}\tshow{eq:tempperturb}
  
\noindent or 

\begin{eqnarray}
\RBAcal & = & {\int e^{-\mba } P(\mba|\pi^c_A)d\mba}\nonumber\\
& \equiv & <e^{-\mba}>_{\pi^c_A}
\label{eq:ep}
\end{eqnarray}\tshow{eq:ep}

\noindent Eq. \ref{eq:ep} (see also Eq. \ref{eq:absoluteswitch}) is what we refer to as the exponential perturbation (EP) method. \RBAcal\ may then be simply estimated by:

\begin{equation}
\RBAcal \est \s {i=1} {b} \hat{P}(\mbai|\pia) e^{-\mbai}
\label{eq:epgub}
\end{equation}\tshow{eq:epgub}

\noindent or:

\begin{equation}
\RBAcal  \est  \frac { \s {i=1} {b} H(M_{BA,i} | \pi_A^c) e^{-M_{BA,i}}} {\s {i=1} {b} H(M_{BA,i} | \pi_A^c)}
\label{eq:estep}
\end{equation}\tshow{eq:estep}

\noindent using the fact that  $\hat{P}(\mbai|\pia)$ is given by Eq. \ref{eq:estcanp}.

 Generally an attempt to estimate  the FED via  Eq. \ref{eq:estep} will fail. The reason for this ultimately stems from the fact that the regions which contribute the most to the numerator of Eq. \ref{eq:estep} will be those regions over which $P(\mba | \pib)$ is most significant \cite{note:propto}, which, in the general case illustrated in figure \ref{bimodaldistribution},  is not contained within the regions typically explored in a sampling experiment performed with $\pi^c_A$. As a result systematic errors will be present in ones estimate of \RBAcal .

The second method is based on the identity:

\begin{equation}
e^{-\mba} = \frac {A(\mba)} {A(-\mba)}
\label{eq:expacc}
\end{equation}\tshow{eq:expacc}

\noindent where A is the Metropolis acceptance function (Eq. \ref{eq:metro}). Using this identity in Eq. \ref{eq:overlapidentity} we get:

\begin{equation}
\RBAcal A(-\mba) P(\mba | \pi^c_B) = A(\mba) P(\mba | \pi^c_A)
\end{equation}

\noindent Integrating and rearranging this equation we find that:

\begin{equation}
\RBAcal = \frac {< A(\mba)>_{\pi^c_A}} {< A(-\mba)>_{\pi^c_B}}
\label{eq:ar}
\end{equation}\tshow{eq:ar}

\noindent Eq. \ref{eq:ar} is what is known as the acceptance ratio (AR) method \cite{p:bennett}. Since $A(M_{\alp\al})$ is the acceptance probability of a Monte Carlo move in which one attempts to switch the configurational energies from $\ecal_\al$ to $\ecal_\alp$ whilst keeping the effective configuration \vvec\ constant (we will refer to such a Monte Carlo move as a phase switch (PS), see section \ref{sec:phaseswitch}), we see that the AR method estimates \RBAcal\ by evaluating the expectations of the acceptance probabilities for the phase switches  in the two phases, without actually performing them as Monte Carlo moves.
It follows from Eq. \ref{eq:ar} that \RBAcal\  may then be estimated via the identity:

\begin{equation}
\RBAcal \est \frac  {\s {i=1} {b} \hat{P}(M_{BA,i} | \pi_A^c) A({M_{BA,i}})} {\s {i=1} {b} \hat{P}(M_{BA,i} | \pi_B^c) A({-M_{BA,i}})} 
\label{eq:arest}
\end{equation}\tshow{eq:arest}

\noindent Since the AR method (Eq. \ref{eq:ar}) rests on the overlap identity (Eq. \ref{eq:overlapidentity}), it follows that its estimator (Eq. \ref{eq:arest}) implicitly assumes that some sort of overlap exists between the estimators $\hat{P}(\mba | \pi^c_A)$ and  $\hat{P}(\mba | \pi^c_B)$. It is not immediately clear to what extent overlap is needed, and the insight into this shall be provided later in chapter \ref{chap:estsamp}. For the moment we shall remain content with the fact that in the general case an attempt to estimate \RBAcal\ via Eq. \ref{eq:arest} will fail  due to the absence of overlap between the estimators of the phase-constrained distributions of the two phases (see figure \ref{bimodaldistribution}).

Eq. \ref{eq:ep} and Eq. \ref{eq:ar}  are particular cases of a much more general formula originally derived by Bennett \cite{p:bennett}.  This formula may be simply obtained by multiplying the left and right sides of the overlap identity (Eq. \ref{eq:overlapidentity}) by an arbitrary function $G(\mba)$ and integrating over all values of \mba :

\begin{equation}
\RBAcal {\int G(\mba) P(\mba | \pi_B^c) d \mba } = {\int G(\mba) e^{-\mba} P(\mba | \pi_A^c) d \mba}
\end{equation}

\noindent or:

\begin{equation}
\RBAcal  =  \frac {<G(\mba)e^{-\mba}>_{\pi^c_A}} {< G(\mba)>_{\pi^c_B}}
\label{eq:dual}
\end{equation}\tshow{eq:dual}

\noindent For the choices of $G(\mba)=1$ one obtains the EP formula (Eq. \ref{eq:ep}) and for $G(\mba) = A(-\mba)$ one obtains the AR formula (Eq. \ref{eq:ar}). In general Eq. \ref{eq:dual} will require two separate simulations, one in each phase. For this reason it will be referred to as the dual phase (DP) formula. 

The perturbation methods described thus far (Eq. \ref{eq:ep}, Eq. \ref{eq:ar}, and Eq. \ref{eq:dual}) rest on  simulations in which configurations are sampled according to their canonical probabilities (see Eq. \ref{eq:sampconstrain}). The spectrum of methods that we will now review go beyond this and rest on the employment of  sampling distributions which are different from the distributions with respect to which the expectations are performed (see Eq. \ref{eq:reweightav}).

\subsection{\label{sec:umbrella}Umbrella Sampling and Multicanonical Methods}
\tshow{sec:umbrella}

The underlying idea behind the umbrella sampling method of Torrie and Valleau \cite{p:valleaucard}, \cite{p:torrievalleau}, 
\cite{p:tvb}\nocite{p:tvb2}-\cite{p:torrievalleaucpl}  is that of sampling with a distribution $\pi^m_A$ which is different from the sampling distribution $\pi^c_A$ with respect to which the expectations are evaluated. That is by applying the identity of Eq. \ref{eq:reweightav} to  Eq. \ref{eq:ep} they obtain the following identity:

\begin{eqnarray}
\RBAcal  & = & <e^{-\mba }>_{\pi^c_A}\nonumber\\
& = &  \frac {<e^{-\mba}  \frac {\pi_A^c} {\pi^m_A }>_{\pi^m_A}} {<\frac {\pi_A^c} {\pi^m_A}>_{\pi^m_A}}
\label{eq:umbrellalead}
\end{eqnarray}\tshow{eq:umbrellalead}

\noindent Torrie and Valleau considerably simplify the problem of constructing an alternative sampling distribution by noticing that the multidimensional quantity in Eq. \ref{eq:umbrellalead} can in fact be solely expressed in terms of the statistics of $\mba$, provided that the ratio $\pi^c_A / \pi^m_A$ can be expressed as a function of \mba . By constructing an alternative sampling distribution $\pi^m_A$ by appeal to a weight function $\eta_A(\mba)$:

\begin{equation}
P(\mba | \pi^m_A ) \equaldot P(\mba | \pi^c_A) e^{-\eta_A(\mba)}
\label{eq:introweights}
\end{equation}\tshow{eq:introweights}

\noindent they use this idea to rewrite Eq. \ref{eq:umbrellalead} as:

\begin{eqnarray}
\RBAcal & = & \frac {<e^{-\mba}  \frac {P(\mba | \pi_A^c)} {P(\mba | \pi^m_A)}>_{\pi^m_A}} {<\frac {P(\mba | \pi_A^c)} {P(\mba | \pi^m_A)}>_{\pi^m_A}}\nonumber\\
& = &  \frac {<e^{-\mba}  e^{\eta_A(\mba)}>_{\pi^m_A}} {<e^{\eta_A(\mba)}>_{\pi^m_A}}
\label{eq:torriereweight}
\end{eqnarray}\tshow{eq:torriereweight}

\noindent \RBAcal\ may then be estimated via:

\begin{eqnarray}
\RBAcal & \est & \frac {\s {i=1} {b} e^{-M_{BA,i}} \hat{P}(M_{BA,i}|\pi^m_A) e^{\eta_A(M_{BA,i})} } {\s {i=1} {b} \hat{P}(M_{BA,i}|\pi^m_A)  e^{\eta_A(M_{BA,i})}}\nonumber\\
& = & \frac {\s {i=1} {b} e^{-M_{BA,i}} H(M_{BA,i}|\pi^m_A) e^{\eta_A(M_{BA,i})}} {\s {i=1} {b} H(M_{BA,i}|\pi^m_A)  e^{\eta_A(M_{BA,i})}}
\label{eq:torriereweightest}
\end{eqnarray}\tshow{eq:torriereweightest}

\noindent Eq. \ref{eq:torriereweightest} also directly follows from Eq. \ref{eq:epgub} by noticing that instead of $\hat{P}(M_{BA,i} | \pi_\al^c)$ being given by Eq. \ref{eq:estcanp}, it is now given by:

\begin{equation}
\hat{P}(M_{BA,i} | \pi_\al^c) = \frac {\hat{P}(M_{BA,i}|\pi^m_A) e^{\eta_A(M_{BA,i})}} {\s {j=1} {b} \hat{P}(M_{BA,j}|\pi^m_A)  e^{\eta_A(M_{BA,j})}}
\end{equation}

\noindent As we saw in section \ref{sec:perturbation}, the reason why the EP method (Eq. \ref{eq:ep}) fails to serve as a useful estimator is essentially due to the fact that $P(\mba | \pi^c_A)$ does not {\em contain}  $P(\mba | \pib)$ \cite{note:conta}. The umbrella sampling method overcomes this problem by constructing a sampling distribution $\pi^m_A$ so that $P(\mba | \pi^m_A)$ contains both $P(\mba | \pi^c_A)$ and $P(\mba | \pi^c_B)$.

The construction of $\pi^m_A$ is however, just as difficult a task as that of finding the FED of the two phases \cite{note:mulweightsfree}, and in the original papers 
\cite{p:valleaucard},  \cite{p:torrievalleau}, \cite{p:tvb}\nocite{p:tvb2}-\cite{p:torrievalleaucpl} no scheme was provided for constructing the sampling distribution $\pi^m_A$. Instead they acknowledged that the task of finding an appropriate $\pi^m_A$ was ''tedious'' and suggested  performing several independent umbrella sampling simulations, with the umbrella distributions overlapping at the edges, in order to cover the desired region of \ecs.

Recently the work of Torrie and Valleau has come alive again in the works of Berg and Neuhaus \cite{p:bergPRL}, \cite{note:mucaother} in which umbrella sampling was reinvented under the name of the multicanonical (MUCA) algorithm \cite{note:mucaillu}. This time, however,  an efficient prescription for constructing the umbrella sampling distribution was provided. Within the context of umbrella sampling the  MUCA algorithm  may be thought of as the two fold process:

\begin{itemize}
\item One constructs an estimate for $P(\mba | \pia )$, which we will denote by $\hat{P}(\mba | \pia )$, over an arbitrary range of \mba\ space.
\item One then samples from the MUCA sampling distribution:
\begin{equation}
\pi^m_A(\vvec) \equaldot \frac {\pia(\vvec)} {\hat{P}(\mba (\vvec) | \pia )}
\label{eq:multicansamp}
\end{equation}\tshow{eq:multicansamp}

\noindent via  the acceptance probability of Eq. \ref{eq:acceptancerate2}.
\end{itemize}

\noindent It follows that the probability of observing a macrostate $\mba$ under the MUCA sampling distribution $\pi^m_A$  is given by:

\begin{eqnarray}
P(\mba^* | \pi^m_A) &  \equiv  &\int \delta (\mba(\vvec) - \mba^*) \pi^m_A (\vvec)\nonumber\\
& \equaldot & \int \delta (\mba(\vvec) - \mba^*) \frac {\pi^c_A(\vvec)} {\hat{P}(\mba (\vvec) | \pia )}\nonumber\\
& = & \frac {P(\mba^* | \pia )} {\hat{P}(\mba^* | \pia )}
\label{eq:pmbamc}
\end{eqnarray}\tshow{eq:pmbamc}

\noindent In the case where the estimate is perfect (that is $\hat{P}(\mba | \pia ) = {P}(\mba | \pia )$ ) $P(\mba | \pi^m_A )$ is {\em constant} for all the range of  $\mba$ space. 
Therefore by obtaining a sufficiently accurate estimate of ${P}(\mba | \pia )$ over the desired range of $\mba$ space, one may construct an umbrella (or MUCA) sampling distribution $\pi^m_A$ (via Eq. \ref{eq:multicansamp}) which is {\em flat} (in $\mba$ space) over that range.

 There are several methods of generating these MUCA weights and we will discuss the two simplest procedures.
The first method (which is called the visited states method \cite{p:bergPRL}, \cite{p:lee}) starts off with an initial estimate in which all the MUCA weights ${\eta}_A(M_{BA,i})$ are set to be equal to each other. One then performs a simulation (or iteration) for $t_{mul}$ Monte Carlo steps using the MUCA sampling distribution Eq. \ref{eq:multicansamp}. The histogram $H(M_{BA,i}|\pi^m_A)$ that is subsequently obtained is then used to improve the estimate of the MUCA weights through the identity \cite{p:bergPRL}, \cite{p:lee}\nocite{p:mezei}-\cite{note:transi}:

\begin{equation}
{\eta}_{A,{j+1}} (\mbai) = \hat {\eta}_{A,j} (\mbai) + \ln [H(\mbai| \pi^m_A) + 1]
\end{equation}

\noindent where \seta {{\eta}_{A,j} (\mbai)} denotes the weights of the current iteration, and  \seta {{\eta}_{A,{j+1}} (\mbai)} denotes the weights of the next iteration. Under this updating scheme macrostates which have been visited will have their weights increased, and as a result the probability (see Eq. \ref{eq:introweights}) of visiting these macrostates is {\em reduced} for the next iteration. On the other hand the weights of  macrostates which are not visited at all are left unaffected, so as to increase their chance of being visited (relative to those macrostates which have been visited) in the next iteration. By iterating this procedure, one may eventually obtain an accurate set of MUCA weights \seta {{\eta}_A} over the desired range, which one may then use to construct a MUCA sampling distribution (Eq. \ref{eq:multicansamp}) which is flat over an arbitrary span of \mba\ space.

 The second scheme for constructing the MUCA weights ${\eta}_A$ is a modification of the Wang-Landau method \cite{p:wanglandauprl}\nocite{p:yan}-\cite{note:genwl} . In this method the time for each iteration corresponds to a single Monte Carlo step. That is  one updates the weight {\em  after each Monte Carlo step} ($t_{mul} = 1$) via the update scheme:

\begin{equation}
{\eta}_{A,{j+1}} (\mbai) = \left\{\begin{array}
{r@{\quad:\quad}l}
\hat {\eta}_{A,j} (\mbai) + \ln f & \mbox{if bin \mbai\ is visited} \\
\hat {\eta}_{A,j} (\mbai) & \mbox{otherwise}
\end{array}\right.
\label{eq:wlupdate}
\end{equation}\tshow{eq:wlupdate}

\noindent  The idea is to start of with an f greater than unity, and perform the update scheme of Eq. \ref{eq:wlupdate} until one obtains a flat histogram \cite{note:wlflat}. One then reduces f (but at the same time constraining it to be greater than unity) and repeats the simulation, this time starting off with the set of MUCA weights obtained at the end of the previous simulation. This procedure is carried out until f has reduced to a value close to unity, at which stage one will have obtained a sufficiently accurate set of MUCA weights so as to ensure that  ${P}(\mba | \pi^m_A )$ is flat over the range of \mba\ space containing both ${P}(\mba | \pi^c_A )$ and ${P}(\mba | \pi^c_B )$. Using this estimate one may perform a final simulation (in which the weights ${\eta}_A$ are unmodified) and use  Eq. \ref{eq:torriereweightest} in order to estimate \RBAcal. In chapter \ref{chap:sampstrat} we will illustrate in greater detail the use of the umbrella sampling strategy (constructing the MUCA weights via the Wang-Landau method) in estimating the desired FEDs.

\subsection{\label{sec:ms}Multistage Methods}
\tshow{sec:ms}

The multistage (MS) method may be considered to be a generalisation of the DP (dual phase) methods  \cite{p:kofkecummings}, \cite{p:radmer}. The central idea of this method \cite{p:valleaucard} is to break up the task of evaluating the FED into a series of  {\em independent} tasks of estimating the FEDs between pairs of systems whose overlap is considerably improved with respect to the original pair. The method is based on the construction of a chain of configurational energies:

\begin{equation}
\seta {\ecal_{\la_1} = \ecal_A, \ecal_{\la_2}, \ecal_{\la_3}, ..., \ecal_{\la_n} = \ecal_B} 
\label{eq:hamilchain}
\end{equation}\tshow{eq:hamilchain}

\noindent in which the configurational energy $\ecal_{\la_i}$ has the  associated sampling distribution:

\begin{equation}
\pi_{\la_i}(\avvec)  \stackrel {\mathrm{.}}{=}  e^{-\beta \ecal_{\la_i}(\avvec)}
\label{eq:samphamilofi}
\end{equation}\tshow{eq:samphamilofi}

\noindent By noticing that:

\begin{equation}
\RBAcal = \frac {\z_{\la_2}} {\z_{\la_1}} \frac {\z_{\la_3}} {\z_{\la_2}} ... \frac{\z_{\la_n}} {\z_{\la_{n-1}}}
\label{eq:multistage}
\end{equation}\tshow{eq:multistage}

\noindent where 

\begin{equation}
\z_{\la_i} \equiv \int d\avvec e^{-\beta \ecal_{\la_i}(\avvec)}
\label{eq:partofstage}
\end{equation}\tshow{eq:partofstage}

\noindent and by constructing a chain of \ces\ in such a way that $P(\mba | \pi_{\la_i})$ and $P(\mba | \pi_{\la_{i+1}})$ sufficiently overlap, one may estimate each $\frac {\z_{\la_{i+1}}} {\z_{\la_{i}}}$ via any one of the techniques presented in this review.

The standard (and simplest) way of constructing $\ecal_{\la_i}$ is via the following linear interpolation scheme:

\begin{eqnarray}
\ecal_{\la_i}(\avvec)& =& \la_i \ecal_B (\avvec) + (1-\la_i) \ecal_A (\avvec)\nonumber\\
& = & \ecal_A (\vvec) + \beta^{-1} \la_i \mba 
\label{eq:linearhamiltonian}
\end{eqnarray}\tshow{eq:linearhamiltonian}

\noindent where $\la_1\equiv 0 < \la_2 ....<\la_n \equiv 1$ \cite{note:aboutlin}. 
As an example one may evaluate Eq. \ref{eq:multistage} via the EP method (Eq. \ref{eq:ep}) \cite{note:launcher}:

\begin{eqnarray}
\RBAcal  =  \p {i=1} {n-1} <e^{-\beta [\ecal_{\la_{i+1}}(\avvec) - \ecal_{\la_i} (\avvec)]}>_{\pi_{\la_i}}
\end{eqnarray}

\noindent which, for the case of Eq. \ref{eq:linearhamiltonian}, may be written as:

\begin{equation}
\RBAcal =\p {i=1} {n-1} < e^{-\triangle \la_i \mba(\avvec)}>_{\pi_{\la_i}}\label{eq:multistagezwanzig}
\end{equation}\tshow{eq:multistagezwanzig}

\noindent where

\begin{equation}
\triangle \la_i \equiv \la_{i+1} - \la_i
\end{equation}

\noindent Eq. \ref{eq:multistagezwanzig} is interesting in its own right since it reduces to the well known method of thermodynamic integration \cite{p:frenkelladd}\nocite{p:fleischman}-\cite{p:straatsma} in the limit of sufficiently small \seta {\delela_i}. To see this we note that for sufficiently small  $\triangle \la$ one may expand the exponential in Eq. \ref{eq:multistagezwanzig}  as a power series in $\triangle \la$ so that:

\begin{eqnarray}
\triangle F_{BA} = -\beta^{-1} \ln \RBAcal & = & -\beta^{-1} \ln(\p {i=1} {n-1} <1-\delela_i \mba+O(\delela^2)>_{\pi_{\la_i}})\nonumber\\
& \approx & -\beta^{-1} \s {i=1} {n-1} \ln (1- <\delela_i \mba >_{\pi_{\la_i}})\nonumber\\
& \approx &  \beta^{-1}\s {i=1} {n-1} < \delela_i \mba >_{\pi_{\la_i}}\nonumber\\
& \approx & \int^{\la=1}_{\la=0} d\la <\frac {\partial \ecal_\la} {\partial \la}>_{\pi_\la}
\label{eq:MSTI}
\end{eqnarray}\tshow{eq:MSTI}

\noindent where in expanding the exponential of Eq. \ref{eq:multistagezwanzig} we have neglected powers of order ${\triangle \la}^2$ and higher. This is valid provided that there are a sufficient number of configurational energies linking $\ecal_A$ to $\ecal_B$ (see Eq. \ref{eq:linearhamiltonian}), so as to ensure that the $\delela_i$ are sufficiently small. We make a point to note that though the appearance of Eq. \ref{eq:MSTI} is identical to that of Eq. \ref{eq:thermody}, it  has incorporated within (as do all the other methods that are being reviewed in this section) it the PM formalism of \cite{p:LSMC}. This allows it to be used to {\em directly} estimate the FED between the two phases, rather than having to use it to evaluate the FED between each system and some reference system, as is the case in the original formulation (see Eq. \ref{eq:thermody}).

\subsection{\label{sec:wham}Weighted Histogram Analysis Method}
\tshow{sec:wham}

The weighted histogram analysis method (also called WHAM \cite{p:weightedhistogram}\nocite{p:weightedhistogram2}\nocite{p:souailleroux}-\cite{p:sugita}) is a generalisation of the histogram re-weighting techniques of \cite{p:reweighting} and \cite{p:reweighting2}, and employs a minimum variance estimator which uses the re-weighting of data from several independent simulations in order to calculate the FEDs. The original derivation is very involved and we follow the simpler derivation given in \cite{p:radmer}.

Suppose that one constructs a chain  of configurational energies \seta {\ecal_{\la_i}} linking $\ecal_A$ to $\ecal_B$ (which we take to be the particular case of Eq. \ref{eq:linearhamiltonian}), whose corresponding sampling distributions are denoted by $\pi_{\la_i}$, Eq. \ref{eq:samphamilofi}. The WHAM method is based on the observation that if one performs several independent sampling experiments with the sampling distributions $\pi_{\la_1}$, $\pi_{\la_2}$, ...., $\pi_{\la_n}$, in which  $N_j$ independent samples are obtained for the sampling experiment performed with $\pi_{\la_j}$, then the probability of observing a macrostate $\mba$ within the collection of data, as obtained from {\em all} the simulations, is given by:
\begin{equation}
P(\mba) = \frac {1} {N_T} \s {i=1} {n} N_i P(\mba | \pi_{\la_i})
\label{eq:whamprob}
\end{equation}\tshow{eq:whamprob}

\begin{equation}
N_T = \s {i=1} {n} N_i
\end{equation}

\noindent The underlying idea of the WHAM method is simple. The strategy is to construct a set of distributions \seta {P(\mba | \pi_{\la_i})} which overlap and connect the regions of \ecs\ associated with one phase to those of the other \cite{note:heypath}. With this choice the resulting $P(\mba)$ will contain {\em both} $P(\mba | \pia)$ {\em and} $P(\mba | \pib)$.  To arrive at an expression which will allow one to estimate the partition functions \seta {{\tilde {Z}}_{\la_i}} (up to a multiplicative constant which is the same for all the estimates) one starts off by inverting the expression given in Eq. \ref{eq:whamprob}:

\begin{eqnarray}
P(\mba) & = & \frac {P(\mba | \pi_{\la_j})} {N_T} \s {i=1} {n} N_i \frac {P(\mba | \pi_{\la_i})} {P(\mba | \pi_{\la_j})} \nonumber\\
& = &  \frac {P(\mba | \pi_{\la_j})} {N_T} \s {i=1} {n} N_i \frac {{\tilde {Z}}_{\la_j}} {{\tilde {Z}}_{\la_i}} e^{-(\la_i-\la_j)\mba}
\end{eqnarray}

\noindent or:

\begin{equation}
{\tilde Z}_{\la_j} P(\mba | \pi_{\la_j}) = \frac {N_T P(\mba) e^{-\la_j \mba }} {\s {i=1} {n} \frac {N_i} {{\tilde {Z}}_{\la_i}} e^{-\la_i \mba}}
\label{eq:whamtemp}
\end{equation}\tshow{eq:whamtemp}

\noindent Summing over all bins in Eq. \ref{eq:whamtemp} and using Eq. \ref{eq:whamprob} it is clear that ${\tilde {Z}}_{\la_j}$ may be estimated by:

\begin{eqnarray}
{\tilde {Z}}_{\la_j} & \est &  \s {k=1} {b} \frac {N_T P(\mbak) e^{-\la_j \mbak }} {\s {i=1} {n} \frac {N_i} {{\tilde {Z}}_{\la_i}} e^{-\la_i \mbak}}\nonumber\\
& = & \s {k=1} {b} \{ \frac {\s {m=1} {n} H(\mbak | \pi_{\la_m}) e^{-\la_j \mbak}} {\s {i=1} {n} \frac {N_i} {{\tilde {Z}}_{\la_i}} e^{-\la_i \mbak}}\}
\label{eq:whamcentral}
\end{eqnarray}\tshow{eq:whamcentral}

\noindent The set of equations given by Eq. \ref{eq:whamcentral} form the core of the WHAM method \cite{note:wham}. It is immediately clear from Eq. \ref{eq:whamcentral} that in order to estimate the ratio $\RBAcal = {\tilde {Z}}_{\la_n}/{\tilde {Z}}_{\la_1}$ one must estimate the partition functions (up to a common constant) of {\em all} the sampling distributions associated with the configurational energies \seta {\ecal_{\la_i}} (see Eq. \ref{eq:hamilchain}) linking $\ecal_A$ to $\ecal_B$. In order to estimate the \seta {{\tilde {Z}}_{\la_i}} one must solve Eq. \ref{eq:whamcentral} {\em iteratively}.
One starts off with a set of estimates for $\{{\tilde {Z}}_{\la_i}\}$, which one then feeds into Eq. \ref{eq:whamcentral} to get a new set of estimates. One then feeds these estimates back into Eq. \ref{eq:whamcentral} to get yet another set of even more accurate estimates. One carries out this iteration until the set of estimates have suitably converged. At this point the estimate of  ${\tilde {Z}}_{\la_n}/{\tilde {Z}}_{\la_1}$ will yield an accurate estimate of \RBAcal . The power of the method clearly lies in the enormous scope for parallelizability that exists in the construction of the path linking  phase A to phase B.

\subsection{\label{sec:st}Simulated Tempering}
\tshow{sec:st}

The simulated tempering method \cite{p:nezbedakolafa}\nocite{p:lyubartsev}\nocite{p:lyubartsevgeneral}-\cite{p:simulatedtempering}, like the multistage method, involves the construction of a chain of configurational energies linking $\ecal_A$ to $\ecal_B$ (see Eq. \ref{eq:hamilchain}). The basic idea behind the method is to simulate from a sampling distribution characterised by the following partition function:

\begin{equation}
\z_{ST} = \s {i=1} {n} \z_{\la_i} e^{-\eta_{ST}^{(i)}} = \s {i=1} {n} \int d\avvec e^{-\beta \ecal_{\la_i} (\avvec) - \eta_{ST}^{(i)}}
\label{eq:partforexpanded}
\end{equation}\tshow{eq:partforexpanded}

\noindent where $\z_{\la_i}$ is the partition function associated with the sampling distribution $\pi_{\la_i}$ (see Eq. \ref{eq:samphamilofi} and Eq. \ref{eq:partofstage}) and $\eta_{ST}^{(i)}$ are some weights. We will refer to each $\z_{\la_i}$ as a sub-ensemble. The idea of the method is to construct a $\z_{ST}$ so that the corresponding sampling distribution explores (evenly) {\em all} the regions of \ecs\ 'containing' the two phase-constrained distributions $P(\mba | \pia)$ and $P(\mba | \pib)$ and the regions in between them. One way \cite{p:bennett} of realising Eq. \ref{eq:partforexpanded} is to sample via the sampling distribution:

\begin{equation}
\pi (\vvec) \equaldot  \s {i=1} {n} e^{-\beta \ecal_{\la_i} (\avvec) - \eta_{ST}^{(i)}}
\label{eq:samemuca}
\end{equation}\tshow{eq:samemuca}

\noindent An alternative sampling distribution (the one used in  \cite{p:nezbedakolafa}\nocite{p:lyubartsev}\nocite{p:lyubartsevgeneral}-\cite{p:simulatedtempering}) is one which 'hops' between the sub-ensembles and in which two types of Monte Carlo moves are employed. The first type of move involves the usual particle displacement. Such moves are accepted with the probability given by Eq. \ref{eq:acceptancerate2}, where $\pi_{\la_i}$ is used in place of $\pi$ if the simulation is in sub-ensemble i. The second type of move attempts to switch sub-ensembles. That is this move keeps the effective configuration \avvec\ constant whilst trying to change the sampling distribution from $\pi_{\la_m}$ to $\pi_{\la_{m'}}$ (generally $m'$ is chosen to be an adjacent sub-ensemble of m). In order to satisfy detailed balance (Eq. \ref{eq:detailedbalance}) and in order to yield a sampling distribution with a  partition function given by Eq. \ref{eq:partforexpanded}, such a move must be accepted with probability:

\begin{equation}
P_a(m\rightarrow m' | \pi_{ST})  =  \mbox{Min} \{1, \frac {e^{-\beta \ecal_{\la_{m'}}(\avvec) -\eta_{ST}^{(m')}}} {e^{-\beta \ecal_{\la_m}(\avvec) -\eta_{ST}^{(m)}}} \}
\label{eq:acceptanceST}
\end{equation}\tshow{eq:acceptanceST}

\noindent Accordingly we may write the sampling distribution as:

\begin{equation}
\pi_{ST} (\vvec , i) \equaldot e^{-\beta \ecal_{\la_i}(\vvec)-\eta_{ST}^{(i)}}
\label{eq:ststo}
\end{equation}\tshow{eq:ststo}

\noindent where i is a stochastic variable assuming any integer value between (and inclusive of) 1 and n. Unless otherwise stated, we will assume hereafter that $\pi_{ST}$ corresponds to the sampling distribution given in Eq. \ref{eq:ststo}.

Suppose that $P_m$ denotes the probability of finding the simulation in sub-ensemble m and suppose that $T_m$ denotes  the time spent in the sub-ensemble m:

\begin{equation}
T_m = \s {i=1} {t} \triangle_m (\vvec(i))
\end{equation}

\noindent where \seta {\vvec{(1)}, \vvec{(2)}, ..., \vvec{(t)}}\ denotes the sequence of effective configurations generated by the simulation, and where:

\begin{equation}
\triangle_m (\vvec(i)) \equiv \left\{\begin{array}
{r@{\quad:\quad}l}
1 & \mbox{if $\vvec(i)$ is a configuration generated in sub-ensemble m} \\
0 & \mbox{otherwise}
\end{array}\right.
\label{eq:triangle3}
\end{equation}\tshow{eq:triangle3}

\noindent Under the sampling distribution $\pi_{ST}$, it follows from Eq. \ref{eq:partforexpanded} that since the ratio of the probabilities of the simulation being found in any two sub-ensembles is given by:

\begin{equation}
\frac {P_m} {P_k} = \frac {\z_{\la_m}} {\z_{\la_k}} e^{-(\eta_{ST}^{(m)} - \eta_{ST}^{(k)})}
\end{equation}

\noindent and since the ratio of the probabilities of finding the simulation in two sub-ensembles  is estimated by the ratio of the times spent in the two sub-ensembles:

\begin{equation}
\frac {\z_{\la_m}} {\z_{\la_k}} e^{- (\eta_{ST}^{(m)} - \eta_{ST}^{(k)})} \est \frac {T_m} {T_k}
\end{equation}

\noindent then the \rpf\ of $\z_{\la_m}$ and $\z_{\la_k}$ may be estimated via:

\begin{equation}
\frac {\z_{\la_m}} {\z_{\la_k}} \est \frac {T_m} {T_k} e^{(\eta_{ST}^{(m)} - \eta_{ST}^{(k)})} 
\label{eq:STest}
\end{equation}\tshow{eq:STest}

\noindent It follows from Eq. \ref{eq:STest} that the quantity \RBAcal\ may then be estimated from the ratio of the times spent in the sub-ensembles 1 and n:

\begin{equation}
\RBAcal \est \frac {T_n} {T_1} e^{\eta_{ST}^{(n)} - \eta_{ST}^{(1)}}
\label{eq:RBAST}
\end{equation}\tshow{eq:RBAST}

\noindent In order to arrive at an estimator in terms of the macrovariable \mba\ (as has been formulated in sections \ref{sec:umbrella} and \ref{sec:wham} for the umbrella and WHAM methods respectively) we first note that:

\begin{eqnarray}
P(\mba | \pi_{ST}) & = & \s {i=1} {n} P(\mba , {\la_i} | \pi_{ST})\nonumber\\
& = & \s {i=1} {n} P(\mba | \pi_{\la_i}) . P(\la_i | \pi_{ST})
\end{eqnarray}

\noindent where $P(\mba , {\la_i} | \pi_{ST})$ denotes the joint distribution of observing the macrostate \mba\ {\em and} of being in sub-ensemble i under the sampling distribution $\pi_{ST}$, and where $P(\la_i | \pi_{ST})$ denotes the probability of being in sub-ensemble i under the sampling distribution $\pi_{ST}$. From Eq. \ref{eq:partforexpanded} and Eq. \ref{eq:ststo} it follows that:

\begin{equation}
P(\la_i | \pi_{ST}) = \frac {e^{-\eta^i_{ST}} \z_{\la_i}} {\s {j=1} {n} e^{-\eta^j_{ST}} \z_{\la_j}}\label{eq:stocc}
\end{equation}\tshow{eq:stocc}

\noindent By noting that in the case where the \ce\ is linearly parameterised (Eq. \ref{eq:linearhamiltonian}):

\begin{eqnarray}
\frac {P(\mba | \pi_{ST})} {P(\mba | \pi_A)} &  = & \s {i=1} {n} \frac {P(\mba | \pi_{\la_i})} {P(\mba | \pi_{\la_0})} P(\la_i | \pi_{ST}) \nonumber\\
& = & \frac {\z_{\la_0}} {\s {j=1} {n} e^{-\eta^j_{ST} }\z_{\la_j}} \s {i=1} {n} e^{-\la_i \mba -\eta^i_{ST}} 
\end{eqnarray}

\noindent One may use Eq. \ref{eq:torriereweight} to arrive at an estimator for \RBAcal\ in terms of the macrovariable \mba :

\begin{eqnarray}
\RBAcal & = &  \frac {<e^{-\mba}  \frac {P(\mba | \pi_A^c)} {P(\mba | \pi_{ST})}>_{\pi_{ST}}} {<\frac {P(\mba | \pi_A^c)} {P(\mba | \pi_{ST})}>_{\pi_{ST}}}\nonumber\\
& = & \frac {<\frac {1} {\s {i=1} {n} e^{(1-\la_i)\mba -\eta^i_{ST}}} >_{\pi_{ST}}}
{<\frac {1} {\s {i=1} {n} e^{-\la_i \mba -\eta^i_{ST}}} >_{\pi_{ST}}}
\label{eq:genstest}
\end{eqnarray}\tshow{eq:genstest}

\noindent The estimator in Eq. \ref{eq:genstest} is also valid if, instead of $\pi_{ST}$, one employs the sampling distribution given in Eq. \ref{eq:samemuca}.

 By hopping between the sub-ensembles the simulation is able to explore a wider region of \ecs\ than it would under a sampling experiment performed with any one of the individual sampling distributions $\pi_{\la_m}$. In order to ensure that the simulation visits all sub-ensembles, one must first ensure that a sufficient number of intermediate sub-ensembles have been constructed (so that $P(\mba | \pi_{\la_i})$ overlaps with $P(\mba | \pi_{\la_{i+1}})$). Secondly one must also ensure that the correct weights \seta {\eta_{ST}^{(i)}} have been chosen so as to guarantee that the simulation is able to frequently traverse between the regions of \ecs\ typically associated with phase A and those typically associated with phase B. One way to do this is to  choose the weights so that equal times are spent in all the sub-ensembles. In this case one sets $\eta_{ST}^{(i)} = \ln {\tilde {Z}}_{\la_i} + constant$. However since a-priori the partition functions are not known, it follows that the weights  must be constructed in an iterative fashion (e.g. via the visited states method or the  Wang-Landau method) as is done in the Umbrella Sampling method (see section \ref{sec:umbrella}). Having obtained the weights one may then proceed to estimate the \rpf\ \RBAcal\ by appeal to the estimator in Eq. \ref{eq:RBAST} or Eq. \ref{eq:genstest}.

\subsection{\label{sec:phaseswitch}Phase Switching Method}
\tshow{sec:phaseswitch}

The Phase Switching (PS) method, along with the whole phase mapping (PM) formalism, was originally developed in \cite{p:LSMC} (see \cite{p:brucereview} for a review and see  \cite{p:LSMCprehard}\nocite{p:LSMCpresoft}\nocite{p:solidliquid}-\cite{p:LSMCerrington} for generalisations). In order to motivate the method consider the case where one samples from a sampling distribution which is associated with the following  partition function (which we refer to as the canonical PS partition function):

\begin{equation}
\z_{PS} = \int d\avvec e^{-\beta \ecal_A(\avvec)}  + \int d\avvec e^{-\beta \ecal_B(\avvec)}
\label{eq:partLS}
\end{equation}\tshow{eq:partLS}

\noindent An example of a  sampling distribution associated with such a partition function is:

\begin{eqnarray}
\pi(\vvec) & \equaldot & e^{-\beta \ecal_A(\vvec)} + e^{-\beta \ecal_B(\vvec)}\nonumber\\
& = & e^{-\beta \ecal_A(\vvec)} [1+e^{-\mba(\vvec)}]
\label{eq:pssingle}
\end{eqnarray}\tshow{eq:pssingle}

\noindent From the discussion of section \ref{sec:st} it is clear that Eq. \ref{eq:partLS} may be equivalently realised by what we call the canonical  PS sampling distribution:

\begin{equation}
\pi^c_{PS} (\vvec, \al)  \equaldot  e^{-\beta \ecal_\al (\vvec)} 
\label{eq:PSsamp}
\end{equation}\tshow{eq:PSsamp}

\noindent where $\al$ is a stochastic variable, which assumes one of two values: $\al=A$ or $\al=B$. Eq. \ref{eq:PSsamp} is clearly a special case of the ST sampling distribution $\pi_{ST}$ in which one has only two sub-ensembles corresponding to the two phases and in which the weights \seta {\eta_{ST}} are the same for both the phases. The sampling distribution $\pips$ then accepts a  'switch' between the two systems with the following probability:

\begin{eqnarray}
P_a(A\rightarrow B | \pips) & = & \mbox{Min} \{1, \frac {\pi^c_{PS}(\vvec,B)}  {\pi^c_{PS}(\vvec,A)}\}\nonumber\\
& = & \mbox{Min} \{ 1, e^{-\mba}\}\nonumber\\
\label{eq:psacc}
\end{eqnarray}\tshow{eq:psacc}

\noindent The PS sampling distribution \pips\ (and also the alternative sampling distribution in Eq. \ref{eq:pssingle}) then yield  the following distribution:

\begin{equation}
P(\mba | \pips) = \frac{\z_A P(\mba | \pia) + \z_B P(\mba | \pib)} {\z_A+\z_B}
\label{eq:mbadist}
\end{equation}\tshow{eq:mbadist}

\noindent For the PS sampling distribution \pips\ the \rpf\ is then given by:

\begin{eqnarray}
\RBAcal &  \equiv & \frac {\int d\vvec e^{-\beta \ecal_B (\avvec)}} {\int d\vvec e^{-\beta \ecal_A (\avvec)}}\nonumber\\
& = &\frac {<\triangle_B (\vvec)  >_{\pi^c_{PS}}} {<\triangle_A (\vvec) >_{\pi^c_{PS}}}
\label{eq:ratiooftimes0}
\end{eqnarray}\tshow{eq:ratiooftimes0}

\noindent which may be  estimated by the ratio of the times spent in the two phases (see Eq. \ref{eq:RBAST}):

\begin{equation}
\RBAcal \est \frac {\s {i=1} {t} \triangle_B (\vvec(i))} { \s {i=1} {t}\triangle_A (\vvec(i)) } 
\label{eq:ratiooftimes}
\end{equation}\tshow{eq:ratiooftimes}

\noindent where $\triangle_\gamma (\vvec)$ is given by:

\begin{equation}
\triangle_\al (\vvec(i)) \equiv \left\{\begin{array}
{r@{\quad:\quad}l}
1 & \mbox{if $\vvec(i)$ is a configuration is generated in phase \al } \\
0 & \mbox{otherwise}
\end{array}\right.
\end{equation}

\noindent More generally (see Eq. \ref{eq:genstest}) one may estimate \RBAcal\ (for both Eq. \ref{eq:pssingle} and Eq. \ref{eq:PSsamp}) from the estimator corresponding to:

\begin{equation}
\RBAcal = \frac {<f(\mba)>_{\pi^c_{PS}}} {<f(-\mba)>_{\pi^c_{PS}}}
\label{eq:firstfermi}
\end{equation}\tshow{eq:firstfermi}

\noindent where $f(x)$ is the fermi function:

\begin{equation}
f(x) = \frac {1} {1+e^x}
\label{eq:fermi}
\end{equation}\tshow{eq:fermi
}

\noindent It is clear from Eq. \ref{eq:psacc} that a PS Monte Carlo move (in which one attempts to switch phases whilst keeping the effective configuration \vvec\ constant) is only likely to be accepted within  the $|\mba | \leq O(1)$ regions. From section \ref{sec:overlap} we saw that the vast majority of equilibrium configurations of the parent phase will be mapped (under the operation of the PM) onto configurations of the conjugate phase which are of higher excitation energy. Therefore the probability of visiting configurations for which the two phases are of roughly the same energy (under the operation of the PM) is negligibly small (see also figure \ref{bimodaldistribution}). Even if one considers the more general case of Eq. \ref{eq:partforexpanded} where one  introduces two weights into Eq. \ref{eq:partLS} so that:

\begin{equation}
\z_{ST} = \int d\avvec e^{-\beta \ecal_A(\avvec)-\eta_{ST}^{(A)}}  + \int d\avvec e^{-\beta \ecal_B(\avvec)-\eta_{ST}^{(B)}}
\label{eq:partforls}
\end{equation}\tshow{eq:partforls}

\noindent the problem will not be solved since, in the absence of overlap, these weights will at best allow the switching to take place only in one direction. As we saw in section \ref{sec:st} the simulated tempering method solves this problem by constructing a series of intermediate systems so as to engineer overlap between adjacent  systems. For such pairs of (sufficiently overlapping) systems the weights can be chosen so as to ensure that switching takes place frequently in both directions. 

The way the PS method overcomes the overlap problem is by using a set of weights which are themselves a function on \ecs. That is rather than simulating via a sampling distribution characterised  by the partition function in Eq. \ref{eq:partforls} one instead employs a MUCA sampling distribution associated with the  partition function:

\begin{equation}
\z_{PS} = \int d\avvec e^{-\beta \ecal_A(\avvec) - \eta_{PS} (\mba (\avvec))}  + \int d\avvec e^{-\beta \ecal_B(\avvec)- \eta_{PS} (\mba (\avvec))}
\label{eq:partLSsamp}
\end{equation}\tshow{eq:partLSsamp}

\noindent The sampling distribution (which we call the MUCA-PS sampling distribution) is then given by :

\begin{eqnarray}
\pi^m_{PS} (\vvec, \al) & \equaldot & \pi^c_{PS} (\vvec, \al) e^{-\eta_{PS}(\mba(\vvec))} \nonumber\\
& \equaldot & e^{-\beta \ecal_\al (\vvec)-\eta_{PS}(\mba(\vvec))} 
\label{eq:pssampmuca}
\end{eqnarray}\tshow{eq:pssampmuca}

\noindent which has the associated \mba\ distribution:

\begin{equation}
P(\mba | \pi^m_{PS} ) \equaldot e^{-\eta_{PS}(\mba)} P(\mba | \pi^c_{PS} ) 
\end{equation}

\noindent Eq. \ref{eq:partLSsamp} may be equivalently realised via the sampling distribution:

\begin{equation}
\pi(\vvec) \equaldot e^{-\beta \ecal_A(\vvec) -\eta (\mba)} [1+e^{-\mba(\vvec)}]
\label{eq:psmuca2}
\end{equation}\tshow{eq:psmuca2}

\noindent It is immediately clear that the probability of a PS Monte Carlo move, as dictated by Eq. \ref{eq:pssampmuca}, is given by Eq. \ref{eq:psacc}, {\em since a PS does not change the value of \mba }:

\begin{eqnarray}
P_a(A\rightarrow B | \pi^m_{PS}) & = & \mbox{Min} \{1, \frac {\pi^m_{PS}(\vvec,B)}  {\pi^m_{PS}(\vvec,A)}\}\nonumber\\
& = & \mbox{Min} \{ 1, e^{-\mba}\}\nonumber\\
\label{eq:lateps}
\end{eqnarray}\tshow{eq:lateps}

\noindent On the other hand the probability of accepting a move from a configuration $\vvec$ to a configuration $\vvec '$ is now given by:

\begin{eqnarray}
P_a(\avvec \rightarrow \avvec ' | \pi^m_{PS}) & = & \mbox {Min} \{ 1, \frac {\pi^m_{PS} (\vvec ' , \al)} {\pi^m_{PS} (\vvec, \al)}\nonumber\\
& = &\mbox {Min} \{ 1, \frac {e^{-\beta \ecal_\gamma(\avvec ') -\eta_{PS} (\mba(\avvec ')) } } {e^{-\beta \ecal_\gamma(\avvec ) -\eta_{PS} (\mba(\avvec )) }} \}
\label{eq:switchLS}
\end{eqnarray}\tshow{eq:switchLS}

\noindent Therefore the role of the MUCA weight function $\eta_{PS}(\mba)$ is not, as is the case in the ST method, to aid the simulation (directly) in switching between phases, but instead to {\em guide} the simulation to regions of \ecs\  from which the simulation can easily switch phases.

In order to determine \RBAcal\ one applies the reweighting formula (Eq. \ref{eq:reweightav}) to Eq. \ref{eq:ratiooftimes0} (so as to remove the bias of the MUCA weights in Eq. \ref{eq:partLSsamp}):

\begin{equation}
\RBAcal =  \frac {<\triangle_B (\vvec) e^{\eta_{PS}(\mba(\vvec))} >_{\pi^m_{PS}}} {<\triangle_A (\vvec) e^{\eta_{PS}(\mba(\vvec))} >_{\pi^m_{PS}}}
\label{eq:psformu}
\end{equation}\tshow{eq:psformu}

\noindent The corresponding estimator is then given by:

\begin{equation}
\RBAcal \est \frac {\s {i=1} {t} e^{+\eta_{PS} (\mba(\vvec(i)))} \triangle_B(\vvec(i))} {\s {i=1} {t} e^{+\eta_{PS} (\mba(\vvec(i)))} \triangle_A(\vvec(i))}
\label{eq:psest}
\end{equation}\tshow{eq:psest}

\noindent Similarly the  more general estimator given in Eq. \ref{eq:firstfermi}, valid for both Eq. \ref{eq:pssampmuca} and Eq. \ref{eq:psmuca2}, is now replaced by:

\begin{equation}
\RBAcal = \frac {<f(\mba)e^{\eta_{PS}(\mba)}>_{\pi^m_{PS}}} {<f(-\mba)e^{\eta_{PS}(\mba)}>_{\pi^m_{PS}}}
\end{equation}

\begin{figure}[tbp]
\begin{center}
\rotatebox{270}{
\hidefigure{\includegraphics[scale=0.6]{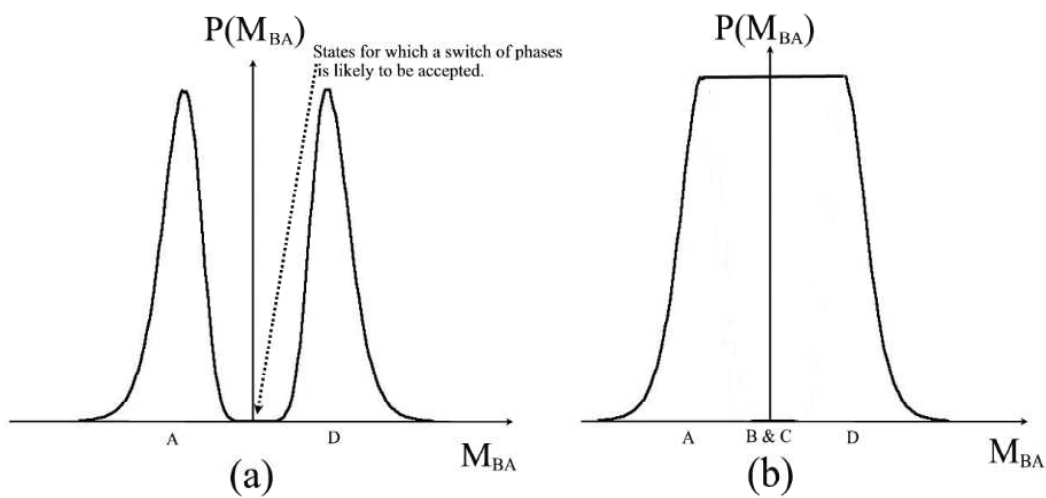}}
}
\end{center}
\caption{{Schematic of probability distribution of $\mba$}}

(a) The canonical distribution $P(\mba | \pi^c_{PS})$

(b) The multicanonical distribution $P(\mba | \pi^m_{PS})$

In order for the simulation to be able to reach those regions characterised by $\mba\sim 0$, one must employ MUCA weights (through Eq. \ref{eq:pssampmuca}). See also figure \ref{phase_switch_fig}.
\begin{center}
{\bf{------------------------------------------}}
\end{center}

\tshow{***schematicpdf}

\label{schematicpdf}
\end{figure}

\begin{figure}[tbp]
\begin{center}
\rotatebox{270}{
\hidefigure{\includegraphics[scale=0.6]{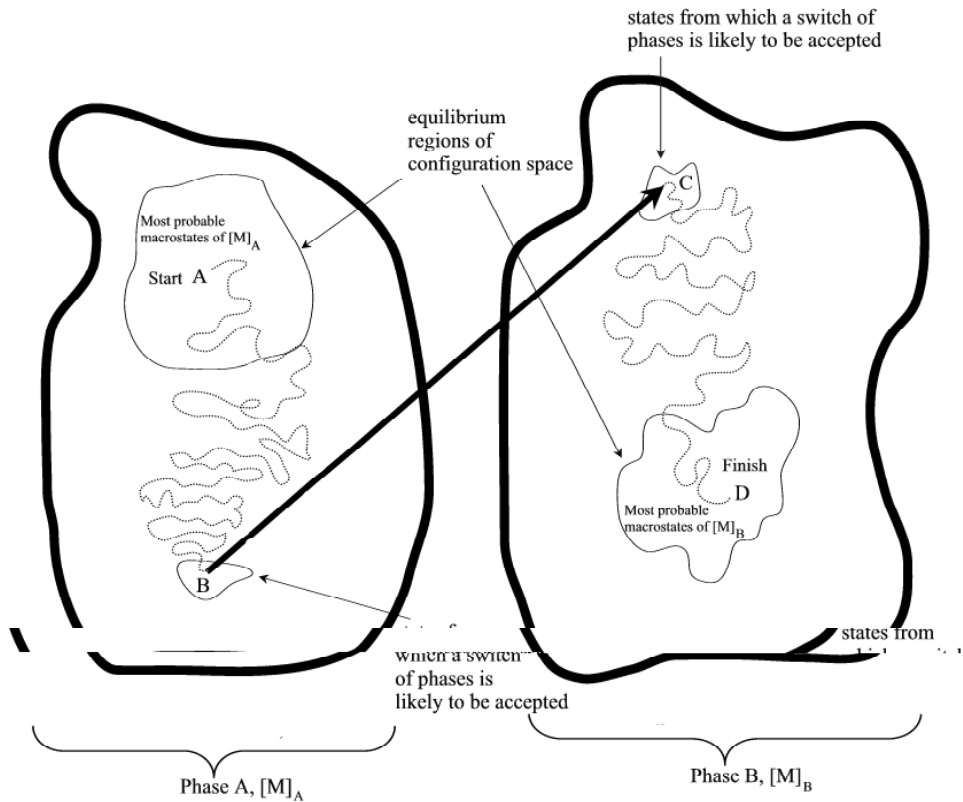}}
}
\end{center}
\caption{Schematic of the PS algorithm in \rvec\ space}

The diagram shows a simulation starting at a typical ('equilibrium') configuration of phase A . The employment of the weights \seta {\eta}  in Eq. \ref{eq:pssampmuca} means that the simulation performs a random walk (in \mba\ space) to the region B. For these configurations $\mba\sim 0$ and therefore a PS (via Eq. \ref{eq:lateps}) has a chance of being accepted.  An accepted  PS then takes the simulation to region C, which just corresponds to a switching of reference configurations (see Eq. \ref{eq:rintermsofv}). Then under the influence of the weights the algorithm performs a random walk in \mba\ space from region C to region D, which is  the equilibrium region of (absolute) \cs\ of  phase B. The corresponding points A, B, C, and D are also shown in figure \ref{schematicpdf}.\newline
The utility of the macrovariable \mba\ (as was mentioned in the discussion of Eq. \ref{eq:partrsestimator2}) is that it serves as a guiding parameter for the simulation. In the case of the PS method, it is used to guide the simulation to regions of (absolute) \cs\ from which a switch of phases has a non-negligible chance of being accepted.
\begin{center}
{\bf{------------------------------------------}}
\end{center}
\tshow{phase_switch_fig}
\label{phase_switch_fig}
\end{figure}

\noindent Figure \ref{schematicpdf} (a) shows a schematic for $P(\mba | \pi^c_{PS})$ and (b) shows a schematic for  $P(\mba | \pi^m_{PS})$. It is clear from (a) that a canonical PS sampling distribution $\pi^c_{PS}$ initiated in one of the two phases will remain stuck in that phase, since the probability of visiting the $\mba \sim 0$ regions is negligibly small. In order to obtain the estimator of the canonical distribution shown in (a), one would first have to perform a MUCA simulation as shown in (b) and then reweight the data appropriately (see \cite{note:canpsprob} for details). The essential feature of the MUCA distribution $P(\mba | \pi^m_{PS})$ is that it contains {\em both} the phase constrained distributions $P(\mba | \pia)$ and $P(\mba | \pib)$. 
Figure \ref{phase_switch_fig} shows a schematic of the PS procedure.

\subsection{\label{sec:fg}Fast Growth Method}
\tshow{sec:fg}

The fast growth (FG) method \cite{p:jarprl} rests on a result called the Fluctuation Theorem. This theorem has been proved for a variety of non-equilibrium processes  \cite{p:jarprl}, \cite{p:crooks}\nocite{p:crookspre}-\cite{p:crookspre2}, \cite{p:gallavotticohenfirst}\nocite{p:gallavotticohensecond}\nocite{p:firstfluctuation}\nocite{p:kurchan}\nocite{p:depken}-\cite{p:zon}. The particular formulation  that we will draw on is set out in \cite{p:jarprl}, \cite{p:crooks}\nocite{p:crookspre}-\cite{p:crookspre2}.

Central to the Fluctuation Theorem is a non-equilibrium process that we will now describe. One starts of  by constructing a configurational energy $\ecal_\lambda$ which is a function of a field parameter $\lambda$ (for example see Eq. \ref{eq:linearhamiltonian}). The field parameter $\lambda$ takes any value between $\la_1$ and $\la_n$, and the set \seta{\ecal_{\lambda}}\ forms a chain of configurational energies linking $\ecal_A$ to $\ecal_B$ (see Eq. \ref{eq:hamilchain}).
 The $A\rightarrow B$ non-equilibrium process, which takes us from phase A to phase B, involves the switching of the field parameter $\la$  from  an initial value of $\la_1 =0$ to $\la_n = 1$ in a series of discrete steps at some predefined, but arbitrary, set of times \seta {t} = \seta {t_1, t_2, ..., t_{n-1}}.

The  implementation procedure may be realised as follows. The initial point $\vvec(1)$ is sampled with respect to the canonical distribution $\pi_{\la_1}=\pi^c_A$. One then increments (at time $t_1$) the field parameter $\la$ from $\la_1$ to $\la_2$; in doing so one performs a (temperature scaled) amount of work (which we will simply refer to as work for the rest of this thesis)  $\delta W_{BA}(t_1) = \beta \{\ecal_{\la_2} (\avvec(1)) - \ecal_{\la_1}(\avvec(1))\}$ on the system. The system is then equilibrated via the sampling distribution $\pi_{\la_2}$ until a time $t_2$ yielding the configuration $\avvec(2)$. At this point one increments \la\ from $\la_2$ to $\la_3$ so as to yield the work increment $\delta W_{BA}(t_2) = \beta \{\ecal_{\la_3} (\avvec(2)) - \ecal_{\la_2}(\avvec(2))\}$. One continues this process until the value of the field parameter has reached its terminal value, $\la_n$, at time $t_{n-1}$.

 As a result one obtains the path:

\begin{eqnarray}
t & \rightarrow & t_1, t_2, ..., t_{n-1}\nonumber\\
\avvec & \rightarrow &\avvec(1), \avvec(2), ..., \avvec(n-1) \equiv \seta {\vvec} \nonumber \\
\la (t) &\rightarrow & \la_1, \la_2, ..., \la_n
\label{eq:path}
\end{eqnarray}\tshow{eq:path}

\noindent The net work for the $A\rightarrow B$ process is then given by:

\begin{equation}
\wba (\seta {\vvec}) = \s {i=1} {n-1} \delta \wba (t_i) = \beta \s {i=1} {n-1} [\ecal_{\la_{i+1}} (\avvec({i})) - \ecal_{\la_{i}}(\avvec({i}))]
\label{eq:totalwork}
\end{equation}\tshow{eq:totalwork}

\noindent In the case where one employs the linear parameterisation as given in Eq. \ref{eq:linearhamiltonian}, Eq. \ref{eq:totalwork} may be written as:

\begin{equation}
\wba = \s {i=1} {n-1} \triangle \la_i \mba (\vvec(i))
\label{eq:linwork}
\end{equation}\tshow{eq:linwork}

\noindent where:

\begin{equation}
\triangle \la_i = \la_{i+1} - \la_i
\end{equation}

\noindent A non-equilibrium process taking the simulation from phase B to phase A may be similarly defined. For the sake of notational convenience it is instructive to think of the $B\rightarrow A$ process as an $A\rightarrow B$ process, which is performed backwards in time (that is the sequence of events is reversed \cite{p:crookspre}, \cite{p:crookspre2}), and with the simple modification that the initial configuration, now corresponding to $\vvec (n-1)$ at time $t_{n-1}$, is sampled from the distribution $\pi_{\la_n} = \pib$. That is the path is constructed  as follows. At time $t_{n-1}$ one decrements \la\ from its initial value of $\la_n$ to $\la_{n-1}$, thus performing an amount of  work $\delta W_{AB}(t_{n-1}) = \beta [\ecal_{\la_{n-1}}(\vvec(n-1)) - \ecal_{\la_{n}}(\vvec(n-1))$ . One then proceeds to equilibrate the system so as to obtain the configuration $\vvec(n-2)$ at the time $t_{n-2}$, at which point \la\ is further decremented from $\la_{n-1}$ to $\la_{n-2}$, thus incurring a work $\delta W_{AB}(t_{n-2}) = \beta [\ecal_{\la_{n-2}}(\vvec(n-2)) - \ecal_{\la_{n-1}}(\vvec(n-2))$. This procedure is repeated iteratively until time $t_1$ at which point \la\ is decremented from $\la_2$ to $\la_1$. It is clear from this that the (temperature reduced) work increment performed at $t_i$, when \la\ is changed from $\la_i$ to $\la_{i+1}$, is given by:

\begin{equation}
\delta W_{AB} (t_i) =\beta [\ecal_{\la_i}(\vvec_i) - \ecal_{\la_{i+1}}(\vvec_i)]
\end{equation}

\noindent The net work for the resulting path:

\begin{eqnarray}
t & \rightarrow & t_{n-1}, t_{n-2}, ..., t_{1}\nonumber\\
\avvec & \rightarrow &\avvec(n-1), \avvec(n-2), ..., \avvec(1) \equiv \seta {\vvec} \nonumber \\
\la (t) &\rightarrow & \la_n, \la_{n-1}, ..., \la_1
\end{eqnarray}

\noindent is given by:

\begin{eqnarray}
W_{AB} (\seta {\vvec}) & = & \s {i=1} {n-1} \delta W_{AB} (t_i)\noindent\\
& = & \beta \s {i=1} {n-1} [\ecal_{\la_i}(\vvec_i) -\ecal_{\la_{i+1}}(\vvec_i)]\nonumber\\
& = & -\beta \s {i=1} {n-1} [ \ecal_{\la_{i+1}}(\vvec_i) - \ecal_{\la_i}(\vvec_i)]\nonumber\\
& = & -\beta \wba (\seta {\vvec})
\end{eqnarray}

\noindent Suppose now that $P(\wba | \pi^c_A)$ corresponds to the probability of obtaining $\wba$ for the $A\rightarrow B$ process and suppose that $P(\wba | \pi^c_B)$ \cite{note:whywrite} is the corresponding quantity  for the $B \rightarrow A $ process (hereafter whenever we mention $\pi^c_\al$ in the context of the FG method we will in fact be referring to the $\al\rightarrow\alp$ process, in which the initial distribution of the configurations is given by $\pi^c_\al$). The fluctuation theorem, which we have also proved in appendix \ref{app:fgproof},   asserts that \cite{p:crooks}\nocite{p:crookspre}-\cite{p:crookspre2}:

\begin{equation}
\z_B P(\wba | \pi_B^c)=\z_A e^{-\wba} P(\wba | \pi_A^c)
\label{eq:crooksfluct0}
\end{equation}\tshow{eq:crooksfluct0}

\noindent  or:

\begin{equation}
\RBAcal = \frac {e^{-\wba} P(\wba | \pi_A^c)}{P(\wba | \pi_B^c) } 
\label{eq:crooksfluct}
\end{equation}\tshow{eq:crooksfluct}

\noindent In the special case of zero equilibration (which is {\em equivalent} to changing  \la\ directly from $\la_1$ to $\la_n$ in a single step):

\begin{eqnarray}
\wba = \delta W_{BA}(t_1) & =& \beta \{\ecal_{\la_n}(\avvec(1)) - \ecal_{\la_1}(\avvec(1))\}\nonumber\\
& = & M_{BA} (\vvec(1))
\label{eq:instantwork}
\end{eqnarray}\tshow{eq:instantwork}

\noindent  Eq. \ref{eq:crooksfluct} reduces to the overlap identity Eq. \ref{eq:overlapidentity}. For this reason we will refer to all the methods based on the identity Eq. \ref{eq:overlapidentity} as the zero equilibration, or elementary, methods.
Since Eq. \ref{eq:crooksfluct} is simply a generalisation of Eq. \ref{eq:overlapidentity} we will also refer to this formula as the overlap identity. It immediately follows that we may generalise Eq. \ref{eq:ep} to:

\begin{equation}
\RBAcal = <e^{-\wba}>_{\pi^c_A}
\label{eq:fgep}
\end{equation}\tshow{eq:fgep}

\noindent and Eq. \ref{eq:ar} to:

\begin{equation}
\RBAcal = \frac {<A(\wba)>_{\pi_A^c}} {<A(-\wba)>_{\pi_B^c}}
\label{eq:FGAR}
\end{equation}\tshow{eq:FGAR}

\noindent Eq. \ref{eq:dual} may be replaced by the more general 'dual-phase' (DP) formula \cite{note:whaov}:

\begin{equation}
\RBAcal  =  \frac {<G(\wba)e^{-\wba}>_{\pi^c_A}} {< G(\wba)>_{\pi^c_B}}
\label{eq:freebennett}
\end{equation}\tshow{eq:freebennett}

\noindent Generally $P(\wba | \pi^c_A)$ and $P(\wba | \pi^c_B)$ will also face an overlap problem in the sense described in section \ref{sec:overlap}. However the FG method does have a way of getting around this; we will postpone our discussion of this point until chapter \ref{chap:sampstrat}.

\subsection{\label{sec:pfg}Path Sampling Fast Growth Methods}
\tshow{sec:pfg}

All the simulations mentioned until now have been discussed in the context of the  sampling of individual configurations \vvec . Let us now consider a simulation which, instead of jumping between configurations, jumps between paths \seta {\vvec} (such as those produced in the FG method, section \ref{sec:fg}). Such a simulation can be though of as comprising of a two fold procedure. In the first stage the simulation generates a path \seta {\vvec}. In the second stage a decision is made whether to accept or reject the new path. The idea behind the path sampling formulation of the FG method  \cite{p:sun} is to express it in terms of the notion of the sampling of paths, in the way that has just been described.

Suppose that ${\cal{P}}^c_{\al\rightarrow \alp} (\seta {\vvec})$ denotes the underlying distribution of the paths produced in the $\al \rightarrow \alp$ FG process as described in section \ref{sec:fg}. It was proved in appendix \ref{app:fgproof} that \cite{p:crookspre}, \cite{p:crookspre2}:

\begin{equation}
\frac {{\cal P}^c_{A\rightarrow B} (\seta {\vvec})}{{\cal P}^c_{B\rightarrow A} (\seta {\vvec})} = \frac {Z_B} {Z_A} \exp \{ W_{BA} (\seta {\vvec})\}
\label{eq:fundafluo}
\end{equation}\tshow{eq:fundafluo}

\noindent We will now show that the FG method described in section \ref{sec:fg} may be interpreted as  a path sampling simulation in which paths are {\em generated} according to ${\cal{P}}^c_{\al\rightarrow \alp} (\seta {\vvec})$ in the $\al \rightarrow \alp$ process, and subsequently {\em accepted} with probability 1. To be more specific, suppose that the current state of the simulation is  described by the path \seta {\vvec} and suppose that \seta {\tilde {\vvec}} corresponds to the path that has just been generated. Then it is clear that if one is to obtain  a set of paths distributed according to ${\cal{P}}^c_{A\rightarrow B}(\seta {\vvec})$ then  the acceptance probability of moving from the path \seta {\vvec} to \seta {\tilde {\vvec}} (in the $A\rightarrow B$ process) is given by (see Eq. \ref{eq:acceptancerate}):

\begin{equation}
P_a(\seta{\vvec}\rightarrow \seta{\tilde{\vvec}} | \pia )   =  \mbox{Min} \seta {1, \frac {P_G(\seta{\vvec} |\seta{\tilde{\vvec}} ) {\cal{P}}^c_{A\rightarrow B}(\seta {\tilde{\vvec}})} {P_G(\seta{\tilde{\vvec}}|\seta{\vvec}) {\cal{P}}^c_{A\rightarrow B}(\seta {\vvec})}}
\label{eq:bigacc}
\end{equation}\tshow{eq:bigacc}

\noindent where $P_G(\seta{\tilde{\vvec}} | \seta {\vvec})$ denotes the probability of the simulation generating a path $\seta{\tilde{\vvec}}$ given that the previous path was $\seta {\vvec}$. In the procedure described in section \ref{sec:fg} the path $\seta{\tilde{\vvec}}$ is constructed from the initial  configuration $\tilde{\vvec}(1)$, which itself is obtained from ${\vvec}(1)$, the initial configuration of $\seta {\vvec}$, by equilibrating the system for a fixed amount of time via \pia .
Using the notation of appendix \ref{app:fgproof} it is clear that $P_G(\seta{\tilde{\vvec}} | \seta {\vvec})$ is given (for the $A\rightarrow B$ process) by:

\begin{equation}
P_G(\seta {\tilde{\vvec}} | \seta {\vvec}) = P_S(\vvec(1) \rightarrow \tilde{\vvec}(1) | \pia)P_{A\rightarrow B} (\seta {\tilde{\vvec}} | \tilde{\vvec}(1))
\label{eq:gensin}
\end{equation}\tshow{eq:gensin}

\noindent where $P_S(\vvec(1) \rightarrow \tilde{\vvec}(1) | \pia)$ denotes the probability of making a transition from the configuration $\vvec(1)$ to the configuration $\tilde{\vvec}(1)$ when sampling from \pia\ for a fixed amount of time and where $P_{A\rightarrow B} (\seta {\tilde{\vvec}} | \tilde{\vvec}(1))$ denotes the probability of obtaining a path $\seta {\tilde{\vvec}}$ via the $A\rightarrow B$ FG process of section \ref{sec:fg}, given an initial configuration of $\tilde{\vvec}(1)$.  Since from Eq. \ref{eq:startfg} we know that:

\begin{equation}
\frac {P_S(\vvec(1) \rightarrow \tilde{\vvec}(1) | \pia)} {P_S(\tilde{\vvec}(1) \rightarrow {\vvec}(1) | \pia)} = \frac {\pia (\tilde{\vvec}(1))} {\pia ({\vvec}(1))}
\end{equation}

\noindent it immediately follows that (for the $A\rightarrow B$ process):

\begin{eqnarray}
\frac {P_G(\seta {\tilde{\vvec}} | \seta {\vvec})} {P_G(\seta {{\vvec}} | \seta {\tilde{\vvec}})} & = & \frac {P_S(\vvec(1) \rightarrow \tilde{\vvec}(1) | \pia)} {P_S(\tilde{\vvec}(1) \rightarrow {\vvec}(1) | \pia)} \frac {P_{A\rightarrow B} (\seta {\tilde{\vvec}} | \tilde{\vvec}(1))} {P_{A\rightarrow B} (\seta {\vvec} | \vvec(1))}\nonumber\\
& = & \frac {\pia (\tilde{\vvec}(1))} {\pia ({\vvec}(1))} \frac {P_{A\rightarrow B} (\seta {\tilde{\vvec}} | \tilde{\vvec}(1))} {P_{A\rightarrow B} (\seta {\vvec} | \vvec(1))}\nonumber\\
& = & \frac {{\cal P}^c_{A\rightarrow B}(\seta {\tilde {\vvec}})} {{\cal P}^c_{A\rightarrow B}(\seta { {\vvec}})}
\label{eq:generation}
\end{eqnarray}\tshow{eq:generation}

\noindent where we have use the fact that:

\begin{equation}
{\cal{P}}^c_{A\rightarrow B} (\seta {\vvec}) = \pi^c_A(\vvec(1)) P_{A\rightarrow B} (\seta {\vvec} | \vvec(1))
\end{equation}

\noindent Eq. \ref{eq:bigacc} then becomes:

\begin{equation}
P_a(\seta{\vvec}\rightarrow \seta{\tilde{\vvec}} | \pia ) = 1
\label{eq:fgacc2}
\end{equation}\tshow{eq:fgacc2}

\noindent Therefore we  infer that the FG method described in section \ref{sec:fg} may be thought of as a path sampling experiment in which the paths  are generated according to the mechanism described in section \ref{sec:fg} and in which moves between old and new paths are {\em accepted with probability 1}. 

The benefit of the path sampling interpretation of the FG method is that it allows us to generalise the PS method (Eq. \ref{eq:PSsamp}) so as to be applicable within the framework of the FG method. To see this we recall that the canonical PS sampling distribution $\pips$ (Eq. \ref{eq:PSsamp}) realises the \mba\ distribution given in Eq. \ref{eq:mbadist}.  It is not hard to show that the path sampling distribution:

\begin{equation}
{\cal{P}}^c_{PS}(\seta {\vvec}, \al) \equaldot \z_\al {\cal{P}}^c_{\al\rightarrow \alp} (\seta {\vvec})
\label{eq:pfgps}
\end{equation}\tshow{eq:pfgps}

\noindent in which $\al$ is a stochastically sampled variable, realises the following \wba\ distribution:

\begin{equation}
P(\wba | \pi^c_{PS}) = \frac{\z_A P(\wba | \pia)  +  \z_B P(\wba | \pib)} {\z_A + \z_B}
\label{eq:fgpst}
\end{equation}\tshow{eq:fgpst}

\noindent where we have used $P(\wba | \pi^c_{PS})$ to denote $P(\wba | {\cal{P}}^c_{PS})$ in order bring out the links with the zero equilibration cases.  Eq. \ref{eq:pfgps} is essentially the path sampling generalisation of Eq. \ref{eq:PSsamp}. The implementation of Eq. \ref{eq:pfgps} involves the employment of an additional Monte Carlo move which allows one to {\em switch between processes}. That is suppose that  $\zeta_{A\rightarrow B}$ labels the $A\rightarrow B$ FG process and  $\zeta_{B\rightarrow A}$ labels the $B\rightarrow A$ FG process, in which paths are generated according to Eq. \ref{eq:generation}. Then in this notation $P(\wba | \pi^c_\al)\equiv P(\wba | \zeta_{\al\rightarrow\alp})$.
As with the original FG method,  the FG phase-switch (FG-PS)  procedure involves (for the $\zeta_{\al\rightarrow \alp}$ process) generating paths in the manner described by Eq. \ref{eq:gensin} and subsequently accepting with probability 1 (see Eq. \ref{eq:fgacc2}). On top of this one introduces an additional Monte Carlo move in which one attempts to switch between the $\zeta_{\al\rightarrow \alp}$ and $\zeta_{\alp\rightarrow \al}$ processes whilst leaving the path $\seta{\vvec}$ unperturbed; the acceptance probability for such a move is given by:

\begin{eqnarray}
P_a(\zeta_{A\rightarrow B} \rightarrow \zeta_{B\rightarrow A}) & = & \mbox{Min} \{ 1,  \frac {\z_B {\cal{P}}^c_{B\rightarrow A}(\seta {\vvec})} {\z_A {\cal{P}}^c_{A\rightarrow B}(\seta {\vvec})}\}\nonumber\\
& = &  \mbox{Min} \{ 1, e^{-\wba}\}
\label{eq:sttops}
\end{eqnarray}\tshow{eq:sttops}

\noindent where we have appealed to Eq. \ref{eq:fundaflu}. In the case where $P(\wba | \pi^c_A)$ and $P(\wba | \pi^c_B)$ partially overlap about the $\wba \sim 0$ regions, the FG-PS method allows one to sample {\em all} the paths that contribute non-negligibly to the estimator of the FED. One may then proceed to estimate \RBAcal\  via:

\begin{equation}
\RBAcal = \frac {<f(\wba)>_{\pi^c_{PS}}} {<f(-\wba)>_{\pi^c_{PS}}}
\end{equation}

\subsection{Looking Forward}

We are now in a position to explain more fully the scope of the study presented here. We have seen that the key obstacle to the problem of determining FEDs is the overlap problem. To deal with this problem one must:

\begin{enumerate}
\item Choose an appropriate global displacement $\avec {D}$  in order to map configurations of one phase onto the other (see Eq. \ref{eq:switchtemp}, figure \ref{phase_in_realspace_withswitch}) and choose a representation \vvec\ (see Eq. \ref{eq:vtoutran})
\item Choose an estimator \cite{p:bennett}.
\item Employ some form of extended sampling strategy \cite{p:iba}.
\end{enumerate}

\noindent In succeeding chapters we take up each of these points in turn. In chapter \ref{chap:tune} we will deal with the first point and will show how the overlap depends on the representation one chooses to work in. Specifically by working in a representation in which the effective configuration \avvec\ corresponds to the normal modes of a crystalline solid we construct a transformation (called the fourier space mapping, FSM) which, under certain conditions, is more efficient than the RSM (see Eq. \ref{eq:RSS}).

In chapter \ref{chap:estsamp} we provide some insight into the second point and show how in the case of partial overlap, which is when there are \wba\ macrostates over which both the estimators $\hat{P}(\wba | \pi^c_A)$ and $\hat{P}(\wba | \pi^c_B)$ are non-zero, and when there are other macrostates for which only one of the estimators is non-zero, then the estimator of  Eq. \ref{eq:freebennett} can be {\em guaranteed} to work (in the sense that the estimate of \RBAcal\ is free of systematic errors) for any $G(\mba)$ simply by restricting the regions of \mba\ space from which the {\em non-negligible} contributions to the expectations come. Furthermore we will also show that there is a family of estimators of the form of Eq. \ref{eq:freebennett} for which no such restrictions are required since {\em by construction} these estimators are guaranteed to be free of systematic errors, provided that there exists some overlap between $\hat{P}(W_{BA} | \pi^c_A)$ and $\hat{P}(W_{BA} | \pi^c_B)$.

The third point is addressed in chapter \ref{chap:sampstrat}. We start off by applying the basic theoretical techniques of umbrella sampling \cite{p:iba}, \cite{p:torrievalleau}, \cite{p:bergPRL} to the problem of phase behaviour, and use the recently developed Wang-Landau \cite{p:wanglandauprl} technique to construct the MUCA weight function needed in order to estimate the FED (see section \ref{sec:umbrella}).  In addition to this we present a new method of overcoming the overlap problem (called the Multihamiltonian (MH) method) which involves simulating several independent sampling distributions simultaneously. Like the WHAM method, the benefit of this method is that it is highly parallelizable. We then proceed to make an investigation of the FG method, and show how it is able to effectively overcome the overlap problem.

Having formulated the FED problem in a strictly classical framework we proceed in chapter \ref{chap:quantum} to consider the quantum formulation as provided by the path integral formalism. The quantum FED problem is even more computationally intensive than its classical counterpart, and we show that the MH method developed in chapter \ref{chap:sampstrat} provides an efficient way of estimating the  relevant quantities. In particular we illustrate its  power with a study of the role of zero-point motion in determining crystal stability.

\chapter{\label{chap:tune}Tuning the Representations}
\tshow{chap:tune}

\section{\label{sec:tuneintro}Introduction}
\tshow{sec:tuneintro}

We saw in sections \ref{sec:statmech} and \ref{sec:samplingstrat} that different phases of a given material may be thought of as corresponding to different basins of attraction of the  configurational energy $E(\rvec)$. For finite-system-constructs of the relevant phases, there may exist several metastable basins of attractions (corresponding to the different phases) and it is ones desire to find the most probable one. In the thermodynamic limit this basin of attraction becomes overwhelmingly more probable than the others, resulting in the corresponding phase being the one that is found in nature. In this thesis we will focus on the case where there are two candidate phases, and our task will be limited to that of determining the more stable out of the two \cite{note:freecomplex}.

Computationally the task of finding the more probable out of the two phases involves implementing a Metropolis simulation in which one visits these two regions in a single simulation. One may then estimate the FED via Eq. \ref{eq:partrsestimator0} and Eq. \ref{eq:rbafree}. However the sequential or pathwise nature of the Metropolis method and the presence of a region of (absolute) configuration space, in between the two phases,  of intrinsically low probability means that a simulation initiated in a given phase will remain trapped in that phase. 

If instead of working  in the \rvec\ representation (in which one attempts to estimate the quantity $R_{BA}$ in Eq. \ref{eq:partrsestimator0}) one chooses to work in the \vvec\ representation (in which one attempts to estimate the quantity \RBAcal\ in Eq. \ref{eq:RBA}), one greatly alleviates the difficulty associated with the problem of estimating the FED of the two phases by bypassing this intermediate region altogether (compare figure \ref{fig:representations} (a) and (b)). The residual difficulty that is left in the associated problem is captured  in the amount by which the two phases overlap in \vvec\ space, and it is this difficulty which must be overcome in order to estimate \RBAcal . 

In order to gauge the amount of overlap between these  two regions, one must (by virtue of the overlap identity Eq. \ref{eq:overlapidentity}) observe the  amount of overlap that is present between $P(\mba | \pi^c_A)$ and $P(\mba | \pi^c_B)$.
This overlap in \mba\ space is clearly dependent upon two factors:

\begin{enumerate}
\item[A] The choice of the reference configurations $\Rvec_\gamma$ (see Eq. \ref{eq:rintermsofv}). Different choices of $\Rvec_A$ and $\Rvec_B$ correspond to different choices of $\Dvec$ in Eq. \ref{eq:DLS} (see also Eq. \ref{eq:switchtemp}, Eq. \ref{eq:absoluteswitch}, figure \ref{phase_in_realspace_withswitch} ).
\item[B] The choice of representation \vvec\ \cite{note:bothimpo}, \cite{note:choosingperfect}, which \mba(\vvec) depends on (Eq. \ref{eq:op}).
\end{enumerate}

\noindent The ideal choices of the global translation $\Dvec$ and the representation \vvec\  are ones for which $P(\mba | \pi^c_A)$ and $P(\mba | \pi^c_B)$ (see figure \ref{bimodaldistribution}) collapse onto the distribution:

\begin{equation}
P(M_{\alp\al} | \pi^c_\al) = \delta (M_{\alp\al} + \ln R_{\alp\al})
\label{eq:perfect}
\end{equation}\tshow{eq:perfect}

\noindent Departures from this ideal limit manifest in the  bi-modality of $P(M_{\alp\al} | \pi^c_\al)$, as shown in figure \ref{bimodaldistribution}.
 The more efficient the choice of \Dvec\ and representation \vvec , the closer the distributions $P(\mba | \pi^c_A)$ and $P(\mba | \pi^c_B)$ will be to the ideal limit (Eq. \ref{eq:perfect}).

The general challenge to the problem of estimating the FED is that of tuning the PM so as to maximise the overlap between the two regions of effective \cs\ (see figure \ref{fig:representations} (b)). In this chapter we will investigate this issue in the particular context of the two phases being crystalline solids. We will primarily focus on the role of the representation (issue B). Specifically we will see that a PM as formulated in a Fourier space (normal mode) representation provides some strategic advantages over the real-space representations (RSM, Eq. \ref{eq:RSS}) utilised in previous studies  \cite{p:LSMC}, \cite{p:LSMCprehard}\nocite{p:LSMCpresoft}\nocite{p:solidliquid}\nocite{p:LSMCerrington}-\cite{p:voter}.

In order to set the context we will now introduce the model system (the Lennard Jones solid) which has been employed in all the simulations carried out in this work. This will be followed by a section illustrating how the PM is implemented for our model systems, followed by a brief discussion of the role of the global translation vector (issue A) in the mappings between the two phases.

\section{\label{sec:system}The model system}
\tshow{sec:system}

For the work in this thesis our model system will be comprised of particles interacting via the pairwise Lennard-Jones (LJ) interatomic potential:

\begin{equation}
\phi(r_{ij}) = 4\epsilon[(\f {\sigma} {r_{ij}})^{12} - (\f {\sigma} {r_{ij}})^{6}]
\label{lennard}
\end{equation}\tshow{lennard}

\noindent where $\phi(r_{ij})$ corresponds to the interaction energy between particles i and j separated by a distance $r_{ij}= | {\vec {r}}_i - {\vec {r}}_j|$, and where $\vec {r}_i$ and $\vec {r}_j$ are the position vectors of particles i and j respectively.  The overall configurational energy $E(\rvec)$ is then given by:

\begin{equation}
E(\rvec) = \h \s {i=1} {N} \s {j\neq i} {N}  \phi(r_{ij})
\label{eq:fullce}
\end{equation}\tshow{eq:fullce}

\begin{figure}[tbp]
\begin{center}
\rotatebox{270}{
\includegraphics[scale=0.5]{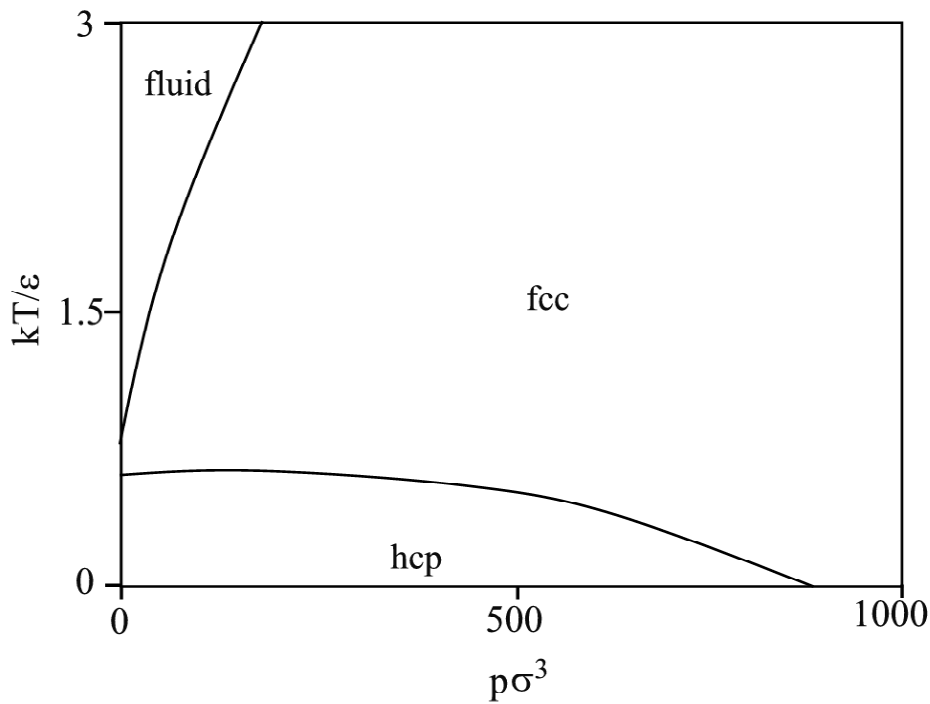}
}
\end{center}
\caption{\textbf{Schematic of classical phase diagram for the LJ potential}}

The figure shows a (scaled temperature versus scaled pressure) schematic for the classical phase diagram of the LJ potential in Eq. \ref{lennard}. In order to determine the hcp, fcc phase boundary, one must determine the more probable of the two phases. The methods utilised and developed in this thesis are able to address this sort of problem. The methods can, without much difficulty, be generalised to the case of the solid-liquid boundary. An initial line of investigation into this has been made in \cite{p:solidliquid}, \cite{p:LSMCerrington}. (See also \cite{note:otherphases}).

\begin{center}
{\bf{------------------------------------------}}
\end{center}

\tshow{fig:lennardcl}
\label{fig:lennardcl}
\end{figure}

\noindent Generally the use of the full configurational energy in Eq. \ref{eq:fullce} in a simulation is prohibitively expensive, and one instead employs some form of approximation whereby the potential is truncated at some distance from the particle \cite{note:truncation}. The ensuing phase diagram (a schematic of which is shown in figure \ref{fig:lennardcl}) is highly sensitive to this truncation radius; the latter has to be chosen carefully if one is to reproduce the true characteristics of the phase under consideration. By analysing the fluctuations in the ground state energies and the harmonic free energy differences as a function of the truncation radius, it was found in \cite{b:andythesis} (where identical systems were employed to those  used in this work) that a truncation radius $r_c$ given by:

\begin{equation}
r_c = 1.5 r_{nn}
\end{equation}

\noindent yielded sufficiently accurate results (where $r_{nn}$ is the nearest neighbour distance). A truncation radius of this magnitude essentially amounts to each particle interacting with both its first nearest neighbour shell (comprising of 12 particles) and its second nearest neighbour shell (comprising of 6 particles) \cite{note:truncationpic}. This truncation radius was also employed for all the simulations used in this thesis, with the exception of those in chapter \ref{chap:quantum} (in which a truncation radius of $r_c = 1.1 r_{nn}$ was employed).

Unless otherwise stated, the system size that we have employed is $N=216$, and the densities are  $\rho \sigma^3 = 1.092$. We will also quote all results in terms of the reduced temperature:

\begin{equation}
T^* = \frac {k T} {\epsilon}
\label{eq:redtemp}
\end{equation}\tshow{eq:redtemp}

\noindent We also make a point here that it is our intention, in this thesis, only to use the LJ system as a testbed for the various methods and not to give definitive results for the LJ phase diagram (which has already been done in \cite{b:andythesis}).

\section{\label{sec:pm}The Phase Mapping for crystalline solids}
\tshow{sec:pm}

Let us now focus out attention on the regions of the phase diagram near the fcc-hcp boundary (see figure \ref{fig:lennardcl}), and let phase label A refer to the fcc structure and phase label B to the hcp structure. In the case of crystalline solids, the reference configuration ($\Rvec_\al$) may most conveniently be chosen to be the ground state configuration (i.e. the lattice sites themselves). The full PM then involves a switch of lattice vectors, accompanied by the mapping of the displacements (of the particles from their lattice sites) of one phase onto the (possibly modified) displacements of particles of the other phase. In the case of the RSM, these displacements are {\em unmodified} on the transition of the phases.

\begin{figure}[tbp]
\begin{center}
\ifpdf
\rotatebox{0}{
\includegraphics[scale=0.6]{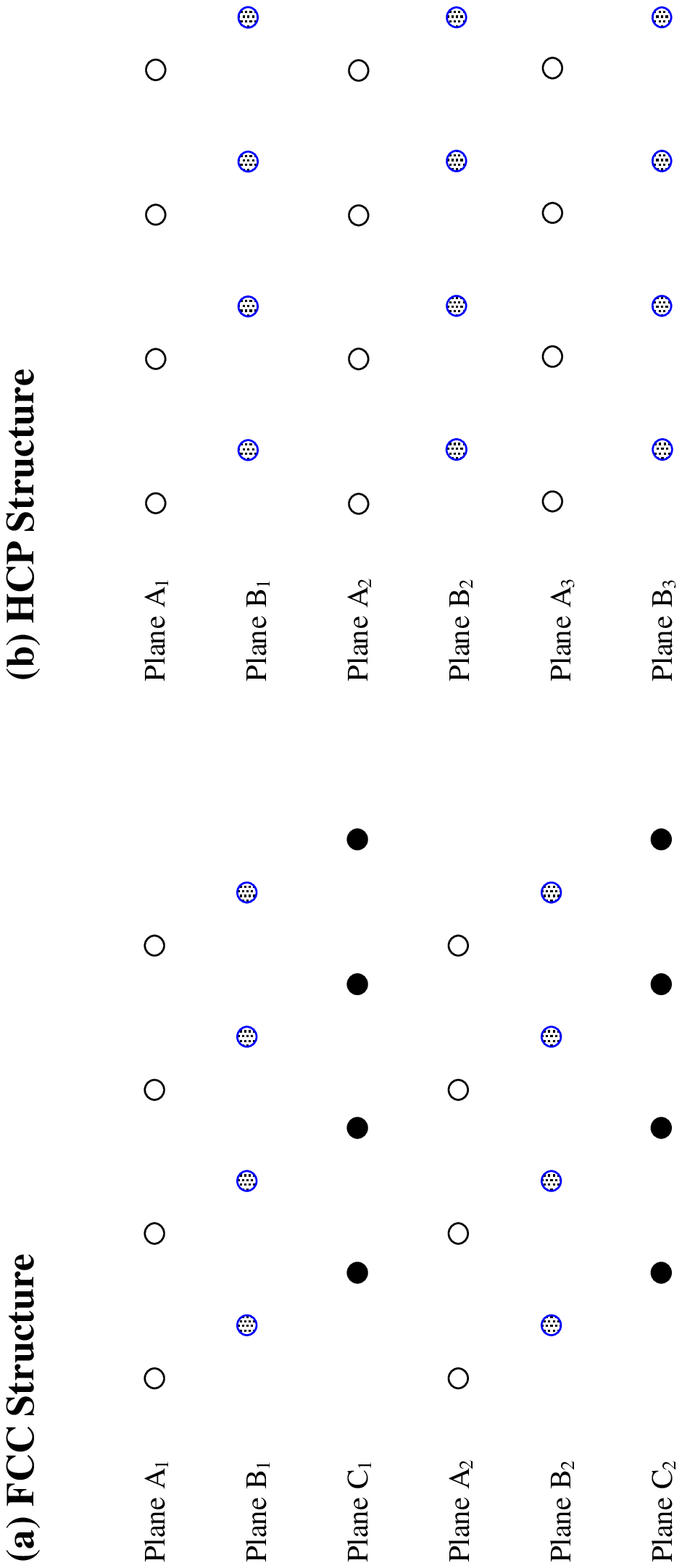}
}
\else
\rotatebox{270}{
\includegraphics[scale=0.6]{fig/fcchcp}
}
\fi
\end{center}
\caption{\textbf{The fcc \& hcp structures}}

The fcc structure may be thought of as comprising of three families of close packed planes, labelled as A, B, and C in figure (a).

The hcp structure, on the other hand, can be represented in terms of the two families of close packed planes (stacked in the ABAB... formation, see figure (b))

\begin{center}
{\bf{------------------------------------------}}
\end{center}

\tshow{fig:hcp}
\label{fig:hcp}
\end{figure}

\begin{figure}[tbp]
\begin{center}
\includegraphics[scale=0.6]{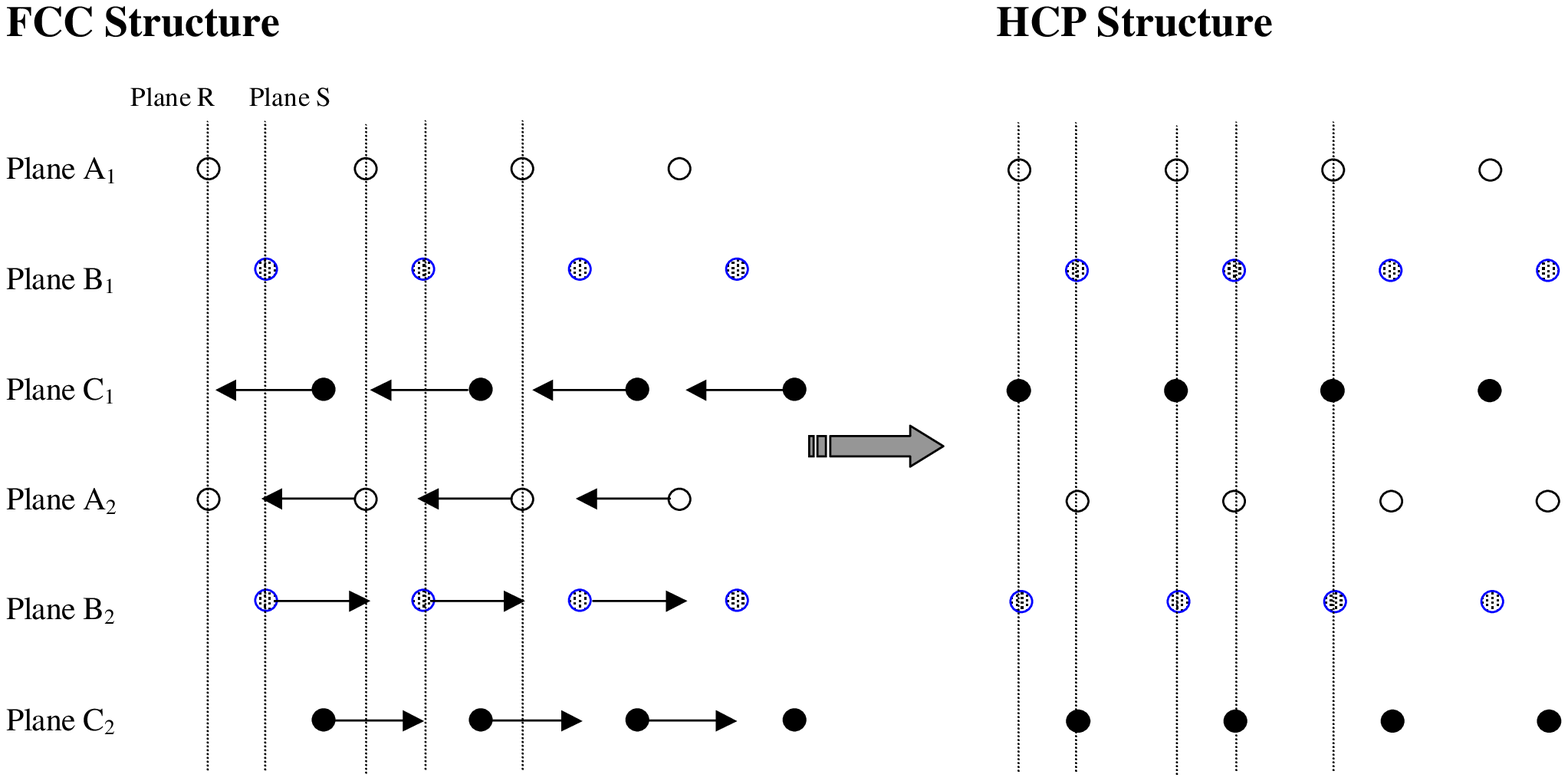}
\end{center}
\caption{\textbf{The PM Transformation}}

The figure showing the fcc structure on the left (characterised by 3 different planes, stacked in a 'ABCABC...' formation) being transformed to the hcp structure (characterised by 2 planes, stacked in the formation 'ABAB...'. See also \fig\  \ref{fig:hcp}). The atoms of plane $C_1$ are made to lie beneath those of plane $A_1$, the atoms of plane $A_2$ underneath those of $B_1$, the atoms of $B_2$ under those of $A_1$, and the atoms of $C_2$ underneath those of plane $B_1$. Note that periodic boundary conditions apply. A consequence of these boundary conditions is that one could, in principle, slide the planes a greater distance than is shown, so as to transform one phase into the other. However we have illustrated only one possibility, the one which is actually used in the simulations. See \cite{p:LSMCpresoft} and \cite{b:andythesis} for more details on the other possibilities.

\begin{center}
{\bf{------------------------------------------}}
\end{center}

\tshow{**latticeswitch}
\label{latticeswitch}
\end{figure}

For the fcc and hcp structures, one may identify families of planes which are common to the reference configurations of both the structures. Whereas the fcc structure may be thought of as being comprised of three families of close packed planes (see figure \ref{fig:hcp} (a)), the hcp structure may be thought of as comprising of two layers of close packed planes (see figure \ref{fig:hcp} (b)). The geometry of the planes are such that they  permit a simple mapping of the lattice structure of one phase onto that of the other (see figure \ref{latticeswitch}).  
In this case the operation of the RSM takes a particularly simple form; one merely slips the planes as shown in figure \ref{latticeswitch}, whilst at the same time preserving the relative positions of the particles within a given plane. For a more detailed illustration of these planes and their corresponding structures, we refer the reader to \cite{b:andythesis} and \cite{b:ashcroftmermin}

The choice that we have made for mapping the lattice sites of one structure onto those of the other does not exhaust the possibilities for $\Dvec$. In fact for a given labelling scheme of the particles, one may choose to map the lattice vector of particle i of structure \al\ onto that of particle j of structure \alp, instead of mapping it onto the same particle of the corresponding phase \cite{note:permlattice}.  This procedure may equivalently be thought of as a permutation of the index labelling the particles under the operation of the PM; there are of the order of N! such permutations. This point was investigated to a limited extent for the RSM in the case of hard spheres by Jackson et. al. \cite{p:LSMCpresoft}. In their work they investigated cases where the planes were displaced a distance greater than that shown in figure \ref{latticeswitch}, in transforming from one phase to the other. They also investigated the cases where the planes of the fcc structure were randomly stacked when forming the hcp structure, and also the case where the displacements of a particle of the fcc structure were mapped onto those of a randomly chosen particle of the hcp structure. All the alternative mappings resulted in greater number of hard spheres overlapping, as compared to the mapping presented in figure \ref{latticeswitch}. In this work we do not investigate this issue any further \cite{note:modemapping}.

\tshow{latticeswitch}

In the case of the solid-liquid phase boundary, an appropriate reference configuration for the liquid phase is simply any typical configuration of the fluid phase \cite{p:solidliquid}. Though we will not have anything more to say about this, we note that an investigation along this direction has been made (for the hard sphere case) in  \cite{p:solidliquid} and more recently (for the soft potential case) in \cite{p:LSMCerrington}.

\section{Formulation of the Fourier Space Mapping}

In section \ref{sec:tuneintro} we mentioned that there are two issues at hand, the first being the choice of a suitable reference configuration $\Rvec_\al$ and the second being the choice of an  appropriate representation \vvec. Given that the lattice sites themselves serve as both natural and convenient choices  for the  reference configurations, we will now concentrate our efforts on finding an optimal representation \vvec\ \cite{note:jarwork}. Specifically we will choose a representation in which \avvec\ corresponds to the normal modes (whose corresponding mapping Eq. \ref{eq:genswitch} we call the Fourier Space Mapping, FSM), and compare the overlap that one obtains in this case to  that of the RSM (Eq. \ref{eq:RSS}). We will show that in the harmonic limit the FSM ensures that the two phases have identical excitation energies, so as to ensure perfect overlap (in the sense of Eq. \ref{eq:perfect}) between the two phases in the effective \cs\ as parameterised by the coordinates \vvec . 
In contrast to this we will also show that for the RSM this overlap will never be perfect \cite{note:cruderep}.

\subsection{\label{sec:fsstran}Constructing the transformation}
\tshow{sec:fsstran}

To motivate the transformation, consider the Taylor expansion of the (excitation) configurational energy (Eq. \ref{eq:expandenergy}) in powers of the displacements \auvec\ with respect to the reference configuration $\aRvec_{\gamma}$ \cite{note:whytaylor}:

\begin{equation}
\ecal_\gamma (\auvec) =E(\Rvec_\al + \uvec) - E(\Rvec_\al) =\ecal^h_\gamma  (\auvec) + \ecal^a_\gamma (\auvec)
\label{eq:energyexp}
\end{equation}\tshow{eq:energyexp}

\noindent where the second term $\ecal^h_\gamma$ denotes the harmonic contributions (containing powers of second order  in the displacement \auvec) and $\ecal^a_\gamma$ denotes the anharmonic contributions (of at least third order in the displacement \auvec). The harmonic contributions may be written as:

\begin{equation}
\ecal^h_\gamma (\auvec) = \frac {1} {2} \auvec^T \avec {K}_\gamma \auvec
\label{eq:energy_har}
\end{equation}\tshow{eq:energy-har}

\noindent where $\avec {K}_\gamma$ is the $3N\times 3N$ dynamical matrix. If $\avec {K}_\gamma^{ij}$ denotes the entry in the i-th row and j-th column of the matrix  $\avec {K}_\gamma$ then:

\begin{equation}
\avec {K}_\gamma^{ij} = \frac {\partial \ecal_\al (\uvec)} {\partial u_i} \frac {\partial \ecal_\al (\uvec)} {\partial u_j} 
\label{eq:dynamatrix}
\end{equation}\tshow{eq:dynamatrix}

 \noindent Since $\avec {K}_\gamma$ is a symmetric matrix (i.e. it is Hermitian) we may (via the Gram-Schmidt orthogonalisation procedure if necessary) construct a set of orthonormal vectors \seta {\avec {e}^j_\gamma} which are the eigenvalue of $\avec {K}_\gamma$. In our case we will take $\avec {e}^j_\gamma$ to be the 3N column vector corresponding to the j-th eigenvector of $\avec {K}_\gamma$. If we set $\avec {T}_\gamma$ (Eq. \ref{eq:vtoutran}) such that:

\begin{equation}
\avec{T}^{ij}_\gamma = \frac {\aevec^{ij}_\gamma} {\sqrt {k^j_\gamma}}
\label{eq:FStran}
\end{equation}\tshow{eq:FStran}

\noindent or:

\begin{equation}
{\mf T}_\al = (\f {1} {\sqrt{k^1_\al}} {\mf e}^1_\al, \f {1} {\sqrt{k^2_\al}} {\mf e}^2_\al, ..., \f {1} {\sqrt{k^{3N}_\al}} {\mf e}^{3N}_\al)
\label{har_notran}
\end{equation}\tshow{har-notran}

\noindent where $\aevec^{ij}_\gamma$ is the i-th component of the j-th eigenvector of $\avec {K}_\gamma$ and where $k^i_\al$ is the eigenvector of ${\mf e}^i_\al$, then from Eq. \ref{eq:vtoutran}

\begin{equation}
\auvec = \s {m} {} v_m \frac{\aevec^{m}_\gamma} {\sqrt {k^m_\gamma}} 
\label{eq:uintermsofmodes}
\end{equation}\tshow{eq:uintermsofmodes}

\noindent The summation in Eq. \ref{eq:uintermsofmodes} is performed over the 3N components  \seta {v_m} (which we refer to as the fourier coordinates) of the column vector $\avvec$. Substituting Eq. \ref{eq:uintermsofmodes} into Eq. \ref{eq:energy_har} one finds that:

\begin{eqnarray}
\auvec ^T \avec {K}_\gamma \auvec & = & \s {i} {} \s {j} {}\s {m} {} \s {n} {} \frac {v_m \aevec^{im}_\gamma} {\sqrt {k_\gamma^m} }  \avec{K}_\gamma^{ij} \frac {v_n \aevec^{jn}_\gamma}  {\sqrt{k_\gamma^n}}\nonumber\\
& = & \s {i} {} \s {m} {} \s {n} {} \frac{v_m \aevec^{im}_\gamma } {\sqrt{k_\gamma^m}} . \sqrt {k_\gamma^n} \aevec^{in}_\gamma v_n\nonumber\\
& = &   \s {m} {} \s {n} {} v_m v_n \frac {\sqrt{k^n_\gamma}} {\sqrt{k^m_\gamma}}\delta_{mn} \nonumber\\
& = & \s {m} {} v^2_m
\end{eqnarray}

\noindent where we have invoked the orthonormality of the eigenvectors:

\begin{equation}
\s {i} {} \aevec^{im}_\gamma . \aevec^{in}_\gamma = \delta_{mn}
\end{equation}

\noindent In other words, by choosing an appropriate representation \avvec\ (which we call the fourier representation)  in which the normal modes (or fourier coordinates) of one crystalline solid are mapped onto those of the other, one may cast $\ecal^h_\gamma$ into a form which contains no phase  labels:

\begin{equation}
\ecal^h_\gamma (\avvec)= \h \vvec^T . \vvec = \frac {1} {2} \s {m} {} v^2_m
\label{eq:einv}
\end{equation}\tshow{eq:einv}

\noindent Using the fact that $[{{\mf {T}}_\gamma}^{-1}]^{ij} = \sqrt {k_\gamma^i} \aevec_\gamma^{ji}$ or

\begin{equation}
[{\mf T}_\al]^{-1} = 
\left( \begin{array}{c}
\sqrt{k_\al^1}[{\mf e}_\al^1]^T\\
\sqrt{k_\al^2}[{\mf e}_\al^2]^T\\
.\\
.\\
.
\end{array} \right)
\label{eq:fsinv}
\end{equation}\tshow{eq:fsinv}

\noindent we may use Eq. \ref{eq:stran} to write down the matrix elements of $\avec {S}_{BA}$:

\begin{equation}
{\avec {S}}^{ij}_{BA} = \s {m} {} \frac{\sqrt{k^m_A}} {\sqrt {k^m_B}}  \aevec^{im}_B  \aevec^{jm}_A  
\label{eq:fsstran}
\end{equation}\tshow{eq:fsstran}

\noindent This matrix has the associated determinant:

\begin{equation}
{\det \avec {S}}_{BA} = \frac {{\det \avec {T}}_B} {{\det \avec {T}}_A} = \p {m} {} \sqrt{\frac {k^m_A} {k^m_B}}
\label{eq:detS}
\end{equation}\tshow{eq:detS}

\noindent The expression in Eq. \ref{eq:FEDexpansion} for the FED may then be written as:

\begin{equation}
\triangle F_{BA} = \triangle E^0_{BA} + \triangle F^h_{BA} + \triangle F^a_{BA}
\label{eq:FED2}
\end{equation}\tshow{eq:FED2}

\noindent where 

\begin{equation}
{\triangle F}^h_{BA} = -\frac {1} {\beta} \ln \det \avec {S}_{BA} = \frac {1} {2\beta} \s {m} {} \ln (\frac {k^m_B} {k^m_A})
\label{eq:harfree}
\end{equation}\tshow{eq:harfree}

\noindent is the harmonic contribution to the overall FED (Eq. \ref{eq:FED2}). Since the harmonic contributions to the excitation energy (Eq. \ref{eq:energy_har}, \ref{eq:einv}) are equal if they share the same fourier coordinates, we see that only the anharmonic contributions to the energy will survive in the evaluation of  $\mba$:

\begin{eqnarray}
\mba (\avvec) & = & \beta[\ecal_B (\avvec) - \ecal_A(\avvec)]\nonumber\\
& = & \beta [\ecal^a_B(\avvec) - \ecal^a_A(\avvec)]
\label{eq:harM}
\end{eqnarray}\tshow{eq:harM}

\noindent As a result the third term in Eq. \ref{eq:FED2}:

\begin{equation}
\triangle F^a_{BA} = -\beta^{-1} \ln {\cal R}^a_{BA}
\label{eq:anharfree}
\end{equation}\tshow{eq:anharfree}

\noindent reflects the purely anharmonic contributions to the FED.

\begin{figure}[tbp]
\begin{center}
\includegraphics[scale=0.75]{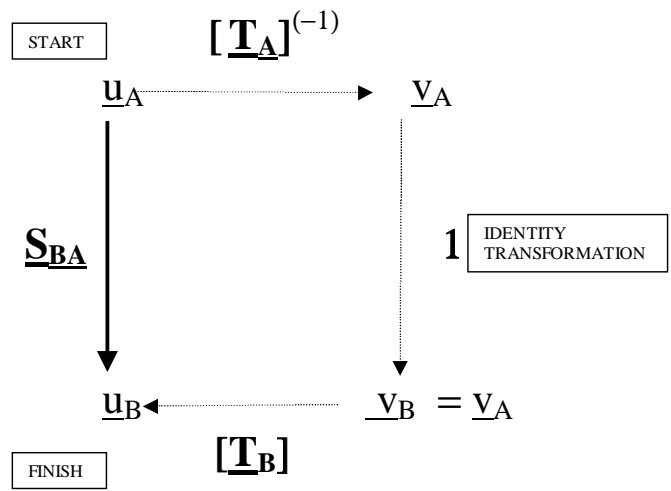}
\end{center}
\caption{\textbf{Diagrammatic Representation of the FSM}}
In order to transform (within the harmonic transformation) the configuration $\mf u_A$ (of phase A) onto a configuration $\mf u_B$ (of phase B) so that  both configurations are excited above their respective ground states by the same amount (see Eq. \ref{eq:energy_har}), we first transform from $\mf u_A$ to $\mf v_A$, which represents the configuration in what we call fourier space. We do this using the transformation $[\mf T_A]^{-1}$, given by \eq\ \ref{eq:fsinv}. We then force phase B to share the same set of fourier  coordinates as phase A by setting $\mf v_B$ equal to $\mf v_A$. We do this via the identity transformation. This ensures (by appeal to Eq. \ref{eq:einv}) that (within the harmonic approximation) the two phases have the same excitation energies above their respective ground states. Finally we transform back to real space via the transformation $[\mf T_B]$. The net transformation is given by ${\avec {S}}^{}_{BA}$ \cite{note:permutation}.

\begin{center}
{\bf{------------------------------------------}}
\end{center}

\label{concept_pic}
\end{figure}

One may realise the associated mapping, which we call the fourier space mapping (FSM),  within the framework of the \auvec\ representation through the operation in Eq. \ref{eq:genswitch} where $\avec {S}_{\tilde{\gamma}\gamma}$ is given by Eq. \ref{eq:fsstran}. Figure \ref{concept_pic} shows a schematic of the conceptual procedure involved in the mapping.

For systems with periodic boundary conditions, additional considerations must be taken into account in  constructing the FSM. In appendix \ref{app:technicality} we outline the necessary  modifications which must be incorporated into the transformation ${\avec {S}}_{BA}$ in order to accommodate these constraints. Generally what one finds is that the use of periodic boundary conditions means that three of the eigenvectors of the dynamical matrix $\avec {K}_\gamma$ (Eq. \ref{eq:dynamatrix}) will have zero eigenvalues. Suppose that  $\aevec^1_\gamma$, $\aevec^2_\gamma$, and $\aevec^3_\gamma$ correspond to these null eigenvectors. The findings of appendix \ref{app:technicality} are that one may simply omit these coordinates in the evaluation of the relevant quantities. For example the displacements \uvec\ are now given by:

\begin{equation}
\auvec = \s {m=4} {3N} v_m \frac{\aevec^{m}_\gamma} {\sqrt {k^m_\gamma}} 
\label{eq:temptune1}
\end{equation}\tshow{eq:temptune1}

\noindent and the transformation matrix ${\avec {S}}_{BA}$ (Eq. \ref{eq:fsstran}) is now replaced by:

\begin{equation}
{\avec {S}}^{ij}_{BA} =  \s {m=4} {3N} \sqrt{\frac {k^m_A}{k^m_B}} \aevec^{im}_B \aevec^{jm}_A 
\label{eq:temptune2}
\end{equation}\tshow{eq:temptune2}

\noindent We will assume that all subsequent summations over the fourier coordinates \seta {v_m} will be of the form of that employed in Eq. \ref{eq:temptune1} and Eq. \ref{eq:temptune2}, i.e. summations which exclude the null modes.

\tshow{concept-pic}
\newpage

\subsection{Summary}

Summarising, we have constructed a transformation $\avec {S}_{\tilde{\gamma}\gamma}$ (Eq. \ref{eq:fsstran}) called the fourier space mapping (FSM) which maps a configuration $\rvec_\gamma$ of phase \al\ onto a configuration $\rvec_\alp$ of phase \alp\ so as to ensure that  the two phases are of identical excitation energies in the harmonic limit, thus guaranteeing perfect overlap. As the harmonic contributions to the FED are already known exactly via Eq. \ref{eq:harfree}, the utility of this transformation will not lie in the overwhelmingly harmonic regime but will instead lie within the anharmonic regime. In particular, since the overlap can be arbitrarily improved simply by reducing the temperature ($\ov\rightarrow 1$ as $T\rightarrow 0$), one might expect that the problem of estimating the anharmonic contributions to the FED might become controllably small in the $T\rightarrow 0$ limit. We will see that this is not quite the case.

Before discussing the limitations of the FSM we will first focus on the efficiency with which it overcomes the overlap problem, and we will use the RSM for comparison.
In order to compare the efficiency with which the FSM and RSM tackle the overlap problem, we will, in the next section, investigate the issue of the overlap between the two \pc s of $\mba$ via analytic techniques. Specifically we will show how, in the harmonic limit, the overlap problem vanishes for the FSM whereas it tends to a constant value for the RSM (in the sense that $P(\mba | \pi^c_\gamma)$ assumes a stationary form), thereby serving as the most extreme illustration of the dependence of the overlap problem on the representation. We will also use these results to outline some of the basic limitations that the FSM faces in estimating the anharmonic FEDs

\section{\label{sec:limitingFSSRSS} Analytic results}
\tshow{sec:limitingFSSRSS}

 Since a probability distribution is completely characterised by its cumulants, one way to obtain insights into the behaviour of the overlap problem is to focus on the cumulants of $P(\mba | \pi^c_\gamma)$. Of particular importance are the first two cumulants, since it is these which correspond to the mean and variance of the distribution, and since it is these which will be most important in indicating the amount of overlap that will be present \cite{note:meanvarspread}. To define the cumulants let us expand \Ralcal\ (using Eq. \ref{eq:ep}) as the exponential of a power series in $M_{\alp\al}$ \cite{b:kendall}:

\begin{equation}
\Ralcal = <\exp \{ -\mal\}>_{\pi^c_\al} = \exp \{ \s {n=0} {\infty} (-1)^n \frac {\omega_n} {n!} \}
\label{eq:cumulant}
\end{equation}\tshow{eq:cumulant}

\noindent where $\omega_n$ is the n-th cumulant. The first three cumulants are then given by:

\begin{equation}
\omega_1 \equiv <\mal >_{\pi^c_\al}
\label{eq:firstcumulant}
\end{equation}\tshow{eq:firstcumulant}

\noindent and

\begin{equation}
\omega_2 \equiv <\mal^2>_{\pi^c_\al}  - <\mal>^2_{\pi^c_\al}
\label{eq:secondcumulant}
\end{equation}\tshow{eq:secondcumulant}

\noindent and:

\begin{equation}
\omega_3 \equiv 2 <\mal>^3_{\pi^c_\al} - 3 <\mal >_{\pi^c_\al} <\mal^2>_{\pi^c_\al} + <\mal^3 >_{\pi^c_\al}
\label{eq:thirdcumulant}
\end{equation}\tshow{eq:thirdcumulant}

\noindent The FED (Eq. \ref{eq:FEDexpansion}) may be cast into a simple form using these cumulants:

\begin{equation}
\triangle F_{BA} = \triangle E^0_{BA} - \beta^{-1} [\ln \det S_{BA} - \omega_1 +\frac {1} {2} \omega_2 + ...]
\label{eq:FEDcumu}
\end{equation}\tshow{eq:FEDcumu}

\noindent In the next two sections we will analyse the behaviour of these cumulants for the RSM and the FSM in the harmonic limit ($T\rightarrow 0$).

\subsection{\label{eq:fssanalytic}Fourier Space Mapping}
\tshow{eq:fssanalytic}

Drawing on anharmonic perturbation theory one may expand the \ce\ $\ecal_\al(\vvec)$ as a power series in $\vvec$ in which the contributions of successive orders become increasingly smaller. As the harmonic limit is approached one may discard all but the terms which scale, upon integration, with lowest order of T, thereby considerably simplifying the analysis of the cumulants. The results of the theory  (appendix \ref{app:fssperturb}) may be summarised as follows. In the limit $T\rightarrow 0$ (or $\beta \rightarrow \infty$), the cumulants of $P(M_{\alp\al} | \pi^c_\al)$ scale in the following way with temperature:

\begin{equation}
\lim_{\beta\rightarrow\infty} \omega_n \sim \left\{\begin{array}
{r@{\quad:\quad}l}
\beta^{-\frac {n} {2}} & \mbox{\ if n is even}\\
\beta^{-\frac {[n+1]} {2}} & \mbox{\ if n is odd}
\end{array}\right.
\label{eq:cs}
\end{equation}\tshow{eq:cs}

\noindent From Eq. \ref{eq:FEDcumu} and Eq. \ref{eq:cs} one observes that in the limit $T\rightarrow 0$ one may write a cumulant approximation expression for the anharmonic contributions to the FED as:

\begin{equation}
\lim_{\beta\rightarrow\infty} \triangle F^a_{BA} = \beta^{-1} [\omega_1 - \frac {\omega_2} {2}]
\label{eq:anharFapprox}
\end{equation}\tshow{eq:anharFapprox}

\subsection{\label{eq:rssanalytic}Real Space Mapping}
\tshow{eq:rssanalytic}

The behaviour of the FSM in the low temperature regime is in sharp contrast to the RSM in which all the temperature-scaled cumulants tend to a constant value, indicating that the overlap of the phase-constrained distributions of $\mba$ assume a constant value in this limit. To see this we once again appeal to anharmonic perturbation theory and expand the configurational energy as a power series of in \vvec . The results have been worked out in appendix \ref{app:rssperturb} and may be summarised as follows:

\begin{equation}
\lim_{\beta\rightarrow\infty} \omega_1 = \h \s {i} {} \kappa_i
\label{eq:rsscumu1}
\end{equation}\tshow{eq:rsscumu1}

\noindent and:

\begin{equation}
\lim_{\beta\rightarrow\infty} \omega_2 = \h \s {i} {} \kappa_i^2
\label{eq:rsscumu2}
\end{equation}\tshow{eq:rsscumu2}

\noindent where \seta {\kappa} are the eigenvalues of the matrix with elements:

\begin{equation}
{\mf W}_{m \tilde{m}} = \s {i} {} \frac {k^i_\alp} {\sqrt{k^m_\al} \sqrt{k^{\tilde{m}}_\al} } [{{\mf e}^i_\alp}^T . {\mf e}^m_\al] [{{\mf e}^i_\alp}^T . {\mf e}^{\tilde{m}}_\al] - \delta_{\tilde{m} m} {\mf {1}}
\end{equation}

\noindent More generally one may conclude that:

\begin{equation}
\lim_{\beta\rightarrow\infty} \omega_n \sim O(1)
\label{eq:rsscumugenscale}
\end{equation}\tshow{eq:rsscumugenscale}

\noindent so that {\em all} the cumulants contribute to the FED (Eq. \ref{eq:FEDcumu}) at arbitrarily low T. The fact that the temperature-reduced cumulants (Eq. \ref{eq:rsscumugenscale}) tend to a constant value in the harmonic limit translates to the fact that the  overlap between $P(\mba | \pi^c_A)$ and $P(\mba | \pi^c_B)$ tends to a constant amount in this limit \cite{note:constov}.

\section{\label{sec:fsssimulation}Some numerical results}
\tshow{sec:fsssimulation}

In section  (in section \ref{sec:numov})  we will start by investigating  (numerically) the overlap of the \pc s for both the FSM and the RSM in the low temperature and high temperature regimes \cite{note:oppositecon}. We will find that at sufficiently low temperatures the overlap associated with the FSM is, as expected, better than the RSM. However for higher temperatures it is the RSM which has the better overlap.
In section \ref{sec:numfee} we  will then proceed to estimate the anharmonic FEDs \cite{note:prioranhar} in the low and high temperature regimes. We will find that for sufficiently low temperatures, the FSM does not require any extended sampling  in order to arrive at an estimate of the FED which is free of systematic errors. This is in contrast to the RSM, which will, in the most general case, require  extended sampling. On the transition to higher temperatures we find that both the RSM and the FSM require extended sampling in order to overcome the overlap problem. We will then end this section with discussion of the relative efficiencies of the two methods.

\subsection{\label{sec:numov}Overlap}
\tshow{sec:numov}

To start off with, let us consider the behaviour of the overlap problem for the FSM and the RSM in the low temperature regime. In considering the issue of the overlap, we recall (see section \ref{sec:overlap}) that the difference in the free energies of two phases manifests itself as an asymmetry (about the origin) in the position at which the two phase-constrained distributions $P(\mab | \pi^c_A)$ and $P(\mab | \pi^c_B)$ intersect. For systems which have a small FED (as is the case for the systems employed here), this asymmetry will be ever so slight.
An illustration of this for the RSM is shown in figure \ref{fig:twopeak}. 
In characterising the overlap, the observed (approximate) symmetry allows us to focus our  attention on the behaviour on a single phase-constrained distribution. The amount of overlap (which can, in an approximate way, be measured by the peak to peak distance) may then be gauged by  measuring the distance of the peak of this phase-constrained distribution from the origin. The smaller this distance is, the greater will be the overlap between the two distributions \cite{note:implicitassumption}.

\begin{figure}[tbp]
\begin{center}
\ifpdf
\rotatebox{90}{
\includegraphics[scale=0.6]{fig/comparisonofsinglephasedistributions}
}
\else
\rotatebox{0}{
\includegraphics[scale=0.6]{fig/comparisonofsinglephasedistributions}
}
\fi
\end{center}
\caption{${P}(\mab | \pi^c_A)$ and ${P}(\mab | \pi^c_B)$ for the RSM}
The phase-constrained  probability distributions $P(\mab | \pi^c_A)$ and $P(\mab | \pi^c_B)$, as obtained for the RSM. An approximate symmetry is exhibited: each distribution is a mirror reflection of the other about the origin. A similar symmetry is also observed for the FSM. 

$T^*=0.8$ (see Eq. \ref{eq:redtemp}).
\begin{center}
{\bf{------------------------------------------}}
\end{center}

\label{fig:twopeak}
\tshow{fig:twopeak}
\end{figure}

\begin{figure}[tbp]
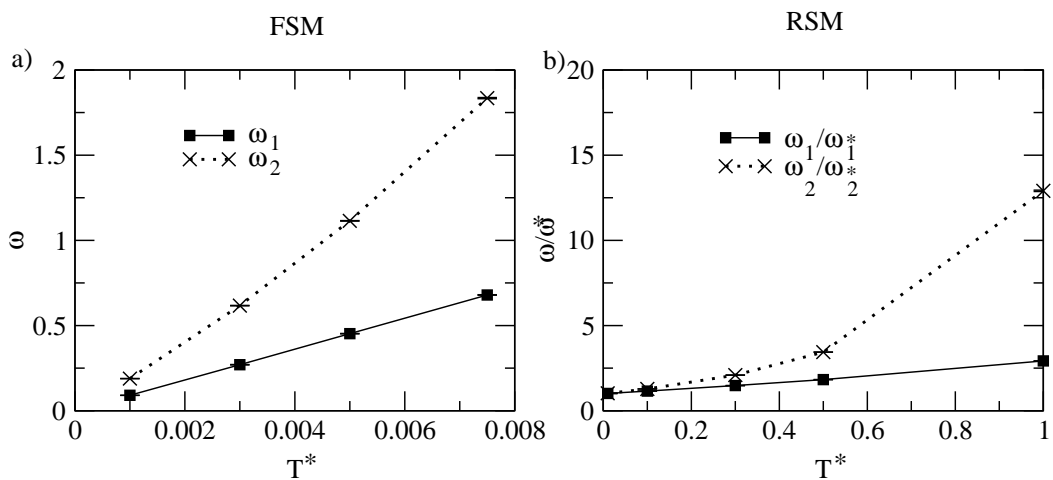

\begin{center}
\ifpdf
\rotatebox{90}{
\includegraphics[scale=0.6]{fig/moments}
}
\else
\rotatebox{0}{
\includegraphics[scale=0.6]{fig/moments}
}
\fi
\end{center}
\caption{Temperature-dependence of the cumulants for the FSM and the RSM}

a) shows the cumulants $\omega_1$ and $\omega_2$ for the FSM for low temperatures. These cumulants are given by Eq. \ref{eq:firstcumulant} and Eq. \ref{eq:secondcumulant}.

 b) shows the scaled cumulants $\frac {\omega_1} {\omega_1^*}$ and $\frac {\omega_2} {\omega_2^*}$ for the RSM, where ${\omega_1^*}$ and $\omega_2^*$ correspond to the values obtained from theory (Eq. \ref{eq:rsscumu1} and Eq. \ref{eq:rsscumu2}).

The temperatures employed in (a) were the highest temperatures at which the simulation results agreed with the predictions (Eq. \ref{eq:cs}) of leading order anharmonic perturbation theory.

\begin{center}
{\bf{------------------------------------------}}
\end{center}

\label{fig:moments}\tshow{fig:moments}

\end{figure}

\begin{figure}[tbp]
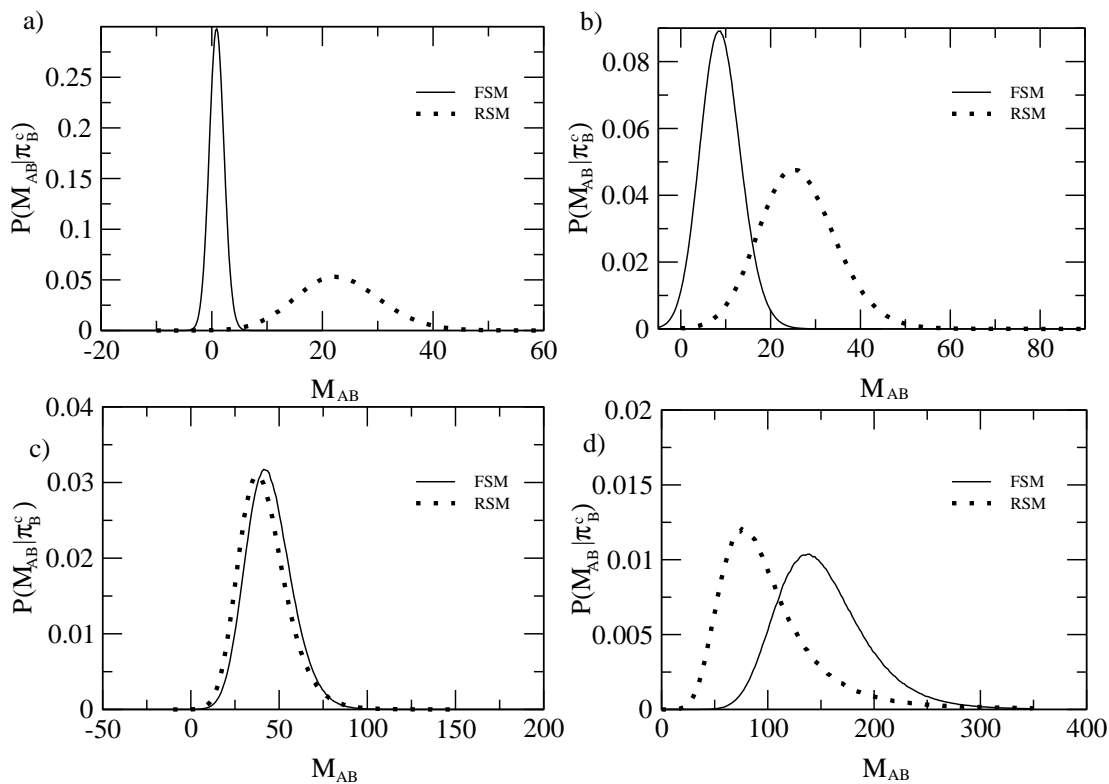

\begin{center}
\ifpdf
\rotatebox{90}{
\includegraphics[scale=0.6]{fig/comparisonoffslsandrsls}
}
\else
\rotatebox{0}{
\includegraphics[scale=0.6]{fig/comparisonoffslsandrsls}
}
\fi
\end{center}
\caption{Comparison of the FSM and RSM with varying temperature}

The evolution of ${P}(M_{AB} | \pi^c_B)$ (for the RSM and FSM) with temperature. The temperatures correspond to (a)$T^*=0.01$, (b)$T^*=0.1$, (c)$T^*=0.5$, (d)$T^*=1.5$. 

Note that in (a) and (b) the RSM essentially assumes its low temperature limiting form \cite{note:checklowtemplimit} (see also figure \ref{fig:moments} (b)).

\begin{center}
{\bf{------------------------------------------}}
\end{center}

\label{fig:comparisonofFSLSandRSLS}\tshow{fig:comparisonofFSLSandRSLS}

\end{figure}

Figure \ref{fig:moments} shows the scaling of the first two cumulants of $P(\mab | \pi^c_B)$  with temperature. In accordance with Eq. \ref{eq:cs}, linear scaling is observed for the FSM. Moreover we see from figure \ref{fig:moments} (b) that the corresponding cumulants for the RSM tend to the limiting values as predicted by Eq.  \ref{eq:rsscumu1} and Eq. \ref{eq:rsscumu2}. Figure \ref{fig:comparisonofFSLSandRSLS} investigates the overlap for a range of temperature by looking at the phase-constrained distribution $P(\mab | \pib)$. One immediately observes that as the temperature is reduced, the overlap of the FSM becomes considerably better than that of the RSM. Whereas in the case of the RSM $P(\mab | \pi^c_B)$ tends to a limiting (stationary) form (see figures \ref{fig:comparisonofFSLSandRSLS} (a) and (b)) \cite{note:checklowtemplimit}, in the case of FSM the corresponding distribution tends to the ideal limit of the delta function (Eq. \ref{eq:perfect}), which in this case is centred on the origin \cite{note:deltacentre}.

To understand the low-temperature behaviour of the FSM we note that the method only probes the anharmonic effects (see Eq. \ref{eq:harM}) since by construction the harmonic contributions to the configurational energy cancel out in the two phases. Since these anharmonic contributions vanish as the harmonic limit is approached, the observed behaviour  is in accordance with what is expected. The RSM, on the other hand, does not 'fold out' the harmonic contributions. For this representation \RBAcal\ must assume a constant value ($\RBAcal = \det  {\avec {S}}_{BA}$) in the harmonic limit. From the overlap identity (Eq. \ref{eq:overlapidentity}) we see that for this to be the case the ratio $P(\mab |\pia) / P(\mab | \pib)$ must approach a stationary value in this limit. One way for this to be achieved is for both $P(\mab | \pia)$ and $P(\mab | \pib)$ to tend to stationary non-singular distributions. This is precisely what is observed.

So far we have analysed the behaviour of the FSM in the limit $T\rightarrow 0$ limit. Let us now discuss the behaviour of the transformation as the temperature is raised ( so as to make the anharmonic effects more prominent).
 From figure \ref{fig:comparisonofFSLSandRSLS} (c) and (d) we see that, though at low temperatures the overlap of the FSM is better than that of the RSM, at high temperatures the situation is reversed; the overlap of the RSM is better than that of the FSM. This can be understood by first noting that the FSM is a global transformation. That is, whereas for the RSM a single particle perturbation in phase \al\ corresponds to a single particle perturbation in phase \alp\  (Eq. \ref{eq:RSS}), in the case of the FSM a single particle perturbation of phase \al\ manifests itself as a global perturbation in which all the particles of \alp\ are perturbed (Eq. \ref{eq:genswitch}, Eq. \ref{eq:fsstran}). Therefore the anharmonic corrections to the energy induced by the exploration of a particle of phase \al\ into the anharmonic regions of the configurational energy $\ecal_\al (\vvec)$ will, under the operation of the FSM, propagate on a global level in phase \alp . In contrast these anharmonic effects only propagate on a local level under the operation of the RSM.
As a result one finds that in the highly anharmonic regimes $M_{\alp\al}$ will be considerably  amplified for the FSM as compared to the RSM.
In addition to this one finds that for the RSM which we have employed (see section \ref{sec:pm}), intra-planar correlations are preserved \cite{p:LSMCprehard} on the transition from one phase to the other, a fact which is true even in the anharmonic regime. The FSM, on the other hand, will preserve little correlations in the anharmonic limit, since the anharmonic effects effectively contaminate the transformation. The net result of these two effects is that the RSM eventually becomes more efficient than the FSM on the transition to sufficiently high temperatures.

An important point to note is that the deviations seen in the distributions of the FSM from the ideal limit (Eq. \ref{eq:perfect}), obtained on increasing the temperatures, are {\em not} due to the increasing prominence of the {\em intrinsic} anharmonic effects but are instead due to the inefficiency of the representation  \cite{note:choosingperfect}. To see this we note that in the harmonic limit the FSM maps configurations of phase \al\ onto configurations of phase \alp\ which are of the same effective  temperature. On increasing the temperature the contamination of the FSM transformation by anharmonic effects results in configurations of phase \al\ being mapped onto configurations of \alp\ which are effectively hotter than the typical configurations of phase \alp. As a result the anharmonic corrections to the total excitation energy will be {\em amplified}, under the operation of the FSM,  over those corrections that are intrinsically present in phase \alp\ at that temperature, so that $M_{\alp\al}$ is not truly representative of the intrinsic anharmonic effects. Therefore, based on an observation of $P(\mab | \pi^c_B)$, one may naively conclude that the anharmonic effects are greater than they really are. 
This idea is supported by figure \ref{fig:comparisonofFSLSandRSLS} (b) where, despite the fact that at $T^*=0.1$  $P(\mab | \pib)$  has assumed its low temperature (harmonic) limiting form for the RSM (see also figure \ref{fig:moments} (b)), the corresponding distribution for the FSM exhibits a significant departure from the ideal limit (Eq. \ref{eq:perfect}). In fact from figure \ref{fig:moments} (a) we see that departure (for the FSM) from the linear scaling predictions (Eq. \ref{eq:cs}) of leading order anharmonic perturbation theory are observed at a temperature  two orders of magnitude {\em lower} than that for which departures from the {\em harmonic} predictions are observed for the RSM. From  figure \ref{fig:anharfree} we see that the anharmonic contributions to the FED are smaller than what one would expect based on the observation of $P(\mab | \pib)$ in figure \ref{fig:comparisonofFSLSandRSLS}.

\subsection{\label{sec:numfee}Estimating the FEDs}
\tshow{sec:numfee}

\begin{figure}[tbp]
\begin{center}
\ifpdf
\rotatebox{90}{
\includegraphics[scale=0.6]{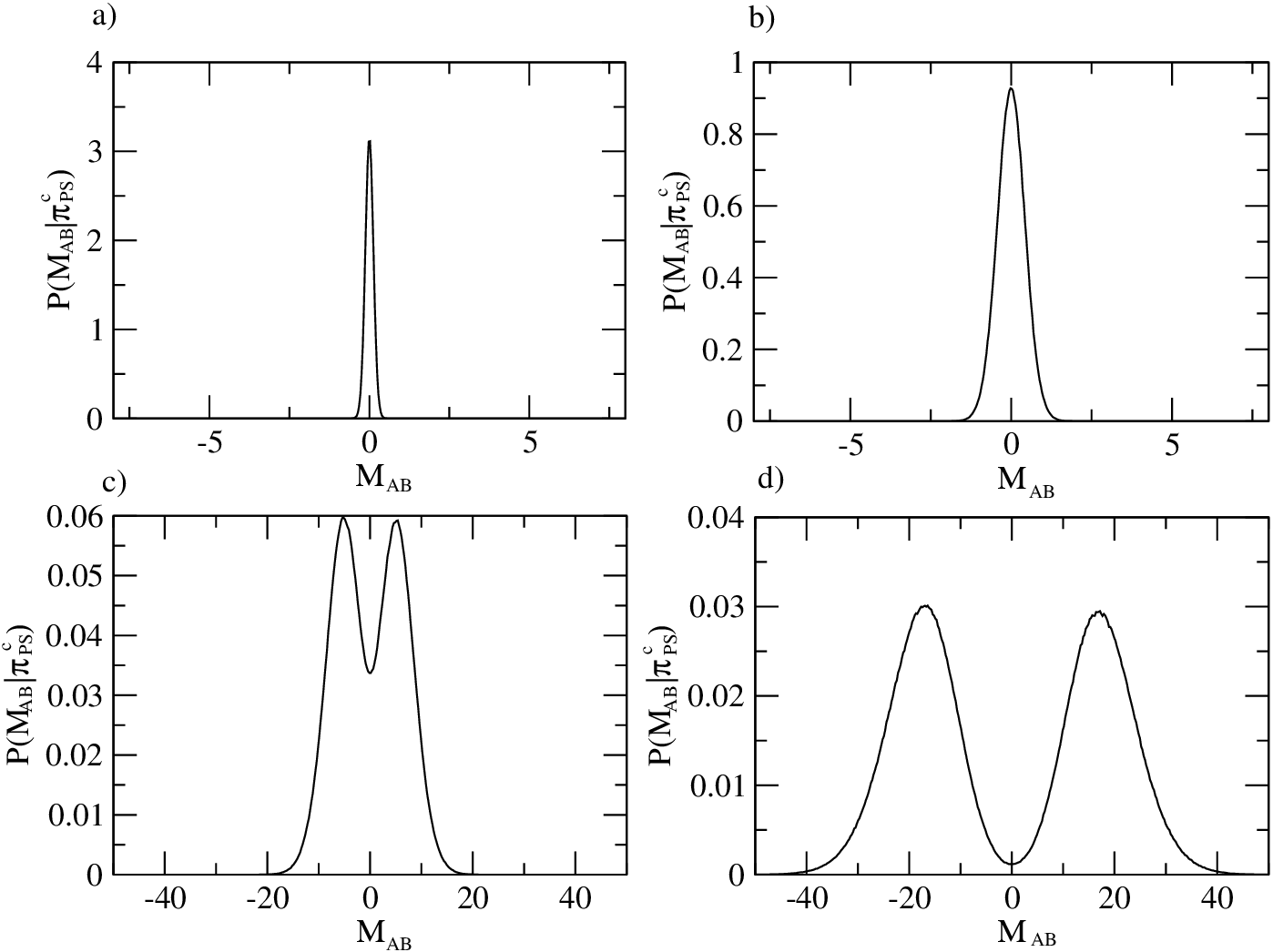}
}
\else
\rotatebox{0}{
\includegraphics[scale=0.6]{fig/ctm_switch_diff_temps}
}
\fi
\end{center}
\caption{The probability density function ${P}(M_{AB} | \pi^c_{PS})$ for the FSM as a function of temperature}
a)$T^*=0.00009$, b)$T^*=0.001$, c)$T^*=0.06$, d)$T^*=0.2$

\begin{center}
{\bf{------------------------------------------}}
\end{center}

\label{fig:ctmswitch}\tshow{fig:ctmswitch}
\end{figure}

Figure \ref{fig:ctmswitch} shows the full \mab\ probability distributions for the FSM-PS method (see section \ref{sec:phaseswitch}); {\em no} form of extended sampling was employed here. It is clear that on the transition to sufficiently low temperatures, the FSM-PS method is no longer plagued with being constrained to the phase which it is initiated in \cite{note:overlapRSS} signalling the absence of an overlap problem since the {\em full} (effective) \cs\ associated with  {\em both} the phases is visited. The overlap problem (for this estimator at these temperatures) is effectively cured.

\begin{figure}[tbp]
\begin{center}
\ifpdf
\rotatebox{90}{
\includegraphics[scale=0.6]{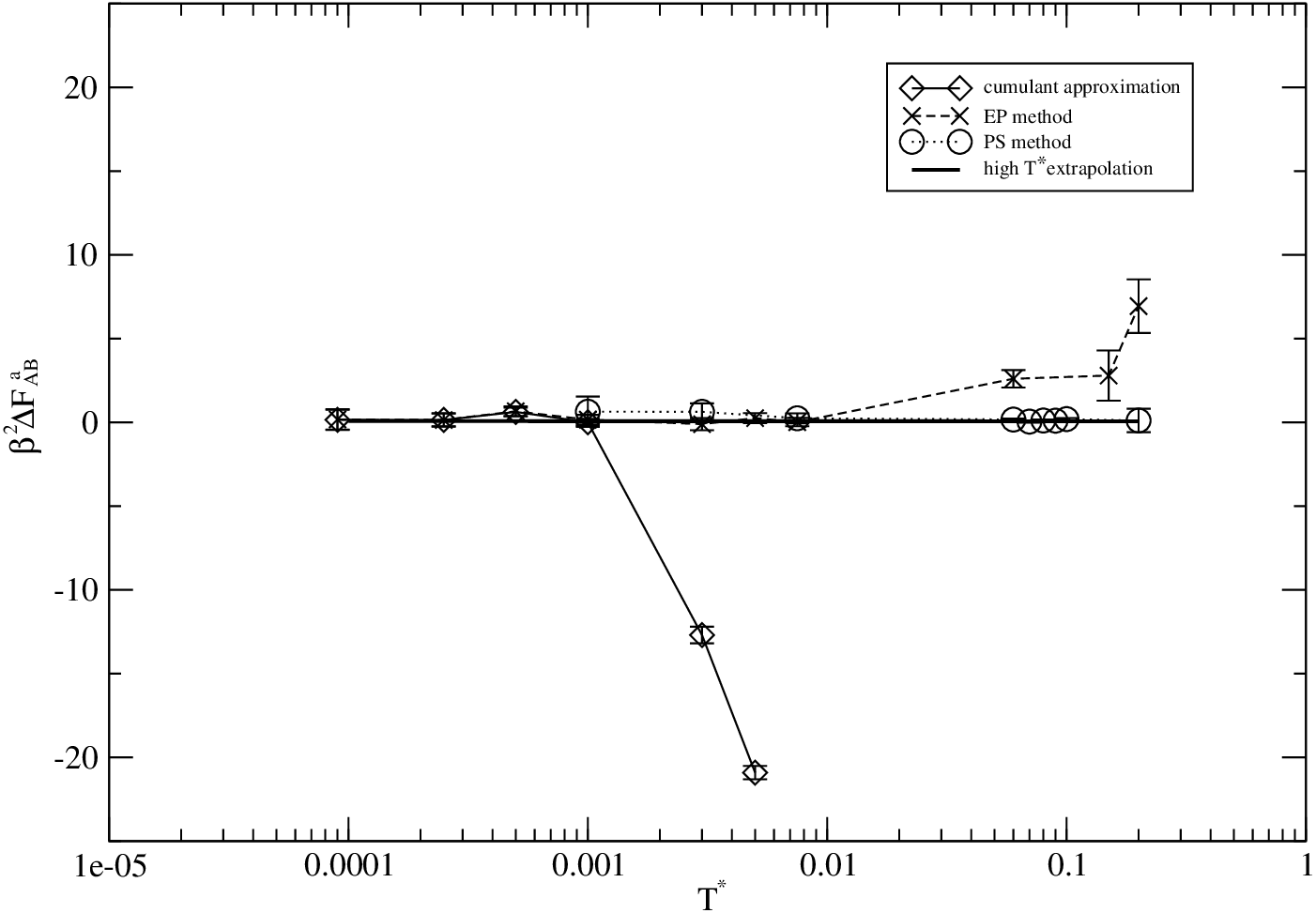}
}
\else
\rotatebox{0}{
\includegraphics[scale=0.6]{fig/anharmonicfreeenergy}
}
\fi
\end{center}
\caption{$\beta^2 \triangle F^a_{AB}$ versus $T^*$}
The temperature-scaled anharmonic contributions to the FED, $\beta^2 \triangle F^a_{AB}$, estimated (without the use of \es ) via the FSM-PS method (Eq. \ref{eq:ratiooftimes0}), the cumulant approximation (Eq. \ref{eq:anharFapprox}), and the FSM-EP formula (Eq. \ref{eq:ep}). 

We have plotted $\beta^2\triangle F^a_{AB}$ instead of $\triangle F^a_{AB}$ since, from perturbation theory, we know this quantity should be constant at sufficiently low temperatures. From perturbation theory we also know that at higher temperatures contributions to $\triangle F^a_{AB}$ appear which scale as $\beta^{-3}$. For this reason we have use a linear extrapolation to compare the high temperature results of figure \ref{fig:ctmswitchfree} to those obtained here \cite{note:fedexpan}.

$\triangle F^a_{AB}$ is in units of $k/\epsilon$.

\begin{center}
{\bf{------------------------------------------}}
\end{center}

\label{fig:anharfree}\tshow{fig:anharfree}
\end{figure}

\begin{figure}[tbp]
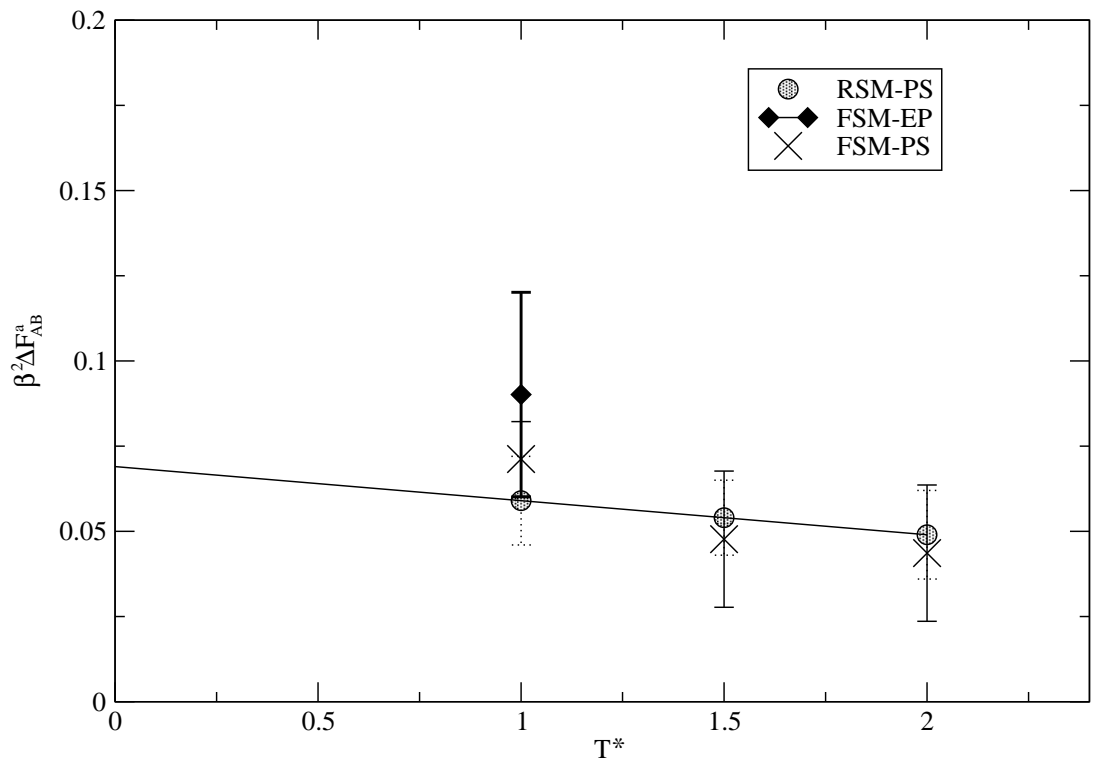

\begin{center}
\ifpdf
\rotatebox{90}{
\includegraphics[scale=0.6]{fig/harmonic_switch_free}
}
\else
\rotatebox{0}{
\includegraphics[scale=0.6]{fig/harmonic_switch_free}
}
\fi
\end{center}
\caption{$\beta^2 \triangle F^a_{AB}$ as obtained from the MUCA FSM-PS and RSM-PS methods}
Note that for the temperatures investigated, MUCA weights had to be employed in order to ensure that the simulation was able to visit the $M_{AB} \sim 0$ regions. 

The straight line denotes a linear extrapolation based on the results of the RSM-PS simulations \cite{note:fedexpan}.

$\triangle F^a_{AB}$ is in units of $k/\epsilon$.

\begin{center}
{\bf{------------------------------------------}}
\end{center}

\label{fig:ctmswitchfree}\tshow{fig:ctmswitchfree}
\end{figure}

Figure \ref{fig:anharfree} shows the estimates of the anharmonic contributions to the temperature-scaled FED for the range of temperatures shown in Figure \ref{fig:ctmswitch}. The significant feature is the agreement of the three estimators (the FSM-EP estimator Eq. \ref{eq:estep}, the FSM cumulant approximation Eq. \ref{eq:anharFapprox}, and the FSM-PS estimator Eq. \ref{eq:ratiooftimes}) at low temperatures and the disagreement between them at high temperatures. The FSM-PS estimator indicates that the anharmonic contributions are unresolvably small throughout. This  conclusion is consistent with what one would expect based on the extrapolation of FSM-PS measurements at higher temperatures (see figure \ref{fig:ctmswitchfree}). Moreover since the FSM-PS method visits (for the range of temperatures investigated in figure \ref{fig:anharfree}) the regions of the effective \cs s associated with {\em both} the phases (as is clear from figure \ref{fig:ctmswitch}) we expect that it should be free of systematic errors. Accordingly we will use the results of the FSM-PS estimator as the benchmark (albeit a rather uninteresting one) for the other methods.

Let us start by discussing the low temperature limit. The ability of the FSM in overcoming the overlap problem (on the transition to sufficiently low temperatures) for {\em all} the estimators is clearly evident from figure \ref{fig:anharfree}. However contrary to initial expectations this does not mean that the task of {\em resolving} the anharmonic contributions to FED becomes any easier. To see this we note that for the FSM  $\omega_1 \sim T$ and $\omega_2 \sim T $ (see Eq. \ref{eq:cs}), with all higher orders vanishing at a higher rate. As a result the error in ones estimate of the mean ($\omega_1$) of $P(\mab | \pi^c_B)$ is proportional to $T^{\frac {1} {2}}$, since this quantity itself is proportional to the standard deviation $\sqrt {\omega_2}$ of $P(\mab | \pi^c_B)$. The fact that $\omega_1$ decays faster, with decreasing temperature, than does its error is indicative of a signal to noise problem that is present on the transition to lower temperatures. That is, in the region where the overlap problem is overcome the (small) anharmonic contributions are entirely masked by the residual noise in the transformation in a way which is not cured by going to (still) lower temperatures.

As the temperature is increased what is observed in figure \ref{fig:anharfree} is the eventual departure of the estimates of both the cumulant approximation and the FSM-EP methods from the estimates of the FSM-PS method. The first estimator to depart from the benchmark line is the cumulant approximation (Eq. \ref{eq:anharFapprox}),  signalling the increasing importance of the higher order cumulants in the expansion of Eq. \ref{eq:FEDcumu} \cite{note:cumudir}. Upon inclusion of all the cumulants (which is simply done by estimating \RBAcal\  via the FSM-EP method, Eq. \ref{eq:cumulant}) one does indeed estimate the quantity $\beta^2 \triangle F^a_{BA}$ correctly, since the results of the FSM-EP method and the FSM-PS method coincide. On increasing the temperature further, the estimates of \RBAcal\ via the FSM-EP method  also begin to depart from those of the FSM-PS method. The reason is that now systematic errors are arising from the fact that $\hat{P}(\mab | \pi^c_A)$ and $\hat{P}(\mab | \pi^c_B)$ do not completely overlap. 
This is clearly the case in figure \ref{fig:ctmswitch} (d). 
We note that unlike the cumulant approximation, which {\em underestimates} the FED \cite{note:cumudir}, the FSM-EP method {\em overestimates} the desired quantity when systematic errors begin to set in \cite{note:whyover}.

The decreasing amount of overlap between $P(\mab | \pi^c_A)$ and $P(\mab | \pi^c_B)$  obtained on increasing the temperature means that eventually even the FSM-PS estimators will not be free of systematic errors without the use of some form of extended sampling strategy (see section \ref{sec:review} and chapter \ref{chap:sampstrat}). For the systems investigated here, the maximum temperature at which the FSM-PS method could successfully be implemented without the use of extended sampling was T=0.2 (figure \ref{fig:ctmswitch}). Beyond this extended sampling was required. Figure \ref{fig:ctmswitchfree} shows the temperature-scaled anharmonic FEDs obtained  with the MUCA  \cite{note:mulps} sampling distribution and shows clear agreement between the RSM and FSM methods. The results are consistent with those of \cite{p:LSMCpresoft}. In particular {\em the anharmonic contributions act so as to favour the hcp (B) phase.}

The MUCA extended sampling strategy (and indeed all the other extended sampling strategies, to be discussed in chapter \ref{chap:sampstrat}) allows one to tackle the overlap problem {\em irrespective} of the representation. The choice of representation then manifests itself in the residual statistical errors in the estimate of the FED. In comparing these statistical errors for the FSM and the RSM, we first note from figure \ref{fig:comparisonofFSLSandRSLS} that for all temperatures of up to T=0.5, the FSM will require a narrower region of \mab\ space to be reweighted (as compared to the RSM) within the MUCA approach \cite{note:howmuca}. This has two consequences for the simulation. Firstly this will translate to a smaller statistical error in the estimate of \RBAcal\ for a given number of Monte Carlo steps since the system will fluctuate over a narrower region of \mab\ space. Secondly this will correspond to a reduced computational effort in the task of constructing the multicanonical weights. However on top of this one must also give consideration to the differences in computational effort (i.e. the time for each Monte Carlo step) between the FSM and RSM. In order to  understand this latter issue more fully, we note that in a  MUCA simulation  the macrovariable $\mab$ will have to be evaluated for each Monte Carlo step. Suppose the simulation is in phase $\al$ and suppose that one employs a short ranged potential. Then a single particle perturbation will, for the RSM, require a {\em local} re-evaluation of the configurational energies of both $\al$ and $\alp$ in order to compute the new value of $\mab$. 
The number of computational steps needed for such a task is $O(N^0)=O(1)$. For the FSM the re-evaluation of $\ecal_\al$ will also be local in nature. 
However since the FSM (see Eq. \ref{eq:fsstran}) induces a {\em global} rearrangement of the atoms of phase $\alp$ (so that the calculation of $\ecal_\alp$ is an $O(N)$ calculation) the number of computational steps needed for the reevaluation of \mab\ for the FSM will be $O(N)$. This significant advantage that the RSM holds over the FSM vanishes when long ranged potentials are employed, in which case {\em both} methods will involve $O(N^2)$ calculations.

\section{Summary}

In this section we have clearly illustrated both analytically and numerically the dependence of the overlap on the representation. We have shown that adopting a {\em fourier} representation of the displacements allows one to cure the overlap problem at sufficiently low temperatures. This is in sharp contrast to the RSM, for which the overlap of the two distributions tends to a limiting form. The main benefit of the FSM over the RSM is that for sufficiently low temperatures, \es\ will not be needed in order to arrive at an estimate of the FED which is free of systematic errors. 

However our expectations of being able to estimate the FED via the FSM with increasing ease are not fulfilled due to the presence of a signal to noise problem which gets {\em worse} as the temperature decreases, even though the overlap between the \pc s {\em improves}. Furthermore  by being a transformation which is global in nature, one must expend a considerably greater amount of computational effort in dealing with the FSM than is required for the RSM.

In tackling the problem of estimating FEDs in the most general cases, one must not only give consideration to the choice of  representation but one must also give consideration to the choices of estimators and the choices of  \es\ strategies. The importance of the choice of estimator has already been illustrated to a certain extent in figure \ref{fig:anharfree}, where we have seen that the PS estimator (Eq. \ref{eq:ratiooftimes}) is better than that of the EP method (Eq. \ref{eq:estep}). In chapter \ref{chap:estsamp} we will address these issues in greater depth. In chapter \ref{chap:sampstrat} we will then proceed to discuss the use of extended sampling strategies in the task of estimating FEDs.

\chapter{\label{chap:estsamp}Estimators}
\tshow{chap:estsamp}

\section{\label{sec:estintro}Introduction}
\tshow{sec:estintro}

Imperative to the understanding of the FED problem is the appreciation of the distinction that must be made between statistical and systematic errors (see section \ref{sec:metrop}).  Whereas statistical errors may be reduced to a desired level simply by running the simulation for a sufficient duration of time, systematic errors in general can not be controlled in this way. In the context of FED calculations the origin of these systematic errors is the partial overlap \cite{note:partialmean} between  $\hat{P}(\mba | \pi^c_A)$ and $\hat{P}(\mba | \pi^c_B)$ (see section \ref{sec:overlap}). As we have discussed in chapter \ref{chap:review} these systematic errors may be minimised through efficient choices of PM (that is choices of the global configuration space displacement $\Dvec$ and representation $\vvec$). In the case where it is possible to construct an efficient PM so as to yield at least some overlap between $\hat{P}(\mba | \pi^c_A)$ and $\hat{P}(\mba | \pi^c_B)$, it is possible to eliminate the systematic errors that arise in ones estimate of \RBAcal\ by constructing an appropriate estimator. In this chapter we will investigate this issue in two stages. In the first part  we will show how  one may restrict the regions of \mba\ space which contribute to the relevant expectations (which appear in the estimators) so as to yield an estimate of \RBAcal\ which is free of systematic errors. 
We will then show that an alternative strategy to this is that of employing estimators which are unrestricted, in the sense just described, and which are instead {\em designed} to have their most significant contributions originating from those regions of \ecs\ over which both $\hat{P}(\mba | \pi^c_A)$ and $\hat{P}(\mba | \pi^c_B)$ overlap \cite{note:limworkoverlap}. This latter idea has been studied (and understood) in a different way in \cite{p:bennett}, \cite{p:kofkecummings}, \cite{p:radmer}, \cite{p:lusinghkofke}\nocite{p:lukofke}\nocite{p:lukofke2}-\cite{p:lukofke3}. At the heart of our insight is the appreciation that only within the region of overlap does the estimator in Eq. \ref{eq:overlapest} yield an estimate which is free of systematic errors. Since ultimately {\em all} estimators may be derived directly from Eq. \ref{eq:overlapest}, it follows that the successful estimators will be those which pool together the estimates of \RBAcal\ made by Eq. \ref{eq:overlapest} {\em within} the region of overlap.

\begin{figure}[tbp]
\begin{center}
\rotatebox{0}{
\includegraphics[scale=0.5]{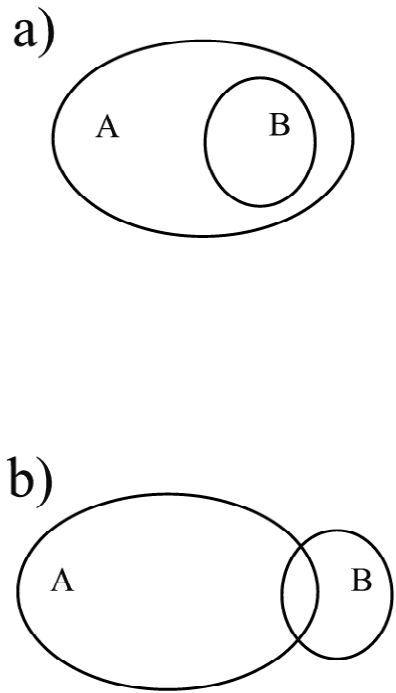}
}
\end{center}
\caption{Schematic of the way in which the two phases can overlap in (effective) configuration space}

a)Here the regions of (effective) configuration space  typically explored by phase B are a {\em subset} of those typically explored by phase A.

b)In the second type of overlap there are regions typically explored by each phase that are {\em not} visited by the other phase. When we refer to 'partial overlap' we will have the case shown in (b) in mind.

In the whole of this thesis we will only concern ourselves with cases where the overlap is of the type shown in (b).

\begin{center}
{\bf{------------------------------------------}}
\end{center}

\tshow{pic:twooverlap}
\label{pic:twooverlap}
\end{figure}

Consider figure \ref{pic:twooverlap}. In the most general case, there will be two types of overlap that one encounters when one attempts to estimates the FED. In the first case (see figure \ref{pic:twooverlap} (a)) the regions of \ecs\ of one of the phases forms a subset of that of the other phase. This type of situation typically arises in the calculation of the chemical potential (via the insertion method), where one attempts to determine the FED between an N particle system (A) and an N+1 (B) particle system (see \cite{p:kofkecummings} for an excellent discussion). In this case it has been argued \cite{p:kofkecummings} that the EP estimator (Eq. \ref{eq:ep}, in which A, the N particle system, is the parent phase) will yield an estimate of the FED which is free of systematic errors. The basic idea is that a sampling experiment performed in phase A will capture {\em all} the regions of \ecs\ relevant to phase B \cite{note:kofex}.

The latter type of overlap (see figure \ref{pic:twooverlap} (b)) appears when one attempts to determine the FED between {\em different} phases of a {\em single} system (see section \ref{sec:overlap}). In regards to the estimator we have already seen in section \ref{sec:fsssimulation} (in particular figure \ref{fig:anharfree}) how the estimator of the EP method (Eq. \ref{eq:estep}) fails in this case, due to the fact that {\em a single} \pc\ fails to capture {\em all} the important regions of \ecs\ which contribute to the FED. One must instead employ estimators which involve the sampling of  the regions of \ecs\ associated with {\em both} phases \cite{p:bennett}.

These  estimators, which involve the simulation of both phases,  may be broadly categorised into two groups: the phase-constrained estimators and the phase-switching (PS) estimators. The phase-constrained estimators involve expectations with respect to sampling distributions  confined to the phase in which they are initiated, whereas the phase-switching estimator involves a sampling distribution  which actually switches between the phases. As we will now show, one must in general take explicit steps so as to ensure that the phase-constrained estimators are free of systematic errors. We will also show that, for a particular subgroup of the phase-constrained estimators, no such steps are needed since these estimators are, by construction, free of systematic errors even in the case of partial overlap. In the case of the PS estimator, we show that for partial overlap the method can be guaranteed to be free of systematic errors simply by appropriately weighting the two phases (in a way as prescribed in the simulated tempering method \cite{p:nezbedakolafa}\nocite{p:lyubartsev}\nocite{p:lyubartsevgeneral}-\cite{p:simulatedtempering}, section \ref{sec:st}) so as to increase the probability with which the simulation visits the phase with the smaller partition function (greater free energy). In order to keep the discussion as general as possible we will formulate our arguments within the context of the FG method.

\section{Phase-constrained estimators}

\subsection{\label{sec:elimsys}Eliminating systematic errors via restricted expectations}
\tshow{sec:elimsys}

\begin{figure}[tbp]
\begin{center}
\rotatebox{270}{
\includegraphics[scale=0.5]{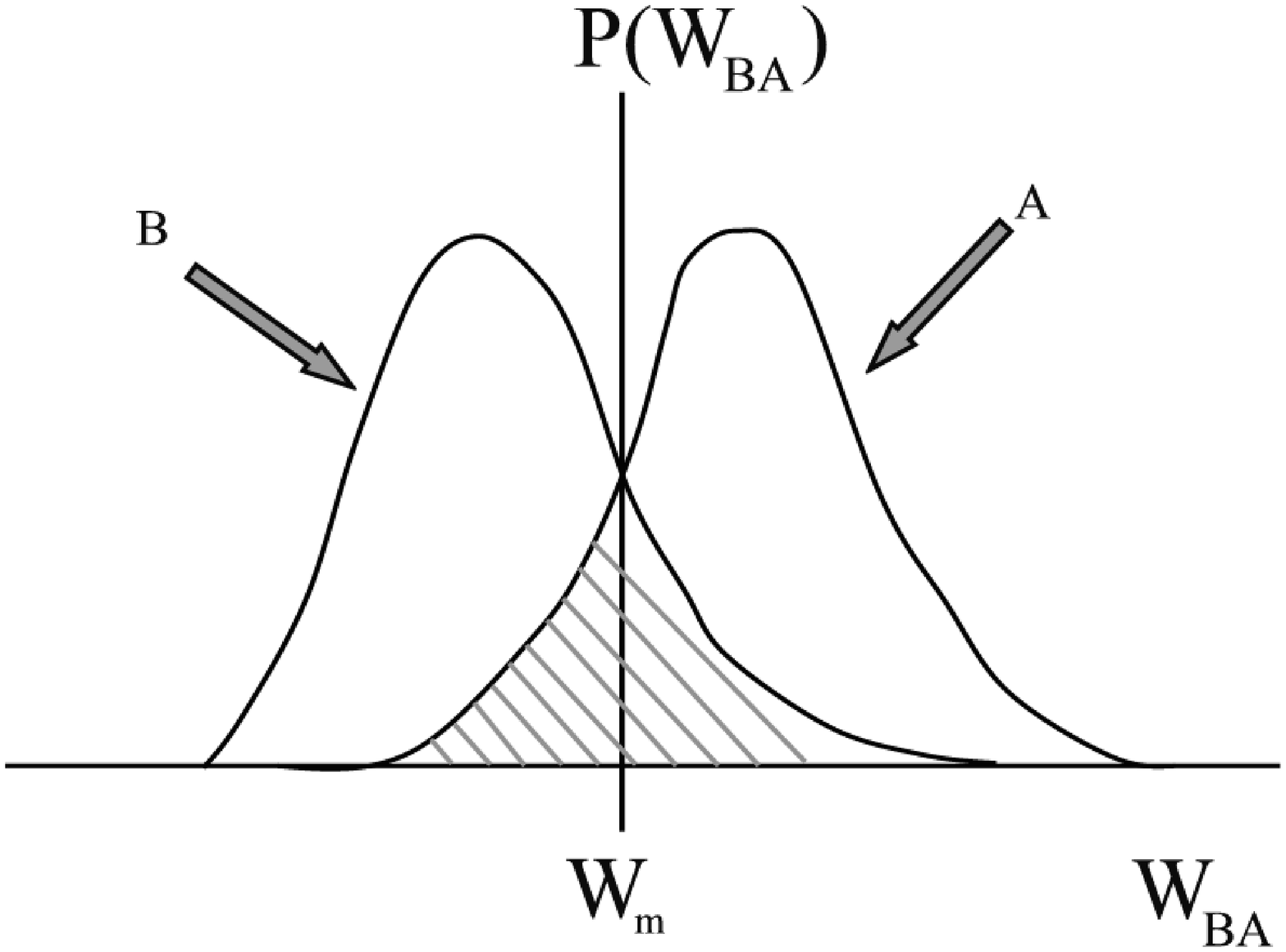}
}
\end{center}
\caption{Schematic of ${P}(\wba|\pi^c_A)$ and ${P}(\wba|\pi^c_B)$ in the case of partial overlap}

Using the same arguments as those employed in section \ref{sec:overlap} it is easy to show that the point at which the two distributions intersect is given by $W_m = -\ln \RBAcal$. 

\begin{center}
{\bf{------------------------------------------}}
\end{center}

\tshow{pic:overlapping}
\label{pic:overlapping}
\end{figure}

Suppose that $\hat{P}(\wba | \pia)$ and $\hat{P}(\wba | \pib)$ partially overlap \cite{note:partialmean} in the manner shown in figure \ref{pic:overlapping}. Using the same arguments employed in  section \ref{sec:overlap}, it follows that the point at which they intersect is given by:

\begin{equation}
W_m = -\ln \RBAcal = \beta \triangle F_{BA}
\label{eq:intersect}
\end{equation}\tshow{eq:intersect}

\noindent 

We will now proceed to show how, in the case of partial overlap, the FED may be estimated from any estimator (see Eq. \ref{eq:freebennett}) in a way which is free of systematic errors, merely by restricting the range of \wba\ space from which the non-negligible contributions originate.
Let us first consider the overlap identity (Eq. \ref{eq:crooksfluct}). Multiplying by an arbitrary non-zero function $G(\wba)$ and then integrating both sides over the restricted range $W_{BA}^0 \leq W_{BA} \leq W_{BA}^1$  we arrive at a formula which we call the restricted dual phase perturbation (RDP) formula:

\begin{equation}
\RBAcal  \int_{W_{BA}^0}^{W_{BA}^1} G(\wba) P(W_{BA} | \pi_B^c) d W_{BA} = \int_{W_{BA}^0}^{W_{BA}^1} G(\wba) e^{-W_{BA}} P(W_{BA} | \pi_A^c) d W_{BA}
\end{equation}

\noindent or

\begin{equation}
\RBAcal = \frac {<G(\wba) Y(\wba) e^{-\wba}>_{\pi^c_A}} {<G(\wba) Y(\wba) >_{\pi^c_B}}
\label{eq:dualphaserestrict}
\end{equation}\tshow{eq:dualphaserestrict}

\noindent where:

\begin{equation}
Y(\wba)=\left\{\begin{array}
{r@{\quad:\quad}l}
1 & \mbox{$W_{BA}^0 \leq W_{BA} \leq W_{BA}^1$}\\
0 & \mbox{otherwise}
\label{eq:res}
\end{array}\right.
\end{equation}\tshow{eq:res}

\noindent Eq. \ref{eq:dualphaserestrict}  may then be used to estimate \RBAcal\ via:

\begin{equation}
\RBAcal  \est  \frac {\s {i=1} {b} \hat {P}(\wbai | \pi^c_A) G(\wbai) Y(\wbai) e^{- \wbai}} {\s {i=1} {b} \hat {P}(\wbai | \pi^c_B) G(\wbai) Y(\wbai)}
\label{eq:estdualphaserestrict}
\end{equation}\tshow{eq:estdualphaserestrict}

\noindent where:

\begin{equation}
\hat {P} (\wbai | \pi^c_\al ) = \frac {H(\wbai | \pi^c_\al)} {\s {i=1} {b} H(\wbai | \pi^c_\al)}
\label{eq:probestfg}
\end{equation}\tshow{eq:probestfg}

\noindent $W_{BA}^0$ and $W_{BA}^1$ are in principle arbitrary. In practise, however, they are not if one is to arrive at an estimator which will yield an estimate of \RBAcal\ which is free of systematic errors.  
In order to obtain the necessary insights it is instructive to derive Eq. \ref{eq:estdualphaserestrict} directly from the {\em estimator} for \RBAcal\ associated with the overlap identity (Eq. \ref{eq:crooksfluct}) itself:

\begin{equation}
\RBAcal \est {\hat{\cal R}}_{BA} =  \frac {\hat {P} (W_{BA,i} | \pi_A^c) e^{- W_{BA,i}}} {\hat {P} (W_{BA,i} | \pi_B^c)}
\label{eq:crooksest}
\end{equation}\tshow{eq:crooksest}

\noindent where $ {\hat{\cal R}}_{BA}$ is an estimate for \RBAcal . Rearranging Eq. \ref{eq:crooksest} and multiply both sides by\newline $G(\wbai) Y(\wbai)$ one obtains:

\begin{equation}
 {\hat{\cal R}}_{BA} G(\wbai) Y(\wbai) {\hat {P} (W_{BA,i} | \pi_B^c)} = G(\wbai) Y(\wbai) {\hat {P}(W_{BA,i} | \pi_A^c)} e^{-\wbai}
\end{equation}

\noindent Summing both sides over all the bins and rearranging leads to Eq. \ref{eq:estdualphaserestrict}.

The necessary restrictions that are needed become apparent when one notices that, implicit in this derivation, is the assumption that the histograms $H(W_{BA,i} | \pi_A^c)$ and $H(W_{BA,i} | \pi_B^c)$ of the bins \wbai\  over which the summations are performed are {\em simultaneously} non-zero. This requirement stems from Eq. \ref{eq:crooksest}, which itself assumes that both $\hat{P}(\wbai | \pi^c_A)$ and $\hat{P}(\wbai | \pi^c_B)$ are non-zero. However the regions of \ecs\ over which the estimators  $\hat{P}(\wbai | \pi^c_A)$ and $\hat{P}(\wbai | \pi^c_B)$ of the phase-constrained distributions are {\em both} non-zero is precisely what we defined (in section \ref{sec:overlap}) to be the region of overlap.
 {\em In other words the widest choice of} $ W_{BA}^0 \leq  W_{BA} \leq  W_{BA}^1$ {\em should correspond directly to the region over which the estimators} $\hat{P}(\wba | \pia)${\em  and }$\hat{P}(\wba | \pib)$ {\em of the \pc s overlap (i.e. the shaded region in figure \ref{pic:overlapping}).}

\begin{figure}[tbp]
\begin{center}
\rotatebox{270}{
\includegraphics[scale=0.5]{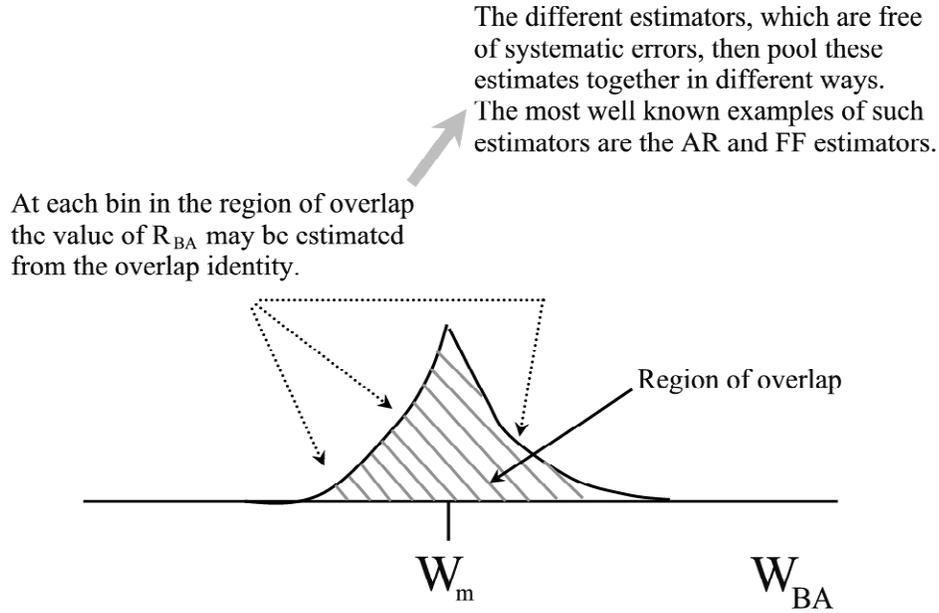}
}
\end{center}
\caption{Schematic illustrating the principle workings of Eq. \ref{eq:dualphaserestrict}}

This figure illustrates the fact that the phase-constrained estimators whose estimates of \RBAcal\ are free of systematic errors  
 are those whose {\em non-negligible} contributions solely come from the region of overlap.

\begin{center}
{\bf{------------------------------------------}}
\end{center}

\tshow{pic:estov}
\label{pic:estov}
\end{figure}

The key point is that within the overlapping region each bin has, associated with it, an estimate of \RBAcal\ given by Eq. \ref{eq:crooksest}. One may then pool these estimates together in different ways;  the result is the array of  different estimators whose form is most generally given by Eq. \ref{eq:dualphaserestrict}. This idea is illustrated in figure \ref{pic:estov}. As we will show in the next section the acceptance ratio (AR) and  fermi function (FF) are prime examples of estimators which pool the estimates in this way and which {\em do not} require any restrictions (see Eq. \ref{eq:res}) to be imposed. The phase switch (PS) method is another such method, which accounts for all the regions of \ecs\ which contribute non-negligibly to the FED by actually switching phases (in the case of zero equilibration, see section \ref{sec:phaseswitch}) or more generally switching between processes (in the case of the arbitrary equilibration FG method, see section \ref{sec:pfg}). The PS method will be discussed later in section \ref{sec:fgps}.

Some flexibility does exist in setting the range $ W_{BA}^0 \leq  W_{BA} \leq  W_{BA}^1$. Specifically the range $ W_{BA}^0 \leq  W_{BA} \leq  W_{BA}^1$ can be widened so as to include \wba\ macrostates originating  from outside the region of overlap, {\em provided} that they contribute negligibly to the relevant estimators. Since from the overlap identity (Eq. \ref{eq:crooksfluct}) we know that:

\begin{equation}
P(\wba | \pi^c_B) \propto e^{-\wba } P(\wba | \pi^c_A)
\label{eq:propcrooks}
\end{equation}\tshow{eq:propcrooks}

\noindent it follows that regions which are negligible to the estimator of the numerator of Eq. \ref{eq:estdualphaserestrict} are also negligible to the denominator, and vice-versa, which makes them easily identifiable. For example, in the case where $G(\wba)=1$, these  regions are those over which $P(\wba | \pib)$ is negligible. This includes all the parts  of $P(\wba | \pia)$ which do not overlap with $P(\wba | \pib)$, which means that in the task of estimating \RBAcal\ from Eq. \ref{eq:estdualphaserestrict} one may remove the upper restriction $W_{BA}^1$ (when $G(\wba)=1$). The sole purpose of this is merely to simplify the evaluation of the relevant estimators (see section \ref{sec:numiest}). The important point is that the regions which {\em do} contribute {\em non-negligibly} to these summations are limited to the regions of overlap.

\subsection{\label{sec:chooseg}Eliminating systematic errors via $G(\wba)$}
\tshow{sec:chooseg}

We have seen in the previous section how one may construct an estimator based on a given $G(\wba)$ which is free of systematic errors merely by restricting the expectation, so as to ensure that the \wba\ macrostates which contribute {\em non-negligibly}  to the estimator come from within the overlapping region (see figure \ref{pic:overlapping}, figure \ref{pic:estov}). Since the most significant contributions to the numerator and the denominator of Eq. \ref{eq:dualphaserestrict} come from the {\em same} regions of \ecs\ (by virtue of Eq. \ref{eq:propcrooks}) we see that an alternative strategy is to {\em construct} a $G(\wba)$ so as to ensure that the non-negligible contributions originate from the regions of \wba\ space over which  $\hat{P}(\wba | \pi^c_A)$ and $\hat{P}(\wba | \pi^c_B)$ overlap.
In this case the restrictions imposed on the expectations in Eq. \ref{eq:dualphaserestrict} may be lifted.

In order to facilitate our analysis let us define a set of weight functions (not to be confused with MUCA weights) for the estimator of \RBAcal .  In the case of the DP estimators (Eq. \ref{eq:freebennett}) let us define a weight function $w_n(\wba)$ as:

\begin{equation}
w_n(\wba) = G(\wba) e^{-\wba} \hat{P}(\wba | \pia)
\label{eq:weinum0}
\end{equation}\tshow{eq:weinum0}

\noindent and a weight function $w_d(\wba)$ as:

\begin{equation}
w_d(\wba) = G(\wba) \hat{P}(\wba | \pib)
\label{eq:weiden0}
\end{equation}\tshow{eq:weiden0}

\noindent These weight functions essentially measure the contribution of a macrostate \wba\ in the numerators and the denominator of the estimator of Eq. \ref{eq:freebennett}. Since from the overlap identity (Eq. \ref{eq:crooksfluct}) we know that:

\begin{equation}
w_n(\wba) = \RBAcal w_d(\wba)
\label{eq:propwei}
\end{equation}\tshow{eq:propwei}

\noindent we will only concentrate on $w_n$ in the following analysis of the DP estimators.

In the case of the EP estimator, which is obtained by setting $G(\wba)=1$ in Eq. \ref{eq:freebennett}, we will depart from the definitions given in Eq. \ref{eq:weinum0} and Eq. \ref{eq:weiden0} and instead define the weights in accordance to the contributions of macrostates to the numerator and denominator of the corresponding estimator:

\begin{equation}
\RBAcal  \est  \frac { \s {i=1} {b} H(W_{BA,i} | \pi_A^c) e^{-W_{BA,i}}} {\s {i=1} {b} H(W_{BA,i} | \pi_A^c)}
\label{eq:estepfg}
\end{equation}\tshow{eq:estepfg}

\noindent That is we define the weights as:

\begin{eqnarray}
w_n(\wba) &  = & e^{-\wba} \hat{P}(\wba | \pia)\nonumber\\
&  \propto & \hat{P}(\wba | \pib)
\label{eq:epn}
\end{eqnarray}\tshow{eq:epn}

\noindent and:

\begin{equation}
w_d(\wba) = \hat{P} (\wba | \pia)
\label{eq:epd}
\end{equation}\tshow{eq:epd}

\noindent In this case one must separately analyse $w_n$ and $w_d$ since they are no longer proportional, as is the case in Eq. \ref{eq:propwei}.

Let us now motivate the construction of estimators in which no restrictions of the form of Eq. \ref{eq:res} are needed. From Eq. \ref{eq:propcrooks} it is clear that the choice  $G(\wba)=1$ in Eq. \ref{eq:freebennett} results in the most significant contributions originating from the regions where ${P}(\wba |\pi^c_B)$ is most significant. On the other hand choosing $G(\wba) = e^{\wba}$ (this merely corresponds to performing the EP method in the other phase) results in the contributing regions being those over which ${P}(\wba |\pi^c_A)$ is most significant. For reasons mentioned in section \ref{sec:elimsys}, both these choices can only be guaranteed to yield estimates of \RBAcal\ which are free of systematic errors (in the case of partial overlap) by imposing the restrictions mentioned in the previous section. One may, then, naively expect that the construction of an interpolation $G(\wba)=[1+e^{\wba}]/2$, which leads to the following formula:

\begin{equation}
\RBAcal = \frac {<[1+e^{-\wba}]>_{\pi^c_A}} {<[1+e^{+\wba}]>_{\pi^c_B}}
\label{eq:invfermi}
\end{equation}\tshow{eq:invfermi}

\noindent to lead to a more useful estimator of \RBAcal. This is not in fact the case. To see this we first notice (from Eq. \ref{eq:weinum0}) that the weight function $w_n(\wba)$ is given by:

\begin{equation}
w_n (\wba) = [1+ e^{-\wba}] \hat{P}(\wba | \pia)
\end{equation}

\noindent so that:

\begin{equation}
w_n (\wba) \approx \left\{\begin{array}
{r@{\quad:\quad}l}
\hat{P}(\wba | \pia) & \mbox{\ for $\wba > 0$}\\
\RBAcal \hat{P}(\wba | \pib) & \mbox{\ for $\wba < 0$}
\end{array}\right.
\label{eq:weighthypo}
\end{equation}\tshow{eq:weighthypo}

\noindent It is immediately apparent from Eq. \ref{eq:weighthypo} and figure \ref{pic:overlapping} that the regions of \wba\ space which contribute significantly are those regions of \wba\ space spanned by {\em both} $\hat{P}(\wba | \pia)$ and  $\hat{P}(\wba | \pib)$, and is not simply limited to the regions of \wba\ space over which the two \pc s overlap, as one might originally expect. Therefore Eq. \ref{eq:invfermi} has not got the desired property that we are looking for, namely the property of having the non-negligible contributions coming solely from the region of overlap.

Now let us examine the choice of $G(\wba) = A(-\wba)$, which leads to the acceptance ratio (AR) formula (Eq. \ref{eq:FGAR}). As we will now see, the AR formula is a prime example of an estimator which does not require restrictions to be imposed on its corresponding estimator in order to guarantee that it is free of systematic errors.
In order to see this, let us establish the AR formula in a slightly more general way. To do this we first re-write the overlap identity (Eq. \ref{eq:crooksfluct}) as:

\begin{equation}
G(\wba-C) P(\wba | \pi^c_B) \RBAcal = e^{-C} G(\wba - C) e^{-[\wba - C]} P(\wba | \pi^c_A)
\end{equation}

\noindent where C is an arbitrary constant. Integrating both sides over \wba\ and rearranging gives:

\begin{equation}
\RBAcal = e^{-C} \frac {<G(\wba-C) e^{-[\wba - C]}>_{\pi^c_A}} {<G(\wba-C)>_{\pi^c_B}}
\label{eq:freebennettgen}
\end{equation}\tshow{eq:freebennettgen}

\noindent Following our earlier definitions (see Eq. \ref{eq:weinum0} and Eq. \ref{eq:weiden0}) we define the weight functions as:

\begin{equation}
w_n(\wba) = G(\wba-C) e^{-[\wba-C]} \hat{P}(\wba | \pia)
\label{eq:weinum}
\end{equation}\tshow{eq:weinum}

\noindent and $w_d(\wba)$ as:

\begin{equation}
w_d(\wba) = G(\wba-C) \hat{P}(\wba | \pib)
\end{equation}

\noindent The interrelation between the weights may now be written more generally as:

\begin{equation}
w_n(\wba) = e^C \RBAcal w_d(\wba)
\label{eq:accdecom5}
\end{equation}\tshow{eq:accdecom5}

\noindent Once again it suffices to focus ones attention on only one of these weight function (which in the following analysis will be $w_n$). 

The constant C in Eq. \ref{eq:freebennettgen} is important in that it directly affects the statistical and systematic errors associated with the corresponding estimator. We will return to the optimal choice of C later. If one substitutes  $G(\wba) = A(-[\wba-C])$ into Eq. \ref{eq:freebennettgen} one obtains  a generalisation of the  AR formula, Eq. \ref{eq:FGAR}:

\begin{equation}
\RBAcal = e^{-C} \frac {<A(\wba-C)>_{\pi_A^c}} {<A(-[\wba-C])>_{\pi_B^c}}
\label{eq:FGARgen}
\end{equation}\tshow{eq:FGARgen}

\noindent whose weight function $w_n(\wba)$ is given by:

\begin{equation}
w_n(\wba)  = A(\wba - C) \hat{P}(\wba | \pia)
\end{equation}

\noindent It then follows that:

\begin{equation}
w_n(\wba) = \left\{\begin{array}
{r@{\quad}l}
e^C \RBAcal \hat{P}(\wba | \pib) \mbox {\  for $\wba > C$}\\
e^C \RBAcal \hat{P}(\wba | \pib) = \hat{P}(\wba | \pia)  \mbox {\  for $\wba = C$}\\
\hat{P}(\wba | \pia)  \mbox {\  for $\wba < C$}
\end{array}\right.
\label{eq:accdecom3}
\end{equation}\tshow{eq:accdecom3}

\begin{figure}[tbp]
\begin{center}
\rotatebox{0}{
\includegraphics[scale=0.6]{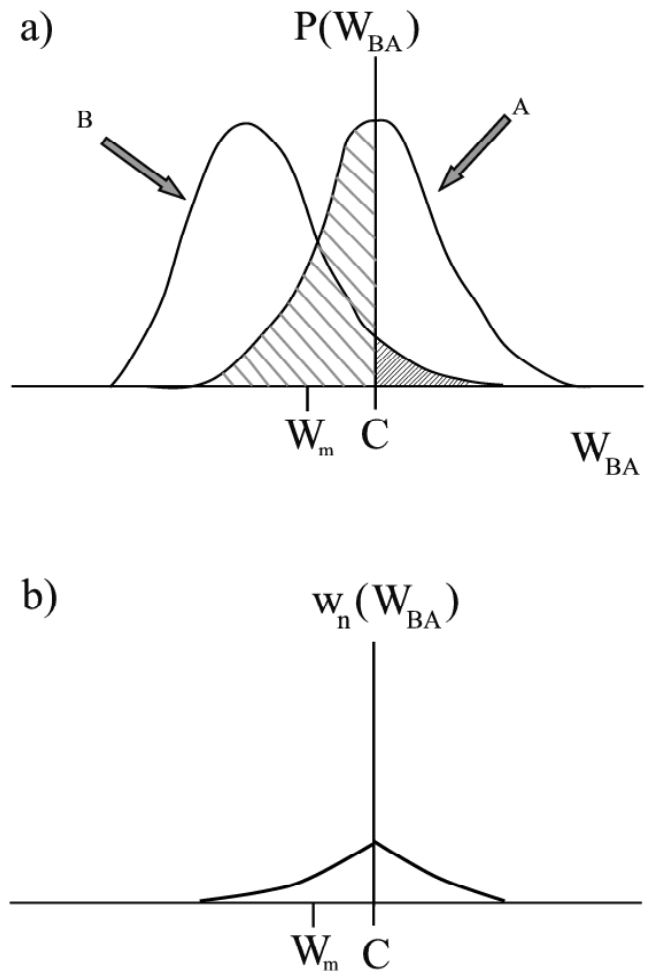}
}
\end{center}
\caption{
The weight function $w_n(\wba)$ for a given C.
}
Figure (a) shows the portions of the distributions which contribute to the estimate of the FED when C is displaced from $W_m$. 

Figure (b) shows the resulting weight function $w_n(\wba)$ for the AR method (see Eq. \ref{eq:accdecom3}). 

\begin{center}
{\bf{------------------------------------------}}
\end{center}

\tshow{pic:overlappingc}
\label{pic:overlappingc}
\end{figure}

\noindent Referring to figure \ref{pic:overlappingc} (a) and (b), it is clear from Eq. \ref{eq:accdecom3} that if C lies within the overlapping region, then the regions which contribute most significantly to the estimators of the expectations appearing in the numerator and denominator of Eq. \ref{eq:FGARgen} are those  over which $\hat{P}(\wba | \pi^c_A)$ and $\hat{P}(\wba | \pi^c_B)$ overlap. 
In the case where the FEDs are small, it suffices to use the original AR formula Eq. \ref{eq:FGAR}, since $W_m \approx 0$.

Therefore the AR method is only free of systematic errors for choices of C lying within the overlapping regions. If the value of C lies outside this region, it is not hard to see that the regions of \ecs\ which contribute the most significantly to Eq. \ref{eq:FGARgen} will no longer  be contained entirely within the region of overlap. As a result restrictions will have to be imposed on the corresponding estimator (for reasons mentioned in section \ref{sec:elimsys}) in order to guarantee that the associated estimate of \RBAcal\ is free of systematic errors.
We will now discuss another estimator which, like the AR method, has its most significant contributions originating from the region of overlap and which is 
unique in that it is the estimator for which the statistical variance is a minimum.

\subsubsection{Minimising the statistical errors : Bennett's fermi function estimator}

The AR formula is but one of a family of estimators for which the regions which contribute most significantly correspond to  the overlapping regions.
The question that one may now proceed to ask is which, out of the family of these estimators, is the one whose corresponding estimator for $\triangle F_{BA}$ is of minimum statistical variance \cite{note:statest}. The task of finding a minimum-variance estimator has been tackled by Bennett \cite{p:bennett}. We will now present his estimator within the more general context of the fast growth (FG) method. Consider the following choice for G:

\begin{equation}
G(\wba -C) = f(-[\wba -C] )
\label{eq:gmin}
\end{equation}\tshow{eq:gmin}

\noindent where f is the fermi function. Substitution of Eq. \ref{eq:gmin} into Eq. \ref{eq:freebennettgen} yields \cite{p:bennett}, \cite{p:radmer},  \cite{p:shirts}\nocite{p:meng}-\cite{note:bderi}, \cite{note:lala}:

\begin{equation}
\RBAcal = e^{-C} \frac {<f(\wba - C)>_{\pi^c_A}} {<f(-[\wba - C])>_{\pi^c_B}}
\label{eq:fermiform}
\end{equation}\tshow{eq:fermiform}

\noindent where we have used the fact that:

\begin{equation}
\frac {f(x)} {f(-x)} = e^{-x}
\end{equation}

\begin{figure}[tbp]
\begin{center}
\rotatebox{0}{
\includegraphics[scale=0.5]{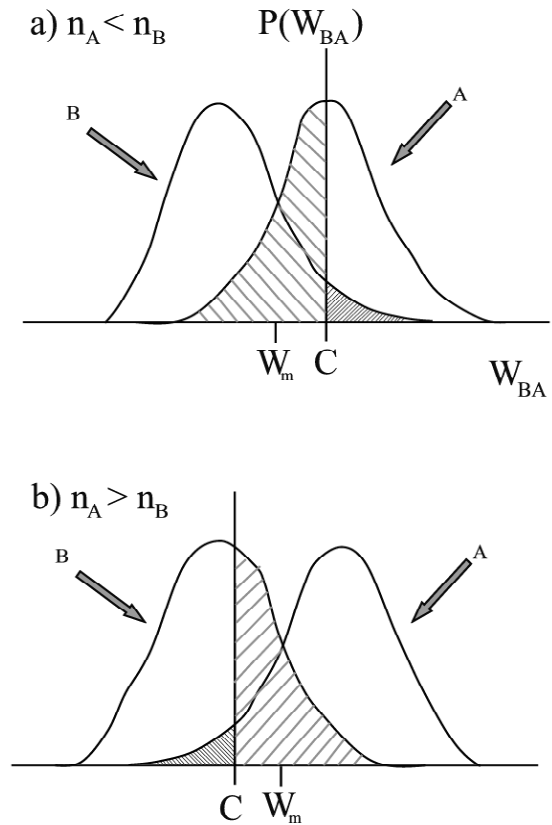}
}
\end{center}
\caption{Schematic illustrating the idea behind Bennett's prescription for C (Eq. \ref{eq:daconst})}

According to Bennett \cite{p:bennett} the optimal estimator is given by Eq. \ref{eq:fermiform}, where C is given by Eq. \ref{eq:daconst}. This choice of C has a physical interpretation. In short, when phase A is sampled less well than phase B, then the choice of C ensures that one includes an increasing proportion of the distribution of phase A in estimating the FED in order to compensate for the increased statistical errors associated with the estimator $\hat{P}(\wba | \pia)$. This is illustrated schematically in figure (a). On the other hand when phase B is less well sampled, then the opposite is done, so that one includes a greater proportion of  $\hat{P}(\wba | \pib)$. This is shown in figure (b).

\begin{center}
{\bf{------------------------------------------}}
\end{center}

\tshow{pic:bennettadjust}
\label{pic:bennettadjust}
\end{figure}

\noindent  Bennett showed that the choice of G and of C which lead to a minimum variance estimator of  $\triangle F_{BA}$ is that of Eq. \ref{eq:fermiform} in which C is set to be:

\begin{equation}
C = W_m - \ln \frac {n_A} {n_B}
\label{eq:daconst}
\end{equation}\tshow{eq:daconst}

\noindent and where $n_\al$ are the number of independent data samples obtained in phase \al. Eq. \ref{eq:daconst} has the following simple physical interpretation. When $n_A<n_B$, then $C>W_m$ so that (with respect to the case $C=W_m$) increasing amounts of the tail of $\hat{P}(\wba | \pia)$ are included in the contributions that come from the overlap region, whilst a smaller proportion of  the tail of $\hat{P}(\wba | \pib)$ is included. The reason  for this is that the statistics of the tail of  $\hat{P}(\wba | \pia)$ will be a lot worse than that of  $\hat{P}(\wba | \pib)$, and therefore it makes sense to take contributions from a larger proportion of its tail  and a smaller proportion of the tail of  $\hat{P}(\wba | \pib)$. When $n_A>n_B$ the opposite is true. That is since C is now less than $W_m$, a larger proportion of the tail of  $\hat{P}(\wba | \pib)$ is taken into account, whereas a smaller proportion of the tail of $\hat{P}(\wba | \pia)$ contributes, thus balancing the fluctuations contributed  by  $\hat{P}(\wba | \pia)$ and $\hat{P}(\wba | \pib)$ in the estimate of the FED. This is illustrated in figure \ref{pic:bennettadjust}

In order to estimate the FED from Eq. \ref{eq:fermiform} and Eq. \ref{eq:daconst} Bennett's prescription involves iteratively solving the set of equations:

\begin{equation}
\hat{W}_m = -\ln \{ \frac {\s {i=1} {B} H(\wbai | \pi^c_A) f(\wbai - C)} {\s {i=1} {B} H(\wbai | \pi^c_B) f(-[\wbai - C])} \} + \ln \frac {n_A} {n_B} + C
\label{eq:recurse1}
\end{equation}\tshow{eq:recurse1}

\noindent and:

\begin{equation}
C = \hat{W}_m - \ln \frac {n_A} {n_B}
\label{eq:recurse2}
\end{equation}\tshow{eq:recurse2}

\noindent where $\hat {W}_m$ is the estimate for $W_m$ (see Eq. \ref{eq:intersect}). That is one starts off with an arbitrary estimate of $\hat{W}_m$, say unity. One then uses Eq. \ref{eq:recurse2} to calculate a value of C, which one then substitutes into Eq. \ref{eq:recurse1}. From this one obtains a new estimate $\hat{W}_m$ which one then substitutes back into Eq. \ref{eq:recurse2} to get yet another value of  C. This value of C is then fed back into Eq. \ref{eq:recurse1} and one continues this procedure until convergence is obtained. That is the process is carried out in an iterative fashion until the value of $\hat{W}_m$ obtained from Eq. \ref{eq:recurse1} for a particular value of C also agrees with Eq. \ref{eq:recurse2}. At this point $\hat{W}_m$ yields (in the case of partial overlap between  $\hat{P}(\wba | \pia)$ and $\hat{P}(\wba | \pib)$) a minimum variance unbiased estimate of the true value of $W_m$:

\begin{equation}
W_m \est \hat{W}_m
\end{equation}

\noindent Like the generalised AR formula (Eq. \ref{eq:FGARgen}), no restrictions are required (when C does not differ too greatly from $\hat{W}_m$) in order to ensure that the associated estimate is free of systematic errors (in the case of partial overlap). To see this we first note that the weight function  $w_n(\wba)$ is given by:

\begin{eqnarray}
w_n (\wba) & = & \frac {\hat{P}(\wba | \pi^c_A)} {1+e^{\wba - C}}\nonumber\\
& = & \frac {\hat{P}(\wba | \pi^c_A)} {1+ \frac {1} {e^C \RBAcal} \frac {\hat{P}(\wba | \pi^c_A)} {\hat{P}(\wba | \pi^c_B)}}\nonumber\\
& = & \frac {\hat{P}(\wba | \pi^c_A) \hat{P}(\wba | \pi^c_B)} {\frac {\hat{P}(\wba | \pi^c_A)} {e^C \RBAcal} + \hat{P}(\wba | \pi^c_B)}
\label{eq:fweight1}
\end{eqnarray}\tshow{eq:fweight1}

\noindent where in going from the first to the second line we have employed the overlap identity (Eq. \ref{eq:crooksfluct}).

 From Eq. \ref{eq:fweight1} we see that (provided C does not differ too greatly from $W_m$, so that $e^C \RBAcal \sim 1$)  the regions of \wba\ space for which the weight function $w_A(\wba)$ is most significant are those regions over which $\hat{P}(\wba | \pi^c_A)$ {\em and} $\hat{P}(\wba | \pi^c_B)$ overlap (the shaded region of figure \ref{pic:overlapping}). 
Therefore like the generalised AR formula (Eq. \ref{eq:FGARgen}), the fermi function (FF) formula (Eq. \ref{eq:fermiform}) is free of systematic errors (provided C does not differ too greatly from $W_m$).

Let us now analyse the case where C differs significantly from $W_m$. If $e^C \RBAcal$ is considerably different from unity then  it follows that the weights $w_n(\wba)$ may be approximated by:

\begin{equation}
w_n(\wba) \propto \left\{\begin{array}
{r@{\quad:\quad}l}
\hat{P}(\wba | \pia) & \mbox{\ if $e^C \RBAcal >> 1$}\\
\hat{P}(\wba | \pib) & \mbox{\ if $e^C \RBAcal << 1$}
\end{array}\right.
\end{equation}

\noindent Therefore if C is too large, then the \wba\ macrostates contributing to the estimators of Eq. \ref{eq:fermiform} correspond to  the regions of \wba\ space for which $\hat{P}(\wba |\pi^c_A)$ is significant. In this case the estimator of Eq. \ref{eq:fermiform} will not yield an estimate free of systematic errors, since the contributing regions no longer come from the regions of overlap, and the necessary steps outlined in section \ref{sec:elimsys} will need to be taken. Likewise if C is too negative, then the important regions will be those for which $\hat{P}(\wba |\pi^c_B)$ is significant, once again leading to systematic errors. From this analysis we also see that in the limit of the number of independent samples obtained in the two phases ($n_A$ and $n_B$) becoming very disparate, the Bennett prescription for constructing the optimal C, given by Eq. \ref{eq:daconst}, will lead to systematic errors for the reasons that we have just mentioned. In his paper \cite{p:bennett} however, Bennett does advocate the use of an equal number of independent samples in each phase, so that $C = W_m$. This choice leads to the contributions coming from the regions of overlap, resulting in the estimator of Eq. \ref{eq:fermiform} yielding estimates which are free  of systematic errors. Further insight into Bennett's approach (Eq. \ref{eq:fermiform} and Eq. \ref{eq:daconst}) can be obtained by noticing the links that exist between the method and the task of estimating the overlap parameter \ov\ (Eq. \ref{eq:overlap}). We refer the interested reader to appendix \ref{app:bennett} for the relevant discussion.

\section{\label{sec:fgps}Phase switch estimator}
\tshow{sec:fgps}

In section \ref{sec:pfg} we saw how the PS method could be generalised so as to be applicable within the framework of the (arbitrary equilibration) FG method. In order for the method to work $\hat{P}(\wba | \pia)$ and $\hat{P}(\wba | \pib)$ should be non-zero for the  $\wba\sim 0$ regions. It is only in this case that a simulation initiated in either of the phases will be able to reach the $\wba\sim 0$ regions, from which it will have a non-negligible chance of switching phases. In the general case where the FED differs significantly from $0$, the estimators $\hat{P}(\wba | \pia)$ and $\hat{P}(\wba | \pib)$ will lie to one side of the axis (see figure \ref{pic:overlapping}). In this case one might find that the $\wba \sim 0$ regions are not visited, thus preventing the simulation from switching phases. As we will now show,\ {\em provided $\hat{P}(\wba | \pia)$ and $\hat{P}(\wba | \pib)$ overlap}, one may make a slight modification to the method so as to allow it to switch phases. The basic idea is to weight the two phases so as to increase the probability of the simulation visiting the phase with smaller partition function (or larger free energy). We are essentially performing the ST tempering for the case of two sub-ensembles, within the more general context of the FG method.

We recap that in its most general form the \wba\ distribution of the PS method may be written as:

\begin{equation}
P(\wba | \pips) = P(\wba , \zeta_{A \rightarrow B} | \pips) + P(\wba ,  \zeta_{B \rightarrow A} | \pips)
\label{eq:psand}
\end{equation}\tshow{eq:psand}

\noindent where:

\begin{equation}
P(\wba , \zeta_{\al \rightarrow \alp} | \pips) = P(\wba | \pi^c_\al) P(\zeta_{\al \rightarrow \alp}, \pips)
\label{eq:psand2}
\end{equation}\tshow{eq:psand2}

\noindent In the particular version employed in Eq. \ref{eq:mbadist}, we have $P(\zeta_{\al \rightarrow \alp}, \pips) = \z_{\al}$. In general, this quantity may be arbitrary. It then follows that the acceptance probability for switching phases (strictly  processes) is given by:

\begin{eqnarray}
P_a(\zeta_{A\rightarrow B} \rightarrow \zeta_{B\rightarrow A} | \pi^c_{PS}) & = & \mbox{Min} \{ 1, \frac {P(\wba ,  \zeta_{B \rightarrow A} | \pips)} {P(\wba ,  \zeta_{A \rightarrow B} | \pips)}\}\nonumber\\
& = & \mbox{Min} \{ 1, \frac {e^{-\wba}} {\RBAcal} \frac {P(\zeta_{B\rightarrow A}|\pips)} {P(\zeta_{A\rightarrow B}|\pips)}
\label{eq:fgps0}
\end{eqnarray}\tshow{eq:fgps0}

\noindent Since \RBAcal\ is unknown a-priori, it is convenient to factor it out of the acceptance probability of Eq. \ref{eq:fgps0}. Therefore we conveniently write $P( \zeta_{\al \rightarrow \alp} |\pips)$ as:

\begin{equation}
P(\zeta_{\al \rightarrow \alp} | \pips) = \frac {w_\al \z_\al} {w_\al \z_\al + w_\alp \z_\alp}
\label{eq:fgprob}
\end{equation}\tshow{eq:fgprob}

\noindent where $w_\al$ are some arbitrary  weights,  which {\em are} known a-priori. Substituting Eq. \ref{eq:fgprob} into  Eq. \ref{eq:fgps0} we obtain:

\begin{equation}
P_a(\zeta_{A \rightarrow B}\rightarrow \zeta_{B \rightarrow A} | \pips) =   \mbox{Min} \seta {1, e^{-W_{BA} + \ln \{w_B / w_A\}}}
\label{eq:estfgsw}
\end{equation}\tshow{eq:estfgsw}

\noindent or:

\begin{equation}
P_a(\zeta_{\al \rightarrow \alp}\rightarrow \zeta_{\alp \rightarrow \al} | \pips) = \left\{\begin{array}
 {r@{\quad:\quad}l}
\mbox{Min}\{1, e^{-\wba + C}\} & \mbox{\ if \al =A}\\
\mbox{Min}\{1, e^{\wba - C}\} & \mbox{\ if \al =B}
\end{array}\right.
\label{eq:fgar}
\end{equation}\tshow{eq:fgar}

\noindent where:

\begin{equation}
C = \ln \frac {w_B} {w_A}
\label{eq:fermipsconst}
\end{equation}\tshow{eq:fermipsconst}

\noindent Running the argument in reverse it follows that if one adopts the PS acceptance probability given in Eq. \ref{eq:estfgsw}, then the absolute probability of finding the simulation in the $\palalp$ process is given by Eq. \ref{eq:fgprob}.  Therefore if one implements a FG-PS simulation in which the phase (or process) switching probabilities are given by Eq. \ref{eq:fgar} then \RBAcal\ may be estimated via:

\begin{eqnarray}
\RBAcal & = & \frac {w_A P(\zeta_{B \rightarrow A} | \pips)} {w_B P(\zeta_{A \rightarrow B} | \pips)} \nonumber\\
& = & e^{-C} \frac {<\delta_{\al, B}>_{\pips}} {<\delta_{\al, A}>_{\pips}}
\label{eq:rbafgweight}
\end{eqnarray}\tshow{eq:rbafgweight}

\noindent where:

\begin{equation}
\delta_{\al,\alp} = \left\{\begin{array}
 {r@{\quad:\quad}l}
1 & \mbox{\ if simulation in \palpal\ process}\\
0 & \mbox{\ otherwise}
\end{array}\right.
\end{equation}

\noindent Eq. \ref{eq:rbafgweight} merely expresses the fact that \RBAcal\ may be estimated by the weighted ratio of the times spent in the two processes. An alternative expression for \RBAcal\ may also be found which expresses Eq. \ref{eq:rbafgweight} as an expectation over the macrostates \wba . By substituting Eq. \ref{eq:psand2} into Eq. \ref{eq:psand} and by appealing to the overlap identity (Eq. \ref{eq:crooksfluct}) one  finds that:

\begin{equation}
P(\wba,  \zeta_{\al \rightarrow \alp} |\pips) = \left\{\begin{array}
 {r@{\quad:\quad}l}
\mbox{$[1 + e^{-\wba + C}]^{-1} P(\wba | \pips)$} & \gamma=A\\
\mbox{$[1 + e^{\wba - C }]^{-1} P(\wba | \pips)$} & \gamma=B
\end{array}\right.
\label{eq:tempchest}
\end{equation}\tshow{eq:tempchest}

\noindent Using Eq. \ref{eq:psand2} and Eq. \ref{eq:tempchest} we see that Eq. \ref{eq:rbafgweight} may instead be written as:

\begin{eqnarray}
\RBAcal & = & e^{-C}\frac {\int P(\wba , \zeta_{B \rightarrow A} | \pips) d\wba} { \int P(\wba , \zeta_{A \rightarrow B} | \pips) d\wba}\nonumber\\
& = & e^{-C} \frac {<\int [1 + e^{\wba - C }]^{-1} P(\wba | \pips) d\wba>_{\pips}} {<\int [1 + e^{-\wba + C }]^{-1} P(\wba | \pips) d\wba>_{\pips}}
\end{eqnarray}

\noindent or:

\begin{equation}
\RBAcal = e^{-C} \frac {<f(\wba - C)>_{\pips}} {<f(-[\wba - C])>_{\pips}}
\label{eq:psfermi}
\end{equation}\tshow{eq:psfermi}

\noindent The close resemblance of this estimator with that of the FF method (Eq. \ref{eq:fermiform}) is striking. However the estimator of Eq. \ref{eq:psfermi} is markedly different from that of Eq. \ref{eq:fermiform} in one respect. To see this consider the weight function for the numerator and denominator of the estimator of Eq. \ref{eq:psfermi}:

\begin{eqnarray}
w_n(\wba) & = & f(\wba - C) \hat{P}(\wba | \pips) = \hat{P}(\wba , \zeta_{B \rightarrow A} | \pips)\nonumber\\
& \propto & \hat{P}(\wba | \pib)
\label{eq:wpsn}
\end{eqnarray}\tshow{eq:wpsn}

\noindent and:

\begin{eqnarray}
w_d(\wba) & = & f(-[\wba - C]) \hat{P}(\wba | \pips) = \hat{P}(\wba , \zeta_{A \rightarrow B} | \pips)\nonumber\\
& \propto & \hat{P}(\wba | \pia)
\label{eq:wpsd}
\end{eqnarray}\tshow{eq:wpsd}

\noindent We notice that whereas in the case of the  DP estimators (Eq. \ref{eq:freebennett}) the weights $w_n$ and $w_d$ are directly proportional to each other (see Eq. \ref{eq:accdecom5}), in  the case of the PS estimator they are not. For the DP methods, the contributions to the estimators of the expectations appearing in the numerator and denominator come from the same region of \ecs , though the sampling distributions actually employed are different. In the case of the PS method one employs the {\em same} sampling distribution for the two expectations; though  now the contributions to the two expectations come from different regions of \ecs . Whereas the DP methods can prevent the appearance of systematic errors by ensuring that the non-negligible contributions come from the region of overlap, the PS method avoids systematic errors by actually switching phases and separately sampling each phase. As with the DP methods the correct choice of C must be made in order for the PS method to work.

In order to address the choice of C we note that if the weights $w_A$ and $w_B$ are the same for the two phases (so that C=0, Eq. \ref{eq:fermipsconst}), as is the case in the original PS formulation (see section \ref{sec:phaseswitch} and section \ref{sec:pfg}), and if the region over which the two \pc s overlap (see figure \ref{pic:overlapping}) is sufficiently displaced from the origin, then it is clear that the PS sampling distribution will not be able to successfully switch between the phases {\em in both directions}. The way to remedy this is to choose a C which lies within the region of overlap (see figure \ref{pic:overlapping}). In particular if one chooses:

\begin{equation}
C=W_m
\label{eq:enablefgps}
\end{equation}\tshow{eq:enablefgps}

\noindent then from Eq. \ref{eq:rbafgweight} one finds that:

\begin{equation}
\frac {<\delta_{\al, B}>_{\pips}} {<\delta_{\al, A}>_{\pips}} = 1
\label{eq:equaltime}
\end{equation}\tshow{eq:equaltime}

\noindent so that the simulation spends an equal time in the two phases. By setting C as prescribed in Eq. \ref{eq:enablefgps} what one does is to effectively bias the phase with the larger free energy, so as to increase the probability with which the simulation visits it, as compared to the case where C is set to unity.

\section{\label{sec:numiest}Numerical results}
\tshow{sec:numiest} 

We saw earlier in section \ref{sec:elimsys} how any estimator can, in principle, be modified by imposing appropriate 'restrictions' so as to guarantee that the resulting estimate of the FED is free of systematic errors when there is some overlap between the estimators of the phase-constrained distributions $P(\wba | \pia)$ and $P(\wba | \pib)$.  In this section we illustrate the application of these restrictions in the case of the EP estimator and compare the resulting statistical errors to those of the PS, AR, FF, and EP methods.

In order compare the estimators we used the same simulation setup as that used to obtain the data of figure \ref{fig:anharfree} \cite{note:mbacon}. This represents the rather uninteresting case of estimating \RBAcal\ when its assumes a value of approximately unity. However it is useful for the reason that, since the value of \RBAcal\ hardly changes for the range of temperatures investigated, one may effectively probe the behaviour of the statistical errors purely as a function of the overlap; the overlap being  changed simply by varying  the temperature.

We start by recalling (see section \ref{sec:elimsys}) that in the case of $G(\wba)=1$ the form of Eq. \ref{eq:dualphaserestrict} may be simplified by discarding the upper limit $W^1_{BA}$. Furthermore since \RBAcal\ (and therefore $W_m$) varies negligibly over the range of conditions investigated here (see the results of the FSM-PS method in figure \ref{fig:anharfree}), it is convenient to set $W^0_{BA}$ to $W_m$. The result is that in the case of zero equilibration Eq. \ref{eq:dualphaserestrict} reduces to:

\begin{equation}
\RBAcal = \frac {<e^{-\wba}H(\wba-W_m)>_{\pi^c_A}} {<H(\wba-W_m)>_{\pi^c_B}}
\label{eq:restbz}
\end{equation}\tshow{eq:restbz}

\noindent where:

\begin{equation}
H(x) =\left\{\begin{array}
{r@{\quad:\quad}l}
1 & \mbox{$x>0$}\\
0 & \mbox{otherwise}
\end{array}\right.
\end{equation}
 
\noindent Using the fact that $W_m \approx 0$ for the conditions investigated here, we see that in the case of zero equilibration Eq. \ref{eq:restbz} simplifies to:

\begin{equation}
\RBAcal = \frac {<e^{-\mba}H(\mba)>_{\pi^c_A}} {<H(\mba)>_{\pi^c_B}}
\label{eq:restbz0}
\end{equation}\tshow{eq:restbz0}

\noindent We note that though Eq. \ref{eq:restbz0} does omit some of the region of (effective) configuration space over which the two phase-constrained distributions overlap, and therefore has greater statistical errors than it would if all the regions of overlap were included, its advantage lies in its simplicity and in the fact that Eq. \ref{eq:restbz0} may be used for the spectrum of overlaps investigated without requiring one to modify the restrictions as the overlap changes. It is for this reason that we will use Eq. \ref{eq:restbz0} in our comparison of the estimators.

\begin{figure}[tbp]
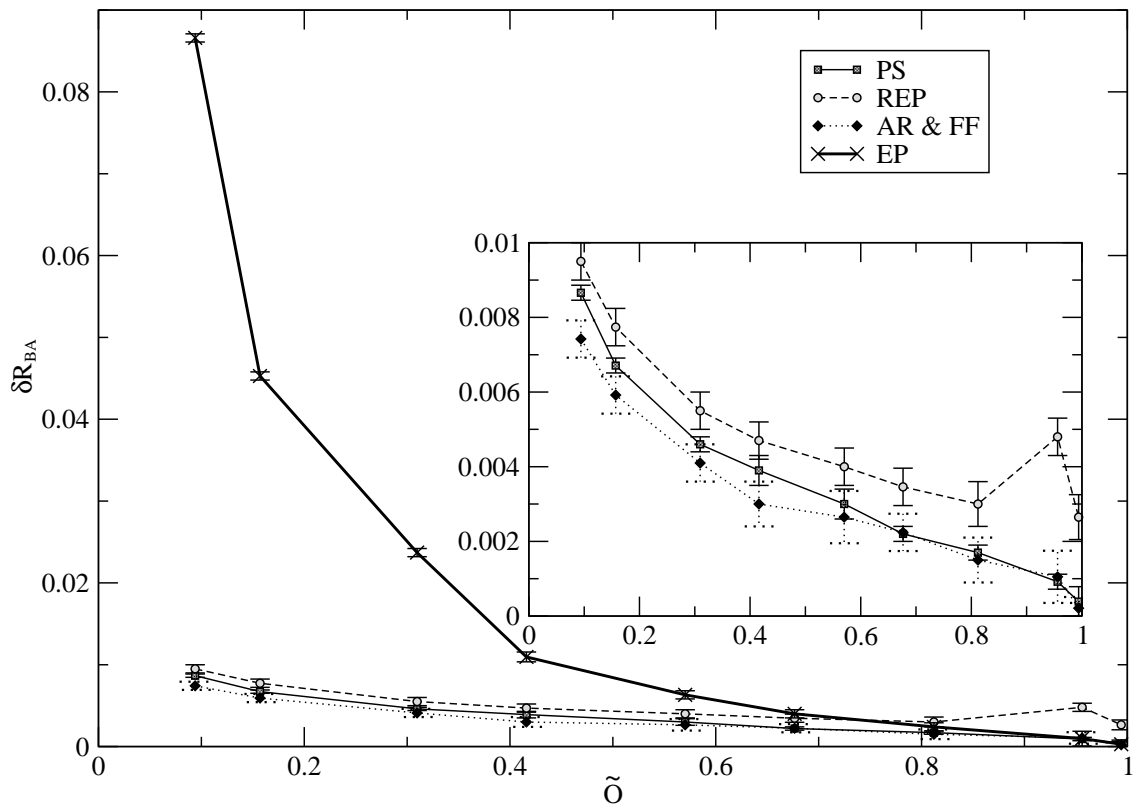

\begin{center}
\ifpdf
\rotatebox{90}{
\hidefigure{\includegraphics[scale=0.6]{fig/overlap_vs_errors_final}}
}
\else
\rotatebox{0}{
\hidefigure{\includegraphics[scale=0.6]{fig/overlap_vs_errors_final}}
}
\fi
\end{center}
\caption{
Errors in \RBAcal\ as a function of the overlap parameter \ov
}

The same amount of computational resources were allocated to all the simulations. The range of the temperatures employed here were the same as those in figure \ref{fig:anharfree}, and the FSM was employed. We tested 5 estimators: the AP, EP, FF, REP, and PS estimators. C=0 for all the estimators. The results of the AR and FF estimators were identical, and are denoted by a single line. \ov\ is given by Eq. \ref{eq:estov}.

All the simulations were zero-equilibration FG simulations (i.e. where $\wba=\mba$).

\begin{center}
{\bf{------------------------------------------}}
\end{center}

\label{fig:overlaperror}
\tshow{fig:overlaperror}
\end{figure}

Figure \ref{fig:overlaperror} shows the statistical errors (and the associated errors of the errors) in the estimates of \RBAcal\ for the different estimators as a function of the overlap parameter \ov . In comparing the different estimators we  once again use the PS estimator as the benchmark. The first observation that we make is that the AR and the FF estimators (the latter of which is not shown in figure \ref{fig:overlaperror} since its results were identical to those of the AR method) yielded statistical errors of roughly the same size as those of the PS method for the whole range of overlaps $\ov$ investigated. The EP estimator, on the other hand, yielded a markedly different behaviour to that of the  PS estimator. 

For high values of overlap \ov\ the EP estimator clearly yields roughly the same statistical errors as those associated with the PS estimator, whereas for low overlaps the errors of the EP estimator are significantly greater. This may be qualitatively understood by noting that for high overlaps, the distributions $P(\mba | \pia)$ and $P(\mba | \pib)$ sufficiently overlap so as to ensure that the statistics of the regions which contribute to {\em both} the numerator and denominator of Eq. \ref{eq:estep} are good under a sampling experiment constrained to a single phase. In the case of the PS method, the statistics of the macrostates relevant to the numerator and the denominator (Eq.\ref{eq:ratiooftimes}) are good because the method (by switching phases) separately visits the regions of \ecs\ associated with the two phases. 

As \ov\ decreases, the two \pc s increasingly separate (see figure \ref{fig:ctmswitch}) until the point is reached where the main body of $P(\mba|\pib)$ resides in the tail of  $P(\mba | \pia)$. Under these conditions even though systematic errors will not be present for the EP estimator, the statistical errors will be greater than those associated with the  PS estimator since now the macrostates which contribute to the numerator of Eq. \ref{eq:estep}   will be visited with a small probability, even though they contribute significantly to the estimate of the FED. On the other hand the PS method visits macrostates with probability in direct proportion to their contribution to the relevant estimator (see Eq. \ref{eq:mbadist}). As a result the statistical errors of the PS estimator in this case will be lower than that of the EP method. For overlaps even lower than this, the systematic errors will begin to set in for the EP estimator, since the regions associated with {\em both} phases will not be visited (as is required) by a simulation constrained to a single phase, even though systematic errors will not be present for the PS method (see also section \ref{sec:fsssimulation}). For the experiments conducted here, it was found that systematic errors begin to set in for the EP estimator for overlaps below $\ov =0.3$.

In contrast to the EP estimator, the estimator of the REP formula (Eq. \ref{eq:restbz0}) does not suffer from any systematic errors for the whole range of overlaps investigated, and its statistical errors were only marginally greater than those of the AR, PS, and FF estimators. The main reason for the increased statistical errors is because the restriction, as imposed in Eq. \ref{eq:restbz0}, excludes some of the overlapping region. That is some of the 'useful' contributions are unnecessarily discarded, resulting in slightly higher statistical errors. It is also for this reason that at high overlaps the error of the EP estimator falls {\em below} that of the  REP estimator. We stress that   this property of the REP estimator is merely an artifact of the particular version of the restrictions we employ in  Eq. \ref{eq:restbz0} and that, upon inclusion of all the region of overlap, the statistical errors should fall to roughly those  of the other methods.  

We finally note that in between the high and low overlap regimes, there is a range of overlaps for which the EP method is free of systematic errors and yet for which the statistical errors are greater than those of the REP method, 
despite the fact that the particular version of the REP that we use discards some of the data originating from the overlap region. In figure \ref{fig:overlaperror} this roughly corresponds to the range $0.3<\tilde{O}<0.8$. To understand this we recall that as the overlap is increased, systematic errors will disappear for the EP method on the onset of the main body of ${P}(\mba | \pib)$ being contained in the tail of ${P}(\mba | \pia)$. However in this regime the statistical errors in the estimate of \RBAcal\ will be large since the statistics of the regions over which ${P}(\mba | \pib)$ is significant will be poor, since this is contained in the tail of  ${P}(\mba | \pia)$, thus offsetting any gains it has over the REP method. This drawback of the EP method will reduce as the overlap increases, until eventually the EP method becomes more efficient than the version of the REP formula that we use.

From this numerical study it is clear that statistical considerations lead one to the conclusion that in the case of partial overlap (see figure \ref{pic:twooverlap} (b) and \ref{pic:overlapping}), estimators which involve the accumulation of data from both the phases (e.g. the PS, AR, FF, and REP estimators) are preferable to those that involve estimators which use the data acquired from a single phase (namely EP method). Out of these estimators, the PS, FF, and AR method are preferable (to the REP method) since one does not need to expend the additional effort of restricting these estimators (as is done in Eq. \ref{eq:dualphaserestrict}). In comparing the AR, FF, and PS methods we note that the PS method requires a single simulation to estimate \RBAcal , whereas the FF and AR methods both require two separate simulations. This latter property can be viewed as an advantage for both groups of methods. On the one hand it affords the FF and the AR method an avenue for parallelisation which is not available to the PS method, since the two phase-constrained simulations may be performed independently. On the other hand the ability of the PS method to estimate \RBAcal\ from data extracted from a single simulation makes it, in some sense, tidier. Another important difference is that the adjustment of  C, so as to yield an estimate of \RBAcal\ which is free of systematic errors when there is partial overlap, have to be made {\em before} the simulation is run in the case of the PS method. In the case of the FF and AR  methods, these adjustments are made {\em after} the simulation is run, when one is trying to estimate \RBAcal\ from the data already obtained; this  may be easily automated. Since a-priori we do not know where the region of overlap is, it is clear that in this case the FF and AR methods have an advantage over the PS method. This, however, is not a significant advantage since one may run  two short simulations, one in each phase, in order to roughly determine the point $M_m$ where  $P(\mba | \pia)$ and $P(\mba | \pib)$ intersect, thereby yielding an appropriate value of C (see Eq. \ref{eq:enablefgps}).

\section{Conclusion}

We saw in section \ref{sec:overlap} that central to ones ability to estimate the FED is the concept of overlap between the estimators  $\hat{P}(\wba | \pi^c_A)$ and $\hat{P}(\wba | \pi^c_B)$ of the phase-constrained distributions. For systems characterised by a (effective) configuration space structure as shown in figure \ref{pic:twooverlap} (b), successful estimators based on the sampling of the \pc s will have their most significant contributions originating from the region of overlap \cite{p:bennett},  \cite{p:kofkecummings}, \cite{p:radmer}, \cite{p:torrievalleau}, \cite{p:lusinghkofke}. The way that we have realised this idea is by appreciating that all estimators are based on the overlap identity (Eq. \ref{eq:crooksfluct}). Since the corresponding estimator of the overlap identity, Eq. \ref{eq:crooksest}, is itself only valid within the region of overlap, we see that estimators which are free of systematic errors can {\em only} have their  non-negligible contributions coming from this region. In a sense one may think of these estimators as {\em pooling} together the estimates of \RBAcal , as made by Eq. \ref{eq:crooksest}, from  within the region of overlap (see figure \ref{pic:estov}). 

An alternative, and equally suitable,  strategy to the phase-constrained simulations is the PS strategy \cite{p:LSMC} in which one actually switches between phases (or more generally between $A\rightarrow B$ and $B\rightarrow A$ processes in the case of the arbitrary equilibration FG method). In this case the method overcomes the problem presented by partial overlap by actually switching  between the phases (or processes) thereby sampling each phase (process) separately.

Generally however, the scope for refinement of the estimator is limited. In the absence of overlap one must resort to some sort of extended sampling strategy \cite{p:iba} in which one {\em engineers} overlap by forcing the simulation to visit regions of \ecs\ which it would not otherwise sample (under the influence of the canonical sampling distribution Eq. \ref{eq:sampconstrain}). This, after the choice of representation (chapter \ref{chap:tune}) and the choice of estimator (the present chapter), forms the final part of the overall strategy of tackling the overlap problem. This topic of discussion will form the core of the next chapter.

\chapter{\label{chap:sampstrat}Sampling Strategies}
\tshow{chap:sampstrat}

\section{Introduction}

In the case where one is unable to construct a PM which ensures some overlap between the two \pc s, one must {\em engineer} overlap by refining the sampling strategy \cite{p:iba}.  Studies until now have focused on methods which fall into one of three broad categories:

\begin{enumerate}
\item They sample from some form of \es\ distribution \cite{p:iba} and extract the FED via an appropriate reweighting scheme (see Eq. \ref{eq:reweightav}). The \es\ strategy involves the employment of a non-canonical sampling distribution so as to allow the simulation  to visit  wider regions of \ecs\ than it normally would under the canonical distributions. Such methods include Umbrella sampling \cite{p:valleaucard}, \cite{p:torrievalleau}, \cite{p:tvb}\nocite{p:tvb2}-\cite{p:torrievalleaucpl},  PS method \cite{p:LSMC}, \cite{p:LSMCprehard}\nocite{p:LSMCpresoft}\nocite{p:solidliquid}-\cite{p:LSMCerrington}, Simulated Tempering \cite{p:nezbedakolafa}\nocite{p:lyubartsev}\nocite{p:lyubartsevgeneral}-\cite{p:simulatedtempering},  and the Weighted Histogram Analysis Method \cite{p:radmer}, \cite{p:weightedhistogram}\nocite{p:weightedhistogram2}\nocite{p:souailleroux}-\cite{p:sugita}. 

\item The fine tuning of the Fast Growth (FG) Method \cite{p:jarprl}, \cite{p:crooks}\nocite{p:crookspre}\nocite{p:crookspre2}-\cite{p:hummerszabo}, \cite{p:sun}. By making the incremental perturbations to the configurational energy (which constitute the work elements of the process, Eq. \ref{eq:totalwork}) sufficiently small and by choosing sufficiently long equilibration times between these work elements, this method allows for the engineering of overlap between $P(\wba | \pi^c_A)$ and $P(\wba |\pi^c_B)$ (see section \ref{sec:asfg}).

\item They split the calculation of \RBAcal\ (Eq. \ref{eq:RBA}) into many small and separate FED calculations, between pairs of systems whose \pc s overlap considerably better than that exhibited by the original pair of systems. These methods are generally referred to as the multistage (MS) methods \cite{p:valleaucard}, \cite{p:rahmanjacucci}, \cite{p:moodyray}, \cite{p:torrievalleau}, \cite{p:tvb}\nocite{p:tvb2}-\cite{p:torrievalleaucpl}, \cite{p:jorgensen}. In the limit of an infinite number of stages we arrive at the thermodynamic integration method (Eq. \ref{eq:MSTI}) \cite{p:frenkelladd}\nocite{p:fleischman}-\cite{p:straatsma}.
\end{enumerate}

In this chapter we will study these three strategies in the following manner. First we will deal with point 1 by  showing how the EP, AR, and PS methods can be made to work by appealing to the MUCA \es\ strategy, as described in section \ref{sec:umbrella}. The generalisation of this strategy to the case of an arbitrary estimator (Eq. \ref{eq:dual}) is straightforward.
Following this we construct a new way of estimating the FED in which one employs a series of parallel simulations (as is the case for  the WHAM method, section \ref{sec:wham}). In certain limiting cases this new method may be thought of as a realisation of both the FG method (point 2) and the MS method (point 3). 
In the final part we illustrate point 2 by applying the FG method to the model systems under consideration here (see section \ref{sec:system}), and show  how it overcomes the overlap problem by means of the fine-tuning of the relevant parameters (see sections \ref{sec:fg} and \ref{sec:asfg}).

\section{\label{sec:mul}The Multicanonical strategy}
\tshow{sec:mul}

The MUCA strategy \cite{note:wheremuca} is a {\em serial} strategy (used in the case of zero-equilibration FG simulations, \wba=\mba) and involves, as we saw in section \ref{sec:umbrella} and \ref{sec:phaseswitch} (see figure \ref{schematicpdf}), the 'warping' of the canonical distribution so as to produce the necessary bridging distribution \cite{note:welimit}. 
Suppose that $\pi^c$ and $\pi^m$ denote the canonical and the MUCA sampling distributions respectively. By accepting moves via \cite{note:chanham}:

\begin{equation}
P_a (\vvec \rightarrow \vvec' | \pi^m)= \mbox {Min} \{ 1, \frac {e^{-\beta \ecal(\avvec ') -\eta (\mba(\avvec ')) } } {e^{-\beta \ecal(\avvec ) -\eta (\mba(\avvec )) }} \}
\end{equation}

\noindent one realises:

\begin{equation}
P(\mba | \pi^m) \equaldot P(\mba | \pi^c) e^{-\eta(\mba)}
\label{eq:mucasampch5}
\end{equation}\tshow{eq:mucasampch5}

\noindent In order for $P(\mba | \pi^m)$ to be {\em flat} over the desired regions of \ecs\ one sets:

\begin{equation}
e^{-\eta(\mba)} \propto \frac {1} {P(\mba | \pi^c)}
\label{eq:mulch4}
\end{equation}\tshow{eq:mulch4}

\noindent In this section we will use the Wang-Landau \cite{p:wanglandauprl}  method to obtain the weights (see section \ref{sec:umbrella}),and  we will focus (our discussion) on three estimators, the EP, AR, and PS estimators. Generalisation to the general estimator of Eq. \ref{eq:dual} is straightforward.

\subsection{The Exponential Perturbation estimator}

In section \ref{sec:perturbation} we saw that it was the failure of a simulation constrained to a single  phase to account for the typical configurations of {\em both} phases which ultimately led to the failure of the EP estimator, even in the case of partial overlap. That is a sampling experiment performed in phase A (via \pia ) only samples the macrostates for which the weights $w_d(\mba)$ (see Eq. \ref{eq:epd}) are non-negligible, and fails to capture all the regions of \ecs\ for which $w_n(\mba)$ (see Eq. \ref{eq:epn}) is non-negligible.

The way this problem is remedied (see \cite{p:torrievalleau} and section \ref{sec:umbrella}) is by constructing  a MUCA distribution  $P(\mba | \pi^m_A)$ which contains  $P(\mba | \pi^c_A)$ and  $P(\mba | \pi^c_B)$, so that the simulation visits the regions of \ecs\ associated with {\em both}  phases.

\subsection{The Acceptance Ratio estimator}

Consider the use of the AR formula (Eq. \ref{eq:ar}) in the absence of overlap (see figure \ref{bimodaldistribution}). In this case the MUCA strategy involves the construction of two separate MUCA distributions $\pi^m_A$ and $\pi^m_B$. $\pi^m_A$ has  to sample all the macrostates \mba\ which contribute non-negligibly to $<A(\mba)>_{\pia}$ and  $\pi^m_B$ has to sample those which  contribute non-negligibly to $<A(-\mba)>_{\pib}$. As we saw in section \ref{sec:chooseg} these correspond to the regions of overlap. However, unlike the case where the estimators $\hat{P}(\mba | \pia)$ and $\hat{P}(\mba | \pib)$ overlap, in the case where they do not overlap it is not clear a priori where these regions are. To determine them one may plot a graph of the weight function $w_n(\mba)$ versus \mba , as one constructs the MUCA weights. Once the MUCA distributions $\hat{P}(\mba | \pi^m_A)$ and  $\hat{P}(\mba | \pi^m_B)$ are wide enough so as to contain all the regions over which  $w_n(\mba)$ is non-negligible \cite{note:wnb} one may then proceed to estimate \RBAcal\ via:

\begin{equation}
\RBAcal \est \frac {\s {i=1} {b} A(\mbai) \hat{P}(\mbai | \pi^c_A)} {\s {i=1} {b} A(-\mbai)\hat{P}(\mbai | \pi^c_B)}
\label{eq:arestmuca}
\end{equation}\tshow{eq:arestmuca}

\noindent where:

\begin{equation}
\hat{P}(\mbai | \pi^c_\al) =  \frac {H(\mbai | \pi^m_\al)e^{\eta_\al(\mbai)}} {\s {i=1} {b} H(\mbai | \pi^m_\al)e^{\eta_\al(\mbai)}}
\end{equation}

\subsection{The Phase Switch  estimator}

We saw in section \ref{sec:phaseswitch} that for the PS method \RBAcal\ may be evaluated by appeal to Eq. \ref{eq:firstfermi}. This identity corresponds to a PS simulation in which the probability of switching phases is given by Eq. \ref{eq:psacc}. In this case it is clear that a PS will have a non-negligible chance of being accepted only around the  $\mba \sim O(1)$ regions, and therefore the MUCA sampling distribution should ensure that these regions are visited by the simulation. In this case (see section \ref{sec:phaseswitch}) a suitable MUCA distribution $P(\mba | \pi^m_{PS})$ is one which is  flat and which  contains both  $P(\mba | \pi^c_A)$ and $P(\mba | \pi^c_B)$ \cite{p:LSMC}, \cite{p:LSMCprehard}\nocite{p:LSMCpresoft}\nocite{p:solidliquid}-\cite{p:LSMCerrington}. If $H(\mbai | \pi^m_{PS})$ denotes the number of data entries falling in bin \mbai\ under the MUCA-PS sampling distribution $\pi^m_{PS}$, then the estimator for \RBAcal\ is given by \cite{note:ch4ps}:

\begin{equation}
\RBAcal \est \frac {\s {i=1} {b} f(\mbai) \hat{P}(\mbai | \pips)} {\s {i=1} {b} f(-\mbai) \hat{P}(\mbai | \pips) }
\label{eq:psestfer}
\end{equation}\tshow{eq:psestfer}

\noindent where:

\begin{equation}
\hat{P}(\mbai | \pips) = \frac {H(\mbai | \pi^m_{PS}) e^{\eta_{PS}(\mbai)}} {\s {i=1} {b} H(\mbai | \pi^m_{PS}) e^{\eta_{PS}(\mbai)}}
\label{eq:estcanps}
\end{equation}\tshow{eq:estcanps}

\noindent where $\eta_{PS}(\mba)$ is the associated MUCA weight function. Like the EP method, the essential feature of the MUCA distribution ${P}(\mba | \pi^m_{PS})$ is that it contains both $P(\mba | \pia)$ and $P(\mba | \pib)$, so as to account for {\em all} the regions of (effective) configuration space over which the weights $w_n(\mba)$ (see Eq. \ref{eq:wpsn}) and  $w_d(\mba)$ (see Eq. \ref{eq:wpsd}) are significant.

\subsection{\label{sec:numeri4}Numerical results}
\tshow{sec:numeri4}

\begin{figure}[tbp]
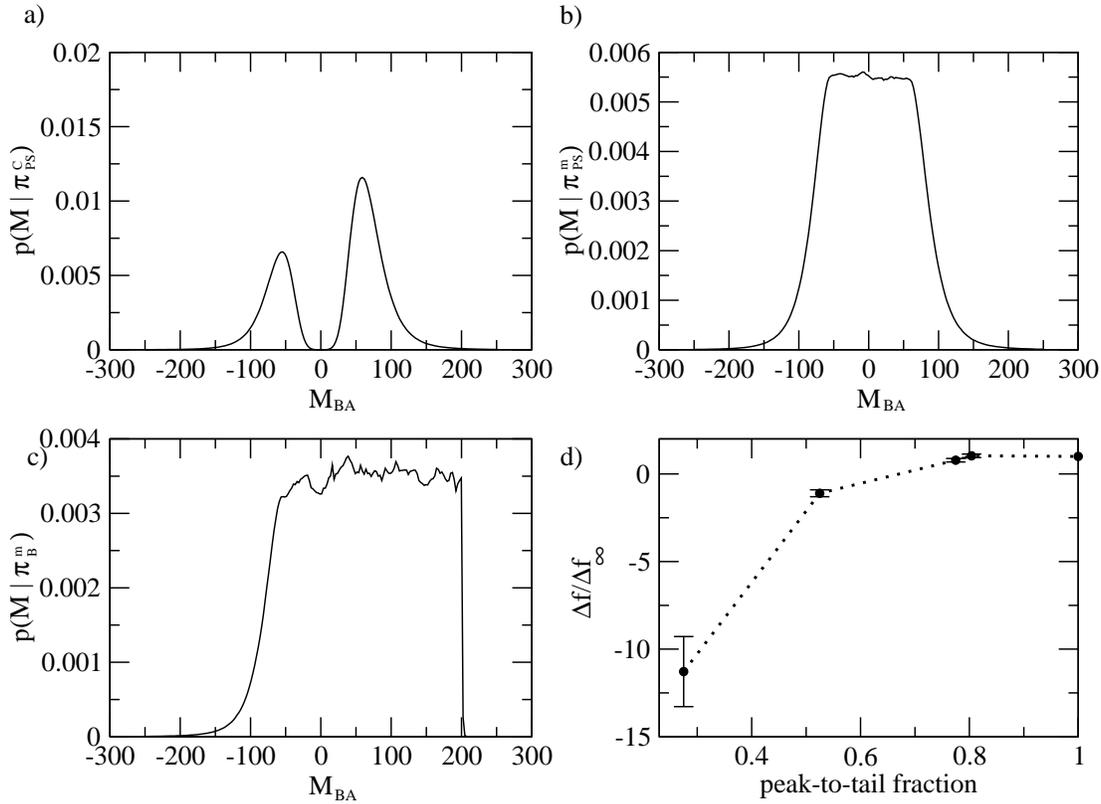

\ifpdf 
\rotatebox{90}{
\includegraphics[scale=0.6]{fig/lsvsbzandfree_mba}
}
\else
\rotatebox{0}{
\includegraphics[scale=0.6]{fig/lsvsbzandfree_mba}
}
\fi
\caption{MUCA Strategies}

a)The canonical probability distribution ${P}(\mba | \pips)$ for the PS method 

b)The MUCA probability distribution ${P}(\mba | \pi^m_{PS})$ for the PS method

c)The full MUCA probability distribution for the EP method {\em initiated in phase B} (left hand peak in (a)) \cite{note:oppest}. The MUCA distribution has been constructed in a way which ensures that a simulation initiated in phase B is able to visit {\em all} the regions of \ecs\ associated with phase A (right peak in (a)). 

d)The convergence of the FED per particle for a series of MUCA simulations (initiated in phase B) as an increasing proportion of the conjugate distribution (right  hand peak in (a)) is included in the MUCA probability distribution. The horizontal axis measures, as a fraction of the distance from the maximum of the left hand peak in (a) to the tail of the right hand peak in (a), the distance up to which the weights satisfy the  relation in Eq. \ref{eq:mulch4}.

$T^*=1.0$, RSM.

\begin{center}
{\bf{------------------------------------------}}
\end{center}
\label{fig:LSvsBZandfree}
\tshow{fig:LSvsBZandfree}
\end{figure}

Figure \ref{fig:LSvsBZandfree} shows an illustration of the application of the MUCA strategy to the EP and the PS methods. Figure (b) shows the MUCA-PS distribution which allows switching between the two phases. Figure (a) shows an estimate of the canonical distribution $P(\mba | \pips)$ as obtained from Eq. \ref{eq:estcanps}. Since the two peaks in figure \ref{fig:LSvsBZandfree} (a) do not overlap, one may (by virtue of Eq. \ref{eq:mbadist}) think of these peaks as  effectively corresponding  to (scaled versions of) the \pc s $P(\mba | \pia)$ and  $P(\mba | \pib)$. From this figure it is clear that the (canonical) probability of the simulation visiting the $\mba \sim O(1)$  regions is negligible, and it is for this reason that one must sample from the distribution shown in figure \ref{fig:LSvsBZandfree} (b) in order for the simulation to be able to switch phases. Figure \ref{fig:LSvsBZandfree} (c) shows the MUCA distribution (for a simulation initiated in phase B) that is required in order to ensure that the EP estimator (Eq. \ref{eq:ep}) is free of systematic errors. Figure  \ref{fig:LSvsBZandfree} (d) shows the convergence of the FED per particle as the MUCA-EP distribution $P(\mba | \pi^m_B)$ is extended so as to include increasing proportions of the distribution associated with phase A . It is clear that convergence is obtained in the limit of the MUCA distribution $P(\mba | \pi^m_B)$ containing the whole of $P(\mba | \pi^c_A)$.

The underlying feature of the form of the two MUCA distributions (Figure \ref{fig:LSvsBZandfree} (b) and (c)) is that they both 'contain' the two canonical distributions $P(\mba | \pia)$ and  $P(\mba | \pib)$. The difference lies  in the {\em way} in which they achieve this. In the case of the EP method, one employs a single sampling distribution $\pi^m_B$. In the case of the PS method, one employs either $\pi^m(\vvec,A)$ or $\pi^m(\vvec,B)$, depending on which phase the simulation is in. This difference manifests itself in the {\em range} of $\mba$ space over which multicanonicalisation must be performed. Whereas in the case of PS method one explicitly constructs the weights (via Eq. \ref{eq:mulch4}) over the region of $\mba$ space lying {\em between} the maxima of the  two peaks, in the case of the EP method one performs enhancement on the whole region between the maximum of the peak of the parent phase (left hand peak in Figure  \ref{fig:LSvsBZandfree} (a)) and the tail (and not merely the peak) of the conjugate distribution (right hand peak of Figure  \ref{fig:LSvsBZandfree} (a)).

The reason for this can be understood as follows. When the PS method `switches phases', it switches the sampling distributions to that which would naturally lead to the exploration of the conjugate phase, even without the aid of MUCA weights. As a consequence the role of multicanonicalisation is merely to ensure that the $\mba\sim O(1)$ regions are accessible to simulations initiated in {\em either} of the phases. This entails the peak-to-peak reweighting, which is evident in the MUCA sampling distribution shown in  figure \ref{fig:LSvsBZandfree} (b). In the case of the EP method, the canonical sampling distribution (which in our case is $\pib$ and is associated with the $\mba<0$ regions) is fixed and is ill-suited to sampling of the regions of \ecs\ associated with  the conjugate phase (which in our case is phase A and corresponds to the $\mba>0$ regions). As a consequence the MUCA weights must not only take the simulation to the $\mba\sim O(1)$ regions, but must also  {\em force} the simulation to visit the entire region of \ecs\ relevant to the conjugate phase ($\mba>0$ regions), since  the sampling distribution \pib\  will typically try to direct the simulation back in the direction of $\mba$ space associated with the parent phase ($\mba<0$ regions). 
It is for this reason that the EP method requires the additional construction of multicanonical weights (Eq. \ref{eq:mulch4}) over the regions  spanning from the maximum of the distribution of the conjugate phase to its tail (compare figure \ref{fig:LSvsBZandfree} (b) with (c)).

\begin{figure}[tbp]
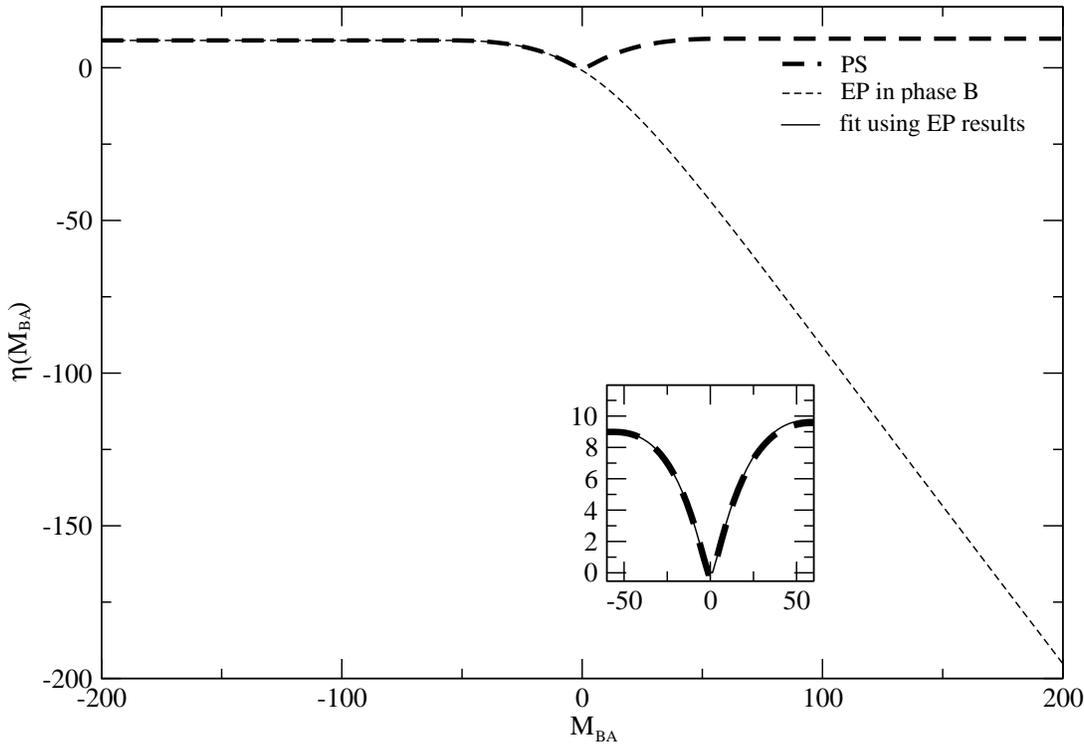

\ifpdf 
\rotatebox{90}{
\includegraphics[scale=0.6]{fig/multicanweightscollated2}
}
\else
\rotatebox{0}{
\includegraphics[scale=0.6]{fig/multicanweightscollated2}
}
\fi
\caption{Comparison of the EP and PS - MUCA weights}
The MUCA weights as a function of $\mba$ for both the MUCA-PS method and for a MUCA-EP simulation initiated in phase B. The inset compares the actual MUCA weight function for the PS method $\eta_{PS}(\mba)$ with $\eta_B(\mba) +\ln [1+e^{\mba}]$. The agreement between the two is in accordance with Eq. \ref{eq:weightsrelation}.

$T^*=1.0$, RSM.

\begin{center}
{\bf{------------------------------------------}}
\end{center}

\label{fig:compareweights}\tshow{fig:compareweights}
\end{figure}

This difference manifests itself in the MUCA weights. 
Figure \ref{fig:compareweights} shows a comparison of the MUCA weights for the two methods. It is clear that for  the $\mba < 0$ regions, the MUCA weights for the EP and PS methods are the same; the reason for this is that  the canonical distribution associated with these regions is \pib\ for both methods. 
This property holds until the $\mba \sim 0$ regions. For the $\mba > 0$ regions, the profiles of the two weight functions diverge. In the case of the EP method the weights decrease as $\mba$ increases, whereas the weights of the PS method increase before levelling off. The reason for this is due to the switching of the phases that takes place in the PS method. That is for the $\mba>0$ regions, the probability of a switch of phases being accepted will be unity. 
 On switching phases the PS simulation will naturally explore the $\mba>0$ regions, even without the aid of MUCA weights. The presence of weights in the $\mba>0$ region is merely to guarantee that the simulation is, once it has jumped from phase B to phase A, able to come back at a later time from the $\mba>0$ regions to the $\mba\sim O(1)$ regions, so as to allow the simulation to switch back to phase B, which will then allow it to naturally explore the $\mba<0$ regions once again. 
 On the other hand in the EP method the simulation  will have to be forced to visit the $\mba>0$ regions of (effective) configuration space; figure \ref{fig:compareweights} clearly illustrates this.

On the transition to larger system sizes, the differences seen in the MUCA distributions of the EP and PS methods become less noticeable. The reason for this lies in the ways the means and the spreads of the peaks scale with the system size. Since for each peak the mean will scale as N, whereas the standard deviation (which measures the spread) scales as $\sqrt{N}$, we see that the additional amount of \mba\ space which will require reweighting in the case of the EP method, over that of  the PS method, 
will become smaller as a fraction of the peak-to-peak distance (which scales as N), on the transition to larger system sizes.

 For finite systems there will also be a difference between the MUCA-EP distributions associated with a simulation constrained to phase A  as compared with one constrained to phase B. In the case of a MUCA-EP simulation constrained to phase A, the MUCA reweighting will now need to be performed from the {\em maximum} of the right hand peak in  figure \ref{fig:LSvsBZandfree} (a) (which is now the peak corresponding to the parent phase) to the {\em tail} of the {\em left}  hand peak (which now corresponds to the conjugate phase). This asymmetry that appears, due to the peak to tail reweighting, will disappears on the transitions to larger system sizes, for the same reasons cited above. 

The intrinsic similarity of the MUCA distributions indicates a connection between the MUCA weights of the different methods. For example by appeal to the overlap identity (Eq. \ref{eq:overlapidentity}) we see that:

\begin{equation}
\RBAcal P(\mba | \pi^m_B) e^{\eta_B(\mba)} \propto e^{-\mba} P(\mba | \pi^m_A) e^{\eta_A(\mba)}
\end{equation}

\noindent Assuming that the two MUCA distributions are flat, one may infer that for the regions over which the two distributions overlap \cite{note:muloverlap}:

\begin{equation}
\eta_B(\mba) = -\mba + \eta_A(\mba) + constant
\label{eq:mulab}
\end{equation}\tshow{eq:mulab}

\noindent Similarly the fact that the MUCA-EP and the MUCA-PS distributions are approximately the same means that one may arrive at a similar identity relating the MUCA weights of  the two methods. To arrive at the result we substitute Eq. \ref{eq:psand2} into Eq. \ref{eq:tempchest} (in which the weights $w_\al$ have been set to be equal) so as to obtain the following relation:

\begin{equation}
P(\mba | \pi^c_\al) \frac {\z_\al} {\z_\al + \z_\alp} = \frac {1} {1+e^{-\mal}} P(\mba | \pips)
\end{equation}

\noindent Using  Eq. \ref{eq:mucasampch5} one obtains:

\begin{equation}
\eta_{PS}(\mba) =\left\{\begin{array}
{r@{\quad}l}
\eta_A(\mba) + \ln[1+e^{-\mba}] + L_1\\
\eta_B(\mba) + \ln[1+e^{\mba}] + L_2
\end{array}\right.
\label{eq:weightsrelation}
\end{equation}\tshow{eq:weightsrelation}

\noindent where $L_1$ and $L_2$ are some constants. This identity also naturally follows from Eq. \ref{eq:pssingle}. Since these constants do not affect the simulation in any way (since it is only the relative values of these constants that matter) we may, without loss of generality, set these two constants to zero. The inset in figure  \ref{fig:compareweights} shows a plot of $\eta_{PS}(\mba)$ and a plot of $\eta_B(\mba) + \ln[1+e^{\mba}]$, where both $\eta_{PS}(\mba)$ and $\eta_B(\mba)$ have been estimated via simulation. The clear agreement between the two curves verifies that the relation in Eq. \ref{eq:weightsrelation} does indeed hold.

So far we have noted that the EP and PS method are different in two respects. Firstly they require different ranges of \mba\ space to be reweighted, and secondly the MUCA weights are different. However these differences mask the underlying similarity of the two methods. The first difference, that is the difference in the range of \mba\ space which requires reweighting, vanishes on the transition to sufficiently large system sizes. The second difference merely arises from the fact that the canonical sampling distributions are different. However since the MUCA distributions are the same, and since the weight of macrostates are proportional (compare Eq. \ref{eq:epn} to Eq. \ref{eq:wpsn} and Eq. \ref{eq:epd} to Eq. \ref{eq:wpsd}), we see that the methods are essentially {\em identical}. This equivalence between the two methods may be most readily expressed through the corresponding estimators:

\begin{eqnarray}
\RBAcal & = & \frac {\s {i=1} {b} f(\mbai) H(\mbai | \pi^m_{PS}) e^{\eta_{PS}(\mbai)}} {\s {i=1} {b} f(-\mbai) H(\mbai | \pi^m_{PS}) e^{\eta_{PS}(\mbai)}}\nonumber\\
& = & \frac {\s {i=1} {b} f(\mbai) H(\mbai | \pi^m_{PS}) e^{\eta_{A}(\mbai)} [1+e^{-\mbai}]} {\s {i=1} {b} f(-\mbai) H(\mbai | \pi^m_{PS}) e^{\eta_{A}(\mbai)}[1+e^{-\mbai}]}\nonumber\\
& \approx & \frac {\s {i=1} {b} H(\mbai | \pi^m_A)e^{-\mbai} e^{\eta_A(\mbai)}} {\s {i=1} {b} H(\mbai | \pi^m_A) e^{\eta_A(\mbai)}} 
\label{eq:estequi}
\end{eqnarray}\tshow{eq:estequi}

\noindent where we have used the fact that the MUCA distributions $P(\mba | \pi^m_B)$ and  $P(\mba | \pi^m_{PS})$ are the same, so that $H(\mbai | \pi^m_A)\propto  H(\mbai | \pi^m_{PS})$. Eq. \ref{eq:estequi} establishes the equivalence between the PS estimator (Eq. \ref{eq:psestfer}) to that of the EP estimator (Eq. \ref{eq:torriereweightest}) in the MUCA limit.

In the absence of MUCA weights, the two estimators are no longer equivalent. The reason for this is because now the \mba\ distributions are no longer the same, even though the weights remain proportional to each other.
As a result the statistical errors will be different for the two methods in a finite run simulation.

\section{\label{sec:mh}The Multihamiltonian strategy}
\tshow{sec:mh}

\subsection{Theory}

In the previous section we have seen that the MUCA strategy provides an efficient framework (in the case of  zero equilibration) for tackling of the overlap problem. The basic idea is to construct a sampling distribution which contains the two \pc s $P(\mba | \pi^c_A)$ and $P(\mba | \pi^c_B)$ so as  to allow for the construction of a path (in a piecewise but serial manner) from the region of \ecs\ associated with phase A to that of phase B. This allows one to determine the weight of the typical macrostates associated with phase B relative to those of phase A. The important thing to notice is that this path can also  be constructed in a parallel manner (that is piecewise and independent fashion). This is effectively what the MS and WHAM methods do \cite{note:ARMH} and was originally proposed by Geyer \cite{p:geyer}\nocite{p:geyerthom}\nocite{p:hukushima}-\cite{note:pt}. The essential ingredient of all these methods is that the independent simulations {\em overlap} in some region of the \ecs\ that they explore. It is only when they overlap that the data of a simulation obtained with ${\pi}$, say, may be reweighted with respect to $\tilde{\pi}$, which overlaps with $\pi$, so as to yield a set of macrostates whose relative probabilities are in agreement with $\tpi$ and which, at the same time, extend outside the range normally explored by $\tpi$ (see Eq. \ref{eq:reweightav}). In this way one may use the idea of reweighting to estimate the probabilities of the regions of \ecs\ typically associated with phase B in relation to those of phase A.

With this in mind let us proceed to construct a new way of estimating the FED based on the idea of simulating several independent systems. Consider the construction of a chain of \ces , as has been done in Eq. \ref{eq:hamilchain}, whose associated sampling distributions (Eq. \ref{eq:samphamilofi}) overlap in a manner so as to yield a path connecting the two regions of \ecs\ associated with the two phases. Then instead of writing the \rpf\ as has been done in Eq. \ref{eq:multistage}, one may instead choose to write it as:

\begin{eqnarray}
\RBAcal & = & \frac {\int d\vvec e^{-\beta \ecal_{\la_2} (\vvec)}} {\int d\vvec e^{-\beta \ecal_{\la_1} (\vvec)}}. \frac {\int d\vvec e^{-\beta \ecal_{\la_3} (\vvec)}} {\int d\vvec e^{-\beta \ecal_{\la_2} (\vvec)}}.....\frac {\int d\vvec e^{-\beta \ecal_{\la_n} (\vvec)}} {\int d\vvec e^{-\beta \ecal_{\la_{n-1}} (\vvec)}}\nonumber\\
&  = & \frac {\int \p {j=1} {n-1} d\vvec_{j} e^{-\beta\{\ecal_{\la_2}(\vvec_{1})+\ecal_{\la_3}(\vvec_{2}) +.....+\ecal_{\la_{n}}(\vvec_{n-1})\}}} {\int \p {j=1} {n-1} d\vvec_{j} e^{-\beta\{\ecal_{\la_1}(\vvec_{1})+\ecal_{\la_2}(\vvec_{2}) +.....+\ecal_{\la_{n-1}}(\vvec_{n-1})\}}}
\end{eqnarray}

\noindent or:

\begin{equation}
\RBAcal  =  \frac {\int e^{-\beta \hcal {B}{{\Vvec}}}d{\Vvec}} {\int e^{-\beta \hcal {A}{{\Vvec}}}d{\Vvec}}
\label{eq:gen_part}
\end{equation}\tshow{eq:gen_part}

\noindent where {\Vvec} = \seta {\vvec_{1}, \vvec_{2}, ... , \vvec_{n-1}} denotes the collection of the configurations of the  n-1 independent replicas, and where:

\begin{equation}
\hcal {\gamma} {{\Vvec}} = \left\{\begin{array}
{r@{\quad:\quad}l}
\s {j=1} {n-1} \ecal_{\la_{j}}(\vvec_{j}) & \mbox{if \al\ = A}\\
\s {j=1} {n-1} \ecal_{\la_{j+1}}(\vvec_{j}) & \mbox{if \al\ = B}
\end{array}\right.
\label{eq:gen_hamil}
\end{equation}\tshow{eq:gen_hamil}

\noindent By writing \RBAcal\ as has been done in Eq. \ref{eq:gen_part}, a new strategy immediately becomes apparent. That is rather than simulating the actual systems with the \ces\ $\ecal_A$ and $\ecal_B$ (see Eq. \ref{eq:RBA}) one may instead simulate the composite systems described by the extended \ces\  $\hcalone {A}$ and $\hcalone {B}$. If one now generalises the PM operation from the original version:

\begin{equation}
A\leftrightarrow B \mbox{\ \ \ } \vvec \rightarrow \vvec
\end{equation}

\noindent to that of a PM between the composite systems in which the extended configuration {\Vvec}\ is matched for the two systems:

\begin{equation}
A\leftrightarrow B \mbox{\ \ \ } {\Vvec} \rightarrow {\Vvec}
\label{eq:genpm}
\end{equation}\tshow{eq:genpm}

\noindent then it is clear that the array of estimators and techniques used to estimate  Eq. \ref{eq:RBA} may also be used here. The key idea is that the independent sampling distribution \seta {\pi_{\la_1}, \pi_{\la_2}, ..., \pi_{\la_n}}  provides the necessary extended sampling strategy that is needed in order to overcome the overlap problem. The greater the number of \ces\ in the chain,  the greater is the weight of the set of \ces\ \seta{\ecal_{\la_2},\ecal_{\la_3},...,\ecal_{\la_{n-1}}} that the two hamiltonians $\hcalone {A}$ and $\hcalone {B}$ share, and therefore the greater is the overlap between the effective  configuration space of the two composite systems. 

In order to be able to quantify the overlap, it is useful to once again define a macrovariable:

\begin{equation}
\mba({\Vvec})  =  \beta [\hcal {B} {{\Vvec}} -  \hcal {A} {{\Vvec}}]
\label{eq:mhwork}
\end{equation}\tshow{eq:mhwork}

\noindent $\mba$ essentially corresponds to  the (temperature scaled) work (which, as before, we will subsequently refer to as work) incurred in switching (in an instantaneous fashion) between the extended configurational energies $\hcalone {A}$ and $\hcalone {B}$ whilst preserving the extended configuration {\Vvec}. In the case of the linear parameterisation given in Eq. \ref{eq:linearhamiltonian}, Eq. \ref{eq:mhwork} may be written as:

\begin{equation}
\mba({\Vvec})  = \s {i=1} {n-1} \delela_i \mba(\vvec_i)
\label{eq:linmhwork}
\end{equation}\tshow{eq:linmhwork}

\noindent In order to quantify the overlap in a meaningful way  one must be able to relate the probabilities of a macrostate \mba\ associated with one system relative to that of the other. Suppose that:

\begin{equation}
\tpi_\al^c \equaldot e^{-\beta  \hcal {\al} {{\Vvec}}}
\label{eq:mhsamp}
\end{equation}\tshow{eq:mhsamp}

\noindent denotes the sampling distribution of the composite system (which may be  realised by  independently simulating the n-1  sampling distributions \seta {\pi_{\la_i}}).  The procedure of sampling via $\tpi_\al$ will be referred to as the multihamiltonian (MH) method. Suppose that $P(\mba | \tpi_\al^c)$ denotes the probability of obtaining the $\mba$ when sampling with $\tpi^c_\al$ . It immediately follows from the form of Eq. \ref{eq:gen_part} (compare this to Eq. \ref{eq:RBA}) that the probability of obtaining $\mba$ (as defined in Eq. \ref{eq:mhwork}) when sampling with $\tpi^c_A$ relative to that when sampling with respect to $\tpi^c_B$, is simply given by the overlap identity (Eq. \ref{eq:crooksfluct}). To see this more explicitly we observe that:

\begin{eqnarray}
P (\mba | \tpi_A)& = & \p {i=1} {n-1} \frac {1} {\z_{\la_i}} \int \delta(\mba(\Vvec) - \mba) e^{-\beta \sum^{n-1}_{i=1}\ecal_{\lambda_{i}}(\vvec_{i})}  \p {i=1} {n-1} d\vvec_{i}\nonumber\\
& = &  \p {i=1} {n-1} \frac {1} {\z_{\la_i}} \int \delta({\mba}(\Vvec) - \mba) e^{-\beta \sum^{n-1}_{i=1}\ecal_{\lambda_{i+1}}(\vvec_{i})}\times\nonumber \\
&  & e^{\beta [\sum^{n-1}_{i=1}\ecal_{\lambda_{i+1}}(\vvec_{i}) -\sum^{n-1}_{i=1}\ecal_{\lambda_{i}}(\vvec_{i})]} \p {i=1} {n-1} d\vvec_{i} \nonumber\\
& = &  \p {i=1} {n-1} \frac {1} {\z_{\la_i}} \int \delta({\mba}(\Vvec) - \mba) e^{-\beta \sum^{n-1}_{i=1}\ecal_{\lambda_{i+1}}(\vvec_{i})}e^{{\mba}(\Vvec)} \p {i=1} {n-1} d\vvec_{i} \nonumber\\
& = & e^{\mba} \frac {\z_{\la_n}} {\z_{\la_1}}\p {i=2} {n} \frac {1} {\z_{\la_i}} \int \delta({\mba}(\Vvec) - \mba) e^{-\beta \sum^{n-1}_{i=1}\ecal_{\lambda_{i+1}}(\vvec_{i})}\p {i=1} {n-1} d\vvec_{i}\nonumber\\
& = & e^{\mba} \frac {\z_B} {\z_A} P(\mba | \tpi_B)
\label{eq:crooksgeneral}
\end{eqnarray}\tshow{eq:crooksgeneral}

\noindent which is the overlap identity given in Eq. \ref{eq:crooksfluct}. Once again the distribution of $P(\mba | \tpi^c_\al)$ will look something similar to that shown in figure \ref{pic:overlapping} with the point of intersection being located at $\mba = M_m$. We note that this also follows from the fact that the MH method can be viewed as a limiting case of the FG method (see appendix \ref{app:equivalent}). 

The crucial point to realise is that as the number of \ces\ in  Eq. \ref{eq:gen_hamil} increases, the overlap \ov\ increases and tends to unity. We recall that heuristically this may be understood by noting that the greater the number of replicas, the greater is the weight of the set of \ces\ \seta{\ecal_{\la_2},\ecal_{\la_3},...,\ecal_{\la_{n-1}}} that the two hamiltonians $\hcalone {A}$ and $\hcalone {B}$ share (see also figure \ref{pic:OPforreplicas} for an alternative explanation). In the case of the n=2 this set has zero weight. As $n\rightarrow \infty$, the weight of this set dominates over the \ces\ $\ecal_{\la_1}$ and $\ecal_{\la_n}$ which are at the edge of the chain of the \ces\ in Eq. \ref{eq:hamilchain} and which describe the two phases whose FED one is trying to measure. It is these edge \ces\ that give rise to the difference between $\hcalone {A}$ and $\hcalone {B}$ (see Eq. \ref{eq:gen_hamil}). In the limit of the number of \ces\ in the chain tending to infinity, one may write:

\begin{eqnarray}
<\mba ({\Vvec})>_{\tpi_A^c} & = & \s {i=1} {n-1} <\ecal_{\la_{i+1}} (\vvec_{i}) - \ecal_{\la_{i}} (\vvec_{i})>_{\pi^c_i}\nonumber\\
& = & \delela_i \s {i=1} {n-1} <\frac {[\ecal_{\la_{i+1}} (\vvec_{i}) - \ecal_{\la_{i}} (\vvec_{i})]} {\delela_i}>_{\pi^c_i}\nonumber\\
& \approx & \int^{\la=\la_n}_{\la=\la_1} d\la <\frac {\partial \ecal_\la} {\partial \la}>_{\pi_\la^c}
\end{eqnarray}

\noindent Similarly, it follows that:

\begin{equation}
<\mba ({\Vvec})>_{\tpi_B^c} \approx \int^{\la=\la_n}_{\la=\la_1} d\la <\frac {\partial \ecal_\la} {\partial \la}>_{\pi_\la^c}
\label{eq:toti}
\end{equation}\tshow{eq:toti}

\noindent Therefore in the limit of the number of \ces\ in Eq. \ref{eq:gen_hamil} tending to infinity, we find that $<\mba ({\Vvec})>_{\tpi_\al^c}$ approaches the value obtained by thermodynamic integration:

\begin{equation}
\lim_{n\rightarrow \infty} <\mba ({\Vvec})>_{\tpi_\al^c}\rightarrow \int^{\la=\la_n}_{\la=\la_1} d\la <\frac {\partial \ecal_\la} {\partial \la}>_{\pi_\la^c} = -\ln \RBAcal
\end{equation}

\noindent It follows  that in this limit \mba\ must have the distribution corresponding to perfect overlap:

\begin{equation}
P(\mba | \pi^c_\al) = \delta (\mba + \ln \RBAcal)
\label{eq:mhbzlim}
\end{equation}\tshow{eq:mhbzlim}

\noindent Comparing Eq. \ref{eq:gen_part} to Eq. \ref{eq:RBA}  it is immediately apparent that the array of estimators as parameterised by Eq. \ref{eq:dual} are available for the task of estimating FEDs. {\em More generally one could also perform an arbitrary switching FG process in which $\hcalone {\al}$} {\em is gradually switched into $\hcalone {\alp}$} {\em so as to allow an estimate of the FED to be obtained either from the phase-constrained methods Eq. \ref{eq:crooksfluct},}  {\em Eq. \ref{eq:freebennettgen}, } {\em or from the PS formulae (Eq. \ref{eq:psfermi})}. 
For example in the case of zero-equilibration the 
MH version of the PS sampling distribution is given by:

\begin{equation}
\tpips ({\Vvec}, \al) = \frac {{w_\al} e^{-\beta \hcal {\al} {\Vvec}}} {w_A \z_A + w_B \z_B}
\label{eq:mhps}
\end{equation}\tshow{eq:mhps}

\noindent where $w_\al$ is the weight which biases phase \al , and \al\ is a stochastically sampled variable (see section \ref{sec:fgps}). \RBAcal\ may then be estimated either via Eq. \ref{eq:rbafgweight} or Eq. \ref{eq:psfermi} (where \wba = \mba).

Inspection of the MH sampling distribution (Eq. \ref{eq:mhsamp}) may lead one to believe that the MH method is equivalent to the MS method. This is generally true if one can write Eq. \ref{eq:dual} as:

\begin{eqnarray}
\RBAcal & = & \frac {<G(\beta [\ecal_B-\ecal_A])e^{-\beta [\ecal_B-\ecal_A]}>_{\tpia}} {< G(\beta [\ecal_B-\ecal_A])>_{\tpib}}\nonumber\\
& = & \frac {\p {i=1} {n-1} <G(\beta\{\ecal_{\la_{i+1}}-\ecal_{\la_i}\})e^{-\beta\{\ecal_{\la_{i+1}}-\ecal_{\la_i}\}}>_{\pi_{\la_i}}} {\p {i=1} {n-1} <G(\beta\{\ecal_{\la_{i+1}}-\ecal_{\la_i}\})>_{\pi_{\la_i}}}
\label{eq:mhdecouple}
\end{eqnarray}\tshow{eq:mhdecouple}

\noindent which only holds if G(x) is of the exponential form:

\begin{equation}
G(x) = e^{ax}
\end{equation}

\noindent where a is some constant. Therefore the only choice of $G(\wba)$ for which the MS and MH methods are equivalent is the EP (Eq. \ref{eq:ep}) method. In this case the only difference that arises for the two methods is that  in the case of MH estimator one only deals with a single estimate  of the error, where are in the case of the MS method one must combine n-1 such errors in determining the error of  the final value for the FED between the two phases. For other estimators (including the PS estimator) the two strategies are not equivalent. 

We finally note that the potential for parallelising the sampling of  Eq. \ref{eq:mhsamp} is a clear advantage of the MH method. Furthermore, in comparison to the WHAM method, the storage requirements are far less. If one records the value of \mba\ during the course of the simulation (as one would do if one wanted to avoid the systematic errors introduced by employing finite bin width histograms) then the use of the MH method will yield significant gains in regards to this issue, since for this method  one will need to record only a single temporal sequence of \mba 's, as opposed to recording one such sequence for each and every replica.

\subsection{Numerical Results}

\begin{figure}[tbp]
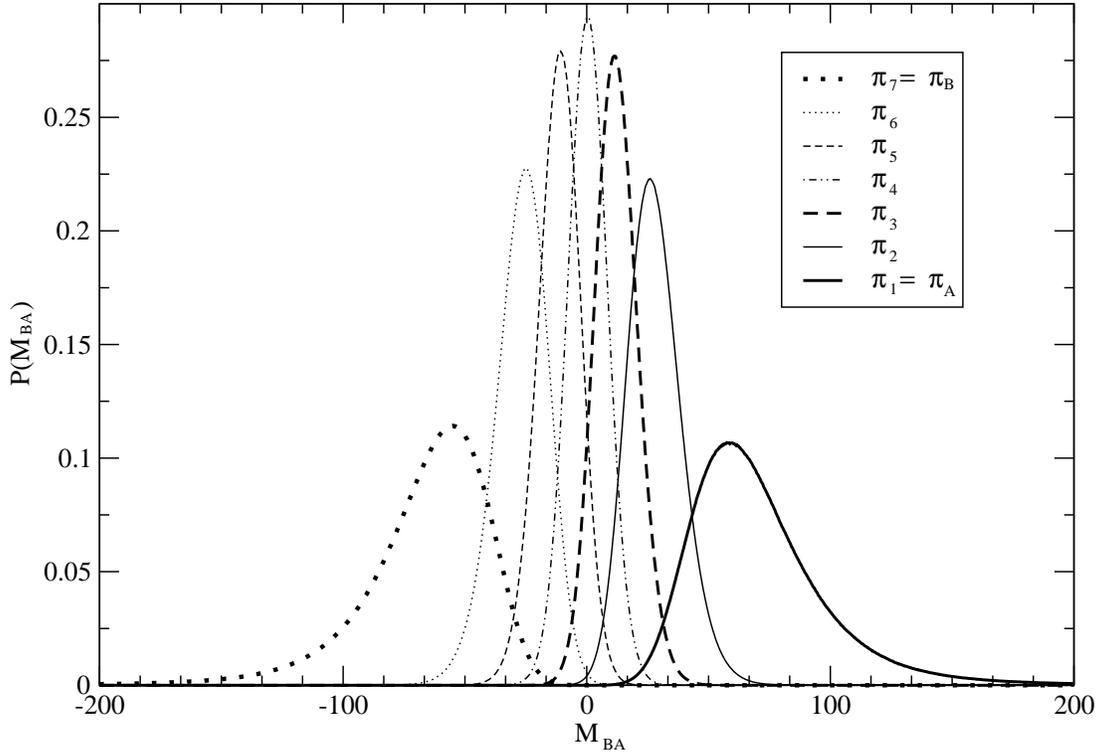

\begin{center}
\ifpdf
\rotatebox{90}{
\includegraphics[scale=0.6]{fig/comparisonofmhlsandmsls_mba}
}
\else
\rotatebox{0}{
\includegraphics[scale=0.6]{fig/comparisonofmhlsandmsls_mba}
}
\fi
\end{center}
\caption{Probability distribution functions ${P}( \mba| \pi_{\la_i} )$}
The figure shows the  probability distributions ${P}( \mba |\pi_i)$ (where $\pi_i=\pi_{\la_i}$, see Eq. \ref{eq:samphamilofi}) for the replicas constituting the chain of \ces\ (Eq. \ref{eq:hamilchain}, Eq. \ref{eq:gen_hamil}) linking the two phases. In this figure we have chosen the linear parameterisation (Eq. \ref{eq:linearhamiltonian}) and set n=7. By constructing a series of independent and overlapping distributions, so as to bridge the regions of \ecs\ associated with the two phases, one overcomes the overlap problem. More specifically, since for some of these distributions $\mba>0$ and for others $\mba<0$, one obtains cancellations in the overall summation given in Eq. \ref{eq:linmhwork}. As a consequence the distribution of $\mba(\Vvec)$, as given by Eq. \ref{eq:linmhwork}, will reside between the peaks associated with the two phases, resulting in improved overlap between $P(\mba | \tpia)$ and $P(\mba | \tpib)$, as compared to that between $P(\mba | \pia)$ and $P(\mba | \pib)$.

See figure \ref{pic:pdfsofMSLS} (a) for the PS distribution for the corresponding composite system described by Eq. \ref{eq:gen_hamil}.

$T^*=1.0$, RSM.

\tshow{pic:OPforreplicas}
\label{pic:OPforreplicas}

\begin{center}
{\bf{------------------------------------------}}
\end{center}

\end{figure}

\begin{figure}[tbp]
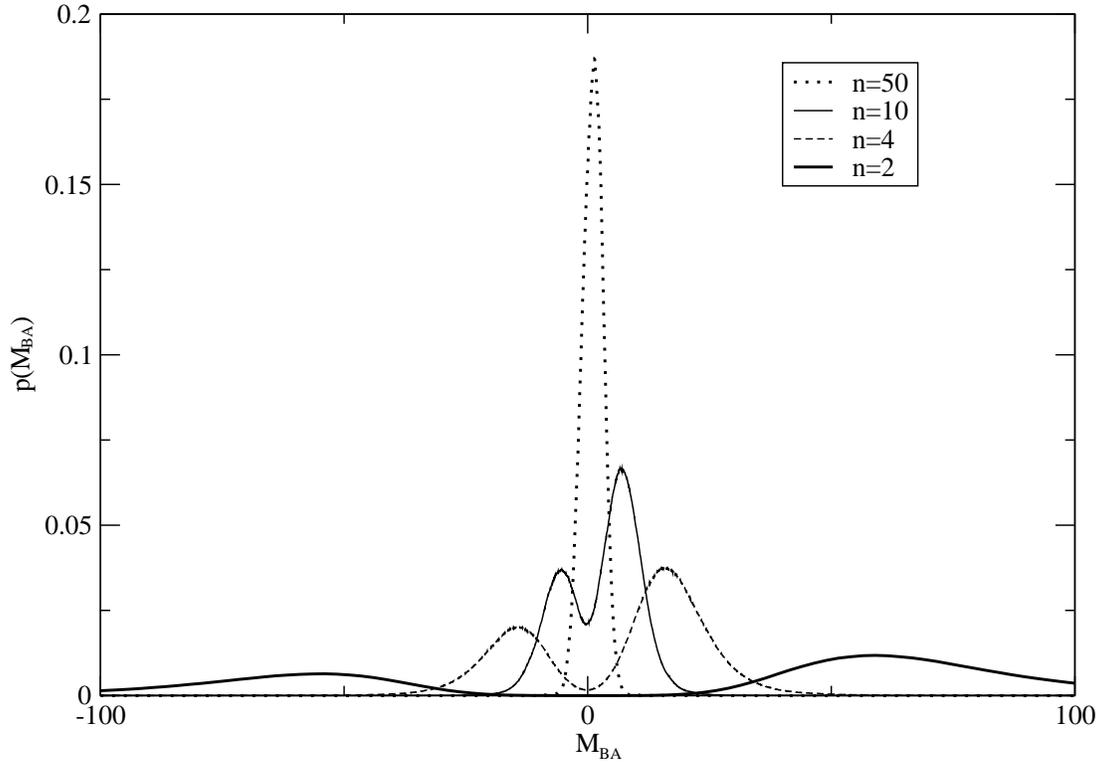

\begin{center}
\ifpdf
\rotatebox{90}{
\includegraphics[scale=0.6]{fig/mhls}
}
\else
\rotatebox{0}{
\includegraphics[scale=0.6]{fig/mhls}
}
\fi
\end{center}
\caption{MH-PS Probability distribution functions of $\mba$ for different n}
{This graph shows the probability distribution ${P}(\mba|\tpips)$ for the MH-PS for n= 4, 10 and 50. The corresponding probability distribution for the original PS simulation (n=2) is shown for comparison. As the number of replicas n is decreased, the probability distribution of \mba\ tends to that of the original (n=2) PS simulation. However as the number of replicas is increased, the two peaks associated with $\tpi_\al^c$ and $\tpi_\alp^c$ increasingly overlap, thereby increasing the chances of a switch being accepted (see Eq. \ref{eq:lateps}, and notice that $M_m\sim 0 $ for the systems investigated here). 

It is clear from the figure that the chance of a switch taking place is negligible for the original PS method unless one employs some form of extended sampling. The MH method offers an alternative extended sampling strategy to that of the MUCA method. The advantage that the method has is that it is highly parallelizable. On the down side, it makes an increasing demand on the memory requirements since one will have to store n-1 replicas of the given system in the computer memory.

{\em We note that it was found that the errors in the estimate of the FED obtained in a finite run were independent of the number of replicas chosen.}

$T^*=1.0$, RSM }
\tshow{pic:MHLS}
\label{pic:MHLS}

\begin{center}
{\bf{------------------------------------------}}
\end{center}

\end{figure}

In this section we illustrate the application of the MH method to the systems investigated in this thesis. Figure \ref{pic:OPforreplicas} shows the distributions of the macrovariable \mba\ for the replicas of the chain of \ces\ (Eq. \ref{eq:hamilchain}, Eq. \ref{eq:gen_hamil}) in which n=7 and in which the \ces\ are  linearly  parameterised as prescribed in  Eq. \ref{eq:linearhamiltonian}. The crucial feature of this figure is the way in which the distributions overlap, so as to provide a continuous path (in \mba\ space) from the region of \ecs\ associated with phase A (right hand most peak in figure \ref{pic:OPforreplicas}) to that of phase B (left hand most peak). Figure \ref{pic:pdfsofMSLS} shows the corresponding MH-PS distribution which employs these replicas in the sampling distribution Eq. \ref{eq:mhps}, and shows how the MH-PS method is able to effectively overcome the overlap problem (notice that the overlap between the phase A and phase B, the right and left hand most peaks respectively, do not overlap at all).
Figure \ref{pic:MHLS} shows the probability distributions of \mba\ for the MH-PS method (Eq. \ref{eq:mhps}) for different n. It is evident that, as the number of replicas n  increases, the overlap increases, with the distribution $P(\mba | \tpips)$ tending to the ideal limit of a delta function centred on $M_m$, which follows from Eq. \ref{eq:mbadist} and Eq. \ref{eq:mhbzlim}.

\begin{figure}[tbp]
\begin{center}
\ifpdf
\rotatebox{90}{
\includegraphics[scale=0.6]{fig/pdf_mhls_msls2}
}
\else
\includegraphics[scale=0.6]{fig/pdf_mhls_msls2}
\fi
\end{center}
\caption{${P}(\mba|\tpi^c_{PS})$ (MH-PS method) corresponding to figure \ref{pic:OPforreplicas}}
The figure  shows the probability distribution ${P}(\mba|\tpips)$ as obtained via the MH-PS method in which the distributions of the composite replicas \seta {\pi_{\la_i}} is given in figure \ref{pic:OPforreplicas}.

n=7, $\delela = 1/6$, $T^*=1.0$, RSM.

\tshow{pic:pdfsofMSLS}
\label{pic:pdfsofMSLS}

\begin{center}
{\bf{------------------------------------------}}
\end{center}

\end{figure}

\section{\label{sec:asfg} The Fast Growth strategy}
\tshow{sec:asfg}

An alternative to the MUCA and the MH strategies is the FG method (see section \ref{sec:fg}). Under this scheme, one performs work on the system so as to morph the \ce\ from that of phase \al\ ($\ecal_\al$) to that of phase \alp\ ($\ecal_\alp$). In the process a path is constructed linking the set of macrostates associated with phase A to those associated with phase B. The key parameters in the method are the increments $\seta {\triangle \la_i}$ of the field parameter \la\ and the equilibration times $\seta {\triangle t_i}$ (where $\triangle t_i= t_{i+1}-t_i$). For simplicity, we will limit ourselves to the case where all the increments are equal , that is $\triangle \la_i = \triangle \la$, and when all the equilibration times are equal, that is $\triangle t_i = \triangle t$. It is only if the equilibration times \delt\ are sufficiently long and the perturbations to the \ce\ (\delela) are sufficiently small that one is able to construct a path (an overlapping sequence of macrostates) connecting the two phases.

One may immediately identify two limiting cases. In the case where $\delt = 0$, one obtains the  zero equilibration  methods, irrespective of $\delela$, as one does in the case where $\delela=1$ \cite{p:jarprl}. In the limit where $\delela \rightarrow 0$, $\delt >0$ (which we will refer to as adiabatic equilibration since such a method takes an infinite time to switch from the \ce\ of phase A to that of phase B), one obtains the thermodynamic integration method (Eq. \ref{eq:MSTI}) \cite{p:jarprl}:

\begin{equation}
\lim_{\delela \rightarrow 0, \delt > 0} <e^{-\wba}>_{\pi^c_A} =  e^{-\beta \int_{\la=0}^{\la=1} d\la <\frac {\partial \ecal_\la} {\partial\la}>_{\pi_\la}}
\label{eq:fgti}
\end{equation}\tshow{eq:fgti}

\noindent Both limits have  undesirable features as they stand. On the one hand the choice of  zero equilibration induces systematic errors (as described in section \ref{sec:perturbation}) which must be overcome by appeal to some form of \es\ strategy (\cite{p:iba}, section \ref{sec:mul}, section \ref{sec:mh}). On the other hand the choice of adiabatic equilibration is time consuming. Furthermore it is not clear how one accounts for the systematic errors induced in making an approximate evaluation of the integral in  Eq. \ref{eq:fgti}.  Generally some intermediate strategy is preferable in which Eq. \ref{eq:dualphaserestrict}, Eq. \ref{eq:psfermi} is used to estimate \RBAcal .

\begin{figure}[tbp]
\begin{center}
\ifpdf
\rotatebox{90}{
\includegraphics[scale=0.6]{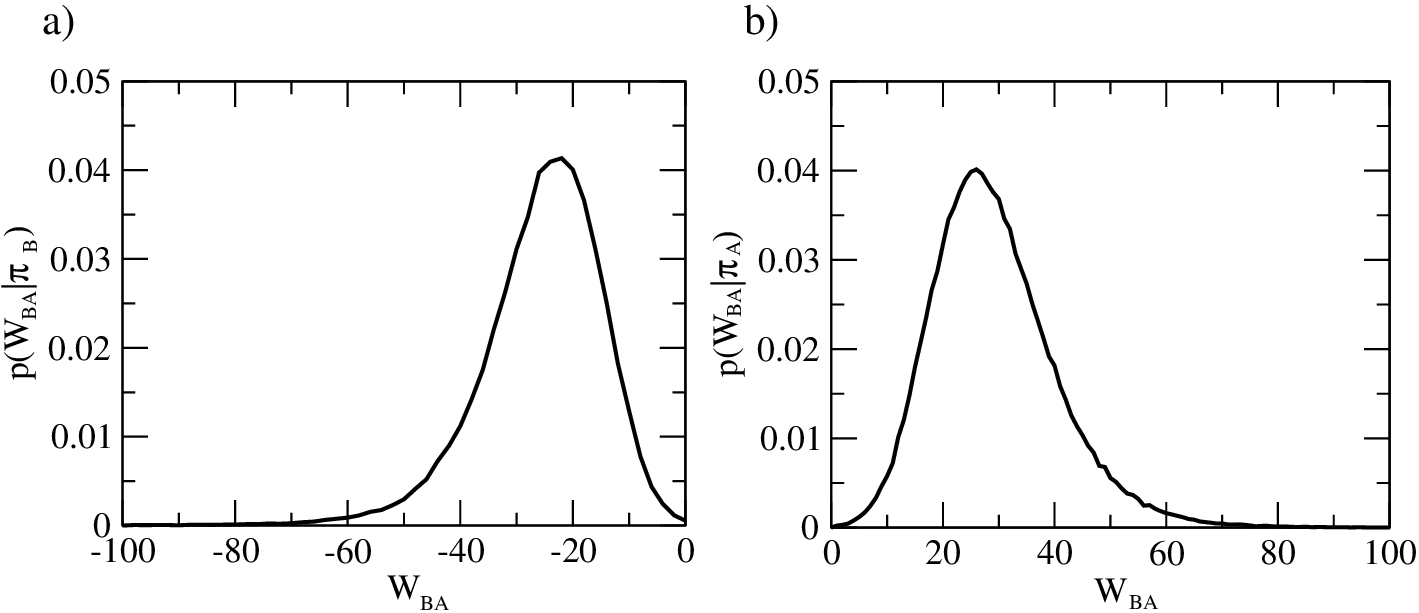}
}
\else
\includegraphics[scale=0.6]{fig/compare_forward_reverse_pdfs}
\fi
\end{center}
\caption{Symmetry of ${P}(\wba | \pia)$ and ${P}(\wba | \pib)$}
The probability distribution of \wba\ for the $A\rightarrow B$ process (in b), in which \la\ is increased from 0 to 1, and for a $B\rightarrow A$  process (in a), in which \la\ is decreased from 1 to 0.

The equilibration time is expressed as a multiple of lattice sweeps. In other words, a single lattice sweep, in which one attempts to sequentially perturb all N particles, corresponds to an equilibration time of $\triangle t = 1$.

$\delt$ = 1, $T^*$ =1.0, n=11, RSM.

\label{fig:forwardreversepdf}
\tshow{fig:forwardreversepdf}

\begin{center}
{\bf{------------------------------------------}}
\end{center}

\end{figure}

We will now focus our attention on such intermediate strategies. Namely we will  investigate the way in which the FG method overcomes the overlap problem for these intermediate strategies through control of the parameters \delela\ and $\delt$  \cite{note:constantFG}. In particular we will examine three variations. In the first case we will  keep \delela\ constant and vary $\delt$. In the second we keep $\delt$ constant and vary \delela . In the third we investigate the case when $\delt /\delela$ is held constant. Since the total time allocated to obtaining each work term $\wba$ is given by $(n-1)\delt =  \delt / \delela$, we see that the last case corresponds to the variation of the parameters  \delela\ and $\delt$ so as to ensure that the amount of time allocated to performing work on the system is held constant. In all cases we will demonstrate the improvement in the overlap by demonstrating  the convergence of the FED estimate as obtained by the EP estimator (Eq. \ref{eq:fgep}) relative to that obtained via the MUCA-PS method. 
Again we illustrate the overlap by examining only one of the phase constrained distributions, which in this case is $P(\wba | \pi^c_A)$, since its conjugate partner $P(\wba | \pi^c_B)$ is roughly symmetrically positioned about the origin (which is where $W_m$, see figure \ref{pic:overlapping}, roughly lies). An example of this approximate symmetry is illustrated in figure \ref{fig:forwardreversepdf}. The particular parameterisation of \seta {\ecal_{\la_i}} that we  employ is the linear parameterisation given by Eq. \ref{eq:linearhamiltonian}.

\subsection{\label{eq:dtc} Keeping \delela\ constant, varying $\delt$}
\tshow{eq:dtc}

\begin{figure}[tbp]
\begin{center}
\ifpdf
\rotatebox{90}{
\includegraphics[scale=0.6]{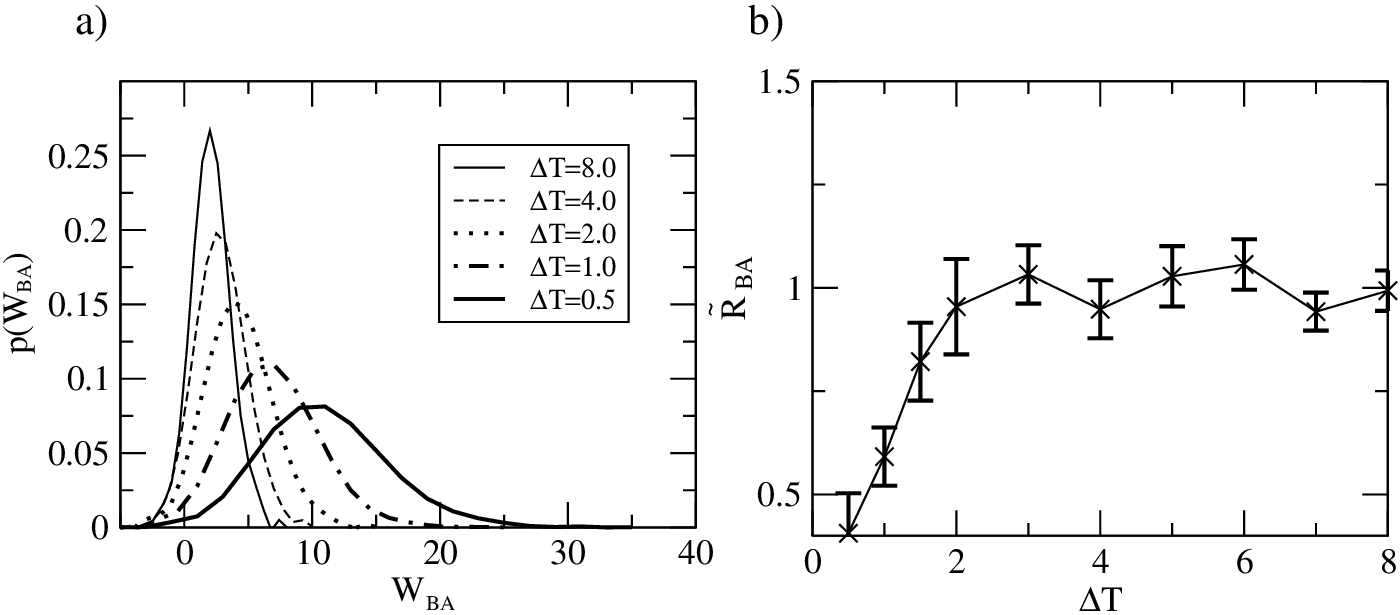}
}
\else
\includegraphics[scale=0.6]{fig/increasingamountofcooling}
\fi
\end{center}
\caption{Convergence of ${\tilde {\cal{R}}}_{BA}$ with increasing $\delt$ , \delela = constant}
Figure (a) shows the distribution ${P}(\wba | \pia)$ as the equilibration times \delt\ is increased whilst \delela\ is maintained at a constant value.

Figure (b) shows the convergence of ${\tilde{\cal R}}_{BA}$ as \delt\ is increased.

${\tilde{\cal R}}_{BA} = {\cal{R}}_{BA}/{\cal{R}}^*_{BA}$, where ${\cal{R}}^*_{BA}$ is the value obtained by the MUCA-PS method and where $\cal{R}_{BA}$ is estimated by the EP estimator (Eq. \ref{eq:fgep}). 

$T^*=1.0$, RSM.

\tshow{fig:increasecool}
\label{fig:increasecool}

\begin{center}
{\bf{------------------------------------------}}
\end{center}

\end{figure}

Figure \ref{fig:increasecool} (a) shows the probability distribution of \wba\ as the increment \delela\ is kept fixed but the equilibration time $\triangle t$ is varied, and figure \ref{fig:increasecool} (b) shows the estimate of ${{\tilde {\cal R}}_{BA}}$ (the normalised value of \RBAcal\ with respect to the corresponding value as obtained by the MUCA-PS simulation) as a function of \delt . 

Figure \ref{fig:increasecool} (a) clearly shows that as the equilibration time $\triangle t$ is increased, the mean of the distribution $P(\wba | \pia)$ and its associated variance both decrease. To understand this we first note that each time a work increment $\delta W_{BA,i}$ is performed, a lag develops in the ensemble of configurations immediately associated with the system $\ecal_{\la_{i+1}}$ after this operation. Namely, when the  switch from a \ce\ $\ecal_{\la_i}(\vvec(i))$ to $\ecal_{\la_{i+1}} (\vvec(i))$ is made, the configuration ${{\vvec(i)}}$ will not be typical of the set of configurations associated with $\pi_{\la_{i+1}}$. Furthermore this lag accumulates as one performs the FG process. A consequence of this is that the distributions of the energies $\seta {\ecal_{\la_{i+1}} ({\vvec(i)})}$ associated with the system immediately after its \ce\ has been incremented from $\ecal_{\la_i}$ to $\ecal_{\la_{i+1}}$ will not be the same as the equilibration distribution which we denote by $\seta {\ecal_{\la_{i+1}}(\vvec(i))}_e$. 
Typically the values within the set $\seta {\ecal_{\la_{i+1}}(\vvec(i))}$ will be higher than  those of $\seta {\ecal_{\la_{i+1}}(\vvec(i))}_e$. Increasing the equilibration time \delt\ decreases this lag, and this is precisely what is observed in Figure \ref{fig:increasecool} (a).

Further insight into the workings of the FG method may be obtained by noticing that the components $\delta \wbai = \delela \mbai$ of the overall work term (\ref{eq:linwork}) can be both positive and negative. Suppose that one performs a FG simulation in which the canonical distribution of the 'intermediate' stages are given by those shown in figure \ref{pic:OPforreplicas}.
 For zero equilibration time (\delt = 0) $\delta \wbai = \delela M_{BA} (\vvec(1))$ for all i, where $M_{BA} (\vvec(1))$ is the starting value of \mba, namely $\mba(\vvec(1))$. Since this corresponds to a value of $\mba$ chosen from $P(\mba|\pia)$ (the right hand peak of \ref{pic:OPforreplicas}) we see that $\delta \wbai$ will be almost always positive. Suppose that $P(\mba | t_i)$ denotes the probability distribution of \mba\ at time $t_i$, when the \ce\ has been incremented from $\ecal_{\la_{i-1}}$ to $\ecal_{\la_i}$ and {\em after} the system has been equilibrated with $\pi^c_{\la_i}$ for a time \delt . Then as the equilibration time \delt\ increases, the distributions  $P(\mba | t_i)$ will shift from the right hand peak in figure \ref{pic:OPforreplicas} towards the left. In the limiting case of $\delt \rightarrow \infty$ one will find that $P(\mba | t_i) \rightarrow P(\mba | \pi_{\la_i})$, so as to yield the collection of distributions in figure \ref{pic:OPforreplicas} \cite{note:supfg}. 
 Therefore we see that the increase of the equilibration time \delt\ will eventually lead to significant cancellations between terms  in Eq. \ref{eq:linwork},  resulting in a decrease (on average) of \wba\  from the value it assumes in the case of  zero equilibration. 
As a result one obtains  improved overlap between $\hat{P}(\wba | \pi^c_A)$ and $\hat{P}(\wba | \pi^c_B)$, resulting in the convergence of \RBAcal\ in the limit of large \delt , as can be seen from figure \ref{fig:increasecool}(b) (note that the convergence is conditional on \delela\ being small enough).

\subsection{Keeping \delt\ constant, varying \delela}

\begin{figure}[tbp]
\begin{center}
\ifpdf
\rotatebox{90}{
\includegraphics[scale=0.6]{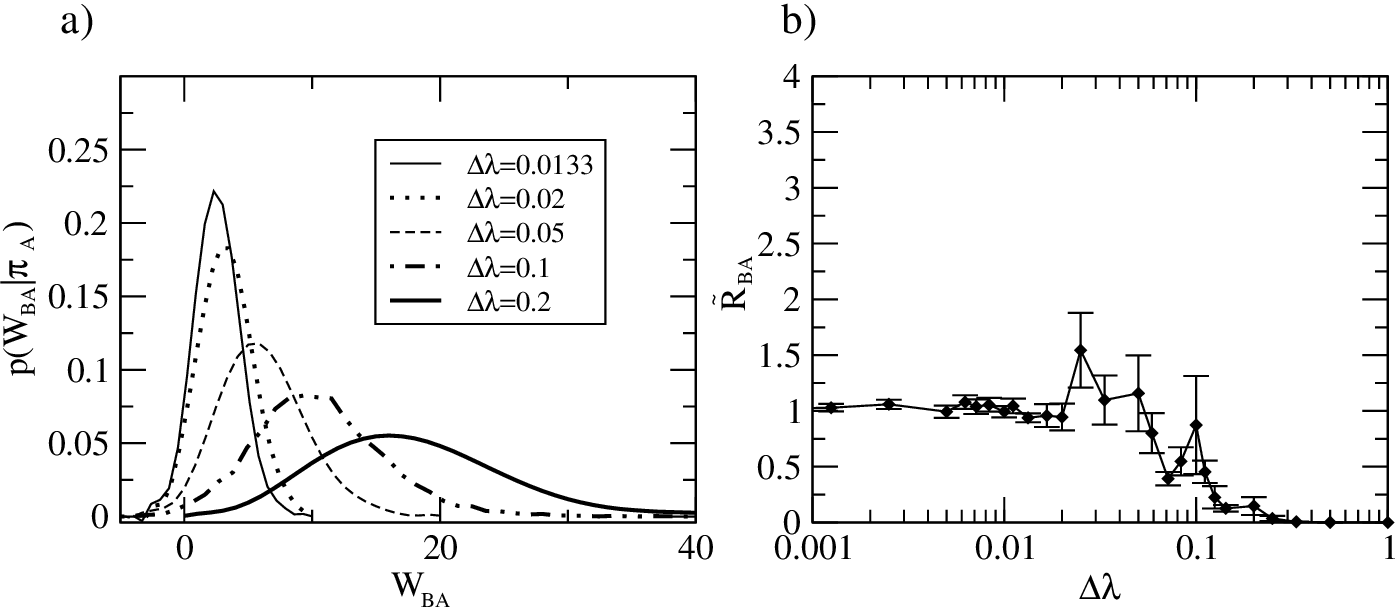}
}
\else
\includegraphics[scale=0.6]{fig/keepingcoolingtimeconstant2}
\fi
\end{center}
\caption{Convergence of ${\tilde {\cal{R}}}_{BA}$ as the equilibration time $\triangle t$ is kept constant but as $\triangle \la$ is varied}

Figure \ref{fig:smallertimestep} (a) shows the variation of ${P}(W_{BA}| \pia)$ as the equilibration time $\triangle t$ is kept fixed, but as the increment $\delela$ is varied. 

Figure \ref{fig:smallertimestep}(b) shows the associated convergence of ${{\tilde {\cal R}}_{BA}}$.

See figure \ref{fig:increasecool} for the definition of $\tilde{\cal{R}}_{BA}$.

$T^*=1.0$, RSM.

\tshow{fig:smallertimestep}
\label{fig:smallertimestep}
\begin{center}
{\bf{------------------------------------------}}
\end{center}
\end{figure}

Figure \ref{fig:smallertimestep} (b) illustrates that as the equilibration time $\delt$ is kept constant, whilst \delela\ is decreased, \RBAcal\ converges as the overlap between $P(\wba | \pia)$ and $P(\wba | \pib)$ increases. In order to understand this consider the following argument. As \delela\ decreases, less equilibration time is needed between successive work increments, until eventually \delt\ matches the equilibration time 'needed' in order for the lag to be absent so that $P(\mba | t_i) =  P(\mba | \pi_{\la_i})$. In this case the distribution of \mba\ at each timeslice will be something reminiscent of what is shown in figure \ref{pic:OPforreplicas}. Subsequent decrease of \delela\ will just correspond to increasing the number of \ces\ in the chain of Eq. \ref{eq:hamilchain}, which will lead to 'better' cancellations in Eq. \ref{eq:linwork}, thus taking $P(\wba | \pib)$ closer to the ideal limit in Eq. \ref{eq:mhbzlim}. Eventually the overlap between the two \pc s will be sufficiently great so as to ensure the convergence of \RBAcal\ even when estimated via the EP method (Eq. \ref{eq:fgep}), as is clearly verified in figure \ref{fig:smallertimestep}.

Let us now analyse the behaviour of the statistical and systematic errors for the FG-EP estimator in the context of the systems that we have studied. Insight into the interplay between statistical and systematic errors may  be obtained from figure \ref{fig:smallertimestep} (b). It is clear from this figure that whereas for large \delela\ systematic errors are present (since $\tilde{\cal R}_{BA} \approx 0$), for small \delela\  they are absent (since $\tilde{\cal R}_{BA} \approx 1$). In between these two limits, one finds that as \delela\ decreases, the systematic errors decrease. The behaviour of the statistical errors, on the other hand, is quite different. In this case the statistical errors are small in both the large \delela\ and small \delela\ limits, and in between these limits there is a transient regime where the statistical errors greatly increase. This is merely an artifact of the EP estimator, and may be understood as follows.

In the case of negligible overlap the weights  $w_n(\wba)$ (see Eq. \ref{eq:epn}) of the macrostates {\em actually sampled} are small in value in comparison to the macrostates which contribute most significantly to the numerator of Eq. \ref{eq:estepfg}. As a consequence the variance of the estimate of \RBAcal\ (see Eq. \ref{eq:estepfg})  will be small, since the estimate of \RBAcal\ will itself be small. As the overlap increases, one eventually enters a regime where  the main body of $\hat{P}(\wba | \pib)$ resides within the tail of $\hat{P}(\wba | \pia)$. In this case, the macrostates which contribute significantly to the numerator of Eq. \ref{eq:estepfg} will originate from the tail of $\hat{P}(\wba | \pia)$, and as a consequence their statistics will be bad, resulting in large statistical errors in the FED estimate, despite the absence of systematic errors. As the overlap improves more of  $\hat{P}(\wba | \pib)$ gets contained in the main body of  $\hat{P}(\wba | \pia)$, and as a result the statistical error in the estimate of the FED improves.

\subsection{Keeping \f{\delt} {\delela} constant}

\begin{figure}[tbp]
\begin{center}
\ifpdf
\rotatebox{90}{
\includegraphics[scale=0.6]{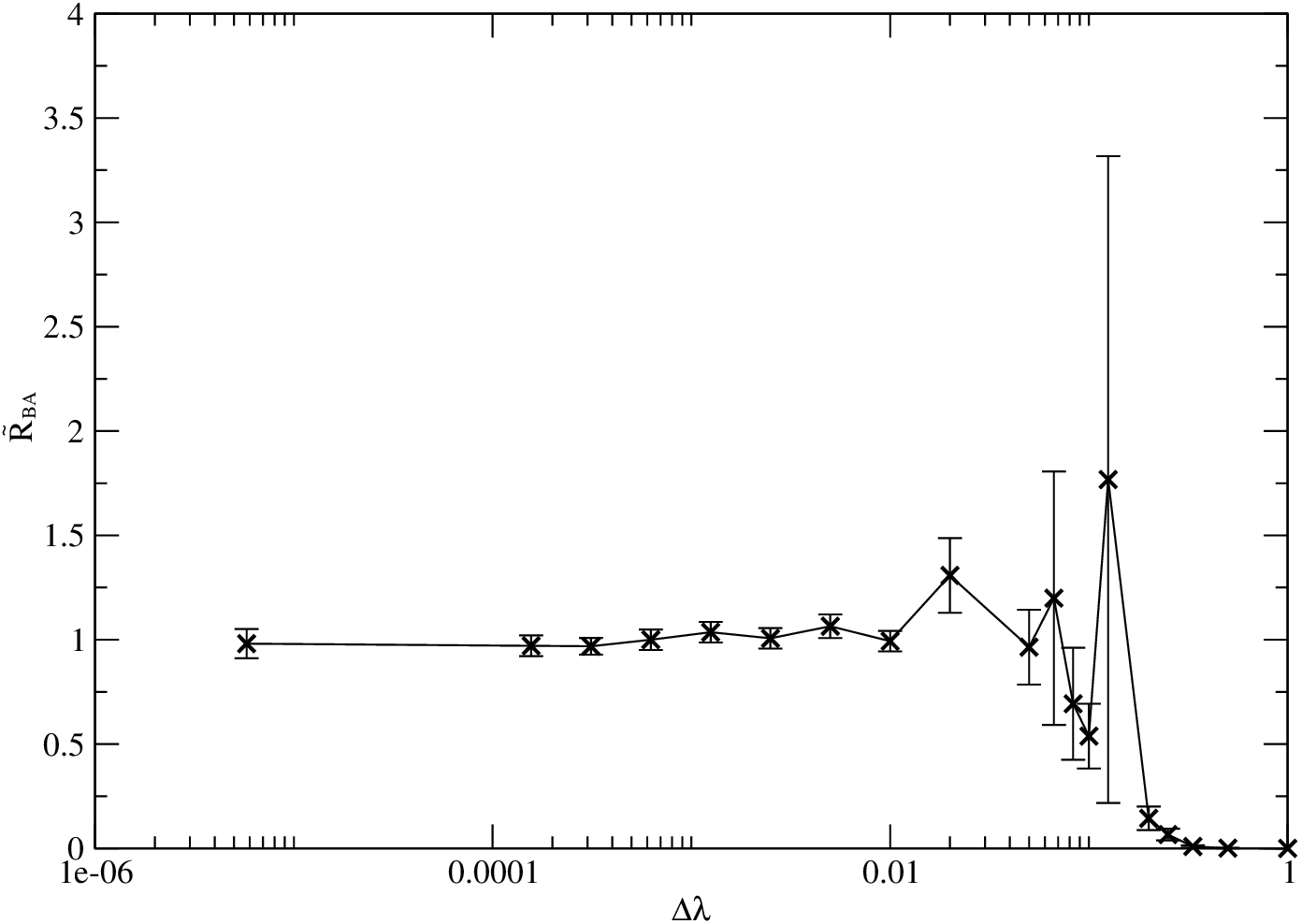}
}
\else
\includegraphics[scale=0.6]{fig/convergenceoffree_keepingworktimeconstant}
\fi
\end{center}
\caption{Convergence of ${\tilde {\cal{R}}}_{BA}$ as a function of \delela\  given $\frac {\triangle t} {\triangle \la}$=constant}

$\delt/ \delela$ corresponds to the time spent obtaining each work term \wba . Therefore the parameters \delt\ and \delela\ are varied so as to keep the total time expended on obtaining each work term \wba\ constant.

See figure \ref{fig:increasecool} for the definition of $\tilde{\cal{R}}_{BA}$.

\tshow{fig:constantworktime}

\label{fig:constantworktime}

\begin{center}
{\bf{------------------------------------------}}
\end{center}

\end{figure}

We also considered the convergence of $\tilde{\cal R}_{BA}$ as \delela\ is varied, given the constraint that the total time spent obtaining each work term $W_{BA}(i)$ is kept constant. Since $\frac {1} {\delela}$ corresponds to the number of times the \ce\  $\ecal_\la$  is 'perturbed'  the constraint of keeping the time ($t_W$) spent obtaining each work term $W_{BA}(i)$  constant corresponds to:

\begin{equation}
\frac {\delt} {\delela} = t_W
\label{eq:deltcon}
\end{equation}\tshow{eq:deltcon}

\noindent It is clear that  if \delela\ is too large, then one will approach the limit of  zero equilibration, resulting in the appearance of the systematic errors described in section \ref{sec:perturbation}. This is exactly what is observed in figure \ref{fig:constantworktime}, since as $\delela\rightarrow 1$ $\tilde {\cal R}_{BA}\rightarrow 0$. As \delela\ is decreased these systematic errors vanish, resulting in \RBAcal\ converging to unity \cite{note:conditionalcon}.

The behaviour exhibited in figure \ref{fig:constantworktime} may be understood as follows. After each work increment a lag develops. This lag can be overcome by appropriately equilibrating the system. However the lag obtained from going from $\la_{n-1}$ to $\la_n$ (as measured by the deviations of \seta {\ecal_{\la_{n-1}}(\vvec(n-1))} from $\seta {\ecal_{\la_{n-1}}(\vvec(n-1))}_e$) can only be removed by performing equilibration {\em after} the \wba\ measurement has been made. Therefore this lag {\em persists} in the value of \wba\ that one obtains. As a consequence if \delela\ is too large, $|\wba|$ will become large, leading to less overlap and eventually to systematic errors.

We finish off this section by noting that it  was observed from the numerical data that the errors seemed to be essentially independent of \delela, provided \delela\ was sufficiently small. In other words our numerical work seems to indicate that the error in ones estimate of \RBAcal\  (based on the estimator of the EP method, Eq. \ref{eq:fgep}) essentially depends on the total time allocated to obtaining each estimate of \wba\ (provided \delela\ is sufficiently small), and not on $\delela$.

\subsection{The choice of estimator}

Our investigations into the way in which the FG method deals with the overlap problem have primarily focused on the EP estimator (Eq. \ref{eq:fgep}). As we have noted in chapter \ref{chap:estsamp} this estimator will require a significantly greater amount of overlap between the phase-constrained distributions than will be the case of the dual-phase (DP) estimators (Eq. \ref{eq:dualphaserestrict}, Eq. \ref{eq:freebennettgen}) or the PS estimator (Eq. \ref{eq:psfermi}) which only require partial overlap \cite{note:partialmean} between the two distributions. Since the application of these estimators is straightforward, we will only illustrate the use of the PS estimator, which has not been formulated before within the context of FG.

\begin{figure}[tbp]
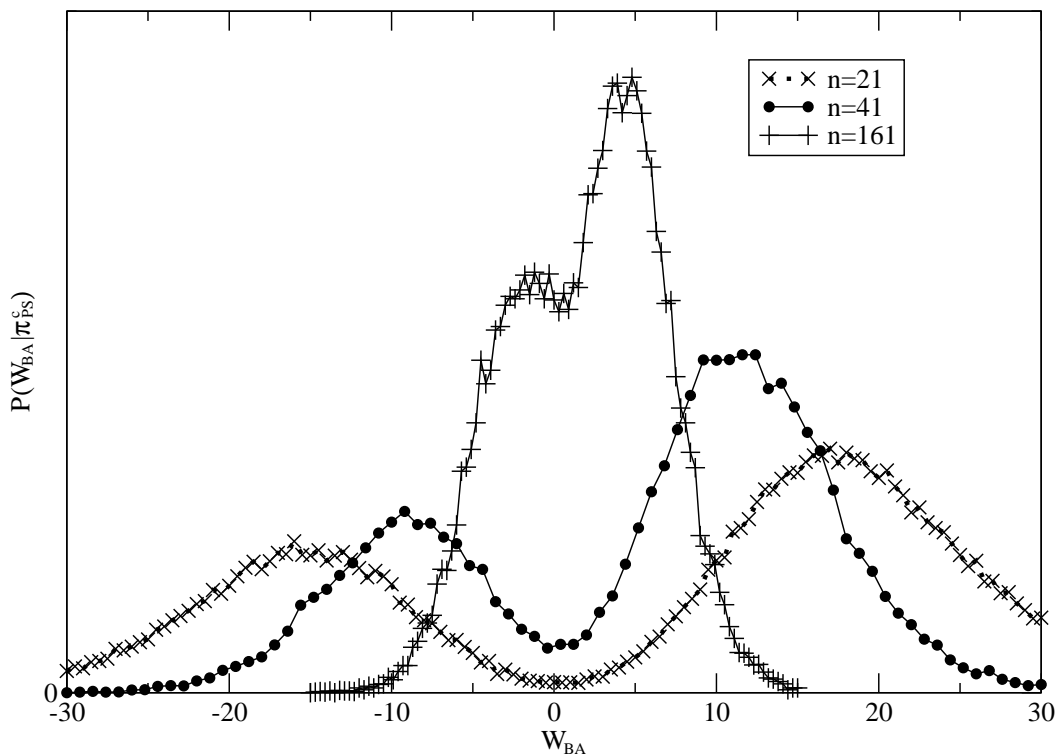

\begin{center}
\ifpdf
\rotatebox{90}{
\hidefigure{\includegraphics[scale=0.6]{fig/fg_switch}}
}
\else
\rotatebox{0}{
\hidefigure{\includegraphics[scale=0.6]{fig/fg_switch}}
}
\fi
\end{center}
\caption{
The PS method as incorporated for the FG method
}

In this figure, we illustrate use of the PS method in conjunction with the FG method (section \ref{sec:pfg}, Eq. \ref{eq:pfgps}). The probability distributions were obtained by keeping \delt\ constant and decreasing \delela . It is clear as \delela\ decreases, the overlap increases, so as to result in the convergence of the two peaks.

$T^*=1.0$, RSM.

\begin{center}
{\bf{------------------------------------------}}
\end{center}
\label{fig:fgswitch}
\tshow{fig:fgswitch}
\end{figure}

Figure \ref{fig:fgswitch} illustrates the application of the FG-PS estimator, as described in section \ref{sec:pfg}, to the systems studied here. In this experiment the equilibration time \delt\ has been kept constant whilst $\delela = 1 / (n-1)$ has been gradually decreased. As we saw earlier in section \ref{eq:dtc} that this leads to increased overlap between $\hat{P}(\wba | \pia)$ and $\hat{P}(\wba | \pib)$. This is also clearly evident in figure \ref{fig:fgswitch}, since the two peaks begin to merge into one as \delela\ decreases, and is reminiscent of what is observed in figure \ref{pic:MHLS}. Until now all FED calculations have been limited to the EP estimator and  figure \ref{fig:fgswitch} (in which the overlap problem has been cured) shows the scope for improvement in using more 'intelligent' estimators.

\section{\label{sec:conc}Conclusion}
\tshow{sec:conc}

Given a choice of representation one obtains two phase constrained distributions $P(\wba | \pia)$ and $P(\wba | \pib)$. In a finite run simulation their estimators $\hat{P}(\wba | \pia)$ and  $\hat{P}(\wba | \pib)$ may or may not overlap. If they overlap then one may choose an estimator (Eq. \ref{eq:dualphaserestrict}, Eq. \ref{eq:freebennettgen}, or Eq. \ref{eq:psfermi}) which yields an estimate of the FED which is free of systematic errors. In the absence of overlap, one must {\em engineer} overlap via one of three possible strategies:

\begin{enumerate}
\item {\em MUCA method:} In this approach MUCA weights are employed to force the simulation to visit the regions of \ecs\ which it would not visit under the canonical sampling distribution and which at the same time contribute non-negligibly to the estimator of the FED. In this way a path is constructed from one region of \ecs\ to the other.

\item {\em MH method:} In this method one employs a series of {\em independent} simulations which overlap in the regions of \ecs\ that they explore, so as to provide a path linking the two phases. Specifically, one engineers overlap by increasing n, the number of configurational energies comprising the composite systems. The benefit of this approach is that it is highly parallelizable, albeit at the expense of additional overheads in terms of memory requirements.

\item {\em FG method:} In this approach the path is constructed by performing non-equilibrium work on the system so as to take it from the regions of \ecs\ associated with one phase to those of the other. One engineers overlap by making the work increments \delela\ sufficiently small and by adequately equilibrating the system between successive work increments.

\end{enumerate}

\noindent These extended sampling strategies may be combined in a straightforward way. For example, {\em the MH method may be incorporated into the framework of the FG method simply by performing work on the hamiltonians $\hcalone {A}$} {\em and $\hcalone {B}$}. Other combinations (such as MUCA and MH) are also possible. In deciding what combinations to use, it is important to bear in mind that both the MUCA and FG are {\em serial} strategies, whereas the MH strategy is a {\em parallel} strategy. Since speedup offered by the MH strategy comes at the expense of additional memory requirements, combinations of this method with either the MUCA or FG methods may be an attractive option.

\chapter{\label{chap:quantum}Quantum Free Energy Differences}
\tshow{chap:quantum}

\section{Introduction}

Our focus until now has been limited to the classical regime of the phase diagram, and  in this chapter we will concentrate our attention on the task of estimating the FEDs within the quantum regimes. Specifically we will use the Path Integral Formalism of statistical mechanics \cite{b:feynmanstat}, \cite{p:fosdickpart}\nocite{p:fosdick}\nocite{p:jordanfosdick}\nocite{p:fosdickreview}\nocite{p:gillanreview}\nocite{p:creutzreview}\nocite{p:bernereview}\nocite{p:ceperleyreview}\nocite{p:schmidtreview}-\cite{b:feynmanhibbs} to map the problem onto that of determining the ratio of two multidimensional integrals of the form used in Eq. \ref{eq:RBA} (see also \cite{p:voter}, \cite{p:fosdickpart}, \cite{p:sscw}\nocite{p:kono}\nocite{p:beck}\nocite{p:pimcreinhardt}\nocite{p:mielke}\nocite{p:barratloubeyre}\nocite{p:vorontsov}\nocite{p:srinivasan}-\cite{note:ackrev}). This will make available to us the spectrum of methods discussed in the previous chapters. For a more rigorous and in depth presentation of the following material, we refer the reader to \cite{b:pathria}, \cite{p:gillanreview}, \cite{p:ceperleyreview}, and \cite{p:tuckermanonline}.

\section{\label{sec:PIMCstat} Path integral formulation of statistical mechanics}

\subsection{Quantum statistical mechanics}

We recap that in the canonical case, classical statistical mechanics \cite{p:vved}\nocite{b:reif}\nocite{b:honerkamp}-\cite{b:pathria} begins with the construct of a system of N particles and of volume V  which is {\em weakly} coupled to a {\em macroscopic} reservoir. Suppose that $\sigma^s$ denotes a state of the system (which we call a microstate of the system) and that $\sigma^r$ denotes a state (or microstate) of the reservoir. Also let us suppose for the moment that the set of states $\seta {\sigma^s,\sigma^r}$ is finite \cite{note:discrete}. The core assumption of classical statistical mechanics is that if the collection of the system and the reservoir is itself considered to be an isolated system of total energy $H^{s+r}$, then this collective system is {\em equally} likely to be in any one of the microstates $(\sigma^s,\sigma^r)$ accessible to it (i.e. of energy  $H^{s+r}$) when the dynamics of this collective system have been averaged out over sufficiently long periods of time.  A consequence of this so called `Equal A Priori Probabilities' assumption and the weak coupling assumption is that one finds that the absolute probability of the system being in microstate $\sigma^s_i$ at any given instant of time is given by:

\begin{equation}
P(\sigma^s_i) = \f 1 Z e^{-\beta H ({\sigma^s_i})}
\label{Gibbs_prob}
\end{equation}\tshow{Gibbs_prob}

\noindent where the partition function Z is given by:

\begin{equation}
Z= \s {i} {} e^{-\beta H( {\sigma^s_i})}
\label{part_fn}
\end{equation}\tshow{part_fn}

\noindent $H$ is the hamiltonian of the system and $\s {i} {}$ denotes a summation over all microstates $\sigma^s_i$ that are available to the system \cite{note:newpb}. The expectation of a general macrovariable (or 'observable', as will be more appropriate in the quantum case) is then given by:

\begin{equation}
<O> = \frac {1} {Z} \s {i} {} {O(\sigma^s_i) e^{-\beta  H({\sigma^s_i})}} 
\label{eq:classicalexp}
\end{equation}\tshow{eq:classicalexp}

In the quantum mechanical case the construct follows a similar procedure (see \cite{b:honerkamp}, \cite{b:pathria}). At the heart of this procedure is the realisation that because the system is coupled to a {\em macroscopic} heat bath, the system will be in one of the eigenstates of the hamiltonian operator of the system, and not in a superposition of states \cite{note:qfur}. Furthermore, in the derivation of Eq. \ref{Gibbs_prob} the exponential comes directly from the entropic properties of the reservoir (\cite{b:reif}\nocite{b:honerkamp}-\cite{b:pathria}). Since the reservoir is classical even in the quantum formulation of statistical mechanics, we deduce that in the quantum mechanical case the probability of finding the system in a microstate $\sigma^s_i$ is once again given by Eq. \ref{Gibbs_prob}. What actually changes is, firstly, what one actually means by a microstate and, secondly, the link that one makes between the observable O and the microstate $\sigma^s_i$. In classical statistical mechanics the observable takes a precise value for each  microstate, since a microstate essentially corresponds to a fixed spatial and momentum configuration of the system. In the quantum mechanical case it will, in the most general case, no longer be the case that the observable takes a definite value for each microstate, since now the microstate $\sigma^s_i$ corresponds to a quantum state (or wavefunction). At best one will only be able to specify the quantum mechanical {\em expectation} of the observable with respect to a given  microstate. Therefore in the quantum mechanical case one replaces  Eq. \ref{eq:classicalexp} by:

\begin{equation}
<O> = \frac {1} {Z} \s {i} {} {<\hat {O}>_{q,\sigma^s_i} e^{-\beta  H({\sigma^s_i})}}
\label{eq:qexp}
\end{equation}\tshow{eq:qexp}

\noindent where $<\hat{O}>_{q,\sigma^s_i}$ denotes the quantum mechanical expectation of the operator $\hat {O}$ with respect to the i-th quantum microstate, or wavefunction, $\sigma^s_i$. By comparison of Eq. \ref{eq:classicalexp} and Eq. \ref{eq:qexp} it is clear that  whereas in classical statistical mechanics one only performs one type of averaging, in the quantum mechanical case one must perform two sorts of averaging. The first average is the quantum mechanical expectation (with respect to a microstate) of the operator $\hat{O}$, and the second is the averaging of this expectation over the quantum microstates accessible to the system. This first averaging, the quantum mechanical expectation, is not something which one performs due to our ignorance of the constituent system, but it is something we {\em have} to do because of the {\em inherent} quantumness of systems. 

To formally develop the theory let us begin by denoting the set of eigenstates of the hamiltonian operator $\hat {H}$ of the system (see \cite{b:sakurai}) by $\{\ket {H_i} \}$. The hermiticity of $\hat H$ will mean that eigenstates of different energy eigenvalues will be orthogonal, though states with the same energy eigenvalue are not necessarily orthogonal. One may, however, employ the Gram-Schmidt orthogonalisation procedure (see \cite{b:kreizig}) to construct a new set of states corresponding to the degenerate eigenvalue which are mutually orthogonal.  Therefore there is no loss in generality if we assume $\{\ket {H_i}\}$ to be a mutually orthonormal set (and therefore correspond to a basis set, see \cite{b:sakurai}). It follows from Eq. \ref{eq:qexp} that the statistical mechanical expectation of an observable O may now be written as:

\begin{equation}
<O> = \f 1 Z \s {i} {} e^{-\beta H_i} \bra {H_i} \hat O \ket {H_i}
\label{q_expect}
\end{equation}

\noindent where $H_i$ denotes the energy eigenvalue associated with the eigenvector $|H_i>$, $\s {i} {}$ denotes a sum over eigenstates, and where the partition function Z is once again given by Eq. \ref{part_fn}. $\bra {H_i} \hat {O} \ket {H_i}$ denotes the {\em quantum mechanical} expectation of the observable O for a given eigenstate $\ket {H_i}$ and accounts for the quantum mechanical properties of the system. As is the case in the classical formula (Eq. \ref{eq:classicalexp}) the weighted summation $\s {i} {} e^{-\beta H_i}$ essentially describes the coupling between the quantum system and the classical reservoir.

Eq. \ref{q_expect} may be written in a more general way as follows:

\begin{eqnarray}
<O> & = & \f 1 Z \s {i} {} \bra {H_i} \hat O e^{-\beta \hat H} \ket {H_i}\\
      & = & \f 1 Z \mbox{Tr}(\hat O \hat \rho) 
\label{trace_expt}
\end{eqnarray}

\noindent where $\hat{\rho}$, the density matrix, is written as:

\begin{equation}
\hat\rho= e^{-\beta \hat H} 
\label{eq:densitymatrix}
\end{equation}\tshow{eq:densitymatrix}

\noindent and where $\mbox{Tr}$ denotes a trace over the matrix elements of the operator $\hat{O}\hat{\rho}$. The partition function may then be written as:

\begin{equation}
Z = \mbox{Tr}(\hat\rho)
\label{eq:quantpart}
\end{equation}\tshow{eq:quantpart}

\noindent The tracing operation performed in Eq. \ref{trace_expt} and Eq. \ref{eq:quantpart} has only been implemented with respect to the orthonormal set corresponding to the energy eigenstates. A-priori these eigenstates, and their associated eigenvalues $H_i$, are not known. Progress is made by noting that the trace is independent of the basis in which it is carried out \cite{b:sakurai}, and therefore one is  free to choose {\em any} representation. A convenient representation is the position representation, in which case the partition function of Eq. \ref{eq:quantpart} for distinguishable (identical but localised) particles simply becomes:

\begin{equation}
Z = \int d\mf r \bra {\mf r} \hat \rho \ket {\mf r}
\label{part_in_pos}
\end{equation} 

\noindent The position representation is useful for the simple  reason that, as we will soon see, it allows one to map the problem of determining Eq. \ref{part_fn} onto that of determining an integral of the form of  Eq. \ref{eq:absolutepart}. This mapping is known as the classical-quantum isomorphism, and is what forms the basis of the path integral computational techniques.

\subsection{The classical-quantum isomorphism \& the path integral}

The partition function, as formulated in Eq. \ref{part_in_pos}, is not fully quantum mechanical in that it ignores exchange. Exchange (see \cite{b:pathria},  \cite{p:gillanreview}, \cite{p:ceperleyreview}) is a quantum mechanical property that arises out of the indistinguishability of identical particles \cite{b:sakurai}, \cite{note:echno}. In order to incorporate this property  into Eq. \ref{part_in_pos}, one rewrites it as:

\begin{equation}
Z = \s {{\cal P}} {} \delta_{\cal P} \int d\mf r \bra {\mf r} \hat \rho \ket {\mf {{\cal P}r}}
\label{eq:gen_QMpart}
\end{equation}\tshow{eq:gen_QMpart}

\noindent where $\mf {{\cal P}r}$ denotes a permutation of the particles, $\s {{\cal P}} {}$ denotes the sum over all such permutations, and $\delta_P$ denotes the sign of the permutation. For bosons $\delta_P = 1$ and for fermions $\delta_P$ assumes  the values 1 and -1 depending on the sign of the permutation (\cite{b:pathria},  \cite{p:gillanreview}, \cite{p:ceperleyreview}).

The expression in Eq. \ref{eq:gen_QMpart} is still not suitable, as it stands, for use in simulation. What remains to be done is to find a way to project the density matrix operator $\hat {\rho}$  onto the position representation $\ket {\rvec}$ so as to ensure that one is left with an expression involving only real numbers. To do this we first decompose our hamiltonian into the sum of a kinetic part $\hat {T}$:

\begin{equation}
\hat {T} = \frac {\hbar^2 \hat{\mf p}^2} {2m}
\label{eq:keop}
\end{equation}\tshow{eq:keop}

\noindent and a configurational part:

\begin{equation}
\hat{E} (\rvec) = E(\rvec)
\end{equation}

\noindent so that:

\begin{equation}
\hat{H} = \hat{T} + E(\rvec)
\end{equation}

\noindent where $\hat {\mf p}$ denotes the momentum operator corresponding to the classical variable $\mf p$, which represents the collective momenta of all the particles, where $\hbar$ denotes Planck's constant, and where m is the mass of the particle. Then the  main obstacle to expressing Eq. \ref{eq:gen_QMpart} in terms of real numbers is the fact that $\hat {T}$ and $E(\rvec)$ are non-commuting operators, which means that there exists no basis in which $\hat {T}$ and $E$ are {\em simultaneously} diagonal \cite{b:sakurai}. Clearly in order to achieve our goal of recasting Eq. \ref{eq:gen_QMpart} as an expression involving only real numbers, we must find a way of separating out the kinetic and the configurational terms in $\hat{\rho}$ so that we can separately diagonalise each contribution; $E$ with respect to the position representation and $\hat {T}$ with respect to  the momentum representation (a representation which it is diagonal in).

The fundamental identity  which allows this to be done is (\cite{p:trotter}\nocite{p:deraedt}\nocite{p:fye}\cite{p:suzukisp}):

\begin{equation}
 e^{-\tau \hat {H}} = e^{-\tau \hat {T}} e^{-\tau E(\rvec)}[1 + O(\frac {\beta^2} {P^3})]
\label{eq:separateexp}
\end{equation}\tshow{eq:separateexp}

\noindent where: 

\begin{equation}
\tau = \frac {\beta} {P}
\label{eq:eftemp}
\end{equation}\tshow{eq:eftemp}

\noindent The Trotter theorem essentially states that in the limit of small $\tau$ one may approximate the operator $e^{-\tau \hhat{}}$ as the product of a 'kinetic' operator $e^{-\tau \hat {T}}$ and a 'configurational' operator $e^{-\tau E}$.

We may now use this (Eq. \ref{eq:separateexp}) to write the partition function as an integral over real numbers. To do this we use the identity $\mf 1 = \int d\rvec \ket{\rvec} \bra{\rvec}$ to re-write Eq. \ref{eq:gen_QMpart} as:

\begin{eqnarray}
Z & = & \s {{\cal P}} {} \delta_{\cal P} \int d\mf r_1 \bra {\mf r_1} e^{-\tau \hat H} e^{-\tau \hat H}....e^{-\tau \hat H} \ket {\mf {Pr}_1} \nonumber \\
  & = & \s {{\cal P}} {} \delta_{\cal P} \int d\mf r_1 d\mf r_2 \bra {\mf r_1} e^{-\tau \hat H}\ket {\mf r_2}\bra {\mf r_2} e^{-\tau \hat H}....e^{-\tau \hat H} \ket {{\mf {Pr}}_1}\nonumber\\
& \vdots & \nonumber \\
  & = & \s {{\cal P}} {} \delta_{\cal P} \int d\mf r_1 ...... d\mf r_P \bra {\mf r_1} e^{-\tau \hat H}\ket {\mf r_2} ...... \bra {\mf r_P} e^{-\tau \hat H}\ket {{\mf {Pr}}_1}\nonumber \\
  & = & \s {{\cal P}} {} \delta_{\cal P} \int \p {i=1} {P}  \bra {\mf r_i} e^{-\tau \hat H}\ket {\mf r_{i+1}} 
\label{eq:part_via_chapman}
\end{eqnarray}\tshow{eq:part_via_chapman}

\noindent where $\mf r_{P+1} = {\mf {{\cal P} r}}_1$ and $\mf r_{i}$ represents the collective displacements of all the particles of replica system i. Applying Eq. \ref{eq:separateexp} to Eq. \ref{eq:part_via_chapman} we see that:

\begin{eqnarray}
\lim_{P\to\infty} \bra {\mf r_i} e^{-\tau \hat H}\ket {\mf r_{i+1}}& = & \bra {\mf r_i} e^{-\tau \hat T} e^{-\tau E} \ket {\mf r_{i+1}}\nonumber \\
 & = &  e^{-\tau E(\mf r_{i+1})} \bra {\mf r_i} e^{-\tau \hat T}  \ket {\mf r_{i+1}}
\label{separation}
\end{eqnarray}\tshow{separation}

\noindent where $E(\mf r_{i})$ denotes the total  \ce\ of replica i. To recast  $\bra {\rvec_i} e^{-\tau \hat {T}} \ket{\rvec_{i+1}}$ we use the identity \cite{b:sakurai}, \cite{note:discretemom}:

\begin{equation}
\mf 1 = \int d\mf p \ket {\mf p} \bra {\mf p}
\label{cont_mom}
\end{equation}

\noindent where $\mf 1$ is the identity operator. Substituting Eq. \ref{cont_mom} into Eq.  \ref{separation} yields:

\begin{eqnarray}
\bra {\mf r_i} e^{-\tau \f {\hbar^2{\hat {\mf p}}^2} {2m} } \ket {\mf r_{i+1}} & = & \int d\mf p \braket {\mf r_i} {\mf p} \bra {\mf p}e^{-\tau \f {\hbar^2{\hat {\mf p}}^2} {2m} }\ket {\mf r_{i+1}}\nonumber \\
 & = & \int d\mf p e^{-\tau \f {\hbar^2{\mf p}^2} {2m}  } \braket {\mf r_i} {\mf p} \braket {\mf p} {\mf r_{i+1}}
\label{kinetic}
\end{eqnarray}\tshow{kinetic}

\noindent which, using the identity \cite{b:sakurai} $\braket {\mf r} {\mf p}  =  \f {1} {(\sqrt{2\pi\hbar})^{3N}} e^{i \f {\mf p . \mf r} {\hbar}}$, becomes:

\begin{eqnarray}
\bra {\rvec_i} e^{-\tau \hat T} \ket{\rvec_{i+1}}  & = & \f {1} {(2\pi\hbar)^{3N}} \int d\mf p e^{i \f {\mf p . (\mf r_i - \mf r_{i+1})} {\hbar}} e^{-\tau \f {\hbar^2 \mf p . \mf p} {2m}}\nonumber \\
  & = & {(\f {1} {4\pi\la_q\tau})^{\f {3N} {2}}\exp\{-\f {1} {4\la_q\tau} (\mf r_{i+1} - \mf r_i)^2\}}
\label{eq:finalkinetic}
\end{eqnarray}\tshow{eq:finalkinetic}

\noindent and where $\la_q$ is given by:

\begin{equation}
\la_q = \frac{\hbar^2} {2m}
\label{eq:debrog}
\end{equation}\tshow{eq:debrog}

\noindent Eq. \ref{eq:finalkinetic} represents the kinetic component appearing in Eq. \ref{separation}. Collating the results of Eq. \ref{eq:part_via_chapman}, Eq. \ref{separation}, and Eq. \ref{eq:finalkinetic} we finally see that the quantum partition function in Eq. \ref{eq:gen_QMpart} may be written as the following limit:

\begin{equation}
Z = \lim_{P\to\infty} Z_P
\label{eq:zmlimit}
\end{equation}\tshow{eq:zmlimit}

\noindent or \cite{note:lowerorder}:

\begin{equation}
Z_P = Z [1+ O(\frac {\beta^2} {P^2})]
\label{eq:scalepart}
\end{equation}\tshow{eq:scalepart}

\noindent where:

\begin{equation}
Z_P = (\f {1} {4\pi\la_q\tau})^{\f {3NP} {2}} e^{-\beta E^0}  \s {{\cal P}} {} \delta_{\cal P} \int d\mf r_1 d\mf r_2 ...d\mf r_P \exp\{-\beta \hcal {} {\{\rvec\}} \}
\label{eq:poly_part}
\end{equation}\tshow{eq:poly_part}

\noindent and:

\begin{equation}
\hcal {} {\{\rvec\}} = \s {i=1} {P} [ \f {P} {4\la_q\beta^2} (\mf r_{i+1} - \mf r_i)^2 + \frac {1} {P} \ecal(\mf r_{i+1}) ]
\label{eq:prim_pot}
\end{equation}\tshow{eq:prim_pot}

\noindent $E^0$ corresponds to the classical groundstate, $\ecal$ denotes the total configurational energy minus the groundstate energy $\ecal(\rvec) = E(\rvec) - E^0$, and  $\rvec_{P+1} = {\mf {\cal P}} \rvec_1$ corresponds to a permutation of the particles in replica $\mbox {1}$. In the absence of exchange  (where particles are localised, so as to make them distinguishable) Eq. \ref{eq:poly_part}  may be simplified to:

\begin{equation}
Z_P = (\f {1} {4\pi\la_q\tau})^{\f {3NP} {2}} e^{-\beta E^0}  \int d\mf r_1 d\mf r_2 ...d\mf r_P \exp\{-\beta \hcal {} {\{\rvec\}} \}
\label{eq:partqm}
\end{equation}\tshow{eq:partqm}

\noindent Eq. \ref{eq:partqm}  represents the partition function of a classical system which is isomorphic to the quantum system of interest. This is generally what is referred to as the classical-quantum isomorphism and forms the starting point, in one form or the other, for the majority of path integral based simulations \cite{note:whypi}.

\subsection{Heuristics of the polymeric system}

\begin{figure}[tbp]
\begin{center}
\rotatebox{270}{
\includegraphics[scale=0.5]{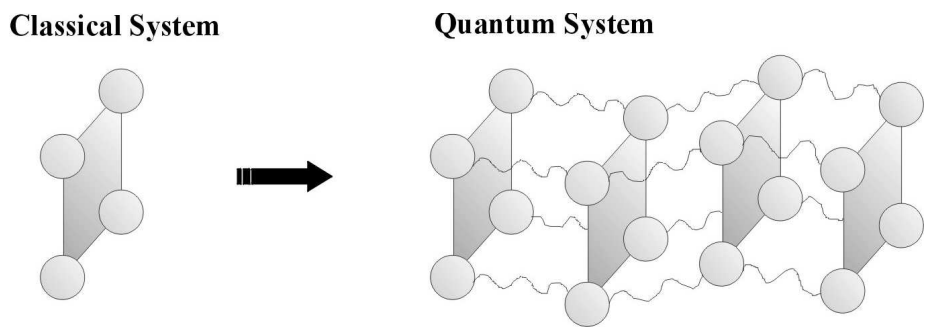}
}
\end{center}
\caption{A schematic of the classical polymer corresponding to the quantum system}
{
By using the path integral formalism of quantum statistical mechanics one finds that an interacting quantum system may be represented by a classical system of polymers, composed of replicas of the given system (see Eq. \ref{eq:poly_part} and Eq. \ref{eq:prim_pot}). The chain linking the {\em same} particle in the adjacent replicas may be thought of as the polymer and a particle in a given timeslice is, in places,  referred to as a bead \cite{note:ceperleyterm}.

Particles within a given replica interact with each other via the classical $\ecal$  giving rise to the configurational action $S_V$ (see Eq. \ref{potential_action}). The quantum effects are controlled by harmonic interactions between the replicas (giving rise to the kinetic action $S_K$, Eq. \ref{kinetic_action}), in which a bead  of a given replica interacts with beads of the adjacent replicas sharing the same bead index.

Strictly the classical-quantum isomorphism mentioned above only holds in the limit of an infinite number of replicas. However in a Path Integral Monte Carlo / Molecular Dynamics simulation, one approximates the quantum system by a finite replica (P) classical polymer. Increasing the number of replicas employed makes the approximation more accurate, but at the expense of increasing the strength of the harmonic interactions between the replicas. For the simulations, this means an increased relaxation time for the polymer and as a result a greater amount of time must be spent (over what one would expect merely by accounting for the increase in the numbers of replicas)  performing the simulation in order to obtain the desired estimates.

\begin{center}
{\bf{------------------------------------------}}
\end{center}
}

\tshow{pic:quantum_switch}
\label{pic:quantum_switch}
\end{figure}

The partition function of Eq. \ref{eq:poly_part} and Eq. \ref{eq:prim_pot} contains all the equilibrium, time independent, information of the quantum system, and serves as the starting point for the Path Integral Monte Carlo (PIMC) methods  \cite{p:fosdick}\nocite{p:jordanfosdick}\nocite{p:fosdickreview}\nocite{p:gillanreview}\nocite{p:creutzreview}\nocite{p:bernereview}\nocite{p:ceperleyreview}-\cite{p:schmidtreview},  
\cite{note:fpimc}, \cite{note:pimd}. Its usefulness lies in the fact that it maps the problem of dealing with an expression involving operators (Eq. \ref{eq:quantpart}) onto one  involving only real numbers. The classical system, represented by Eq. \ref{eq:poly_part} and Eq. \ref{eq:prim_pot}, can be thought of as a system of interacting polymers (see figure \ref{pic:quantum_switch}) with the following action:

\begin{eqnarray}
S_T & \equiv & \beta \hcalone {}\nonumber\\
& = & S_K + S_V
\label{tot_action}
\end{eqnarray}\tshow{tot_action}

\noindent where $S_K$ is the kinetic action:

\begin{equation}
S_K = \f {1} {4\la_q\tau} \s {i=1} {P} (\mf r_{i+1} - \mf r_i)^2
\label{kinetic_action}
\end{equation}\tshow{kinetic_action}

\noindent and $S_V$ is the configurational action:

\begin{equation}
S_V = \s {i=1} {P} \tau  \ecal(\mf  r_{i+1})
\label{potential_action}
\end{equation}\tshow{potential_action}

\noindent Two main parameters control the behaviour of this system of classical polymers. The first is the effective inverse temperature $\tau$ and the second is $\la_q$ (Eq. \ref{eq:debrog}). It is the interplay between these two quantities which determines the strengths of the harmonic interactions in the kinetic action $S_K$ relative to those originating from the  configurational action $S_V$.

This system of interacting polymers is unique (as compared to classical polymers) in that beads of a given polymer interact only with beads of other polymers which are in the same replica, or timeslice \cite{note:ceperleyterm} as we will also call it, labelled by the index i, via the \ce\ $\ecal$ appearing in the configurational action $S_V$.
 In addition to this beads of a given polymer interact, through the term $\f {1} {4\la_q\tau} (\mf r_{i+1} - \mf r_i)^2$ in the kinetic action $S_K$, with the two adjacent beads (of the same polymer) belonging to the two neighbouring  replicas, resulting in a coupling between consecutive replicas (see figure \ref{pic:quantum_switch}). $S_K$ essentially contains the forces which propagate {\em along} a given polymer and $S_V$ contains the forces which give rise to interactions {\em between} polymers.

These polymers are also unique in the sense that they have a special boundary condition, namely that $\rvec_{P+1} = {\mf {\cal P}} \rvec_1$. For the case of distinguishable quantum mechanical particles (${\cal P}=1$), the endpoints of the polymers connect to form loops. Distinguishability of the particles then arises from the fact that one may identify each particle with a given loop. In the presence of exchange the endpoints of  loops coalesce with the starting points of other loops so as to form larger loops, making it impossible to distinguish the exchanging particles since now a single loop may  represent more than one particle. It is in this way that indistinguishability is incorporated into the theoretical framework of the model.

\subsection{\label{sec:tempreg} Temperature regimes in quantum simulations}
\tshow{sec:tempreg}

In the case of quantum systems one may identify three distinct temperature regimes. In the high temperature limit, the system resides within the classical regime where quantum effects may be safely ignored and where the particles are localised to regions in the immediate vicinity of their lattice sites. As the temperature is lowered, one first enters the weak quantum regime where quantum discreteness effects begin to become important. By quantum discreteness we mean those effects arising from the quantisation of energies that accompanies the confinement of particles in their interatomic potential wells. A characteristic of this regime is the increased  amplitudes of vibrations of the particles about their lattice sites (relative to the classical predictions) . This is called the zero point motion and arises from the Heisenberg uncertainty principle. At this point the quantum effects are not strong enough to give rise to exchange, and the particles may, therefore,  still considered to be distinguishable. As the temperature is further reduced one {\em may} enter the strong quantum regime, where  exchange effects can no longer be ignored and where one must explicitly take into account the indistinguishability of the particles. For the Lennard-Jones potential one may easily identity these regimes.

Consider the Lennard-Jones potential given in Eq. \ref{lennard}. The parameter $\epsilon$ roughly measures the well depth, so that one is in the classical regime when the temperature is of the order:

\begin{equation}
kT_c \sim \epsilon
\label{eq:clreg}
\end{equation}\tshow{eq:clreg}

\noindent In this regime the classical effects mask the quantum effects.

As the temperature is further reduced, one eventually enters the weak quantum regime in which the typical particle  energy is less than that of the well depth and is instead of the order of the typical phonon excitation energy:

\begin{equation}
kT_{qw} \sim  h \nu
\label{eq:firstweak}
\end{equation}\tshow{eq:firstweak}

\noindent In this case the quantum zero-point motion effects will be important, but at the same time the exchange effects will not show up in the system. To determine this temperature we note that:

\begin{equation}
\nu \sim \sqrt {\frac {k_{eff}} {m}}
\label{eq:nusp}
\end{equation}\tshow{eq:nusp}

\noindent where m is the mass of the particle and ${k_{eff}}$ is the 'force constant', given by:

\begin{equation}
k_{eff} \sim u^{''}(r=\sigma) \sim \frac {\epsilon} {\sigma^2}
\label{eq:spri}
\end{equation}\tshow{eq:spri}

\noindent Substituting Eq. \ref{eq:nusp} and Eq.\ref{eq:spri} into Eq. \ref{eq:firstweak} one finds that:

\begin{equation}
kT_{qw} \sim h \sqrt {\frac {\epsilon} {m\sigma^2}}
\end{equation}

\noindent It then follows that the difference in the orders between $T_{qw}$ and $T_c$ is given by:

\begin{equation}
\frac {T_{qw}} {T_{c}}\sim \frac {h} {\sqrt{m\epsilon\sigma^2}} = \sqrt {\tilde{D}}
\end{equation}

\noindent where \deb , the De Boer parameter, is given by:

\begin{equation}
\deb = \frac {\hbar^2} {m \epsilon \sigma^2}
\label{eq:pdeb}
\end{equation}\tshow{eq:pdeb}

\noindent As the temperature is further reduced one {\em may} eventually enter a regime where the exchange effects may no longer be ignored. This will happen if the de Broglie wavelength becomes  of the order of the interparticle spacing a:

\begin{equation}
\la \sim a
\label{eq:condexch}
\end{equation}\tshow{eq:condexch}

\noindent where $\la$ is given by:

\begin{equation}
\la = \frac {h} {p}
\label{eq:dbq}
\end{equation}\tshow{eq:dbq}

\noindent where $p$ is the momentum. Since there is a {\em minimum non-zero} value that the total energy of the system can assume it follows that there will be a minimum characteristic value $p^*$ that the absolute value of the momentum can assume. Since for a harmonic oscillator the expectation of the kinetic energy is equal to the expectation of the configurational energy, we see that this  $p^*$ may be crudely estimated by setting:

\begin{eqnarray}
\frac {{p^*}^2} {2m} &  = & \frac {\frac {1} {2} h\nu} {2}\nonumber\\
& = & \frac {h} {4} \sqrt {\frac {\epsilon} {m\sigma^2}}
\end{eqnarray}

\noindent so that:

\begin{equation}
p^* = \sqrt {{\frac {h} {2}} \sqrt{\frac {m\epsilon} {\sigma^2}}}
\end{equation}

\noindent Substituting this in Eq. \ref{eq:condexch}:

\begin{equation}
\frac {\sqrt {2h}} {(\frac {m\epsilon} {\sigma^2})^{\frac{1} {4}}} \sim \sigma
\end{equation}

\noindent Rearranging this equation we get:

\begin{equation}
\tilde{D}\sim 1
\end{equation}

\begin{figure}[htbp]
\begin{center}
\rotatebox{270}{
\includegraphics[scale=0.5]{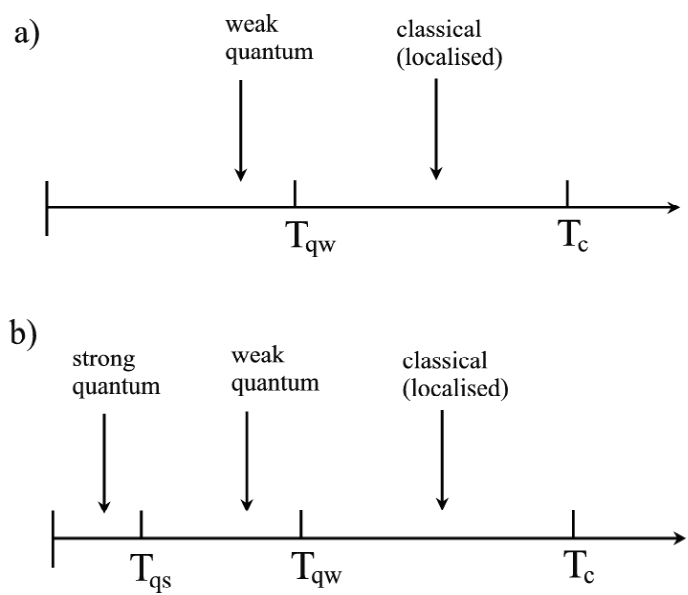}
}
\end{center}
\caption{A schematic showing the different temperature scales that exist}
{
In the {\em weak quantum} regime quantum discreteness is important. By quantum discreteness we mean those effects arising from the quantisation of energies that accompanies confinement of particles in their interactomic potential wells; a regime in which zero point motion is important. In the {\em strong quantum} regime 'exchange effects' become important; particles may no longer be treated as indistinguishable. As an approximate guide, one may say that  if $\tilde{D}$ is small (less than unity) then exchange will not take place \cite{note:nobose}, resulting in the temperature scales shown in (a). If $\tilde{D}$ is large (greater than unity) then exchange will take place and one will have three temperature scales, as shown in (b).

\begin{center}
{\bf{------------------------------------------}}
\end{center}
}

\tshow{pic:tempscales}
\label{pic:tempscales}
\end{figure}

\noindent Therefore if $\tilde{D}$ is sufficiently large (obtained,  for example,  by having a particle of small enough mass) then an additional temperature scale $T_{qs}$ will appear at which point exchange between particles may no longer be neglected. These temperature regimes are shown schematically in figure \ref{pic:tempscales}.

In what follows we address the effects of quantum discreteness but not those of quantum exchange.

\subsection{\label{sec:macest}Estimating macrovariables}
\tshow{sec:macest}

In order  to extract useful information from PIMC simulations, one must find the estimators for the relevant observables within the path integral framework. In the case of estimating thermodynamic quantities, one may derive the estimators merely by taking derivatives of the polymer partition function given in Eq. \ref{eq:partqm}. For example the mean kinetic energy may be derived by using the following relations:

\begin{eqnarray}
<\hat {T}>_P & = & \beta^{-1} m \frac {\partial \ln Z_P} {\partial m}\nonumber\\& = &  \frac {3PN} {2\beta} - \s {m=1} {P} \frac {P m} {2 \beta^2 \hbar^2} <[\rvec_{m+1} - \rvec_{m}]^2>_P\nonumber\\
& = &  \frac {3PN} {2\beta} - \beta^{-1} <S_K>_P
\label{eq:prim_kinetic}
\end{eqnarray}\tshow{eq:prim_kinetic}

\noindent where the subscript P denotes the fact that the expectation is taken with respect to the distribution associated with a P replica polymer (see Eq. \ref{eq:partqm}). Similarly  the mean total  energy of the quantum system may be obtained via the relation:

\begin{eqnarray}
<\hhat>_{P} & = & - \frac {1} {Z_P} \frac{\partial Z_P} {\partial \beta}\nonumber\\
& = & E^0 + <\that>_{P} + <\ecal_q>_P
\label{eq:prim_energy_estimator}
\end{eqnarray}\tshow{eq:prim_energy_estimator}

\noindent where $<\ecal_q>$ is the  \ce\ of the quantum system :

\begin{equation}
<\ecal_q>  =  \beta^{-1} <S_V>_P
\label{eq:prim_potential}
\end{equation}\tshow{eq:prim_potential}

\noindent It must be noted that the estimator in Eq. \ref{eq:prim_energy_estimator} (known as the Barker estimator \cite{p:qbarker}) is not unique. An alternative is the virial estimator \cite{p:herman}. Studies as to the relative efficiencies of these estimators have been made in \cite{p:caoberne}\nocite{p:compest}\nocite{p:compest2}-\cite{p:jankeest}. The general findings are that the more efficient of the two depends on the conditions under which they are used. For example in \cite{p:compest} it was found that for low temperature systems or systems in which the gradients of the \ces\ are high (which is when the quantum effects are typically more significant \cite{note:whysig}) the Barker estimator is preferable. On the other hand it was found that  the virial estimator is preferable at high temperatures or for systems with low gradients of the \ce\ (which is when the quantum effects tend to be less significant). In \cite{p:jankeest}, other findings were also made, one being that as the number of replicas P were increased, the virial estimator eventually became more efficient than the Barker estimator. In this thesis we chose to use  the Barker estimator, mainly due to its simplicity.

The estimators that we have discussed thus far have all been based on finite replica approximations of the quantum partition function Eq. \ref{eq:gen_QMpart}. As a result these estimates will have an associated systematic error (see Eq. \ref{eq:scalepart}).  If $<O>_P$ denotes the finite replica expectation of an operator O \cite{note:ocond} and if $<O>_\infty$ corresponds to the infinite replica estimate, then it follows \cite{p:fye} that:

\begin{equation}
<O>_P = <O>_{\infty} + O(\frac {\beta^2} {P^2})
\label{eq:thermo_var_scale}
\end{equation}\tshow{eq:thermo_var_scale}

\noindent Eq. \ref{eq:thermo_var_scale} provides a clear prescription with which one may proceed to arrive at an estimate which is free of systematic error.
In order to estimate the asymptotic limit $<O>_{\infty}$ what one does is to plot a graph of $<O>_P$ versus $1/P^2$. Provided that P is sufficiently large, so that the corrections in Eq. \ref{eq:scalepart} and Eq. \ref{eq:thermo_var_scale} in which P is raised to a power higher than two may be neglected, the corresponding plot should yield a straight line graph whose intercept gives an estimate for $<O>_{\infty}$ \cite{note:likeclassical}. An illustration of this will be given in section \ref{sec:qres}.

The partition function, with $\hcal {} {\{\rvec\}}$ given by Eq. \ref{eq:prim_pot}, corresponds to what is known as the primitive approximation (PA) in the Quantum Monte Carlo literature \cite{p:creutzreview}\nocite{p:gillanreview}\nocite{p:bernereview}\nocite{p:ceperleyreview}-\cite{p:schmidtreview}. The widespread prevalence of the use of the PA in current literature is due to its underlying simplicity. By exploring more accurate decomposition schemes (section \ref{sec:higherorder}) to that used in Eq. \ref{eq:separateexp}, one may derive what we will refer to as the higher order approximants (HOA). These HOA methods are more accurate than their PA counterparts in that the error terms in equations of the form of Eq. \ref{eq:thermo_var_scale} decay faster than $1/P^2$. However they come at the expense of increased complexity and computational expenditure. In the next section we will briefly discuss the HOA method.

\subsection{Higher Order Approximants \label{sec:higherorder}}
\tshow{sec:higherorder}

The primitive approximation, leading to Eq. \ref{eq:partqm} with $\hcalone {}$ given by Eq. \ref{eq:prim_pot}, is so called due to the fact that the Trotter decomposition given in Eq. \ref{eq:separateexp} are the simplest such breakups of the hamiltonian. Other breakups do exist, and may be written in the form \cite{p:deraedt}\nocite{p:fye}-\cite{p:suzukisp}, \cite{p:libroughton}, and are called the higher order approximants. These higher order approximants allow one to use a smaller number of replicas than the PA methods in order to achieve a desired level of accuracy \cite{note:pimcsys}, since the systematic error associated with these methods decay faster with increasing P. However these gains have to be appropriately balanced against the increased complexity and increased computational cost that accompanies their implementation. One such approximant is based on a Wigner-Kirkwood like expansion \cite{p:ceperleyreview}, \cite{p:libroughton},  \cite{p:wigner}\nocite{p:kirkwood}-\cite{p:takahashi2}). For this method the systematic error in a finite replica approximation scales as $\frac {1} {P^4}$. The method is based on the identity \cite{p:libroughton}:

\begin{equation}
e^{-\tau (A+B)}  = e^{-\tau \frac {A} {2}} e^{-\tau \frac {B} {2}} e^{-\tau^3 \frac {C} {24}}e^{-\tau \frac {B} {2}} e^{-\tau \frac {A} {2}} + O(\tau^5)
\end{equation}

\noindent where 

\begin{equation}
C= [[B,A], A + 2B]
\end{equation}

\noindent and:

\begin{equation}
[A,B]=AB-BA
\end{equation}

\noindent Application of this identity \cite{p:libroughton} once again yields Eq. \ref{eq:zmlimit} and  Eq. \ref{eq:partqm}, where $\hcal {} {\seta {\rvec}}$ is now given by:

\begin{equation}
 \hcal {} {\seta {\rvec}}  = \s {i=1} {P} [ \f {P} {4\la_q\beta^2} (\mf r_{i+1} - \mf r_i)^2 + \frac {1} {P} \ecal (\mf r_{i+1}) + \frac{\beta^2 \la_q} {12P^3} [\nabla_i \ecal(\rvec_i)]^2 ]
\label{eq:HOVeff}
\end{equation}\tshow{eq:HOVeff}

\noindent where:

\begin{equation}
[\nabla_m \ecal(\rvec_m)]^2 \equiv [\frac {\partial \ecal(\rvec_m)} {\partial\rvec_m}]^2 = \s {i} {} \s {j} {} [(\frac {\partial \ecal(s^{(m)}_{ij})} {\partial x^{(m)}_i})^2 + (\frac {\partial \ecal(s^{(m)}_{ij})} {\partial y^{(m)}_i})^2 + (\frac {\partial \ecal(s^{(m)}_{ij})} {\partial z^{(m)}_i})^2]
\end{equation}

\noindent By appropriately differentiating Eq. \ref{eq:poly_part} (with $\hcalone {}$ given by Eq. \ref{eq:HOVeff}) one may also obtain the estimator for the expectation of the kinetic, configurational, and total energies of the system:

\begin{eqnarray}
<\hat{T}>_P & = & \beta^{-1} m \frac {\partial \ln Z} {\partial m}\nonumber\\
& = & \frac {3NP} {2\beta} - \s {i=1} {P} \frac {P} {4\la_q \beta^2} <(\rvec_i - \rvec_{i-1})^2> \nonumber\\
& + & \frac {\la_q} {12} (\frac {\beta} {P})^2 \frac {1} {P} \s {i=1} {P} <(\frac {\partial \ecal(\rvec_i)} {\partial \rvec_i})> 
\end{eqnarray}

\begin{eqnarray}
<\hat{H}>_P & = & -\frac {1} {Z_P} \frac {\partial Z_P} {\partial \beta}\nonumber\\ 
& = & \frac {3PN} {2\beta} - \frac {P} {4\la_q\beta^2} \s {i=1} {P} <(\rvec_{i+1} - \rvec_i)^2>  + \frac {1} {P}  \s {i=1} {P} <\ecal(\rvec_i)>\nonumber\\
& + & \frac {\beta^2 \la_q} {4P^2} <[\nabla_i \ecal(\rvec_i)]^2> + E^0
\label{eq:HOenergy}
\end{eqnarray}\tshow{eq:HOenergy}

\begin{eqnarray}
<\ecal_q>_P & = & <\hcalone{}>_P - <\hat{T}>_P - E^0 \nonumber\\
& = & \frac {1} {P} \s {i=1} {P} \ecal(\rvec_i) + \frac {\la_q} {6} (\frac{\beta} {P})^2 \frac {1} {P} \s {i=1} {P} (\frac {\partial \ecal(\rvec_i)} {\partial \rvec_i})^2
\end{eqnarray}

\noindent It can be shown \cite{p:fye}, \cite{p:libroughton} that within the HOA scheme, the finite replica estimate of the expectation of an observable O, $<O>_P$, will scale towards the asymptotic limit, $<O>_\infty$ in the following way:

\begin{equation}
<O>_P = <O>_{\infty} + O(\frac {\beta^4} {P^4})
\label{eq:thermohoa}
\end{equation}\tshow{eq:thermohoa}

\noindent Similarly it is not hard to show that (Eq. \ref{eq:scalepart}):

\begin{equation}
Z_P = Z [1+O(\frac{\beta^4} {P^4})]
\end{equation}

\noindent As before we note that a finite replica estimate will have a systematic error associated with it. The extrapolation technique described in  section \ref{sec:macest} may then be used to estimate the asymptotic value of the appropriate expectation. That is if one plots a graph of $<O>_P$ against $1/P^4$, then the intercept of the graph on the vertical axis should, provided P is large enough, yield an estimate of $<O>_{\infty}$ which is free of systematic errors.

\subsection{\label{sec:classic_limit} The classical limit}
\label{sec:classical_limit}

In order to develop an intuition for the polymeric-like system described by Eq. \ref{eq:poly_part} it is instructive to observe the emergence of the classical limit out of Eq. \ref{eq:poly_part} by considering the interplay between the kinetic and the configurational actions $S_K$ and $S_V$.
 To do this consider a simulation in which a sufficient number of replicas P have been employed so as to ensure that the quantum effects are modelled to the desired accuracy. Now consider increasing the temperature, so as to reduce $\tau$ \cite{note:reducerep}. The effect of this is to increase the strength of the spring constant $\frac {1} {4\la_q\tau}$ associated with the harmonic-like kinetic term $S_K$, resulting in increased rigidity of the polymers. This increased rigidity has two effects. The first is to make permutations, other than the identity permutation, increasingly unlikely. The second is to make the spatial arrangements of the particles of the various replicas increasingly similar ($\rvec_i\approx\rvec_{i+1}$, so that $\s {i=1} {P} \tau \ecal (\rvec_{i+1}) \approx \beta \ecal (\rvec_j)$ where j denotes any replica). It is clear that what we are seeing is the emergence of classical behaviour, something we would expect on the transition to higher temperatures. That is in this limit Eq. \ref{eq:poly_part} reduces to:

\begin{equation}
Z_P = e^{-\beta E^0} \int d\rvec_1 e^{-\beta \ecal(\rvec_1)} G(\rvec_1)
\end{equation}

\noindent where:

\begin{eqnarray}
G(\rvec_1) & = & (\f {1} {4\pi\la_q\tau})^{\f {3NP} {2}} \int d\mf r_2 d\mf r_3 ...d\mf r_P \exp\{-\f {1} {4\la_q\tau} [(\mf r_{1} - \mf r_2)^2 + (\mf r_{2} - \mf r_3)^2 + ... + (\mf r_{P} - \mf r_1)^2]\}\nonumber\\
& = & [\frac {1} {4\pi\la_q \beta}]^{\frac {3N} {2}}
\end{eqnarray}

\noindent or 

\begin{equation}
Z_P = (\frac {1} {4\pi\la_q \beta})^{\f {3N} {2}} e^{-\beta E^0} \int d\rvec_1 e^{-\beta \ecal(\rvec_1)} 
\label{eq:redcla}
\end{equation}\tshow{eq:redcla}

\noindent Eq. \ref{eq:redcla} is simply the classical partition function.

\section{Quantum FEDs via the  Path Integral Formalism}

By comparing Eq. \ref{eq:partqm} to Eq. \ref{eq:absolutepart}, it is clear that the generalisation of the expression for the ratio of the partition functions (see Eq. \ref{eq:partrsestimator0} and Eq. \ref{eq:RBA}) so as to account for quantum effects is a trivial one (see \cite{p:voter}, \cite{p:fosdickpart}, 
\cite{p:sscw}\nocite{p:kono}\nocite{p:beck}\nocite{p:pimcreinhardt}\nocite{p:mielke}\nocite{p:barratloubeyre}\nocite{p:vorontsov}-\cite{p:srinivasan}). In this section we formulate the quantum version of the real space mapping (Q-RSM) and a  quantum version of the fourier space mapping (Q-FSM) for estimating FEDs of phases.  Though the Q-RSM is quite similar to its classical counterpart, the Q-FSM is quite different, and takes into account the harmonic interactions propagated by the intra-polymer (or inter-replica) interaction term $S_K$ (Eq. \ref{kinetic_action}). In both formulations we will neglect exchange, and we will formulate both methods within the scheme of the PA. Generalisation to the case of HOA methods is straightforward, with the discretisation errors scaling as $1/P^4$ instead of $1/P^2$.

\subsection{Quantum Real Space Mapping}

Limiting ourselves to the case of distinguishable particles, it is clear that if we have two phases A and B and we want to measure the quantum mechanical FEDs then from Eq. \ref{eq:partqm} (see also \cite{p:voter}, \cite{p:fosdickpart}, \cite{p:sscw}\nocite{p:kono}\nocite{p:beck}\nocite{p:pimcreinhardt}\nocite{p:mielke}\nocite{p:barratloubeyre}\nocite{p:vorontsov}-\cite{p:srinivasan})  the ratio of the partition functions is simply determined by:

\begin{equation}
R_{BA} =  e^{-\beta [E^0_B - E^0_A]}\RBAq
\end{equation}

\begin{equation}
\RBAq = \lim_{P\to\infty} \RBAPq
\end{equation}

\noindent where \cite{note:likeclassical}:

\begin{equation}
\RBAPq = \RBAcal [1+O(\frac {\beta^2} {P^2})]
\label{eq:rba_rsm_scale}
\end{equation}\tshow{eq:rba-rsm-scale}

\noindent and:

\begin{equation}
\RBAPq =\f {\int d\mf r_1 ...... d\mf r_P \p {j=1} {P} \triangle_B[\rvec_j]\exp\{-\s {i=1} {P} [ \f {1} {4\la_q\tau} (\mf r_{i+1} - \mf r_i)^2 + \tau \ecal(\mf r_{i+1}) ] \}} {\int d\mf r_1 ...... d\mf r_P  \p {j=1} {P} \triangle_A[\rvec_j] \exp\{-\s {i=1} {P} [ \f {1} {4\la_q\tau} ({\mf r}_{i+1} - \mf r_i)^2 + \tau \ecal({\mf r}_{i+1} )] \} }
\label{quant_free}
\end{equation}\tshow{quant_free}

\noindent Following the derivation of Eq. \ref{eq:RBA} one may map the problem of determining the ratio of the integrals in Eq. \ref{quant_free}, in which a {\em single} hamiltonian is employed, onto that of determining the ratio of two integrals in which the hamiltonians are {\em different}. To do this we express the position of the  particles in terms of the displacements about some reference configuration $\Rvec$, which in the crystalline case is conveniently chosen to be the lattice sites:

\begin{equation}
\rvec_i = \Rvec + \uvec_i
\end{equation}

\noindent where  the subscript i denotes the replica. It then follows that in the $\uvec$ representation the ratio in Eq. \ref{quant_free} may be written as:

\begin{equation}
\RBAP=\f {\int d\uvec_1 ...... d\uvec_P 
e^{-\beta \hcal {B} {\{\uvec\}}}}
{\int d\uvec_1 ...... d\uvec_P 
e^{-\beta \hcal {A} {\{\uvec\}}}}
\label{quant_free_inurep}
\end{equation}\tshow{quant_free_inurep}

\noindent where:

\begin{equation}
 \hcal {\al} {\{\uvec\}} = \s {i=1} {P} [ \f {1} {4\la_q\tau} (\uvec_{i+1} - \uvec_i)^2 + \tau \ecal_\al(\uvec_{i+1}) ]
\label{eq:polyham}
\end{equation}\tshow{eq:polyham}

\noindent This mapping (which we call the quantum RSM, or Q-RSM) is one in which one  {\em simultaneously} maps the displacements of the particles of each and every replica of phase \al\ onto the {\em corresponding} replica of phase \alp :

\begin{equation}
\Rvec_\al, \uvec_i \rightarrow \Rvec_\alp, \uvec_i \mbox{\ \ for all replicas i}
\label{eq:QRSM}
\end{equation}\tshow{eq:QRSM}

\noindent All the simulations performed in this chapter were implemented via the Q-RSM. The crucial feature of the Q-RSM is that since:

\begin{eqnarray}
 \f {1} {4\la_q\tau} (\mf r_{i+1} - \mf r_i)^2 & = &  \f {1} {4\la_q\tau} (\mf R - \mf R + {\mf u}_{i+1} - {\mf u}_i)^2\nonumber \\
& = & \f {1} {4\la_q\tau} ({\mf u}_{i+1} - {\mf u}_i)^2
\end{eqnarray}

\noindent the kinetic part of the action $S_K$ is the same in the two phases under the operation of this mapping. This is a significant advantage of this particular mapping since in the large P limit the kinetic action $S_K$ dominates over the configurational action $S_V$ (see appendix \ref{app:dominate}).

Following the development of earlier chapters one may proceed to define a macrovariable which measures the energy cost of mapping the configuration of the  polymeric system associated with phase A onto that of the polymeric system associated with phase B:

\begin{eqnarray}
\mba(\seta {\uvec}) & = &  \beta [\hcal {B} {\{\uvec\}} -  \hcal {A} {\{\uvec\}}]\nonumber\\
&  = & \frac {\beta} {P} \s {i=1} {P} (\ecal_B(\uvec_{i}) - \ecal_A(\uvec_{i}))
\label{eq:switchcost}
\end{eqnarray}\tshow{eq:switchcost}

\noindent By comparing Eq. \ref{quant_free_inurep} to Eq. \ref{eq:RBA_in_uspace} it is immediately clear that the overlap identity (see Eq. \ref{eq:overlapidentity}) will hold even for our quantum system. If $\pi^q_\al$ denotes the quantum sampling distribution of phase \al :

\begin{equation}
\pi^q_\al \equiv \pi^q_\al(\seta {\uvec}) \equaldot e^{-\beta \hcal {\al} {\{\uvec\}}}
\label{eq:qsamp}
\end{equation}\tshow{eq:qsamp}

\noindent then it immediately follows that:

\begin{equation}
\RBAq = \frac {e^{-\mba} P(\mba | \pi^q_A)} {P(\mba | \pi^q_B)}
\label{eq:qoi}
\end{equation}\tshow{eq:qoi}

\noindent From Eq. \ref{eq:qoi} it clear  that all the discussions of the previous chapters (with the exception of chapter \ref{chap:tune}) also apply to the problem of  estimating the quantum mechanical FED's. In particular the vast array of estimators derived from Eq. \ref{eq:overlapidentity} and the various extended sampling strategies used to overcome the overlap problem may also be used in the quantum case of Eq. \ref{quant_free_inurep}. Later on we will use the PS estimator in conjunction with the MH extended sampling strategy to estimate \RBAq\ for several different values of  P (see section \ref{sec:qfeds}). However before we do this we will derive the quantum version of the FSM. Unlike the Q-RSM, the quantum FSM (Q-FSM) is considerably different in appearance from its classical counterpart.

\subsection{Quantum Fourier Space Mapping}

It is a straightforward exercise to re-write  the expression in Eq. \ref{quant_free_inurep}  in terms of some effective configuration \vvec\  (see Eq. \ref{eq:vtoutran}), corresponding to a PM which matches the fourier coordinates of each and every replica:

\begin{equation}
R_{BA} =  e^{-\beta [E^0_B - E^0_A]} \RBAq
\end{equation}

\noindent and 

\begin{equation}
\RBAq = \lim_{P\to\infty} [\det S_{BA}]^P \RBAPq
\end{equation}

\noindent or

\begin{equation}
\RBAPq = \frac {1} {[\det S_{BA}]^P} \RBAq [1+O(\frac {1} {P^2})]
\end{equation}

\noindent where this time:

\begin{equation}
\RBAPq=\f {\int d\vvec_1 ...... d\vvec_P 
e^{-\beta \hcal {B} {\{\vvec\}}}}
{\int d\vvec_1 ...... d\vvec_P 
e^{-\beta \hcal {A} {\{\vvec\}}}}
\label{qfedvrep}
\end{equation}\tshow{qfedvrep}

\noindent with:

\begin{equation}
\beta \hcal {\al} {\{\vvec\}} = \s {i=1} {P} [ \f {1} {4\la_q\tau} ({\mf T}_\al \vvec_{i+1} - {\mf T}_\al \vvec_i)^2 + \tau \ecal_\al(\vvec_{i+1}) ]
\label{eq:polyham}
\end{equation}\tshow{eq:polyham}

\noindent The relevant macrovariable which quantifies the cost of the mapping then generalises to:

\begin{equation}
M_{BA}(\seta {\vvec}) = \beta [\hcal {B} {\seta {{\mf T}_B \vvec}} - \hcal {A} {\seta {{\mf T}_A\vvec}}]
\label{eq:fsswitchcost}
\end{equation}\tshow{eq:fsswitchcost}

\noindent Though perfectly valid, there are two  problems with the PM as formulated in Eq. \ref{qfedvrep}. The first is that the kinetic action gets modified under the corresponding mapping of configurations. That is even though such a FSM matches the harmonic contribution to the \ce\ of each and every replica, the fact that $S_K$ gets modified means that on the transition to a large number of replicas the cost of making a PM will become energetically expensive, thereby reducing the overlap between the two phases. In this case even if the quantum system becomes harmonic at very low temperatures, the guarantee of curing the overlap problem in the harmonic limit will no longer exist. In fact since the harmonic inter-replica interactions get stronger for larger P and end up dominating the overall action  (see appendix \ref{app:dominate}), and since larger values of P will be needed at lower temperatures, it follows that the cost of the PM, as measured by Eq. \ref{eq:fsswitchcost}, will become greater the lower the temperature.

An alternative formulation reveals itself when we notice that the kinetic action in Eq. \ref{tot_action} is a quadratic function of the displacements $\uvec$. Therefore if the system only explores the harmonic parts of the \ce\ $\ecal$ then the overall action $S_T$ will itself be a quadratic function of the displacements \uvec .  In this case it is possible to define a mapping with ensures perfect overlap between the two systems. The construction of the transformation follows a similar procedure to that used to derive the classical transformation ${\mf S}_{BA}$ (see chapter \ref{chap:tune}). We start by expanding the action in Eq. \ref{tot_action} as a power series in the displacements to yield:

\begin{eqnarray}
S_T & = & \s {i=1} {P} [ \f {1} {4\la_q\tau} (\mf u_{i+1} - \mf u_i)^2 + \tau {\mf u}_{i+1}^T\mf K \mf u_{i+1} ] + O(u^3)\nonumber \\
    & \approx & \s {i=1} {P} [ \f {1} {4\la_q\tau} (\mf u_{i+1}^2 - \mf u_i^2 - 2\mf u_{i+1} . \mf u_i ) +  \tau {\mf u}_{i+1}^T\mf K \mf {u_{i+1}} ]
\label{harmonic_action}
\end{eqnarray}\tshow{harmonic_action}

\noindent where $\mf K$ is the dynamical matrix (see Eq. \ref{eq:dynamatrix}). We may then approximate the total action $S_T$ by:

\begin{equation}
S_T =  \bar{\uvec}^T \Omega \bar{\uvec}
\label{eq:harto}
\end{equation}\tshow{eq:harto}

\noindent where \[ \bar{\uvec} = \left( \begin{array}{c} {\mf u_1}\\ {\mf u_2}\\ \mf .\\ \mf . \\ \mf . \\ {\mf {u_P}} \end{array}\right) \]

\noindent and $\bar{\uvec}_{(\eta -1)(3N) + i} = (\mf u_{\eta})_{i}$ where $\eta$ denotes the replica (assuming the values 1 through to P) and i denotes the component (taking the value 1 through to 3N). It is not hard to show from Eq. \ref{harmonic_action} that:

\begin{equation}
\Omega_{[(\eta -1)(3N)+i][(\nu-1)(3N)+j]} = (\tau \mf K_{i j} + \f {2} {4 \la_q \tau} \mf 1 ) \delta_{\eta \nu} - \f {1} {4\la_q\tau} [\delta_{\eta(\nu-1)} + \delta_{\eta(\nu+1)} ]
\label{Omega_expr}
\end{equation}

\noindent Following the procedure employed in deriving the classical FSM (Eq. \ref{eq:fsstran}), we may immediately write down the transformation matrix for the Q-FSM:

\begin{equation}
{\mf S}_{AB}^{q,ij} = \s {m} {} \sqrt {\frac {k_B^m} {k_A^m}} {\mf {e}}_A^{im} {\mf {e}}_B^{jm}
\label{quan_tran}
\end{equation}\tshow{quan_tran}

\noindent where now ${\mf {e}}_A^m$ and ${\mf {e}}_B^m$ are the normalised 3NP component eigenvectors of $\Omega_A$ and $\Omega_B$ respectively, $k^m$ are the associated eigenvalues, and the summation $\s {m} {}$ is over the non-null eigenvalues of $\Omega$, and where the indices i and j span the values through from 1 to 3NP. In this formulation the 'global' displacement vector of one phase, $\bar{\uvec}^A$ say, is mapped onto that of the other phase via the relation:

\begin{equation}
\bar{\uvec}^B = {\mf {S}}^q_{BA} \bar{\uvec}^A
\label{eq:utan}
\end{equation}\tshow{eq:utan}

\noindent such that the total actions of the two polymers are matched. This transformation may be conceptualised as the following mapping:

\begin{equation}
\mbox{$A\rightarrow B$:\ \ } \Rvec_A \rightarrow \Rvec_B \mbox{\ \ \ \ } \bar{\uvec} \rightarrow {\mf S}^q_{BA} \bar{\uvec}
\end{equation}

\noindent The partition function is now given by:

\begin{equation}
R_{BA} =  e^{-\beta [E^0_B - E^0_A]} \RBAq
\end{equation}

\noindent where

\begin{equation}
\RBAq = \lim_{P\to\infty} [\det S^q_{BA}] \RBAPq
\end{equation}

\noindent where

\begin{equation}
\RBAP=\f {\int d\uvec^B_1 ...... d\uvec^B_P \exp\{-\s {i=1} {P} [ \f {1} {4\la_q\tau} (\uvec^B_{i+1} - \uvec^B_i)^2 + \tau \ecal_B(\uvec^B_{i+1}) ] \}} {\int d\uvec^A_1 ...... d\uvec^A_P \exp\{-\s {i=1} {P} [ \f {1} {4\la_q\tau} (\uvec^A_{i+1} - \uvec^A_i)^2 + \tau \ecal_A(\uvec^A_{i+1} )] \} }
\label{quant_free_invrep}
\end{equation}\tshow{quant_free_invrep}

\noindent and where $\set {\uvec^A}$ and $\seta {\uvec^B}$ are related via Eq. \ref{eq:utan}. The transformation ${\mf S}^q_{BA}$ ensures that the quantity in Eq. \ref{eq:harto} is identical for the two phases.

Unlike the classical FSM, this quantum version of the FSM will not necessarily guarantee an improvement in the overlap as the temperature is reduced. The reason for this is that the presence of zero point motion means that the system may explore the anharmonic regions of the \ce\ even at $T=0$. However in cases where the quantum effects (zero point motion) are not strong enough so as to force the particles to visit the strongly anharmonic regions, then the quantum FSM, as we have formulated here, might serve as a useful tool.

\section{\label{sec:qres}Implementation details and simulation results}
\tshow{{sec:qres}}

\subsection{Motivation}

An archetypal example of a  solid in which  quantum effects are important is ${}^4 He$ \cite{p:ceperleyreview}. In low density solid helium, the atoms have a large zero point motion due to the small atomic mass \cite{p:loubeyre}. Since the atomic interactomic \ce\ is relatively weak as compared to the zero point motion, the lattice expands due to the outward pressure arising from this zero point motion. The result of this is that the solid is destabilised at much lower densities than would be allowed classically \cite{p:runge}. In addition to this one finds that for ${}^4 He$ the result of this zero point motion is that the particles, on average, sit at the relative maximum of a bimodal \ce\ \cite{p:loubeyre}, \cite{p:govpolturak}, \cite{p:gov}, resulting in imaginary frequencies of the dynamical matrix, rendering the harmonic description inaccurate. As the density is increased, the \ce\ eventually comes to dominate the zero point motion, and as a result the crystal develops a single well \ce\   which localises the particles to their lattice sites, resulting in the harmonic description becoming accurate \cite{p:loubeyre}.

As noted before, in addition to zero point motion, a phenomena called exchange arises in the case of indistinguishable particles. In solid ${}^4 He$ this has little effect since the fact that the atoms have no spin to label them means that there is no direct consequence of exchange (see \cite{p:ceperleyreview}). In the case of solid ${}^3 He$ the fact that the atoms have non-zero spin means that exchange plays an important role in determining the magnetic properties (see \cite{p:ceperleyreview}) of the solid.

\begin{figure}[tbp]
\begin{center}
\rotatebox{270}{
\includegraphics[scale=0.5]{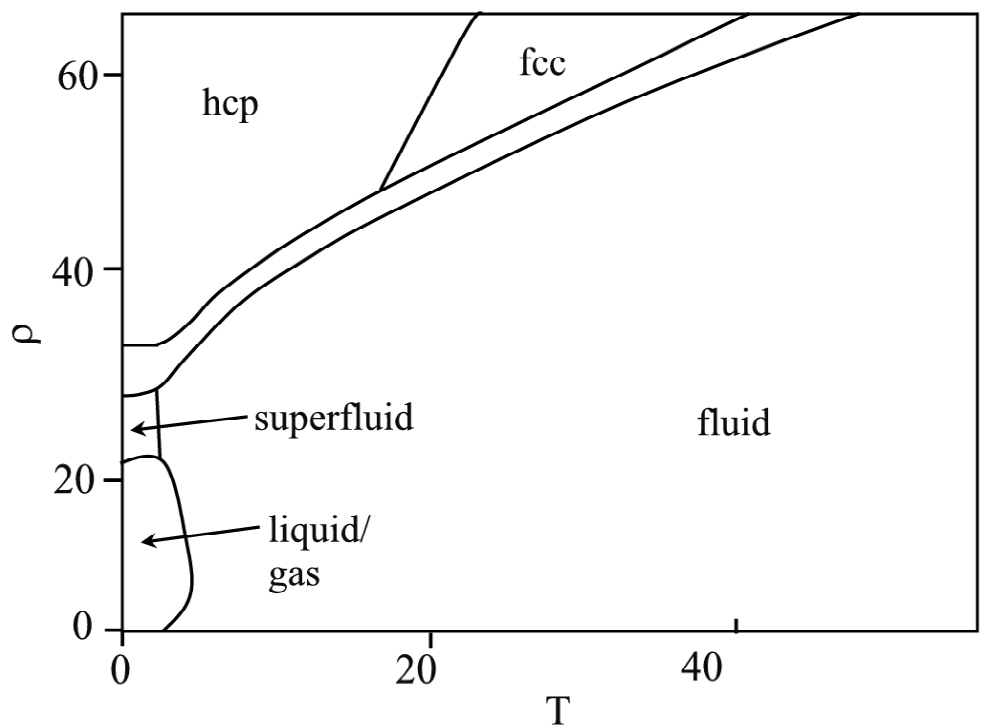}
}
\end{center}
\caption{A schematic of the phase diagram for ${}^4 He$}
This schematic was taken from \cite{p:ceperleysimmons}. The units of $\rho$ are $nm^{-3}$ and the units of the temperature T are Kelvin.

\begin{center}
{\bf{------------------------------------------}}
\end{center}

\tshow{pic:quantumphasediagram}
\label{pic:quantumphasediagram}
\end{figure}

The phase diagram of ${}^4 He$ is shown in figure \ref{pic:quantumphasediagram}. ${}^4 He$ may, at the crudest level, be described by the  Lennard-Jones (LJ) \ce , which provides a reasonable model for  the rare gases  \cite{p:quanlj}, \cite{b:raregas}. However in order to accurately determine the phase diagram one needs to employ more accurate \ces\ (see \cite{p:aziz}\nocite{p:korona}\nocite{p:janzen}-\cite{p:anderson}). In the rest of this chapter we will concern ourselves with the fcc/hcp regimes of the phase diagram (as modelled by the LJ \ce ). Our goal will not be to provide definitive statements about the phase diagram but instead to investigate the methodology developed in this chapter and to get a qualitative feel  for the effect of zero point motion on the relative stability of the fcc (A) and hcp (B) crystalline structures. For detailed studies of the quantum LJ solid, we refer the reader to \cite{p:qpollock}\nocite{p:muser}-\cite{p:chak}.

\subsection{\label{sec:qmodel}The Model System}
\tshow{sec:mode}

As was the case in the classical simulations, the reduced density $\rho\sigma^3$ was set to be $\rho\sigma^3=1.092$. In addition to the parameters $T^*$ and $\rho\sigma^3$, which enter into the classical simulations (see section \ref{sec:system}), one also encounters the  additional parameter $\la_q$ in the case of the quantum simulations. For our LJ systems this  was fixed through the De Boer parameter \cite{p:neumannzoppi},\cite{note:diffdef} \deb\ given by Eq. \ref{eq:pdeb}. In appendix \ref{app:deboer} we clarify the way in which the de Boer parameter enters into the calculations.

Initially simulations employing the same systems as those described in section \ref{sec:system} were implemented in order to estimate the expectation of thermodynamic variables such as the kinetic energy and the \ce, with good accuracy being obtained with both the PA (Eq. \ref{eq:prim_pot}) and the HOA (Eq. \ref{eq:HOVeff}) schemes. However on attempting to estimate the FEDs, it was found that these systems were too large for us to handle with the available computational resources. This meant that we had to restrict our simulations to a system size of N=96, which, from the way the fcc and hcp unit cells were constructed (see figure \ref{fig:hcp}), was the smallest system size that could be used. A consequence of this system size was that in order to fulfil the requirements imposed by the minimum image convention \cite{b:tildesley} particles could only interact with their first nearest neighbour shell (comprising of 12 particles), so that $r_c=1.1r_{nn}$ (see section \ref{sec:system}).

\subsection{Sampling the polymer}

In principle the simulation of a system whose hamiltonian $\hcalone {}$ is described by Eq. \ref{eq:prim_pot} or Eq. \ref{eq:HOVeff} is a straightforward task. One merely performs single particle perturbations, and accordingly accept these moves  with the acceptance probabilities given in  Eq. \ref{eq:sampbol}. In practice however this is not an efficient way to sample this polymeric system. The origins of this lie in the inter-replica coupling term $\frac {1} {4\la_q\tau} (\rvec_{i+1} -\rvec_i)^2$. As the temperature is lowered an increasing number of replicas will need to be employed in order to keep the systematic errors controlled at some prescribed level \cite{note:handrule}. A consequence of this is that the harmonic  inter-replica coupling will get stiffer, resulting in the decrease of the average size of an accepted move.  
Roughly speaking we see that since the kinetic action is a gaussian like term with a prefactor which increases linearly with the number of replicas, the mean square displacement should roughly scale as $1/\sqrt {P}$. 
This leads to a critical slowing down \cite{p:ceperleyreview} of the simulation in this limit, and severely hampers the simulation.

In order to alleviate this problem, one must introduce an additional move to the single particle moves already present in the classical case. This move is a global polymer move \cite{p:ceperleyreview} in which one moves a whole polymer without changing its conformation. Such a move leaves unaffected the kinetic action $S_K$. As a consequence it is only the change in $S_V$ which enters into the acceptance probability of such moves. Both moves are important; on the one hand the global-polymer moves allow faster exploration of the \ce\ $\ecal$, whilst the single particle moves allow the different conformations of the polymer to be explored. For the simulations considered here, it was found that an implementation of the global polymer move for every P single particle moves was optimal, in the sense that the correlation of the underlying data was kept to  a minimum.

\subsection{\label{sec:test} Testing the algorithm}
\tshow{sec:test}

In order to check that the algorithm was functioning correctly, two separate checks were made on the simulation. In the first a harmonic \ce\ was constructed, and the analytic results for the mean total energy $<\hat{H}>$ were compared to that obtained by the simulation. In the second a LJ \ce\ was employed, and the parameters were adjusted so as to get the simulation to explore  a region of the phase diagram in which the quantum effects were non-negligible {\em and} in which only the harmonic regions of the \ce\ were visited. In the latter case \RBAcal\  was also estimated via a MH-PS simulation and compared to the corresponding analytic result.

\subsubsection{Harmonic Potential}

In order to test our algorithm, we considered a system  interacting via the interatomic \ce :

\begin{equation}
\ecal(\rvec) = \frac {1} {2} \s {j\mbox{\ } \epsilon\mbox{\ } nn(i)} {} \s {i} {} \ecal_{ij}
\label{eq:summation_pot}
\end{equation}\tshow{eq:summation_pot}

\noindent where 

\begin{equation}
\ecal_{ij} = \frac {1} {2} K r_{ij}^2
\label{eq:harpot}
\end{equation}\tshow{eq:harpot}

\noindent where $nn(i)$ denotes the set of particles which interact with particle i. It is well known \cite{b:ashcroftmermin} that for a solid interacting via a harmonic \ce , the mean total energy may be obtained exactly from:

\begin{equation}
<\hat{H} > = E^0 + \s {i} {} (\bar{n}_i + \frac {1} {2} )\hbar \omega_i
\label{eq:harmonicenergy}
\end{equation}\tshow{eq:harmonicenergy}

\noindent where:

\begin{equation}
\bar{n}_i = \frac {1} {\exp (\beta\hbar\omega_i) -1}
\label{eq:popu}
\end{equation}\tshow{eq:popu}

\noindent where $\omega_i = \sqrt {\frac {\la_i} {m}}$ and where $\la_i$ are the eigenvalues of the dynamical matrix of $\ecal(\rvec)$, given by:

\begin{equation}
\frac {\partial^2 \ecal} {\partial x_i^\alpha \partial x_k^\beta} = \left\{\begin{array}
{r@{\quad:\quad}l}
\delta_{ik}\delta_{\alpha\beta}\times K \times n(i) - K [1-\delta_{ik}][\delta_{\alpha\beta}] & \mbox{if } k\epsilon nn(i)\\
0 & \mbox{otherwise }
\end{array}\right.
\end{equation}

\noindent where $\s {i} {}$ denotes a summation over the modes, and  where $n(i)$ denotes the number of members in the set $nn(i)$.

\begin{figure}[tbp]
\begin{center}
\ifpdf
\rotatebox{90}{
\includegraphics[scale=0.6]{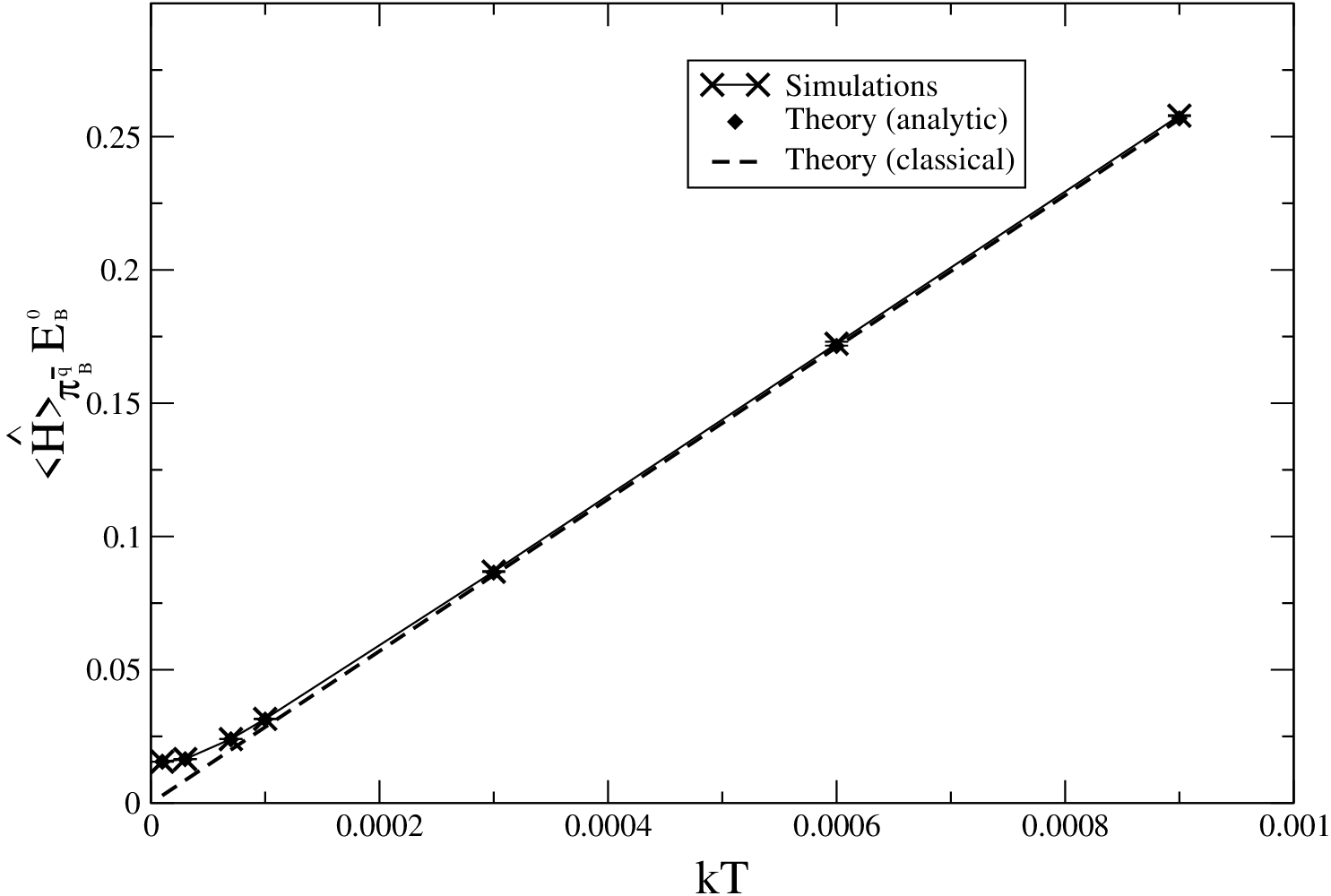}
}
\else
\includegraphics[scale=0.6]{fig/comparisonofanalyticandsimulation}
\fi
\end{center}
\caption{A plot of the analytic results versus those obtained by (HOA) simulation for the mean total excitation energy $<\hat{H}>_{\qpib {HOA}} - E^0_B$ of the quantum system interacting via the harmonic \ce , Eq. \ref{eq:harpot}}

As the temperature is increased the system becomes increasingly classical and hence the relationship between the total excitation energy and the temperature becomes linear, as predicted by the equipartition theorem (Eq. \ref{eq:equipart}). As the temperature is reduced to zero the quantum discretisation effects become increasingly important and the excitation energy becomes constant, due to the presence of zero point motion. It is for this reason that the graph for the quantum system departs from the classical line  in the limit of low temperatures \cite{note:ramirez}.

The range of temperatures were from kT=0.00001 to 0.0009.

$\la_q = 0.5\times 10^{-9}$, K=1.0.

As will all the harmonic calculations via Eq. \ref{eq:harpot},   $<\hat{H}>$ is expressed in units of k.

\begin{center}
{\bf{------------------------------------------}}
\end{center}

\tshow{pic:analyticvssimu}
\label{pic:analyticvssimu}
\end{figure}

Figure \ref{pic:analyticvssimu} shows a graph comparing the estimates of the expectation of the total energy as obtained by simulation with the theoretic values as predicted by classical and quantum statistical mechanics. The curve for the classical theory was obtained from the equipartition theorem \cite{b:reif}\nocite{b:honerkamp}-\cite{b:pathria} which states that for a system of particles interacting via a harmonic \ce , the expectation of the total excitation energy is given by:

\begin{equation}
<\hat {H} > = E^0 + (3N - \nu)kT
\label{eq:equipart}
\end{equation}\tshow{eq:equipart}

\noindent where N is the number of particles (3N represents the number of degrees of freedom) and $\nu$ is the number of constraints. For simulations with periodic boundary conditions $\nu=3$. The curve corresponding to the theoretical predictions as made by quantum statistical mechanics was obtained from Eq. \ref{eq:harmonicenergy}.

The first thing that one notices is that the results of the simulation are in complete agreement with the curve as extracted from Eq. \ref{eq:harmonicenergy}. Comparing the quantum and classical curves, we notice from figure \ref{pic:analyticvssimu}\ that the quantum graph does, as expected, converge onto the classical line on the transition to sufficiently high temperatures. However at lower temperatures the situation is different.
Here the mean total excitation energy  levels off and assumes a constant value, corresponding to the term $\s {i} {} \frac {1} {2} \hbar \omega_i$ arising in Eq. \ref{eq:harmonicenergy}. This is the zero point energy of the system and arises from the inherent motion of the particles, present even at 0 Kelvin. This is purely a quantum phenomenon and arises from the Heisenberg uncertainty principle. In contrast the mean total excitation energy  vanishes in the classical limit, and results in the departure of the quantum system from the classical curve as seen in  figure \ref{pic:analyticvssimu}.

\subsubsection{\label{sublen}Lennard Jones System in harmonic regime}
\tshow{sublen}

The test implemented in the previous section was also implemented on the LJ hcp (B) and fcc system (phase A). The parameters $T^*$ and $\tilde {D}$ were appropriately adjusted so as to ensure that only the harmonic parts of the \ce\ were visited. The eigenvalues  $\la_i$ of the dynamical matrix, obtained via numerical methods, were then used to determine the analytic value for the mean total energy of the system, via Eq. \ref{eq:harmonicenergy}. This was then compared to the result obtained by simulation and used to verify that the simulation was indeed visiting only the harmonic regions of the LJ \ce . In the next stage a check was made in order to verify that the simulation was indeed estimating the FEDs correctly. The (quantum mechanical) analytic values for \RBAcal\ were obtained via the relation \cite{b:ashcroftmermin}:

\begin{equation}
\RBAcal = \frac {\z_B} {\z_A} = \p {i} {} \{\frac{e^{-\frac{{\beta \hbar \omega_i^B}} {2}}}{ {1-e^{-\beta \hbar \omega_i^B}}} . \frac{1-e^{-\beta \hbar \omega_i^A}} {e^{-\frac {\beta \hbar \omega_i^B}{2}}}\}
\label{eq:quanan}
\end{equation}\tshow{eq:quanan}

\noindent where $\omega^\al_i$ denotes the frequency of the i-th mode of phase \al . These results were then compared to those obtained by a simulation employing the HOA scheme (section \ref{sec:higherorder}) using the MH-PS method (section \ref{sec:fgps}). The classical value was obtained from Eq. \ref{eq:harfree}. The results have been tabulated below:

\begin{table}[h]
\begin{center}
\begin{tabular}{|c|c|c|}\hline
   & $<\hat {H}>_{\qpia {HOA}} - E^0_B$ & \RBAq \\\hline\hline
analytic: classical harmonic & 285 & 0.810\\\hline
analytic: quantum harmonic& 2327.39 & 0.671\\\hline
analytic: simulation & $2328.7\pm0.2$ & $0.678\pm 0.006$\\\hline
%\label{table1.0}
\end{tabular}
\end{center}
\caption{The values of \RBAcal\ for a harmonic LJ quantum system}
{ $\tilde{D} = 1.816 \times 10^{-5}$, $T^*$=0.005, P=20}
\label{table:har_free}
\end{table}\tshow{table:har_free}

\begin{figure}[tbp]
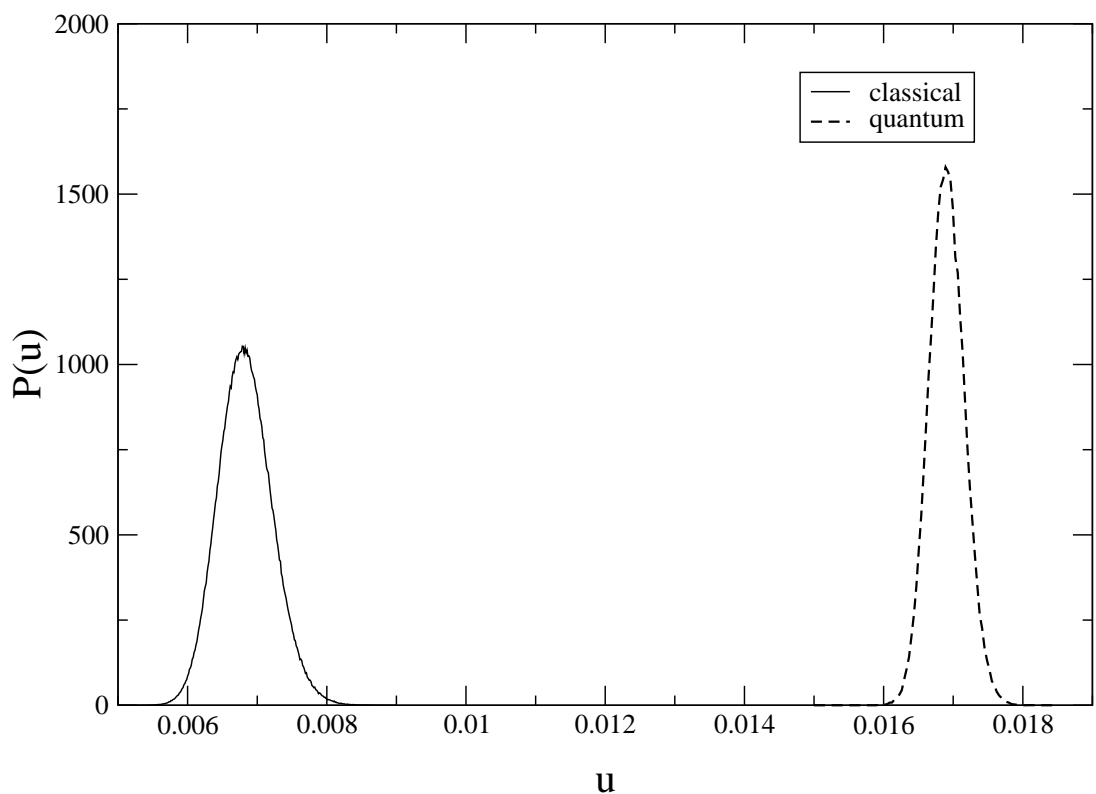

\begin{center}
\ifpdf
\rotatebox{90}{
\includegraphics[scale=0.6]{fig/prob_of_gyrationljhar}
}
\else
\includegraphics[scale=0.6]{fig/prob_of_gyrationljhar}
\fi
\end{center}
\caption{A plot of the distance of a particle from its lattice site, as averaged over all particles (harmonic quantum Lennard-Jones solid)}

$u$ denotes the distance a particle from its lattice site, averaged over all particles of the system. Though the means of the distributions correspond to the mean displacement of a particle from its lattice site, the associate spreads of the displacement of a particles from its lattice sites are $\sqrt{N}=\sqrt{96}\approx 10$ times wider than that shown in the figure above.

It is clear that in the quantum case the zero point motion pushes the particles out to regions  further from the lattice sites than would be the case classically.

$T^*=0.005$, $\tilde{D}=0.00001816$.

\begin{center}
{\bf{------------------------------------------}}
\end{center}

\tshow{pic:rgljh}
\label{pic:rgljh}
\end{figure}

\noindent The first column of table \ref{table:har_free}\ verifies that the system of particles, interacting via the LJ \ce , were indeed exploring only the harmonic regions of the LJ \ce , since the simulation results agree with the analytic values as predicted by harmonic theory (Eq.\ref{eq:quanan}). The analytic values and the simulation results for \RBAcal\ (see column 2 in table \ref{table:har_free})  clearly agree to within the errors, and differ significantly from the classical value. It is clear the effect of the increased amplitudes of vibration arising from the  zero point motion (see figure \ref{pic:rgljh}) act, in this regime of the phase diagram, favourably towards the fcc (A) phase, making it more probable (relative to the hcp (B) phase) than it would be in the classical case. This is expected since in the classical case \cite{p:LSMCpresoft} the increasing amplitudes of vibration obtained on increasing the temperature also favours the fcc (A) phase, {\em within the harmonic regimes}.

\subsection{Scaling of thermodynamic parameters with P}

In practice a PIMC simulation necessarily involves a finite number of replicas. Unlike in section \ref{sec:test}, where we had an analytic check so as to allow us to determine whether a sufficient number of quantum replicas had been employed, one will not have an idea as to the magnitude of the systematic errors arising from the finite replica simulations in the general case. As mentioned before  the only information available to  us  is  the way the associated systematic errors  {\em scale} with the number of replicas. By  plotting the desired expectation as an appropriate power of the inverse of the number of replicas one may hope to obtain the asymptotic value that the expectation assumes in the limit of an infinite number of replicas.

\begin{figure}[tbp]
\begin{center}
\ifpdf
\rotatebox{90}{
\includegraphics[scale=0.6]{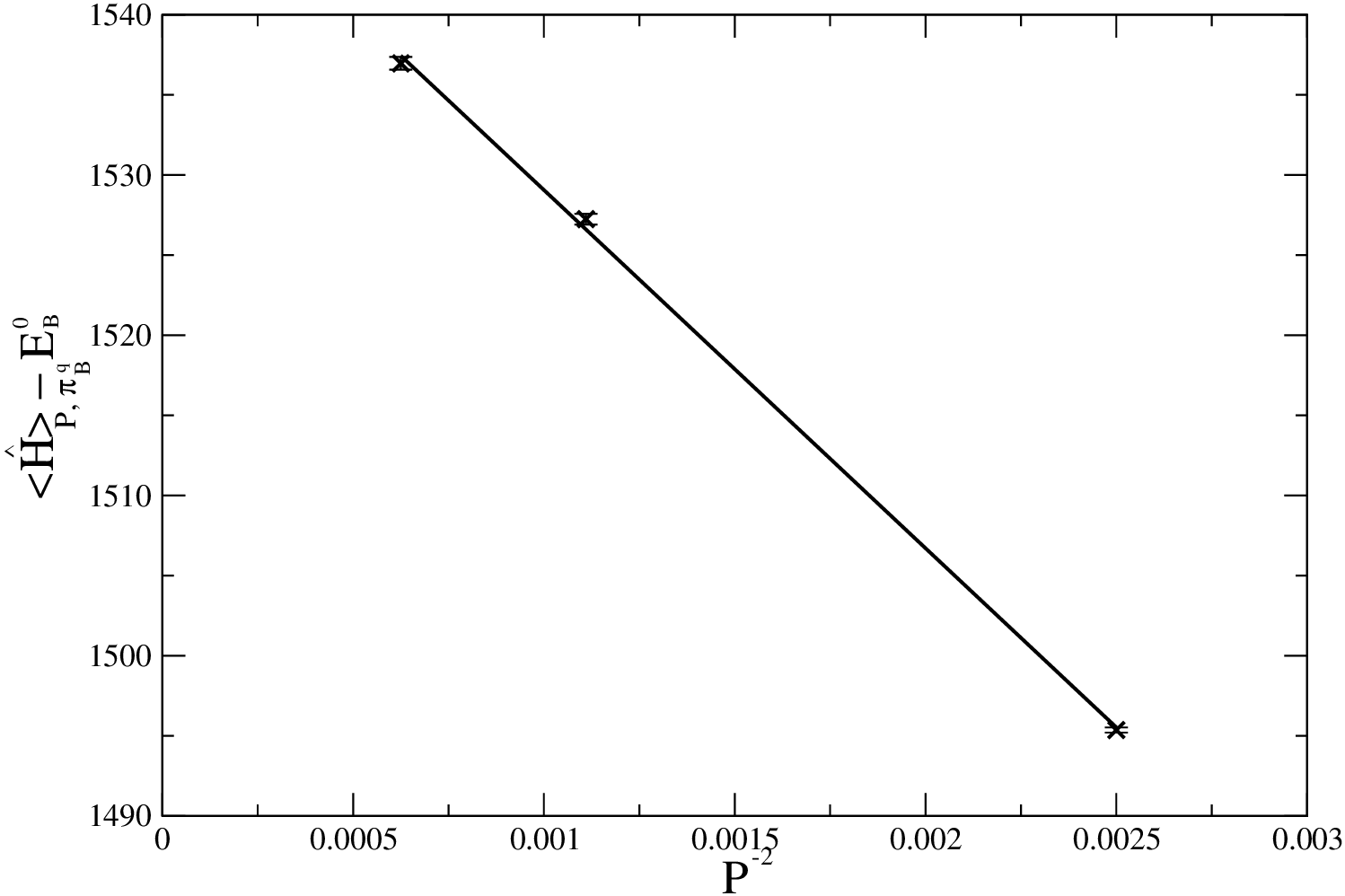}
}
\else
\includegraphics[scale=0.6]{fig/20-30-40reps_harmonic_primitive2}
\fi
\end{center}
\caption{$<\hat{H}>_{P,\qpib {PA}} - E^0_B$ versus $\frac {1} {P^2}$ (PA, harmonic)}
{
The three values of P chosen were P=20, 30, 40. 

The line of fit was given by: $<\hat{H}>_{P,\qpib {PA}}- E^0_B = 1551.4 - 22353 (1/P^2)$.

The analytic value was: $<\hat{H}>=E^0_B+1551.61$. This agrees well with the intercept of 1551.4 (in units of k).

$kT=0.0001$, $\la_q = 0.5\times10^{-9}$, K=1.0. 

}
\begin{center}
{\bf{------------------------------------------}}
\end{center}
\tshow{pic:harmonic_primitive}
\label{pic:harmonic_primitive}
\end{figure}

\begin{figure}[tbp]
\begin{center}
\ifpdf
\rotatebox{90}{
\includegraphics[scale=0.6]{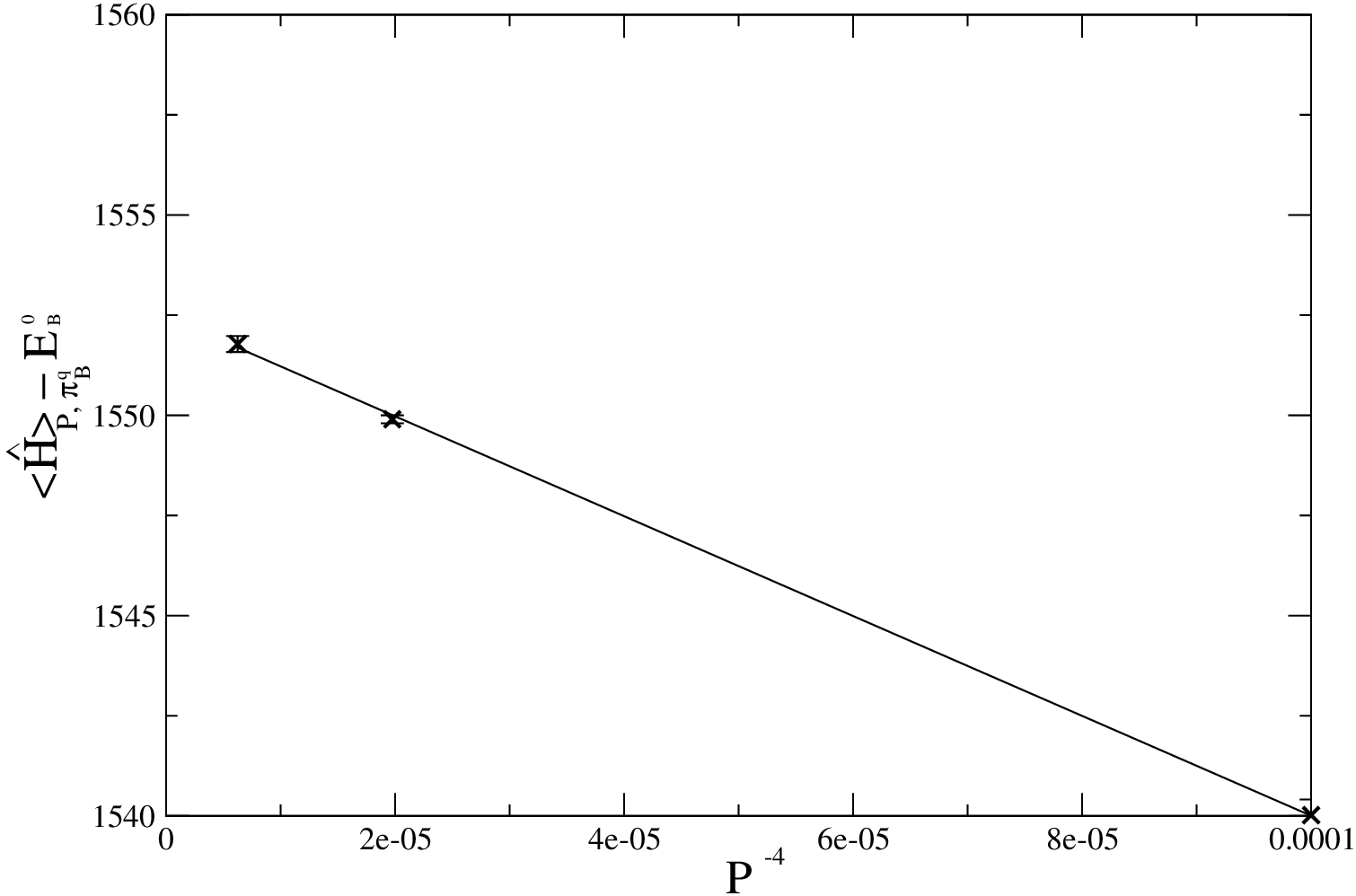}
}
\else
\includegraphics[scale=0.6]{fig/10-15-20reps_harmonic_ho2}
\fi
\end{center}
\caption{$<\hat{H}>_{P,\qpib {HOA}}- E^0_B$ versus $\frac {1} {P^4}$ (HOA, harmonic)}
{
The three replica values that were simulated corresponded to P=10, P=15, and P=20.

The line of fit was given by: $<\hat{H}>_{P,\qpib {HOA}} - E^0_B = 1552.5 - 124630 (1/P^4)$.

The graph shows a plot of $<\hat{H}>_{P,\qpib {HOA}}- E^0_B$ as a function of $\frac {1} {P^4}$ for a system of  particles (in phase B) interacting via the harmonic \ce\ (Eq. \ref{eq:harpot}). The analytic value was  $<\hat{H}> = E^0_B + 1551.6$ (in units of k).

$kT=0.0001$, $\la_q = 0.5\times10^{-9}$, K=1.0.

}

\begin{center}
{\bf{------------------------------------------}}
\end{center}

\tshow{pic:harmonic_HO}
\label{pic:harmonic_HO}
\end{figure}

Specifically we saw in sections \ref{sec:macest} and \ref{sec:higherorder} that the expectations of an observable should have errors which scale as $1/P^2$ (for the PA) and $1/P^4$ (for the HOA). In order to illustrate this scaling, simulations of the harmonic system as described by the \ce\ given in Eq. \ref{eq:summation_pot} and Eq. \ref{eq:harpot} were implemented via both the PA and the HOA methods. The estimates of the asymptotic limits were then extracted via the appropriate graphical extrapolation techniques, and the results were then compared to the analytical value of the total energy. Figure \ref{pic:harmonic_primitive} and figure \ref{pic:harmonic_HO} illustrates that the scaling of the expectation of the total energy does indeed follow that of Eq. \ref{eq:thermo_var_scale} and Eq. \ref{eq:thermohoa} for the PA and HOA methods respectively, yielding the correct value (as obtained analytically) in the asymptotic limit.

Given that both methods accurately determine the asymptotic value of the mean total energy, the question now remains as to which yields a smaller error (for a given computational resource). In addressing this issue, we first note that the number of replicas chosen in  figure \ref{pic:harmonic_primitive} and figure \ref{pic:harmonic_HO} were the {\em minimum} needed in order to be in  the appropriate scaling regimes. Below this range the mean total energies no longer scaled as $1/P^2$ and $1/P^4$ for the PA and HOA methods respectively. From the graphs it is evident that  half the number of replicas were needed in the case of the HOA method than were required for the PA method. However it was also found that the HOA method took roughly double the amount of time to perform a given number of lattice sweeps, as compared to a PA simulation employing the same number of replicas.  In order to understand this we note that since the higher order hamiltonian  of Eq. \ref{eq:HOVeff} has the additional term $[\nabla \ecal(\rvec)]^2$ to the primitive one (Eq. \ref{eq:prim_pot}), and since this term has to be computed over the same number of nearest neighbours for each particle as one would have to do when calculating $\ecal(\rvec)$, the HOA method will require roughly double the time to simulate. As a result we see that a 2P-replica PA simulation will take the same time as a P replica HOA simulation, thus offsetting any gains initially offered by the HOA method.

\begin{figure}[tbp]
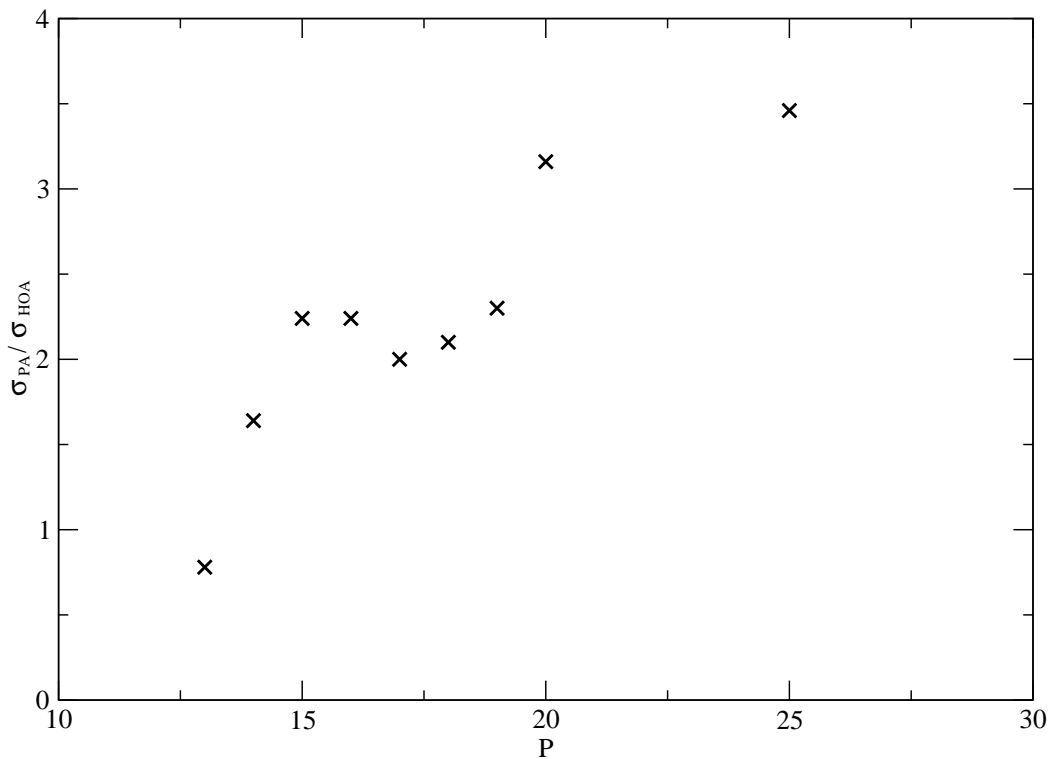

\begin{center}
\ifpdf
\rotatebox{90}{
\includegraphics[scale=0.6]{fig/hoapaerrors2}
}
\else
\rotatebox{0}{
\includegraphics[scale=0.6]{fig/hoapaerrors2}
}
\fi
\end{center}
\caption{Comparisons of the errors of a P replica HOA simulation to a 2P PA one}
For the systems studied here it was found that roughly twice as many replicas were needed for the PA method as compared to the HOA method in order to ensure that the simulation was in the appropriate scaling regime, so as to allow one to arrive at an estimate of observables, free of systematic errors, via the graphical extrapolation technique described in sections \ref{sec:macest} and sections \ref{sec:higherorder}. However the HOA method took twice as long (as compared to a PA simulation employing the same number of replicas) to achieve a given number of sweeps, thereby offsetting the advantages just mentioned (since a 2P-replica PA simulation will take just as long as a P-replica HOA simulation). This graph shows the ratios of the errors of a 2P-replica PA simulation to those of a P-replica HOA simulation, and indicates that the HOA method has a marginal advantage over the PA method. The trend seems to be such that this advantage increases as P increases.

$\tilde {D} = 0.1816$. $T^*=1.5$.

\begin{center}
{\bf{------------------------------------------}}
\end{center}

\tshow{pic:errqm}
\label{pic:errqm}
\end{figure}

All that is left to compare between the two methods is the correlation of the underlying data. Figure \ref{pic:errqm} shows the ratio of the error $\sigma_{HO,P}$ obtained in a P replica HO simulation to the error  $\sigma_{PA,2P}$ obtained from a 2P replica PA simulation, run for the same duration of time \cite{note:duration}. 
Clearly figure \ref{pic:errqm} shows that in regards to correlations, the HO method has a slight advantage over the PA method, and for this reason we employed the HOA method in our attempts at estimating the FEDs. We also note that the trend of the graph indicates that this advantage increases as the number of replicas increases (for the systems studied here).

\subsection{Dependence of $P(\mab | \qpib {PA})$ on P and T}

\begin{figure}[tbp]
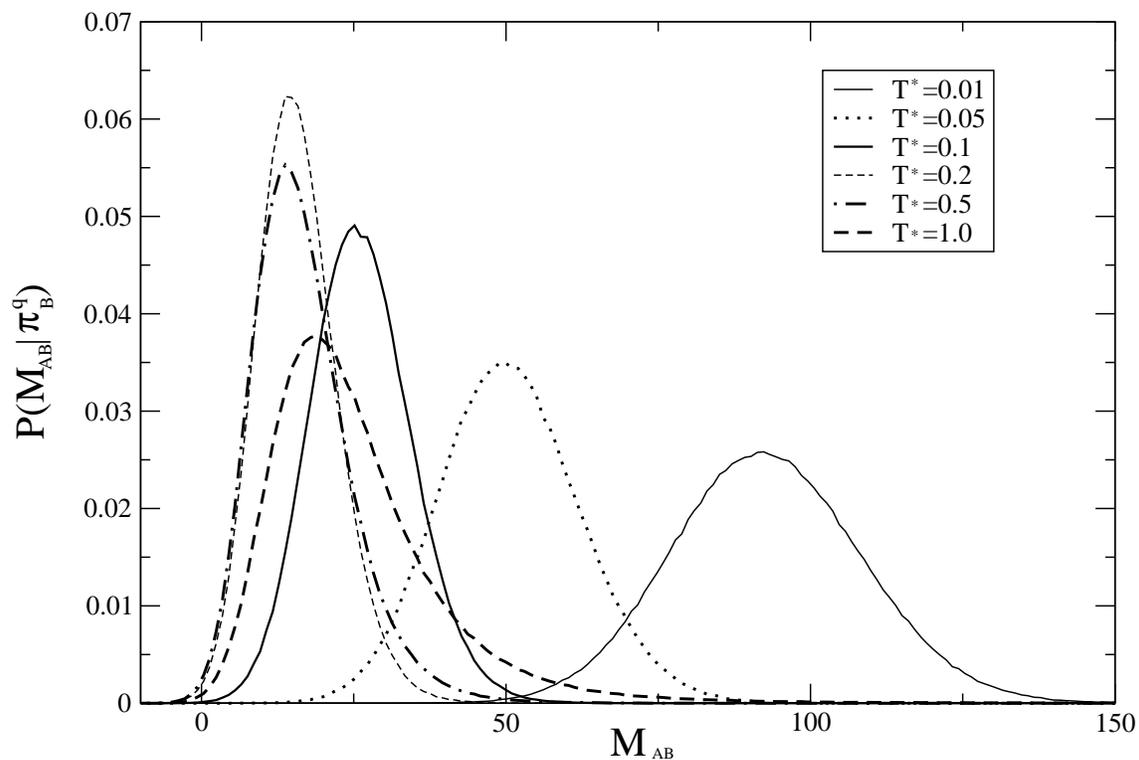

\begin{center}
\ifpdf
\rotatebox{90}{
\includegraphics[scale=0.6]{fig/prob_orderparam_vs_temp}
}
\else
\rotatebox{0}{
\includegraphics[scale=0.6]{fig/prob_orderparam_vs_temp}
}
\fi
\end{center}
\caption{${P}(\mab|\qpib {PA})$ for different T, P constant}
{
 The probability distribution ${P}(\mab|\qpib {PA} )$ was obtained for a selection of temperatures ranging from $T^*=0.01$ to $T^*=1.0$ for the LJ \ce .

$\tilde {D} = 0.001816$, P=10.
}

\begin{center}
{\bf{------------------------------------------}}
\end{center}

\tshow{pic:OP_vs_temp_fixed_rep}
\label{pic:OP_vs_temp_fixed_rep}
\end{figure}

\begin{figure}[tbp]
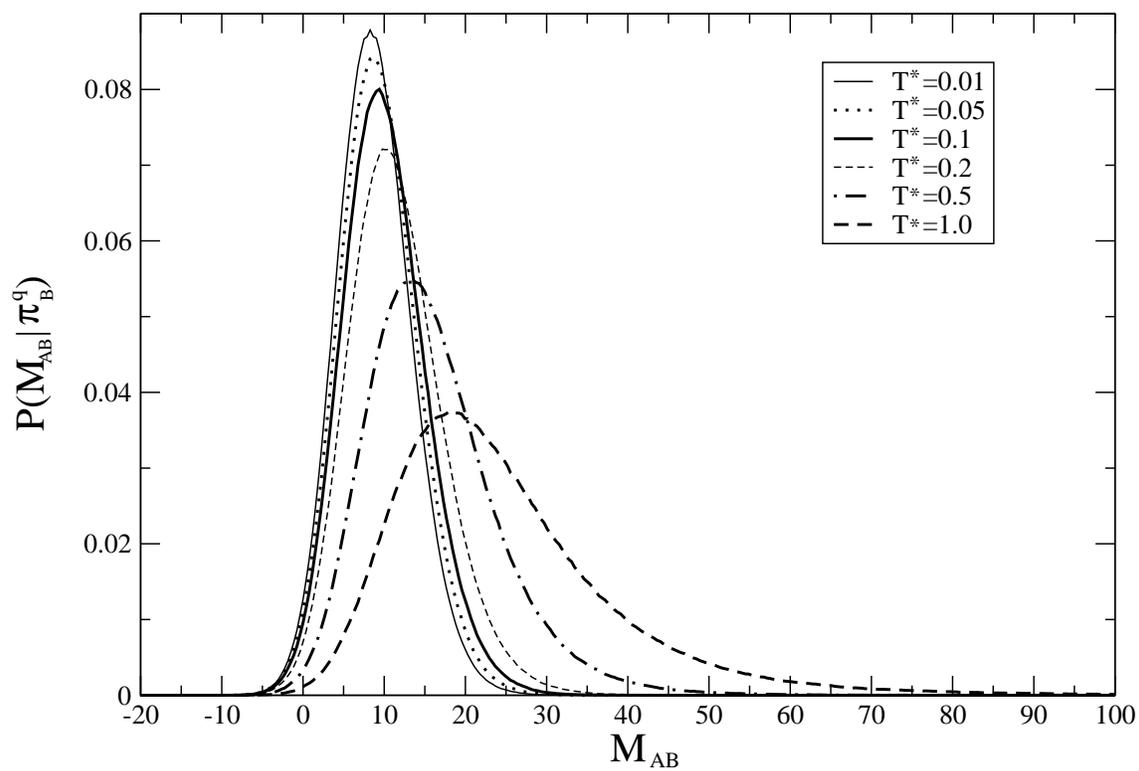

\begin{center}
\ifpdf
\rotatebox{90}{
\includegraphics[scale=0.6]{fig/prob_orderparam_vs_temp_classical}
}
\else
\rotatebox{0}{
\includegraphics[scale=0.6]{fig/prob_orderparam_vs_temp_classical}
}
\fi
\end{center}
\caption{${P}(\mab| \pib)$ for various $T^*$ for the classical simulations}

This figure shows the classical distributions ${P}(\mab| \pib)$ corresponding to those of figure \ref{pic:OP_vs_temp_fixed_rep}.

\begin{center}
{\bf{------------------------------------------}}
\end{center}

\tshow{pic:cwt}
\label{pic:cwt}
\end{figure}

\begin{figure}[tbp]
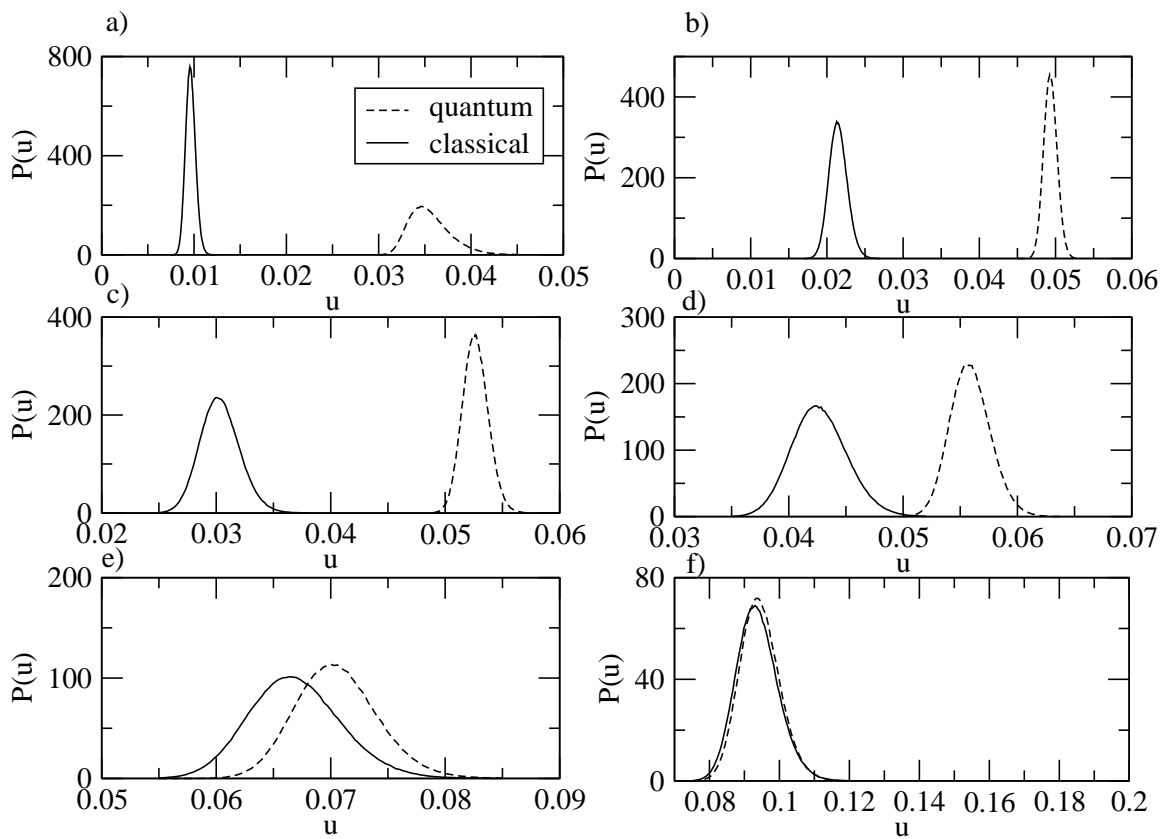

\begin{center}
\ifpdf
\rotatebox{90}{
\includegraphics[scale=0.65]{fig/rg_vstemp}
}
\else
\rotatebox{0}{
\includegraphics[scale=0.65]{fig/rg_vstemp}
}
\fi
\end{center}
\caption{Temperature dependence of the mean distance of a particle from the lattice site}

a)$T^*=0.01$\ b)$T^*=0.05$\  c)$T^*=0.1$\  d)$T^*=0.2$\ e)$T^*=0.5$\ f)$T^*=1.0$\

See figure \ref{pic:rgljh} for an explanation of the way the mean displacement is calculated.

The graph shows the average displacement of the particle in the quantum and classical limits, corresponding to the simulations shown in figures \ref{pic:OP_vs_temp_fixed_rep} and figure \ref{pic:cwt}. It is clear that at low temperatures, the effect of the zero point motion is important and results in a significantly greater mean displacement of the particle from its lattice site than in the corresponding classical case. This results in the peak of ${P}(\mab|\qpib {PA})$ being positioned significantly further out from the origin than the corresponding classical distribution (see figure \ref{pic:OP_vs_temp_fixed_rep} and figure \ref{pic:cwt}). As the temperature increases, the zero point motion becomes less important, resulting in the gradual convergence of the two distributions.

\begin{center}
{\bf{------------------------------------------}}
\end{center}

\tshow{pic:radi}
\label{pic:radi}
\end{figure}

\begin{figure}[tbp]
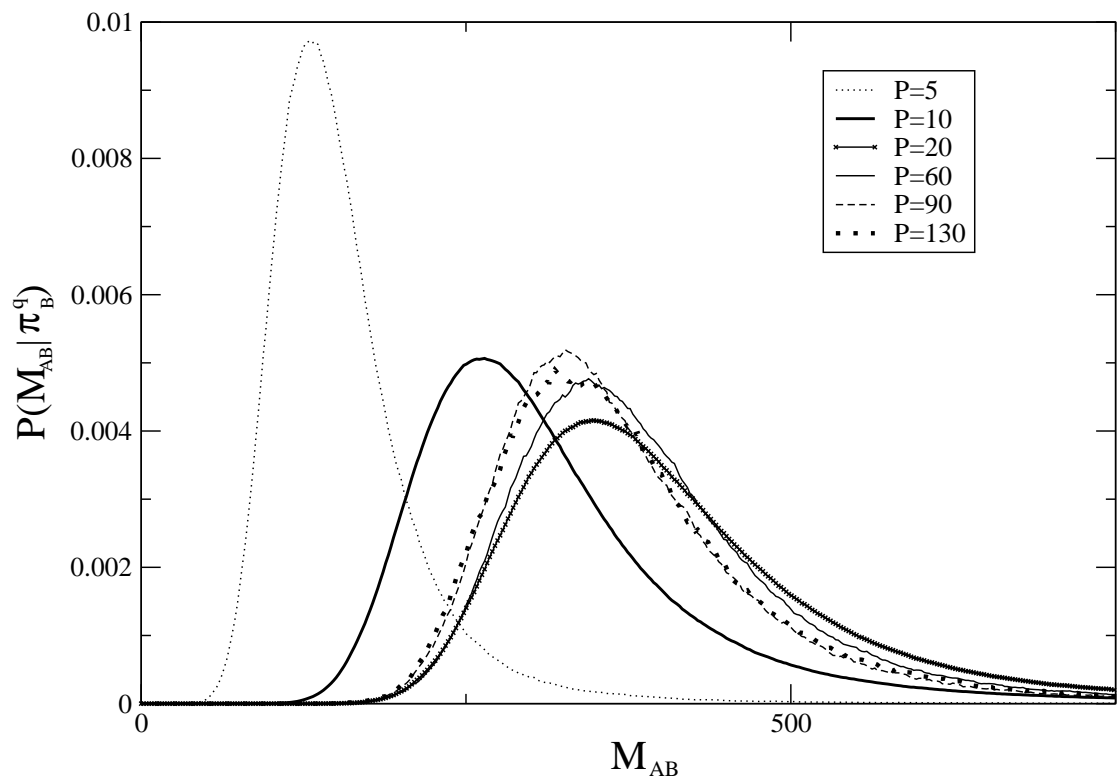

\begin{center}
\ifpdf
\rotatebox{90}{
\includegraphics[scale=0.6]{fig/order_param_vs_replica}
}
\else
\rotatebox{0}{
\includegraphics[scale=0.6]{fig/order_param_vs_replica}
}
\fi
\end{center}
\caption{${P}(\mab|\qpib {PA})$ versus P, for a constant T simulation}
{
The probability distribution ${P}(\mab|\qpib {PA})$ was obtained for a range of replicas ranging from P=10 to P=130, with the temperature being kept fixed at $T^*=0.4$.
 
$\tilde{D}=0.1816$. 

}

\begin{center}
{\bf{------------------------------------------}}
\end{center}

\tshow{pic:order_vs_replica}
\label{pic:order_vs_replica}
\end{figure}

As we have seen before, the statistics of the macrovariable \mab\ essentially contains all the information of the FED between the two phases. Therefore it is instructive to examine the dependence of these distributions on P and T. Figure \ref{pic:OP_vs_temp_fixed_rep} shows the quantum probability distribution $P(\mab | \qpib {PA})$ for different temperatures (and fixed replica number); figure \ref{pic:cwt} shows the classical distributions $P(\mab | \pib)$ for different temperatures; and in figure \ref{pic:order_vs_replica}  $P(\mab | \qpib {PA})$ is plotted (at a given temperature) for different numbers of replicas. 

The first thing that is clear is that for the quantum case  the behaviour of $P(\mab | \qpib {PA} )$ is not monotonic with the variation of the relevant parameters. In figure \ref{pic:OP_vs_temp_fixed_rep} the peak of the distributions initially move in towards the origin with an accompanied decrease of variance. However beyond a certain temperature ($T\sim 0.1$), the mean and variance of the associated distributions start to {\em increase} as the temperature is raised.

The  reason for this observed behaviour is as follows. At sufficiently low temperatures the zero point motion of the particles force the system to wander further away from the lattice sites (see figure \ref{pic:radi}) than would be the case in the classical limit, resulting in the peak of the distribution  being further away from the origin than would be expected in the classical case. This is clearly the case with $T^*=0.01$ (compare the graphs in figures \ref{pic:OP_vs_temp_fixed_rep} and figure \ref{pic:cwt}).
As the temperature increases, the contribution to the energy of the zero point motion remains {\em constant}, since the typical energy has not yet reached that of the phonon excitation energies. From the definition given in Eq. \ref{eq:switchcost} it immediately follows that \mab\ must {\em decrease}. As the temperature is increased even further the thermal excitation contributions to the energy begin to become important. In this case the rate at which \mab\ decreases will itself reduce (since the decrease due to the division by T will be offset by the increase in $\frac {1} {P} \s {i=1} {P} (\ecal_B(\uvec_{i}) - \ecal_A(\uvec_{i})$) , until eventually \mab\ starts to increase. On further increase the system will become classical and the (classical) thermal effects will mask the quantum zero point motion, at which point the difference between the graphs in figure \ref{pic:OP_vs_temp_fixed_rep} and figure \ref{pic:cwt}, arising from the effect of the zero point motion,  will eventually be negligible.

Figure  \ref{pic:order_vs_replica}\ shows the probability distribution $P( \mab | \qpib {PA})$ for different numbers of replicas. Initially as the number of replicas increases the peak moves away from the origin (up to $P\sim 20$). Further increase in the number of replicas leads to the peak  moving closer to the origin, converging (by $P\sim 130$) to the stationary distribution, {\em which is ultimately positioned further away from the origin than the corresponding classical distribution. }
The important thing to note from figures \ref{pic:OP_vs_temp_fixed_rep}\ and \ref{pic:order_vs_replica}\ is that as quantum effects become increasingly important \cite{note:meanwhat}, the peak of $P(\mab|\pib)$ moves away from the origin (relative to the classical distribution) and its width increases. This means that in addition to the fact that the quantum simulation is inherently more time consuming, additional time must also be spent refining the sampling strategy (whether it be increased amounts of multicanonicalisation in the case of MUCA simulations or increased numbers of replicas being employed in the MH method) in order to estimate the  FEDs.

\subsection{\label{sec:qfeds}FEDs}
\tshow{sec:qfeds}

\begin{figure}[tbp]
\begin{center}
\ifpdf
\rotatebox{90}{
\includegraphics[scale=0.6]{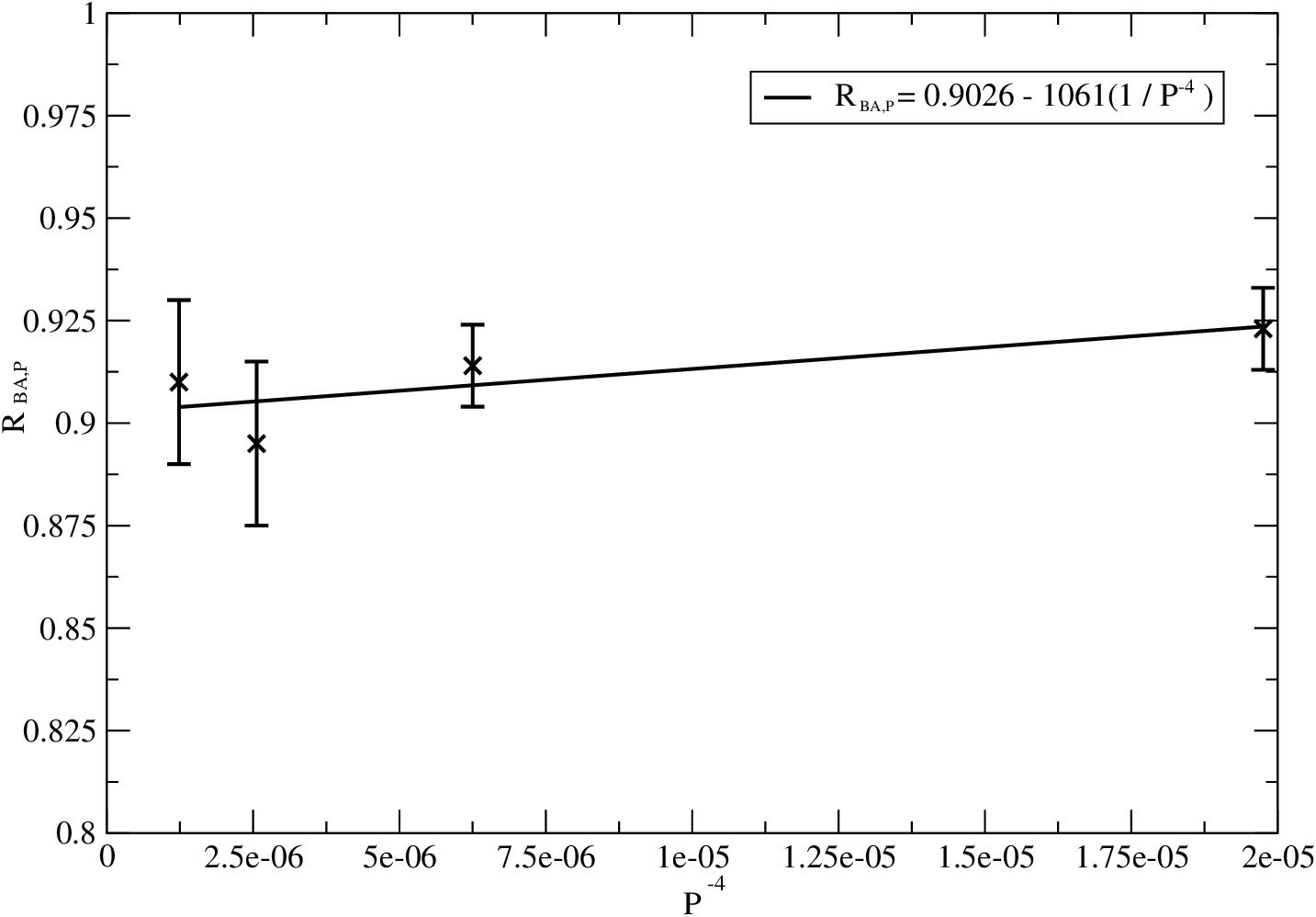}
}
\else
\includegraphics[scale=0.6]{fig/free_energy}
\fi
\end{center}
\caption{$\RBAP$ versus $\frac {1} {P^4}$ (HOA) for the LJ \ce  }
{
The graph shows the plot of the quantity $\RBAP$ against $\frac {1} {P^4}$ for the LJ \ce .

The asymptotic value extracted from the plot: $\RBAP = 0.9 \pm 0.02$.

The classical value:  $\RBAP = 0.844 \pm 0.004$.

$T^*=1.5$.

}

\begin{center}
{\bf{------------------------------------------}}
\end{center}

\tshow{pic:free_energy}
\label{pic:free_energy}
\end{figure}

The primary motivation for developing the path integral machinery in the preceding sections was to use it to estimate quantum FEDs. In fact by formulating it in the way that was  done in Eq. \ref{quant_free_inurep}, we made available to ourselves  the vast  array of tools discussed in the previous chapters that are suitable for tackling this type of problem. In this final section we discuss our attempts at estimating the quantum mechanical FEDs. In estimating the FEDs, our aim was to investigate the role of the zero point motion on the relative stability of the two phases in a regime of the phase diagram in which {\em both} the quantum effects and the anharmonic effects were significant. Figure \ref{pic:free_energy}. shows the estimates of the FEDs, obtained via the MH-PS method, in such a regime.

It is clear from figure \ref{pic:free_energy} that the quantum effects essentially act so as to favour the hcp (B) phase (as compared to the classical case). This can be understood in the context of results obtained in the classical simulations. In \cite{p:LSMCpresoft} the classical LJ system was studied and it was found that the increasing anharmonicity (obtained on increasing the temperature) favoured the hcp (B) phase. This conclusion was arrived at by comparing the simulation results to the harmonic calculations. The same effect is likely to be the cause of the quantum effects favouring the hcp (B) phase. That is the zero point motion pushes the particles into regions further out from the minimum of the \ce\ than they would typically explore in the classical case, making the system more anharmonic. As is the case in the classical systems, this anharmonicity acts in a way which favours the hcp (B) phase. This is in sharp contrast to the {\em harmonic} regions of the quantum phase diagram (see section \ref{sublen}), where the {\em increased} amplitudes of vibration acts so as to favour the fcc (A) phase (as is also the case in the classical regime, see \cite{p:LSMCpresoft}).

\section{Discussion}

The quantum Lennard-Jones phase diagram has not been determined yet via computational techniques and the work here represents a first step in developing the necessary machinery for a move in that direction. The factors limiting our investigation are the following:

\begin{enumerate}
\item The slowing down associated with the simulation of a system of interacting polymers over that of a system of interacting particles.
\item The increasing number (P) of replicas needed as the temperature is reduced. One not only has an increase in computational costs due to the fact that one has to simulate more replicas, but also a critical slowing down associated with the increasing strength of the inter-replica forces.
\item The need to perform graphical extrapolation in order to obtain a single estimate of the expectation of an observable.
\end{enumerate}

\noindent In order to accurately determine the phase diagram (using the same number of processors that we used) at the temperature we chose, one would need to employ a $12\times12\times12$ system. Since the time, associated with keeping the error in the estimate of the FED at some prescribed level, scales roughly as $N^3$ (for short ranged interactions) we see that the simulation of a $12\times12\times12$ system would involve an increase in computational requirement of approximately $[\frac {12^3} {6\times4\times4}]^{3} = 5832$ times. In accordance with Moore's Law, this sort of computational power will be available to us in about 13 years.

However a significant feature of the simulations the way that we have done it (i.e. via the MH route) is the enormous scope for parallelisation. This parallelisability does, in principle, allow us to determine the phase diagram accurately even today, simply by distributing the replicas amongst an increased set of processors \cite{note:extramhq}. In our simulation we employed 256 processors. Therefore to perform the above calculations one would require 1.5 million processors. With the rapid expansion of parallel clusters (e.g. EPCC hpcx) the MH method should make the task of determining the quantum phase diagram a realistic project at a much  earlier time than that predicted above.

\chapter{\label{chap:conclusion}Conclusion}

In order to determine the location of a phase boundary between two phases one must determine at which point in the phase diagram the FED of the two phases is zero. The simplest approach is to tackle the problem via computational techniques (Monte Carlo) whereby one determines the weights of macrostates of one phase relative to those of the other. From this one may then infer the ratio of the partition functions of the two phases.

The problem with this approach is that generally a simulation initiated in a given phase will not visit the regions of (absolute) \cs\ associated with the other phase, since the two phase will in general be separated by a region of configuration space of intrinsically low probability. As a consequence one will not be able to determine the weights of macrostates of one phase {\em relative} to those of the other phase. This is generally referred to as the {\em overlap problem}.

One way to circumvent this problem is to use  the PM formalism \cite{p:brucereview} in which one directly maps the configurations of one phase {\em onto} those  of the other phase. By choosing an 'intelligent' PM one may generate considerable overlap between the two phases. In constructing the PM there are two issues which one must give consideration to. The first is the choice of a reference configurations $\Rvec_\al$ and the second is the choice of coordinate systems (\vvec), or {\em {\bf representation}} as we call it, with which one parameterises the displacements of the particles from the reference configuration. Since the PM matches the  coordinates ($\vvec_B=\vvec_A$), it is clear that the overlap is dependent upon both \Rvec\ and \vvec . The simplest and most straightforward choice of the representation is that in which the coordinates are expressed in terms of the displacements \uvec\ of the particles from the reference configuration $\Rvec_\al$. We call the associated mapping the RSM. Another possible choice is one in which the one parameterises the degrees of freedom in terms of fourier coordinates of the system. This we call the FSM. For the FSM one finds that, in the case of structurally ordered phases, the overlap problem vanishes as the harmonic limit is approached (see chapter \ref{chap:tune}), provided that the reference configurations are chosen to be the ground state configuration (i.e. the lattice sites).

Generally however the scope for refinement of the representation is limited, and one finds that the overlap problem persists. The second strategy that one naturally encounters is that of the {\em {\bf estimator}} which one uses to determine the FED. The choice of the optimal estimator depends on the way in which the regions of (effective) \cs\ associated with the two phases overlaps. In the case where they overlap in the manner shown in figure \ref{pic:twooverlap} (a) the EP estimator (Eq. \ref{eq:ep}), in which one performs a single simulation in phase A, yields an estimate which is free of systematic errors. In the case where they overlap as shown in figure \ref{pic:twooverlap} (b) then one must use estimators which involve simulations in both phases. In this situation one may either choose to use a phase constrained estimator in which one performs two simulations, one in each phase, and in which the non-negligible contributions come from the region of overlap (Eq. \ref{eq:dualphaserestrict} or Eq. \ref{eq:freebennettgen} where G is appropriately chosen), or one may employ the PS estimator (Eq. \ref{eq:psfermi}) in which the sampling distribution actually switches phases. However the validity of the estimates arising from these estimators presupposes some form of overlap in the regions of \ecs\ that the two phases explore. In the most general case, however, there will not be any form of overlap and therefore, like the choice of representation, the scope for refinement of the estimator will be limit.

The final part of the FED problem is that of the {\em {\bf sampling strategy}}. In this case one refines the sampling distribution in order to {\em engineer} overlap. Broadly there are three generic sampling strategies that one may pursue. The first is the MUCA strategy, whereby one introduces corrections to the Boltzmann weights appearing in the acceptance probabilities so as to force the simulation to explore regions of \ecs\ outside those it would normally explore (using the canonical probability distribution). The second is the MH strategy, whereby one simulates several systems independently. By simulating a series of systems in such a way that they overlap in the regions of (effective) configuration space that they explore and which, taken together, connect the regions of configurations space associated with one phase to those regions associated with the other phase, one is able to determine the FED. The advantage of this method is that it is highly parallelisable. The final strategy is the FG method, whereby one performs non-equilibrium work on the system so as to force it from the regions of \ecs\ associated with one phase to those of the other. By ensuring that one performs work in a gradual \cite{note:howgrad}, as opposed to abrupt, fashion, one may generate arbitrary overlap between the two methods. The overall sampling strategy may also involve combinations of these methods (see section \ref{sec:conc}).

\begin{figure}[tbp]
\begin{center}
\rotatebox{270}{
\includegraphics[scale=0.6]{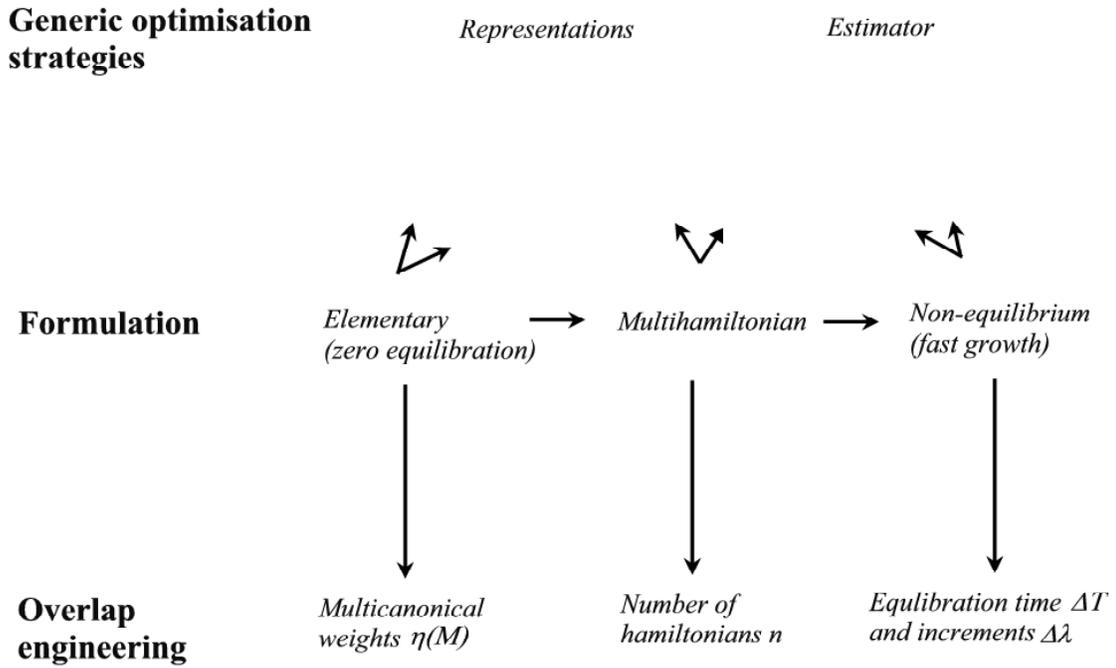}
}
\end{center}
\caption{Overview of the strategies involved in estimating the FED}

The figure above summarises the key components involved in tackling the FED problem. Broadly speaking there are three generic strategies that appear in tackling the overlap problem: the choice of {\em representation}, the choice of {\em estimator}, and the choice of {\em sampling strategy}. Generally the scope for refinement of the representation and estimator is limited, and therefore in order to fully overcome the overlap problem one must refine the sampling strategy.

The horizontal arrows indicate the direction in which one may generalise. That is the elementary (or zero equilibration FG, Eq. \ref{eq:RBA}) formulation may be generalised into the MH one (Eq. \ref{eq:gen_part}). Likewise the MH formulation itself (by defining a non-equilibrium process (see section \ref{sec:fg}) on the hamiltonians \hcalone {A} and \hcalone {B}) may be generalised so as to be incorporated within the FG formulation, Eq. \ref{eq:crooksfluct}.

\begin{center}
{\bf{------------------------------------------}}
\end{center}

\tshow{pic:grand}
\label{pic:grand}
\end{figure}

The key components in tackling the FED problem have been summarised in figure \ref{pic:grand}.

\appendix

\chapter{\label{app:fgproof} Proof of the fluctuation theorem}

\tshow{app:fgproof}

In this section we set out to prove the fluctuation theorem as given by Eq. \ref{eq:crooksfluct}. The original proof was given in \cite{p:crookspre}; we rederive it for the sake of mathematical clarity. The proof that is given here is the particular case of that given in  \cite{p:crookspre} in which \la\ changes discontinuously from $\la_1$ to $\la_2$ at time $t_1$, $\la_2$ to $\la_3$ at time $t_2$, etc for the $A\rightarrow B$ process (and the reverse in the $B\rightarrow A$ process).

\section{Proof of the fluctuation theorem}

We start by deriving a result which is central to the whole procedure. Consider a simulation employing the Metropolis algorithm in which the sequence of configurations $\seta {\sigma_1, \sigma_2, \sigma_3, ...}$ is generated. Under the scheme of the metropolis algorithm the probability of the system going from $\sigma_i$ to $\sigma_{i+1}$ is given by $P_S(\sigma_i \rightarrow \sigma_{i+1})$ (see Eq. \ref{eq:chonetemp}), where $P_S(\sigma_i \rightarrow \sigma_{i+1})$ satisfies  the condition of detailed balance (Eq. \ref{eq:detailedbalance}):

\begin{equation}
\frac {P_S(\sigma_i \rightarrow \sigma_{i+1})} {P_S(\sigma_{i+1} \rightarrow \sigma_{i})} = \frac {\pi(\sigma_{i+1})} {\pi(\sigma_i)}
\label{eq:stosig}
\end{equation}\tshow{eq:stosig}

\noindent where $\pi(\sigma_i)$ is the underlying sampling distribution. Eq. \ref{eq:stosig} may be easily extended to the case of two non-consecutive configurations:

\begin{eqnarray}
P_S(\sigma_1 \rightarrow \sigma_{d}) & = & \int [\p {j=1} {d-1} P_S(\sigma_j \rightarrow \sigma_{j+1})] [\p {j=2} {d-1} d \sigma_j]\nonumber\\
& = & \int [\p {j=1} {d-1} P_S(\sigma_{j+1} \rightarrow \sigma_{j})\times \frac{\pi(\sigma_{j+1})} {\pi(\sigma_1)}] [\p {j=2} {d-1} d \sigma_j]\nonumber\\
& = & \frac {\pi(\sigma_d)} {\pi(\sigma_1)} P_S(\sigma_d \rightarrow \sigma_{1})
\end{eqnarray}

\noindent or:

\begin{equation}
\frac {P_S(\sigma_1 \rightarrow \sigma_{d})} { P_S(\sigma_d \rightarrow \sigma_1)} = \frac {\pi(\sigma_d)} {\pi(\sigma_1)}
\label{eq:startfg}
\end{equation}\tshow{eq:startfg}

\noindent We may now proceed to prove the fluctuation theorem. Suppose that $P_{A\rightarrow B}(\seta {\vvec} | \vvec(1))$ denotes the probability of obtaining a path \seta {\vvec} given an initial configuration of $\vvec(1)$ in the $A\rightarrow B$ process. Then since the initial configuration $\vvec(1)$ is sampled from the distribution $\pia$, it follows that the distribution of the paths is given by:

\begin{equation}
{\cal{P}}^c_{A\rightarrow B} (\seta {\vvec}) = \pi^c_A(\vvec(1)) P_{A\rightarrow B} (\seta {\vvec} | \vvec(1))
\label{eq:fgpsprob}
\end{equation}\tshow{eq:fgpsprob}

\noindent where:

\begin{eqnarray}
P_{A\rightarrow B} (\seta {\vvec} | \vvec(1)) & = & P_S (\vvec(1)\rightarrow \vvec(2) | \pi_{\la_2}) \times P_S(\vvec(2) \rightarrow \vvec(3)| \pi_{\la_3}) .... \nonumber \\
&  & P_S(\vvec(n-2) \rightarrow \vvec(n-1)| \pi_{\la_{n-1}})
\label{eq:prodp}
\end{eqnarray}\tshow{eq:prodp}

\noindent where $P_S(\vvec(i) \rightarrow \vvec(i+1)| \pi_{\la_{i+1}})$ denotes the probability of the system making a transition from the configuration $\vvec(i)$ to $\vvec(i+1)$, between times $t_i$ and $t_{i+1}$, under the sampling distribution $\pi_{\la_{i+1}}$. Similarly the path \seta {\vvec} for the $B\rightarrow A$ process is sampled from the distribution ${\cal{P}}^c_{B\rightarrow A}$ where:

\begin{equation}
{\cal{P}}^c_{B\rightarrow A} (\seta {\vvec}) = \pi^c_B(\vvec(n-1)) P_{B\rightarrow A} (\seta {\vvec} | \vvec(n-1))
\label{eq:fgpsprob}
\end{equation}\tshow{eq:fgpsprob}

\noindent and  where:

\begin{eqnarray}
P_{B\rightarrow A} (\seta {\vvec} | \vvec(n-1)) & = &  P_S (\vvec(n-1)\rightarrow \vvec(n-2) | \pi_{\la_{n-1}}) \times P_S(\vvec(n-2) \rightarrow \vvec(n-3)| \pi_{\la_{n-2}}) .... \nonumber\\
&  & \times P_S(\vvec(2) \rightarrow \vvec(1)| \pi_{\la_{2}})
\label{eq:prodp2}
\end{eqnarray}\tshow{eq:prodp2}

\noindent Since from Eq. \ref{eq:startfg}:

\begin{equation}
\frac {P_S(\vvec(i) \rightarrow \vvec(i+1)| \pi_{\la_{i+1}})}  {P_S(\vvec(i+1) \rightarrow \vvec(i)| \pi_{\la_{i+1}})}= \frac {e^{-\beta E_{\la_{i+1}}(\vvec(i+1))}}{e^{-\beta E_{\la_{i+1}}(\vvec(i))}}
\end{equation}

\noindent it follows that:

\begin{eqnarray}
\frac {{\cal P}^c_{A\rightarrow B} (\seta {\vvec})}{{\cal P}^c_{B\rightarrow A} (\seta {\vvec})}&  = & \frac {\pia (\vvec(1))} {\pib (\vvec(n-1))} \p {i=1} {n-2} \frac {e^{-\beta E_{\la_{i+1}}(\vvec(i+1))}} {e^{-\beta E_{\la_{i+1}}(\vvec(i))}}\nonumber\\
& = & \frac {Z_B} {Z_A} \frac {e^{-\beta E_{\la_1}(\vvec(1))}} {e^{-\beta E_{\la_n}(\vvec(n-1))}} \exp \{ -\beta \s {i=1} {n-2} [E_{\la_{i+1}}(\vvec(i+1)) - E_{\la_{i+1}}(\vvec(i))] \}\nonumber\\
& = & \frac {Z_B} {Z_A} \frac {e^{-\beta E_{\la_1}(\vvec(1))}}  {e^{-\beta E_{\la_n}(\vvec(n-1))}} \exp \{ -\beta [E_{\la_{n-1}}(\vvec(n-1)) - E_2(\vvec(1))] \}\nonumber\\
&  + &  \beta \s {i=2} {n-2} [E_{\la_{i+1}}(\vvec(i)) - E_{\la_i}(\vvec(i))]\}\nonumber\\
& = & \frac {Z_B} {Z_A} \exp \{\beta \s {i=1} {n-1} [E_{\la_{i+1}}(\vvec(i)) - E_{\la_i}(\vvec(i))] \}
\end{eqnarray}

\noindent or:

\begin{equation}
\frac {{\cal P}^c_{A\rightarrow B} (\seta {\vvec})}{{\cal P}^c_{B\rightarrow A} (\seta {\vvec})} = \frac {Z_B} {Z_A} \exp \{ W_{BA} (\seta {\vvec})\}
\label{eq:fundaflu}
\end{equation}\tshow{eq:fundaflu}

\noindent Defining:

\begin{equation}
P(\wba | \pi^c_A) \equiv \int \delta (\wba - \wba(\seta {\vvec})) {\cal P}^c_{A\rightarrow B} (\seta {\vvec})
\end{equation}

\noindent and similarly for $P(\wba | \pi^c_B)$. It follows from Eq. \ref{eq:fundaflu} that:

\begin{eqnarray}
P(\wba^* | \pi^c_A) & = & \int \delta (\wba^* - \wba(\seta {\vvec})) {\cal P}^c_{A\rightarrow B} (\seta {\vvec})\nonumber\\ 
& = & \frac {Z_B} {Z_A} \int e^{\wba (\seta{\vvec})} \delta (\wba^* - \wba(\seta {\vvec})) {\cal P}^c_{B\rightarrow A}(\seta {\vvec})\nonumber\\
& = & \frac {Z_B} {Z_A}  e^{\wba^* } P(\wba ^*| \pi^c_B)
\end{eqnarray}

\noindent or:

\begin{equation}
P(\wba | \pia) = \frac {Z_B} {Z_A} e^{\wba} P(\wba | \pib)
\end{equation}

\noindent It is important to note that in this derivation we relied on Eq. \ref{eq:startfg}. As a consequence it is essential that the equilibration time used to evolve $\vvec(i)$ to $\vvec(i+1)$ in the $A\rightarrow B$ process is the same as that used to equilibrate $\vvec(i+1)$ to $\vvec(i)$ in the $B\rightarrow A$ process. This is consistent with the interpretation of the $B\rightarrow A$ process as being a time-reversal of the $A\rightarrow B$ process, in which the initial configurations are sampled from the distribution $\pib$ instead of $\pia$.

\chapter{\label{app:technicality} Fourier Space Mapping with periodic boundary conditions}
\tshow{app:technicality}

This section primarily deals with the modification that must be made to the FSM (Eq. \ref{eq:genswitch}) in the case where systems with periodic boundary conditions are employed, and is relevant to the discussion of section \ref{sec:fsstran}.

Generally the employment of periodic boundary conditions (in conjunction with a pairwise \ce ) means that there will be three eigenvectors of the dynamical matrix \avec {K} which will be of zero eigenvalue. These eigenvectors correspond to translations of the system. Clearly the fact that they are of zero eigenvalue means that they cannot be incorporated into the framework of Eq. \ref{eq:FStran} and Eq. \ref{eq:uintermsofmodes}. Suppose that  $\aevec^1_\gamma$, $\aevec^2_\gamma$, and $\aevec^3_\gamma$ correspond to the null eigenvectors. In this case we may express the displacements in terms of the fourier coordinates most generally as follows:

\begin{equation}
\auvec = \s {m=1} {3} v_m a^m_\gamma \aevec^{m}_\gamma + \s {m=4} {3N} v_m \frac{\aevec^{m}_\gamma} {\sqrt {k^m_\gamma}} 
\label{eq:utech}
\end{equation}\tshow{eq:utech}

\noindent where $a^1_\gamma$, $a^2_\gamma$, and $a^3_\gamma$ are some arbitrary constants, which are associated with the transformation (see Eq. \ref{eq:vtoutran}):

\begin{equation}
{\mf T}_\al = (a^1_\gamma {\mf e}^1_\al, a^2_\gamma {\mf e}^2_\al, a^3_\gamma {\mf e}^3_\al,\f {1} {\sqrt{k^{3}_\al}} {\mf e}^{3}_\al,..., \f {1} {\sqrt{k^{3N}_\al}} {\mf e}^{3N}_\al)
\label{tran_tech}
\end{equation}\tshow{tran_tech}

\noindent The \rpf\ may then be written as:

\begin{equation}
\RBAcal = \frac {\int d\uvec e^{-\beta \ecal_B (\auvec)}} {\int d\uvec e^{-\beta \ecal_A (\auvec)}} =  C_{\alp\al} \frac {\{\p {i=4} {3N} \frac {1} {\sqrt{k^i_B}}\} \{\int \p {i=4} {3N} dv_i e^{-\beta \ecal_B(\vvec)}\}} {\{\p {i=4} {3N} \frac {1} {\sqrt{k^i_A}}\}\{ \int \p {i=4} {3N} dv_i e^{-\beta \ecal_A(\vvec)}\}}
\label{eq:freetempch3}
\end{equation}\tshow{eq:freetempch3}

\noindent where $C_{\alp\al}$ arises from the centre of mass contributions. Since these should not contribute to the ratio of the partition functions \RBAcal\ we set:

\begin{equation}
C_{\alp\al} = 1
\end{equation}

\noindent The redundancy of the translational degrees of freedom means that one may omit their consideration altogether in mapping the configurations of one phase onto those the other. 
That is the transformation $\avec {T}_\gamma$ (Eq. \ref{eq:FStran}) may be  replaced by a 3N by 3N-3 column vector $\tilde{\avec {T}}_\gamma$ which is given by:

\begin{equation}
\tilde{\avec{T}}^{ij}_\gamma = \frac {\aevec^{i,j+3}_\gamma} {\sqrt {k^{j+3}_\gamma}}
\label{eq:FStran2}
\end{equation}\tshow{eq:FStran2}

\noindent or 

\begin{equation}
\tilde{\mf T}_\gamma = (\f {1} {\sqrt{k^4_\al}} {\mf e}^4_\al, \f {1} {\sqrt{k^5_\al}} {\mf e}^5_\al, ..., \f {1} {\sqrt{k^{3N}_\al}} {\mf e}^{3N}_\al)
\label{har_notran_inv}
\end{equation}\tshow{har-notran-inv}

\noindent The transformation $\tilde{\avec {T}}_\gamma$ now acts on the (3N-3) column vector $\tilde{\avvec}$ where the component $\tilde{v}_i$ is given by $v_{i+3}$. The displacements \uvec\ are then given by:

\begin{equation}
\auvec = \tilde{\avec {T}}_\gamma \tilde{\avvec}
\end{equation}

\noindent or

\begin{equation}
\auvec = \s {m=4} {3N} v_m \frac{\aevec^{m}_\gamma} {\sqrt {k^m_\gamma}} 
\end{equation}

\noindent Likewise the inverse transformation $\avec {T}^{-1}_\gamma$ [(3N-3 by 3N) transformation] may be written as:

\begin{equation}
[\tilde{\avec {T}}^{-1}_\gamma]^{ij} = \aevec_\al^{j,i+3} . \sqrt {k^{i+3}_\gamma}
\end{equation}

\noindent or

\begin{equation}
[\tilde{\mf T}_\al]^{-1} = 
\left( \begin{array}{c}
\sqrt{k_\al^4}[{\mf e}_\al^4]^T\\
\sqrt{k_\al^5}[{\mf e}_\al^5]^T\\
.\\
.\\
.
\end{array} \right)
\label{har_notran_inv}
\end{equation}\tshow{har-notran-inv}

\noindent It then follows that the transformation ${\avec {S}}_{BA}$ (Eq. \ref{eq:genswitch}), which maps the configurations of one phase onto those of the other, may be written as:

\begin{eqnarray}
{\avec {S}}^{ij}_{BA} & = & [\tilde{\avec {T}}_B . \tilde{\avec {T}}^{-1}_A]^{ij}\nonumber\\
& = & \s {m=4} {3N} \avec {T}^{im}_B [\avec {T}^{-1}_A]^{mj}\nonumber\\
& = & \s {m=4} {3N} \sqrt{\frac {k^m_A}{k^m_B}} \aevec^{im}_B \aevec^{jm}_A 
\label{eq:stran2}
\end{eqnarray}\tshow{eq:stran2}

\noindent From Eq. \ref{eq:freetempch3} we see that the  FED may then be written as Eq. \ref{eq:FED2} where now:

\begin{equation}
{\triangle F}^h_{BA} = \frac {1} {2\beta} \s {m=4} {3N} \ln (\frac {k^m_B} {k^m_A})
\label{eq:apharf}
\end{equation}\tshow{eq:apharf}

\noindent and 

\begin{equation}
\triangle F^a_{BA} = -\beta^{-1} \ln {\cal R}^a_{BA} 
\end{equation}

\noindent where 

\begin{equation}
{\cal R}^a_{BA} = \frac {\int  d\tilde{\vvec} e^{-\beta \ecal_B (\tilde{\avvec})}} {\int d \tilde{\vvec} e^{-\beta \ecal_A (\tilde{\avvec})}}
\end{equation}

\noindent Eq. \ref{eq:apharf} is simply obtained from Eq. \ref{eq:freetempch3} by noting that in the harmonic limit $\ecal_A(\vvec) = \ecal_B(\vvec)$, so that the configurational integrals in the numerator and denominator exactly cancel out in this limit.

 The transformation $\avec {S}_{BA}$ in Eq. \ref{eq:stran2}, through the mapping in Eq. \ref{eq:genswitch},  ensures that the effective configuration $\tilde{\vvec}$ are preserved in mapping the displacements \uvec\ of one phase onto those of the other \cite{note:detofS}.

\chapter{\label{app:fssperturb} Perturbation theory for  the Fourier Space Mapping}

\tshow{app:fssperturb}

A probability distribution may be completely characterised by its cumulants (eq. \ref{eq:cumulant}). Therefore an alternative way to investigate the dependence of the overlap on some generic parameter (like the temperature T) is to find the dependence of the cumulants on this parameter. In this appendix we will specifically focus on the FSM, and we will find relations which determine the way the various cumulants of $P( M_{BA} | \pi^c_\al)$ scale with temperature. The primary conclusions of this section will be that in the limit of $T\rightarrow 0$ all the cumulants vanish. 
The discussion of this appendix  is relevant to section \ref{eq:fssanalytic}.

\section{Preliminary Mathematical Properties of Gaussian Integrals}

Let us define the harmonic average of a macrovariable $M(\vvec)$:

\begin{equation}
<M>_h \equiv \f {\int d\vvec M(\avvec) e^{-\f {\beta} {2} \vvec . \vvec}} {\int d\vvec e^{-\f {\beta} {2} \vvec . \vvec}}
\label{gaussian_av}
\end{equation}\tshow{gaussian_av}

\noindent where the limits of integration are implicitly assumed to be from $-\infty$ to $\infty$. Two results which we will frequently use in this appendix and appendix \ref{app:rssperturb} are the following:

\begin{equation}
\f {\int v^{2n} e^{-\frac{\beta}{2} v^2} dv} {\int e^{-\frac{\beta}{2} v^2} dv} = [\f {\beta} {2}]^{-2} \p {i=0} {n-1} (i+0.5)
\label{gauss_1b}
\end{equation}\tshow{gauss-1b}

\noindent and

\begin{equation}
<v_{1}^{q_1} v_{2}^{q_2} ... v_{n}^{q_n}>_h = \left\{\begin{array}
{r@{\quad:\quad}l}
0 & \mbox{\ if any of the $q_i$ are odd} \\
c \beta^{-\f {q} {2}} & \mbox{\ otherwise}
\end{array}\right.
\label{general_average}
\end{equation}\tshow{general_average}

\noindent where $q=\s {i=1} {n} q_i$ and where c is some constant.

\noindent For the sake of notational simplicity, we will, in the following section, denote $M_{\alp\al}(\avvec)$  by $M_{FS}$ when $\avvec$ corresponds to fourier coordinates of the system (defined by the relation in Eq. \ref{eq:vtoutran} when ${\mf {T}}_\al$ is given by Eq. \ref{eq:FStran}) and $M_{\alp\al}(\avvec)$ by $M_{RS}$ when  working with the RSM ($\vvec=\uvec$).

\section{\label{sec:fsmean}Temperature scaling properties of $\omega_n$}
\tshow{sec:fsmean}

In the case of crystalline solids a physical motivation exists for the separation of the harmonic contributions to the excitation energy from the anharmonic ones. Let us start by first considering the Taylor expansion of the excitation energy $\ecal_\gamma$ (Eq. \ref{eq:expandenergy}) in terms of the fourier coordinates \vvec :

\begin{eqnarray}
\exi_\gamma (\avvec)& = &\h \vvec^T . \vvec + \s {ijk} {}  M_{ijk}^\gamma v_i v_j v_k + \s {ijkl} {} N_{ijkl}^\gamma v_i v_j v_k v_l + .....\nonumber\\
& = & H_2 + H^\gamma_3 + H^\gamma_4 + ...
\label{eq:lor_e}
\end{eqnarray}\tshow{eq:lor-e}

\noindent  where $H^\gamma_n$ denotes the summation of all the terms of order n and where $\h \vvec^T . \vvec$ corresponds to the harmonic contributions to the excitation energy (see Eq. \ref{eq:einv}). The expectation of a macrovariable M with respect to the sampling distribution $\pi^c_\al$ may then be written as:

\begin{equation}
<M>_{\pi^c_\gamma} =  \f {1} {\z_\gamma} \int d\avvec M(\avvec) e^{-\beta (H_2 + H^\gamma_3 + H^\gamma_4 + ...)}
\label{new_exp}
\end{equation}\tshow{new_exp}

\noindent where the partition function $\z_\gamma$ may be expanded in the following way:

\begin{eqnarray}
\z_\gamma & = &  \int  d \avvec e^{-\beta \ecal_\gamma(\avvec)} \nonumber\\
 & = &   \int  d \avvec e^{-\beta[H_2 + H_3^\al + H_4^\al+ ...]}\nonumber\\
& = &   \int  d \avvec e^{-\beta H_2}[1-\beta H_3^\gamma -\beta H_4^\al + \f {\beta^2} {2} [H_3^\al]^2 + \beta^2 O(a^8) + \beta O(a^6)]\nonumber\\
& = & Z_h [1 - \beta <H_4^\al>_h + \f {\beta^2} {2} <[H_3^\al]^2>_h + O(\beta^2)]
\label{part_exp}
\end{eqnarray}\tshow{part_exp}

\noindent In Eq. \ref{part_exp} we have used the fact (see Eq. \ref{general_average}) that the integrals of integrands whose overall order of the fourier coordinates $\{v\}$ is odd vanishes. The $O(\beta^2)$ terms in Eq. \ref{part_exp} is what is left over on integrating the  $\beta^2 O(a^8)$ and $\beta O(a^6)$ terms. It then follows that:

\begin{equation}
\f {1} {\z_\gamma} = \f {1} {Z_h} [1 + [\beta <H_4^\al>_h - \f {\beta^2} {2} <(H_3^\al)^2>_h] + O(\beta^{-2})]
\label{inv_part}
\end{equation}\tshow{inv_part}

\noindent Since $[\beta<H_4^\al>_h-\f {\beta^2} {2}<(H^\al_3)^2>_h]\sim \beta^{-1}$ and since we are only interested in the leading order terms in the temperature in Eq. \ref{new_exp} as $\beta\rightarrow\infty$, we see that one may replace $1/\z_\gamma$ appearing in the expectations of Eq.  \ref{new_exp} with $1/Z_h$ in this limit. One may then rewrite Eq. \ref{new_exp} as:

\begin{equation}
<M>_{\pi^c_\gamma} \approx \frac {1} {Z_h} \int d\avvec M(\avvec) G(\vvec) e^{-\beta H_2}
\label{eq:AHA}
\end{equation}\tshow{eq:AHA}

\noindent where 

\begin{equation}
G(\vvec) = 1-\beta H_3^\al + \f {\beta^2} {2} [H^\al_3]^2 -\beta H_4^\al + \beta^2 O(a^8) + \beta O(a^6)
\label{eq:gv}
\end{equation}\tshow{eq:gv}

\noindent In order to analyse the cumulants of $P(\mba | \pi^c_\al)$ let us consider the particular case when $M=\mba$, which we denote by $M_{FS}$. From Eq. \ref{eq:lor_e} we see that one may write:

\begin{equation}
M_{FS} = \beta[\delta H_3^\al + \delta H_4^\al + ...]
\label{eq:taylorfs}
\end{equation}\tshow{eq:taylorfs}

\noindent where 

\[\delta H_n = H_n^\alp - H_n^\al\]

\noindent The expectation of an arbitrary power of the overlap parameter may then be written as:

\begin{equation}
<M^n_{FS}>_{\pi^c_\al} \approx  <G(\vvec)\beta^n (\delta H_3 + \delta H_4 + ...)^n>_h
\label{eq:tempfsexpect}
\end{equation}\tshow{eq:tempfsexpect}

\noindent In evaluating the expectation in Eq. \ref{eq:tempfsexpect} one will obtain a series of terms scaling in different ways with respect to the temperature. Since we are examining the harmonic limit ($T\rightarrow 0$), we are only interested in the terms which are lowest order in T (i.e. highest order in $\beta$). These  originate from the integrals with the lowest overall (even) order of v.  This  means that both $\delta H_3$ and $\delta H_4$ need to be considered in evaluating the expectation of Eq. \ref{eq:tempfsexpect}, since, depending on whether n is even or not, it might be either $\delta H_3$ or $\delta H_4$ which couple to the lowest order terms of $G(\vvec)$ so as to yield the most slowly vanishing term.

Writing Eq. \ref{eq:tempfsexpect} in full and retaining only the lowest order terms, one finds that:

\begin{equation}
<M^n_{FS}>_{\pi^c_\al} \approx \beta^n <[\delta H_3]^n + [\delta H_3]^{n-1} \delta H_4 + \beta H^\al_3 (\delta H_3)^n + [\delta H_3]^{n-1}\delta H_4 \beta H^\al_3>_{h}
\label{eq:mexptwo}
\end{equation}\tshow{eq:mexptwo}

\noindent so that:

\begin{equation}
<M_{FS}^n>_{\pi^c_\al} \sim \beta^n <[\delta H_3]^n>_{\pi^c_\al} \sim \beta ^{-\frac {n} {2}} \mbox{if n is even}
\label{eq:mifeven}
\end{equation}\tshow{eq:mifeven}

\begin{equation}
<M_{FS}^n>_{\pi^c_\al} \sim \beta^n <[\delta H_3]^{n-1} \delta H_4>_{\pi^c_\al} + \beta^n <\beta H^\al_3 (\delta H_3)^n>_{\pi^c_\al} \sim \beta ^{- \frac {[n+1]} {2}} \mbox{if n is odd}
\label{eq:mifodd}
\end{equation}\tshow{eq:mifodd}

\noindent From these relations we see that the mean and the variance of the overlap parameter scale in the following way:

\begin{equation}
\lim_{\beta\rightarrow\infty} <M_{FS} >_{\pi^c_\al} \propto \beta^{-1}
\label{low_ene}
\end{equation}\tshow{low_ene}

\noindent and 

\begin{equation}
\lim_{\beta\rightarrow\infty} <M^2_{FS}>_{\pi^c_\al} - <M_{FS}>^2_{\pi^c_\al} \propto \beta^{-1}
\label{low_var}
\end{equation}\tshow{low_var}

\noindent It is clear from Eq. \ref{eq:mifeven} and  \ref{eq:mifodd} that $\omega_n$ will scale as $<M^n_{FS}>_{\pi^c_\al}$, since all other terms in the cumulants will either scale with the same or higher power of T. Therefore we conclude that for the FSM the cumulants of $P(M_{FS} | \pi^c_\al)$ will scale in the following way:

\begin{equation}
\lim_{\beta\rightarrow\infty}\omega_n \sim \beta^{-\frac {n} {2}} \mbox{\ if n is even}
\end{equation}

\begin{equation}
\lim_{\beta\rightarrow\infty}\omega_n \sim \beta^{-\frac {[n+1]} {2}}\mbox{\ if n is odd}
\end{equation}

\chapter{\label{app:rssperturb} Perturbation theory for  the Real Space Mapping}
\tshow{app:rssperturb}

As in appendix \ref{app:fssperturb}, we examine the temperature dependence of the overlap, as engineered by the RSM, through an investigation of the cumulants (Eq. \ref{eq:cumulant}). We will derive exact expressions for the mean and variances of $P(M_{\alp\al} | \pi^c_\al)$, followed by a general argument to show that $\omega_n$ tends to a constant non-zero value in the limit of $T \rightarrow 0$. The material in this appendix is relevant to the discussion of section \ref{eq:rssanalytic}.

\section{Low temperature limit of $\omega_1$}

Let $\uvec_\al$ collectively denote the displacements of the particles of phase \al\ from the reference configuration ${\mf R}_\al$ (which for the systems employed here correspond to the lattice sites of the crystalline solid, see section \ref{sec:system}), and suppose that $[\vvec_\al]_i$ denotes the i-th component of the fourier coordinates $\vvec_\al$ of phase \al . The RSM ensures that:

\begin{equation}
\uvec_\alp = \uvec_\al
\end{equation}

\noindent Using Eq. \ref{eq:vtoutran} we see that this constraint imposes the following relation between $\vvec_\al$ and $\vvec_\alp$:

\begin{eqnarray}
[\vvec_\alp]_i & = & [{\mf T}_\alp^{-1} . {\mf T}_\al . \vvec_\al ]_i \nonumber\\
& = & \s {j} {} [{\mf T}_\alp^{-1}]^{ij} [{\mf T}_\al \vvec_\al]_j \nonumber\\
& = & \s {m} {} \s {j} {} [{\mf T}_\alp^{-1}]^{ij} {\mf T}_\al^{jm} [\vvec_\al]_m \nonumber\\
& = & \s {j} {} \s {m} {} \frac {\sqrt {k^i_\alp}} {\sqrt {k^m_\al}} {\mf e}^{ji}_\alp {\mf e}^{jm}_\al [\vvec_\al]_m\nonumber\\
& = & \s {m} {} \frac {\sqrt {k^i_\alp}} {\sqrt {k^m_\al}}   [{\mf e}^{i}_\alp]^T . {\mf e}^{m}_\al [\vvec_\al]_m
\end{eqnarray}

\noindent so that :

\begin{equation}
\vvec_\alp . \vvec_\alp = \s {i} {}\s {m} {} \s {\tilde{m}} {} \frac {k^i_\alp} {\sqrt{k^m_\al} \sqrt{k^{\tilde{m}}_\al} } [{{\mf e}^i_\alp}^T . {\mf e}^m_\al] [{{\mf e}^i_\alp}^T . {\mf e}^{\tilde{m}}_\al][\vvec_\al]_m [\vvec_\al]_{\tilde{m}}
\label{eq:rssvrelate}
\end{equation}\tshow{eq:rssvrelate}

\noindent From this we see that  in the harmonic limit (where the excitation energy is given by Eq. \ref{eq:einv}) $M_{\alp\al}$, which we write as $M_{RS}$ to signify the fact that we as using the RSM,  may be written as:

\begin{eqnarray}
M_{RS}(\vvec_\al) & \equiv & \beta [\ecal^h_\alp (\vvec_\alp) - \ecal^h_\al (\vvec_\al)] \nonumber\\
& = & \frac {\beta  }{2} {\mf v}_\al^T  \mf W {\mf v}_\al
\label{eq:harrssm}
\end{eqnarray}\tshow{eq:harrssm}

\noindent where the matrix elements of $\mf W$ are given by:

\begin{equation}
{\mf W}_{m \tilde{m}} = \s {i} {} \frac {k^i_\alp} {\sqrt{k^m_\al} \sqrt{k^{\tilde{m}}_\al} } [{{\mf e}^i_\alp}^T . {\mf e}^m_\al] [{{\mf e}^i_\alp}^T . {\mf e}^{\tilde{m}}_\al] - \delta_{\tilde{m} m} {\mf {1}}
\end{equation}

\noindent We will discard the subscript $\al$ on the variable $\vvec_\al$ since we will assume (for the rest of this section) that $\vvec=\vvec_\al$.

 Because $\mf W$ is symmetric (i.e. Hermitian), we may diagonalise it. Typical diagonalization routines yield eigenvectors, which, in the case of degenerate eigenvalues, may not be orthogonal. In this case one may employ the Gram Schmidt orthogonalisation procedure to construct an orthonormal set amongst these degenerate eigenvectors.
Suppose that $\mf W$ has the eigenvalues \seta{\kappa_i} and suppose that we write $\mf b = \mf N \mf v$ where $\mf N^T \mf W \mf N$ is diagonal. Since $\mf N$ is an orthogonal transformation, we see that $\det N=1$ and $\mf b^T .\mf b=\mf a^T .\mf a$. Then:

\begin{eqnarray}
(\mf v^T .\mf W .\mf v) & = & (\mf b^T (\mf N^T . \mf W . \mf N) \mf b)\nonumber\\
  & = & (\s {i} {} \kappa_i {b_i}^2)
\end{eqnarray}

\noindent and therefore

\begin{eqnarray}
<M_{RS} >_h & = & \beta <\mf v^T .\mf W .\mf v >_h\nonumber\\
& = & \beta \f{\int_{-\infty}^{\infty} (\s {i} {} \kappa_i {b_i}^2)  e^{-\h \beta \mf b^T \mf b}d\mf b}{\int_{-\infty}^{\infty} e^{-\h \beta \mf b^T \mf b}d\mf b}\nonumber\\
& = & \frac{1}{2} \s {i} {} \kappa_i 
\end{eqnarray}

\noindent Alternatively, by directly appealing to Eq. \ref{eq:rssvrelate} and  Eq.\ref{eq:harrssm}, it is not hard to see that:

\begin{equation}
<M_{RS}>_{\al} = \f {1} {2} \s {j} {}(\s {l} {} \f {k^j_\alp} {k^l_\al} [{\mf e^j_\alp}^T . \mf e^l_\al]^2 - 1)
\label{low_temp_lim}
\end{equation}\tshow{low-temp-lim}

\noindent where we have used the fact that $<[\vvec_\al]_m [\vvec_\al]_{\tilde{m}}>_h = \delta_{m\tilde{m}} \frac {\beta^{-1}} {2}$ (see Eq. \ref{gauss_1b}). From Eq. \ref{low_temp_lim} it immediately follows the first cumulant for the RSM may then be written as:

\begin{equation}
\omega_1 = \frac {1} {2} \s {i} {} \kappa_i = \frac {1} {2} \s {j} {}(\s {l} {} \f {k^j_\alp} {k^l_\al} [{\mf e^j_\alp}^T . \mf e^l_\al]^2 - 1)
\label{eq:rsscumu1app}
\end{equation}\tshow{eq:rsscumu1app}

\section{Low temperature limit of $\omega_2$}

In a similar manner, the low temperature limit of the variance of  $M_{RS}$ may be calculated. Using the fact that: 

\begin{eqnarray}
(\mf v^T .\mf W .\mf v)^2 & = & (\mf b^T (\mf N^T . \mf W . \mf N) \mf b)^2\nonumber\\
  & = & (\s {i} {} \kappa_i {b_i}^2)^2
\end{eqnarray}

\noindent we see that:

\begin{eqnarray}
<(M_{RS})^2>_h & = & \beta^2 <(\mf v^T .\mf W .\mf v)^2>_h\nonumber\\
& = &  \beta^2 \f{ \int_{-\infty}^{\infty} (\s {i} {} \kappa_i {b_i}^2)^2  e^{-\h \beta \mf b^T \mf b}d\mf b}{\int_{-\infty}^{\infty} e^{-\h \beta \mf b^T \mf b}d\mf b}\nonumber\\
& = & \beta^2 <(\s {i} {} \kappa_i {b_i}^2)^2>_h\nonumber\\
& = & \beta^2 \s {i} {} <\kappa_i^2 {b_i}^4>_h +  \beta^2 \s {i,j; i \not= j} {} <\kappa_i \kappa_j {b_i}^2 {b_j}^2>_h\nonumber\\
& = & \frac {1} {4} [\s {i} {} 3 \kappa_i^2 + \s {i,j; i\not=j} {}  \kappa_i \kappa_j]
\label{mmmRtwo}
\end{eqnarray}\tshow{low_var}

\noindent Therefore we conclude that:

\begin{eqnarray}
<M_{RS}^2>_h - <M_{RS}>_h^2 & = &  <(\mf v^T .\mf W .\mf v)^2>_h - <(\mf v^T .\mf W .\mf v)>_h^2  \nonumber\\
& = & \frac {1} {2} \s {i} {} {\kappa_i^2}
\label{low_lim_var}
\end{eqnarray}\tshow{low-lim-var}

\noindent Accordingly the second cumulant for the RSM may be written as:

\begin{equation}
\omega_2 = \frac {1} {2} \s {i} {} \kappa_i^2
\label{eq:rsscumu2app}
\end{equation}\tshow{eq:rsscumu2app}

\section{Temperature scaling of $\omega_n$}

Even though in the preceding section we were able to calculate the exact limiting form of $\omega_1$ and $\omega_2$ as $T\rightarrow 0$, extending this to the case of $\omega_n$ becomes tedious. Instead we will follow the presentation in section \ref{sec:fsmean} in order to derive the scaling relation of $\omega_n$ for the general case. In calculating the temperature scaling properties of $\omega_n$ we first note that for the RSM the harmonic terms $H_2$ are not identical in the two phases. Therefore the perturbation expansion of $M_{RS}$ becomes (as compared with Eq. \ref{eq:taylorfs}):

\begin{equation}
M_{RS} \approx \beta [\delta H^\al_2 + \delta H^\al_3 + ....]
\label{eq:taylorrs}
\end{equation}\tshow{eq:taylorrs}

\noindent so that:

\begin{eqnarray}
<M^n_{RS}>_n & \approx & \beta^n <G(\vvec) (\delta H_2 + \delta H_3 + ....)^n>_h\nonumber\\
& = & \beta^n <(1 - \beta H^\gamma_3 + \frac {\beta^2} {2} [H^\gamma_3]^2 - \beta H^\gamma_4 + ...) (\delta H_2 + \delta H_3 + ....)^n>_h
\label{eq:mrsexp}
\end{eqnarray}\tshow{eq:mrsexp}

\noindent It is immediately clear from Eq. \ref{eq:mrsexp} that the temperature scaling properties of $M_{RS}$ will be governed by the leading order term in eq. \ref{eq:taylorrs}, $\delta H_2$, so that:

\begin{equation}
<M_{RS}^n>_{\pi^c_\al} \sim O(1)
\end{equation}

\noindent From this we may infer that the      cumulants will, for sufficiently low temperatures, be independent of the temperature:

\begin{equation}
\lim_{\beta\rightarrow\infty}\omega_n \sim O(1)
\end{equation}

\noindent Therefore in the harmonic limit the distribution of $M_{\alp\al}$, for the RSM, assumes a stationary  form which is not that of the limiting form associated with perfect overlap (Eq. \ref{eq:perfect}).

\chapter{\label{app:errors} Determining Statistical Errors}
\tshow{app:errors}

In this section we discuss the blocking method, which is a way to  determine the error associated with an estimate of the expectation of an arbitrary macrovariable Q (see Eq. \ref{eq:estimateexpectation}) obtained from correlated data. We also illustrate the way in which this blocking method may be used to estimate the error in the FED estimate. For more detailed information on the blocking method we refer the reader to \cite{b:frenkel}, \cite{p:blocking}, \cite{note:lookjanke}.

\section{Errors of averages}

Suppose that we make a series of measurements $Q_i$ (i=1,...,t), sampled from a probability distribution of mean \m\ and variance $\sigma^2$. Suppose that it is also our desire to obtain an unbiased estimate for \m. This can be most simply obtained from the mean of the data set \seta {Q_i}:

\begin{equation}
\hat{\mu}= \f {1} {t} \s {i=1} {t} Q_i
\label{mean_est}
\end{equation}\tshow{mean-est}

\noindent In the case where successive  measurements $Q_i$ are independent, one finds \cite{b:kendall}, \cite{b:grimmett} that the distribution of $\hat{\mu}$ tends to a Normal Distribution with mean \m\ and variance $\sigma^2/t$. That is  ${P}(\hat{\m})\sim N(\mu, \sigma^2/t)$. This is simply  a consequence of the central limit theorem. Therefore in the particular case where the measurements are independent, the ``error'' associated with the  estimate of Eq. \ref{mean_est} is simply given by  $\hat{\sigma}/\sqrt{t}$, where $\hat{\sigma}^2$ is an unbiased estimator of the variance of $Q$, and is given by:

\begin{equation}
\hat{\sigma}^2 = \frac {1} {t-1} \s {i=1} {t} (Q_i - \hat{\mu})^2
\end{equation}

\noindent In the case where the measurements $Q_i$ are  correlated, Eq. \ref{mean_est}\ still yields an unbiased estimate for the mean of the underlying distribution of Q. However the associated error is now no longer given by $\hat{\sigma}/\sqrt{t}$. One method for finding the associated error is the so called blocking method. In this method, the set of data \seta {Q_i} is sectioned into M blocks each containing m data entries. That is block i corresponds to the set \seta {Q_{m(i-1)+1}, Q_{m(i-1)+2},...., Q_{m(i-1)+m}}. Then for each block an estimate $\hat{\mu}_m(i)$ for the mean of the distribution of Q is made, and is given by:

\begin{equation}
\hat{\m}_m(i) = \f {1} {m} \s {j=1} {m} Q_{m(i-1)+j}
\label{block_mean_est}
\end{equation}\tshow{block-mean-est}

\noindent From this we obtain a set of block estimates for the mean of the distribution:

\[ \{ \hat{\m}_m(1),\hat{\m}_m(2),....,\hat{\m}_m(M) \} \]

\noindent Since:

\begin{equation}
\hat{\m} = \f {1} {M} \s {i=1} {M} \hat{\m}_m(i)
\label{mean_from_blocks}
\end{equation}\tshow{mean-from-blocks}

\noindent it follows that the block estimates \seta {\hat{\m}_m(i)} will themselves be distributed with a mean given by $\m$ and a variance given by $\sigma^2_m$, say. The key observation is that for sufficiently large blocksizes successive $\hat{\m}_m(i)$ will independent. In this case the error of the average of the block estimates is simply given by $\hat{\sigma}^2_m / M$. Since the estimator $\hat{\m}$ is precisely this average (see Eq. \ref{mean_from_blocks}) it follows that, for the (sufficiently large m) regimes where successive block estimates are independent, the error in the estimate $\hat{\mu}$ is given by $\hat{\sigma}(\m)$, where:
 
\begin{eqnarray}
\hat{\sigma}^2(\m) & = & \frac {\hat{\sigma}^2_m} {M}\nonumber\\
& = & \frac {m \hat{\sigma}^2_m} {t}
\label{eq:errormu}
\end{eqnarray}\tshow{eq:errormu}

\noindent and where $\hat{\sigma}^2_m$ is obtained from:

\begin{equation}
\hat{\sigma}^2_m = \f {1} {M-1} \s {i=1} {M} ({\hat{\mu}}_m(i)-\hat{\mu})^2
\label{var_blocked_var}
\end{equation}\tshow{var-blocked-var}

\noindent In order to find the regime of blocksizes where successive block estimates become independent,  a simple graphical procedure may be used to estimate the errors in $\hat{\m}$. Since $m\sigma^2_m$ is constant in the regime where the ${\hat{\mu}}_m(i)$ are independent (a result which must hold true since ${\sigma}^2(\m)$ is independent of the blocksize), we see that we may determine the blocksizes m for which the ${\hat{\mu}}_m(i)$ are independent simply by plotting a graph of $m \sigma^2_m$ versus m. As m increases the graph will eventually plateau off, indicating that the block estimates are indeed uncorrelated. From Eq. \ref{eq:errormu} we see that one may then use the value of the plateau (P) to determine the error in the estimate of $\hat{\m}$:

\begin{eqnarray}
\mbox{error in $\hat{\m}$} & = & \sqrt {\f {\hat{\sigma}_m^2} {M}}\nonumber\\
& = & \sqrt {\f {m \hat{\sigma}_m^2} {Mm}}\nonumber\\
& = & \sqrt{\f {P} {t}}
\label{plateau_error}
\end{eqnarray}\tshow{plateau-error}

\noindent For a more mathematically rigorous treatment of the blocking procedure we refer the reader to \cite{p:blocking}.

\section{Errors in the free energy difference}

Though one may estimate the FED by taking an appropriate expectation (Eq. \ref{eq:ep}), the general expression for the FED will involve the ratio of expectations (see for example Eq. \ref{eq:partrsestimator0}, Eq. \ref{eq:psformu}, and  Eq. \ref{eq:freebennett}) . In this case the average of the block estimates of \RBAcal\ is not the same as the estimate of \RBAcal\ obtained from  the whole data set. In this section we show that the blocking method can also be used to estimate the error in \RBAcal . As a specific example we use the PS estimator, in which the ratio of the partition functions is estimated by determining the (unbiased) ratio of the times spent in the two phases (see Eq. \ref{eq:psformu}). For simplicity we consider the case where no weights are employed.

Suppose that we make a series of measurements $\ga_i$ (i=1,....,t) of the ``phase label'' \ga\ which can take on either of two values,  A or B say, during the course of PS simulation. Then the estimator for the probability of being in phase A is given by:

\begin{equation}
\hat{p}(\ga = A)\equiv \f {1} {t} \s {i=1} {t} \delta_{\ga_i , A}
\label{phatal}
\end{equation}\tshow{phatal}

\noindent and the estimator for the probability of being in phase B is given by:

\begin{equation}
\hat{p}(\ga = B)= 1- \hat{p}(\ga = A)
\label{phatalp}
\end{equation}

\noindent It is clear from  Eq. \ref{eq:psformu} that $\ln \RBAcal$ (which is proportional to the FED) may then be estimated by:

\begin{eqnarray}
\hat{D}^{(1)} & = &\ln(\f {\mbox{ Time in B}} {\mbox{ Time in A}})\nonumber\\
& = & \ln(\f {\hat{p}(\ga = B)} {1-\hat{p}(\ga = B)})\nonumber\\
\label{free_d1}
\end{eqnarray}\tshow{free_d1}

\noindent If $\hat {p}(\ga=B)$ is deviated from its true value $p(\ga=B)$ by a small amount $\sigma_B$ (${\sigma_B} / {p(\ga=B)} << 1)$, then it is easy to show, using the approximation $\ln (1+x)\approx x$ valid for small x, that the error $\delta \hat{D}^{(1)}$ in $\hat{D}^{(1)}$ is given by:

\begin{equation}
\delta \hat{D}^{(1)}= \delta\ln (\f {\hat{p}(\ga = B)} {1-\hat{p}(\ga = B)} )= \f {\sigma_B} {p(\ga = B) (1-p(\ga = B))}
\label{error_1}
\end{equation}\tshow{error_1}

\noindent Now let us consider dividing the data of t observations into M blocks, each containing m data points. We may then make a block estimate of $\ln \RBAcal$:

\begin{equation}
\hat{D}_m (i) \equiv \ln(\f {\hat{p}_m (i, \ga = B)} {1- \hat{p}_m(i, \ga = B)})
\label{block_d}
\end{equation}\tshow{block_d}

\noindent where:

\begin{equation}
\hat{p}_m(i, \ga = B) = \frac {1} {m} \s {j=m(i-1)+1} {mi} \delta_{\ga_i , B}
\end{equation}

\noindent The question that we now want to ask is how the block estimates $\hat{D}_m (i)$ in  Eq. \ref{block_d} may be used to determine the error in the estimate given in Eq. \ref{free_d1}. We note that strictly the blocking procedure will only yield the error of the arithmetic average of the block quantities. That is the blocking procedure will estimate the error for the quantity $D^{(2)}$, which is estimated by:

\begin{equation}
\hat{D}^{(2)} = \f {1} {M} \s {i} {} \hat{D}_m (i)
\label{mean_block_d}
\end{equation}\tshow{mean_block_d}

\noindent where $\hat{D}^{(2)}$ also represents an estimator of $\ln \RBAcal$, though it is not an unbiased estimator. We will now show (a result which is expected) that in the large t limit, the distinction between $\hat{D}^{(1)}$ and $\hat{D}^{(2)}$ vanishes, so that we may estimate the error in Eq. \ref{free_d1} simply by using Eq. \ref{var_blocked_var} and Eq. \ref{plateau_error} in which $\hat{\m}_m(i)=\hat{D}_m(i)$ and in which $\hat{\m}$ is replaced by $\hat{D}^{(1)}$. To see this suppose that $\hat{p}_m(i, \ga = B)$ fluctuates about the true value ${p}(\ga = B)$ by an amount $\sigma_{B}(i,m)$:

\begin{equation}
\hat{p}_m(i, \ga=B) = {p}(\ga=B) + \sigma_{B}(i, m)
\end{equation}

\noindent It is clear from Eq. \ref{phatal} that for sufficiently large blocksizes , $\sigma_{B}(i,m)$ will be normally distributed with mean 0 and variance $\sigma^2_B(m)$ say, $\sigma_{B}(i,m)\sim N(0, \sigma^2_B(m))$. Then it follows that:

\begin{eqnarray}
\hat{D}^{(2)} & = & \hat{D}^{(1)} + \f {1} {M} \s {i} {} \f {\sigma_{B}(i,m)} {{p}(\ga = B)(1- {p}(\ga = B) )}\nonumber\\
& = & \hat{D}^{(1)} + \f {1} {M {{p}(\ga = B)(1- {p}(\ga = B) )}} \s {i} {} \sigma_{B}(i,m)
\label{d2_from_d1}
\end{eqnarray}\tshow{d2_from_d1}

\noindent Using the fact that:

\begin{equation}
| \s {i} {} \sigma_{B}(i,m) | \sim \sqrt {M} \sigma_B(m)
\end{equation}

\noindent and using the fact that for sufficiently large blocksizes m:

\begin{equation}
\sigma^2_B(m) \sim \frac {1} {m}
\end{equation}

\noindent we see that:

\begin{equation}
|\hat{D}^{(2)} - \hat{D}^{(1)}| \sim \f {\sqrt{M} \sigma_B (m)} {M {{p}_i(\ga = B)(1- {p}_i(\ga = B) )}} \sim \f {1} {\sqrt{t}}
\end{equation}

\noindent so that the distinction between the two estimators vanishes for sufficiently large t.

\chapter{\label{app:bennett} The overlap parameter and the Fermi function estimator}
\tshow{app:bennett}

In this appendix we bring out a relation that exists between the fermi function (FF) method and the  overlap parameter \ov\ (Eq. \ref{eq:overlap}). We start off by noting that in the case of arbitrary switching FG the overlap parameter may be generalised to:

\begin{eqnarray}
\ov & = &  \int d\wal \frac {2 P(\wal | \pi^c_\al) P(\wal | \pi^c_\alp)} {P(\wal | \pi^c_\al) + P(\wal | \pi^c_\alp)}\nonumber\\
& = & \int d\wal \frac {2P(\wal | \pi^c_\al)} {1+\frac {P(\wal | \pi^c_\al)} {P(\wal | \pi^c_\alp)}}\nonumber\\
& = & \int d\wal \frac {2P(\wal | \pi^c_\al)} {1+\Ralcal e^{\wal}}
\label{eq:overlapfg0}
\end{eqnarray}\tshow{eq:overlapfg0}

\noindent or 

\begin{equation}
\ov  = <\frac {2} {1+\Ralcal e^{\wal}}>_{\pi^c_\al}
\label{eq:overlapfg}
\end{equation}\tshow{eq:overlapfg}

\noindent so that:

\begin{equation}
 <f(\wba - W_m)>_{\pi^c_A} = \h \ov
\end{equation} 

\noindent and 

\begin{equation}
<f(-[\wba - W_m])>_{\pi^c_B} =  \h \ov
\end{equation}

\noindent It is immediately apparent (from Eq. \ref{eq:overlapfg}) that knowledge of the overlap \ov\ translates to direct knowledge of \RBAcal\ \cite{note:alternateO}. The point is that a-priori knowledge of \RBAcal\ (or $W_m$) is not at hand so as to allow an estimation of \ov\ via Eq. \ref{eq:overlapfg}. Consider the case where an equal number of independent samples are obtained in each phase, so that $n_A=n_B$. What the FF method does is to start off with an estimate of \RBAcal , say ${\hat{\cal R}}_{BA}=e^{\hat{C}}$ (where $\hat{C}$ is the estimate of Eq. \ref{eq:daconst}), and use this to obtain separate estimates of the overlap \ov\ from simulations performed in one of the two phases:

\begin{eqnarray}
\ov_A & = & 2<f(\wba - \hat{C})>_{\pia}\nonumber\\
& = & 2 <\frac {1} {1 + {\hat{\cal R}}_{BA} e^{\wba}}>_{\pi^c_A}
\label{eq:estoverlap3}
\end{eqnarray}\tshow{eq:estoverlap3}

\noindent and

\begin{eqnarray}
\ov_B & = & 2<f(-[\wba - \hat{C}])>_{\pib}\nonumber\\
& = & 2 <\frac {1} {1+ \frac {1} {{\hat{\cal R}}_{BA}} e^{-\wba}}>_{\pi^c_B}
\label{eq:estoverlap4}
\end{eqnarray}\tshow{eq:estoverlap4}

\noindent Only if ${\hat{\cal R}}_{BA}$ is an unbiased estimator for  $\RBAcal$  will the two estimates of $\ov_A$ and $\ov_B$ converge to the same value. Therefore what Bennett's recursive prescription (Eq. \ref{eq:recurse1} and Eq. \ref{eq:recurse2}) does is to vary ones estimate of ${\hat{\cal R}}_{BA}$ (through $\hat{C}$) until the estimates of the overlap $\ov_A$ and $\ov_B$  have converged to the same value. At this point one can be sure that the estimate  ${\hat{\cal R}}_{BA}$ reflects the true value of \RBAcal\ since:

\begin{eqnarray}
\RBAcal = e^{-\hat{C}}\frac {<f(\wba - \hat{C})>_{\pia}} {<f(-[\wba - \hat{C}])>_{\pib}} & \est & e^{-\hat{C}} \frac {\hat{\ov}_A} {\hat{\ov}_B}\nonumber\\
& \approx & e^{-\hat{C}}\nonumber\\
& = &  {\hat{\cal R}}_{BA}
\end{eqnarray}

\chapter[Multihamiltonian method from the Fast Growth method]{\label{app:equivalent} Multihamiltonian method as a limiting form of the Fast Growth method}
\tshow{app:equivalent}
 
The MH method can be viewed as a limiting form  of the FG method. The key insight is the observation that as the equilibration time \delt\ increases:

\begin{equation}
\mbox{$\delt \rightarrow \infty$\ \ \ } P(\mba | t_i) \rightarrow P(\mba | \pi_{\la_i}) 
\end{equation}

\noindent where we recall that $P(\mba | t_i)$ denotes the probability distribution of \mba\ at time $t_i$, when the \ce\ has been incremented from $\ecal_{\la_{i-1}}$ to $\ecal_{\la_i}$ and {\em after} the system has been equilibrated with $\pi^c_{\la_i}$ for a time \delt . This stems from the fact that if one perturbs the \ce\ from $\ecal_{\la_{i-1}}(\vvec({i-1}))$ to $\ecal_{\la_i}(\vvec({i-1}))$, and then equilibrates the system for an infinite amount of time, to a configuration $\vvec(i)$, then the ensemble of configurations \seta {\vvec(i)}\ will be Boltzmann distributed with distribution $\pi^c_{\la_i}$. In other words one finds that in the case of adiabatic equilibration ${\cal{P}}^c_{A\rightarrow B}[\vvec(1), \vvec(2), ... \vvec({n-1})]$ assumes the simple form:

\begin{equation}
{\cal{P}}^c_{A\rightarrow B}[\vvec(1), \vvec(2), ... \vvec({n-1})]\propto \p {i=1} {n-1} e^{-\beta \ecal_{\la_i}({\vvec(i)})}
\label{eq:mhlspdf}
\end{equation}\tshow{eq:mhlspdf}

\begin{figure}[tbp]
\begin{center}
\ifpdf
\rotatebox{90}{
\includegraphics[scale=0.4]{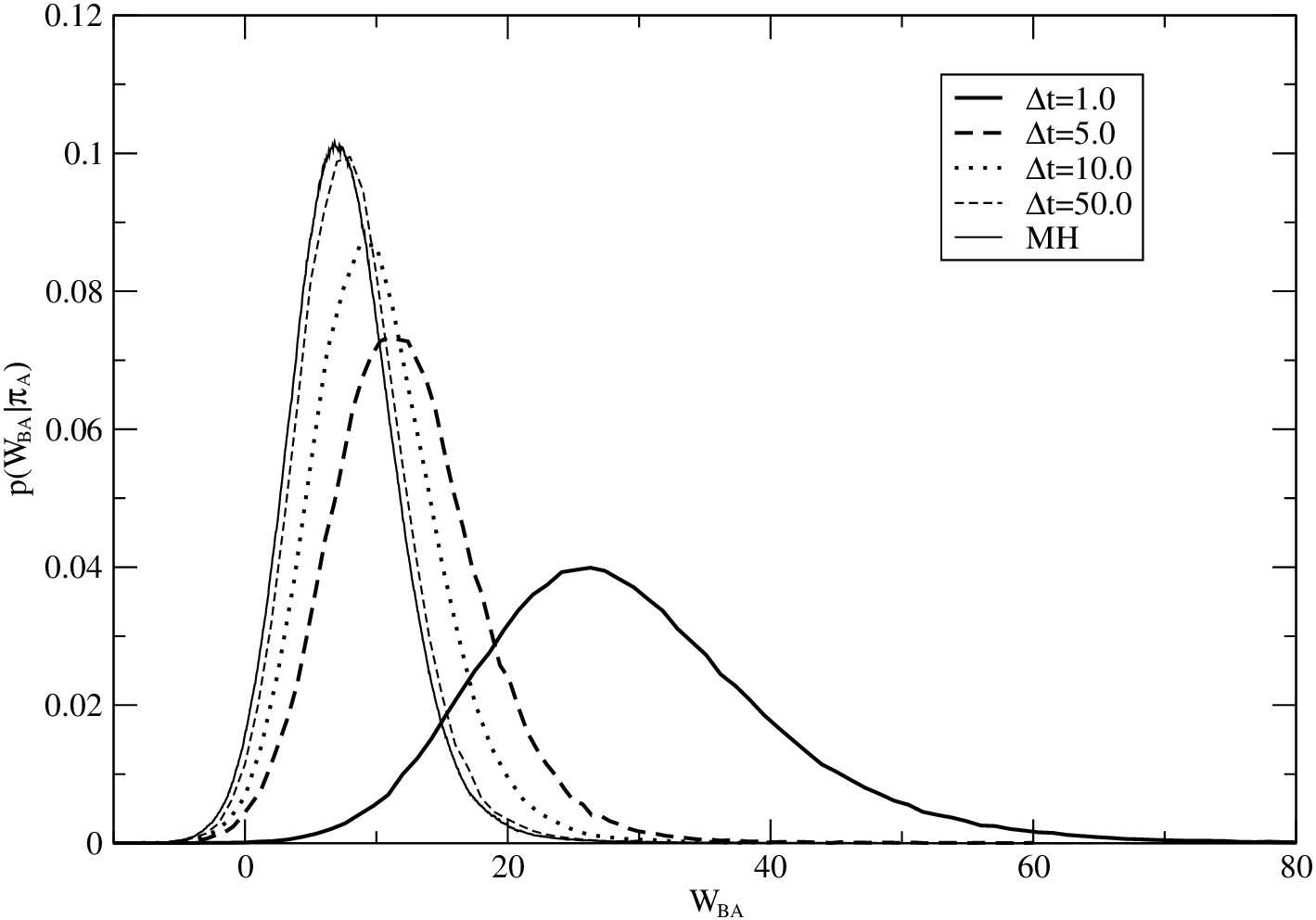}
}
\else
\includegraphics[scale=0.4]{fig/compare_mhbz_jarzynski}
\fi
\end{center}
\caption{$P(\wba| \pia)$ for the MH and the FG methods}
For the MH method \wba = \mba . The figure compares the distributions obtained via the FG method, as the equilibration time \delt\ is increased, with that of the MH method. We see that as increasing equilibration ($\delt$) is allowed between successive work increments $\delta \wbai$ (for the FG method, Eq. \ref{eq:totalwork}), the peaks and widths of the probability distribution reduce, tending asymptotically to the form assumed by the MH method. The number of replicas (n) was 10.

$T^*=1.0$, RSM.

\tshow{pic:comparejar}
\label{pic:comparejar}

\begin{center}
{\bf{------------------------------------------}}
\end{center}

\end{figure}

\noindent This is exactly the sampling distribution of the MH method (see Eq. \ref{eq:gen_hamil}). Therefore the MH method can be viewed (for a given \delela ) as a limiting case of the FG method in which the equilibration time is infinite (i.e. adiabatic equilibration). Figure \ref{pic:comparejar} shows how the distribution $P(\wal | \pib)$ (of the FG method) tends to the limiting form of the distribution of the MH method as the equilibration time (\delt) is increased, and clearly illustrates further the connection between the MH and FG methods that we have just described.

In regards to systematic errors, it follows that if one uses the EP estimator, then the systematic errors associated with the MH method will be less than or equal to those of the FG method since adiabatic equilibration translates to minimum systematic errors. However in the case of the PS method this statement no longer holds if both methods have sufficient overlap so as to ensure that the phase switches can take place. In this case both methods have zero systematic errors since they {\em both} visit all the important regions of \ecs\ which contribute to the estimate of \RBAcal .

Apart from the issue of systematic errors, another difference of the two methods is the way in which they are realised. In the FG method one performs work on a single system as described in section \ref{sec:fg}. In the MH method, one makes use of the form of Eq. \ref{eq:mhlspdf}, which allows one to realise the \wba\ distribution by performing independent simulations in {\em parallel}. This is a significant difference in that it allows considerable speedup of the task of evaluating the FED since one may parallelise the process. This will become especially apparent in chapter \ref{chap:quantum} when we apply the method to the study of quantum FEDs.

\chapter[Quantum Simulations: The details]{\label{app:deboer} Details of the quantum simulations for the Lennard-Jones potential}
\tshow{app:deboer}

In this section we clarify the way the different parameters that enter into the calculations of the hamiltonian of the polymeric systems for the PA (Eq. \ref{eq:prim_pot}) and the HOA (Eq. \ref{eq:HOVeff}). For simplicity we will work in the \rvec\ representation. We recap that for the case of distinguishable quantum particles the PIMC simulation involves the simulation of a system with a partition function given by:

\begin{equation}
Z = \int d\rvec_1 .... d\rvec_n (\frac {Mm} {2\pi \beta \hbar^2})^{\frac {3N} {2}} \exp \{-\beta \hcal {} {\seta {\rvec}}\}
\end{equation}

\noindent where:

\begin{equation}
\beta \hcal {} {\seta {\rvec}} = \s {i=1} {M} \frac {Mm} {2\beta\hbar^2} [\rvec_{i+1}-\rvec_i]^2 + \frac {\beta} {M} \s {i=1} {M} \ecal(\rvec)
\label{app_hamil}
\end{equation}\tshow{app_hamil}

\noindent In the case of the Lennard-Jones \ce\ the distances are measured as units of $\sigma$. Suppose that the superscript $\tilde {}$ over a variable denotes the fact that it is expressed in units of $\sigma$, so that $\tilde{\rvec} = \rvec / \sigma$. Then we may conveniently express all quantities in terms of these scaled variables. Suppose that $x^{(k)}$ denotes the k-th coordinate of a particle, so that $(x^{(1)}, x^{(2)}, x^{(3)}) = (x,y,z)$. If:

\begin{equation}
s_{ij} = \sqrt {(x_i-x_j)^2 + (y_i-y_j)^2 + (z_i-z_j)^2}
\end{equation}

\noindent and if $\ecal (s_{ij})$ denotes the contribution to the overall \ce\ of the interaction between particle i and j, then it follows that:

\begin{eqnarray}
\ecal (s_{ij}) & = & 4 \epsilon [(\frac {\sigma} {s_{ij}})^{12} - (\frac {\sigma} {s_{ij}})^{6} ]\nonumber\\
& = & 4\epsilon [{\tilde{s}_{ij}}^{-12} - {\tilde{s}_{ij}}^{-6} ]\nonumber\\
& = & \epsilon G(\tilde{s}_{ij})
\end{eqnarray}

\noindent where $G(\tilde{s}_{ij})$ is a dimensionless number. If we denote by ${\beta}^*$ the quantity ${\beta}^* = \frac {\epsilon} {kT}$, then it follows that in the case of the PA, Eq. \ref{app_hamil} may be written as:

\begin{equation}
\beta \hcal {} {\seta{\rvec}} = \s {i=1} {M} \frac {M} {2 {{\beta}^*} \deb} [\tilde{\rvec}_{i+1}-\tilde{\rvec}_i]^2  + \frac {\beta ^*} {M} \s {i=1} {M} G[\tilde{\rvec}_i]
\end{equation}

\noindent where:

\begin{equation}
G[\tilde{\rvec}_i] = \h \s {kl} {} G(\tilde{s}^{(i)}_{kl})
\end{equation}

\noindent and where $\tilde{s}^{(i)}_{kl}$ denotes the distance between particles k and l in replica i.

Furthermore:

\begin{eqnarray}
[\frac {\partial \ecal(s_{ij})} {\partial x_i^{(k)}}]^2 & = & [4 \epsilon [6 (\frac {\sigma} {s_{ij}})^6 - 12 (\frac {\sigma} {s_{ij}})^{12}] \frac {x_i^{(k)} -x_j^{(k)}} {s_{ij}^2}]^2\nonumber\\
& = & (\frac {\epsilon} {\sigma})^2 H(\tilde{s}_{ij})
\end{eqnarray}

\noindent where $H(\tilde{r}_{ij})$ is also a dimensionless function. Therefore if the HOA is used then it follows that:

\begin{equation}
\beta \hcal {} {\seta{\rvec}} = \s {i=1} {M} \frac {M} {2 {{\beta}^*} \deb} [\tilde{\rvec}_{i+1}-\tilde{\rvec}_i]^2  + \frac {\beta ^*} {M} \s {i=1} {M} G[\tilde{\rvec}_i] + \frac {{{\beta}^*}^3 \deb} {24 M^3} \s {i=1} {M} [H[\tilde{\rvec}_i])]
\end{equation}

\noindent where $H[\tilde{\rvec}_i]$ is given by:

\begin{equation}
H[\tilde{\rvec}_i] = \s {kl} {} H(\tilde{s}^{(i)}_kl)
\end{equation}

\noindent In using these quantities in the simulation, one must appropriately modify these equations so as to take into account the fact that particles only interact with only the first nearest neighbour shell.

\chapter{\label{app:dominate} Interplay between kinetic and configurational actions}
 
\tshow{app:dominate}

It is well known in the Path Integral Monte Carlo literature \cite{p:ceperleyreview} that on the transition to a large number of replicas, the kinetic action $S_K$ dominates over the configurational action $S_V$. Note however this does {\em not} mean that the configurational action may be neglected on the transition to large number of replicas, since $S_V$ essentially determines {\em where} in configuration space the polymer resides in, where as $S_K$ controls  the magnitude of fluctuations between adjacent replicas within this region of configuration space.

\begin{figure}[tbp]
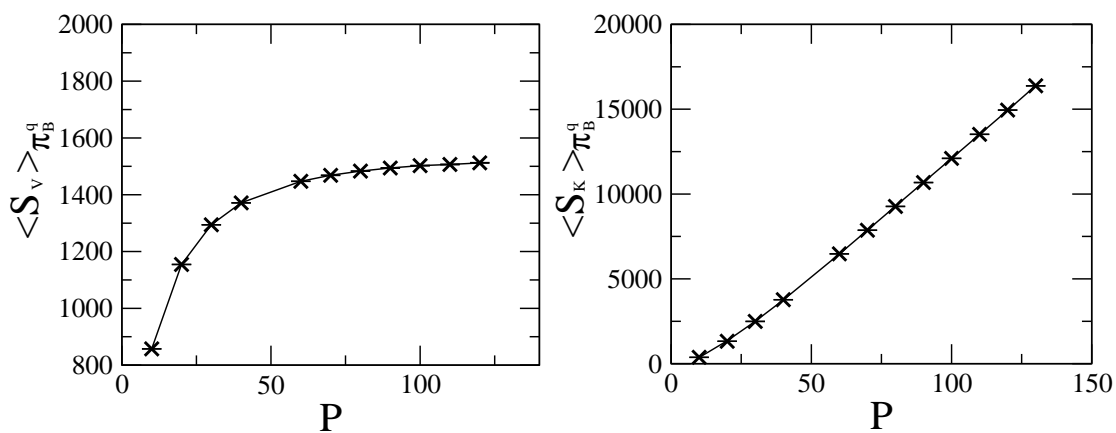

\begin{center}
\ifpdf
\rotatebox{90}{
\includegraphics[scale=0.6]{fig/intra_inter_vs_replica}
}
\else
\rotatebox{0}{
\includegraphics[scale=0.6]{fig/intra_inter_vs_replica}
}
\fi
\end{center}
\caption{Variation of $<S_V>_{\qpib {PA}}$ and $<S_T>_{\qpib {PA}}$ for a (PA) simulation in which the temperature is fixed and the number of replicas is varied}
{
The temperature was fixed at $T^*=0.4$ and the number of replicas employed was varied between the values of P=10 and P=130.

$\tilde {D}=0.1816$, $\rho\sigma^3=1.092$, Q-RSM.
}

\begin{center}
{\bf{------------------------------------------}}
\end{center}

\tshow{pic:inter_intra_vs_replica}
\label{pic:inter_intra_vs_replica}
\end{figure}

In this appendix we provide a simple numerical illustration of this for the LJ systems employed in chapter \ref{chap:quantum}. 
Figure \ref{pic:inter_intra_vs_replica} shows the dependence of $<S_V>_{\qpib {PA}}$ and $<S_K>_{\qpib {PA}}$ on the number of replicas P for a simulation at a fixed temperature. For small number of replicas $<S_K>_{\qpib {PA}}$ starts off assuming a lower value than $<S_V>_{\qpib {PA}}$. As the number of replicas increase both  $<S_K>_{\qpib {PA}}$ and $<S_V>_{\qpib {PA}}$ increase, until eventually $S_K$ comes to dominate over $S_V$. Within this regime $S_K$ scales linearly with P.  $<S_V>_{\qpib {PA}}$ on the other hand, had a positive gradient which decreases as P increases, but never quite reaches zero. As a result  $<S_V>_{\qpib {PA}}$ appears to plateau off, though the plateau is only reached in the $P\rightarrow \infty$ limit. The figure clearly illustrates the dominance of $S_K$ over $S_V$ in the large P limit.

% ======================== The bibliography ===============================

\donarrowly
\bibliographystyle{unsrt}
\addcontentsline{toc}{chapter}{Bibliography}
\bibliography{main}

\begin{thebibliography}{100}
\bibitem{note:partialmean}
We remind the reader that by partial overlap we mean there are \wba\
  macrostates over which both the estimators $\hat{P}(\wba | \pi^c_A)$ and
  $\hat{P}(\wba | \pi^c_B)$ are non-zero, and at the same time when there are
  other macrostates for which only one of the estimators is non-zero. See
  figure \ref{pic:overlapping}.
\bibitem{note:thermominfree}
An exception to this is on the coexistence line, where the two phases have
  equal free energies, and therefore are equally probable. Therefore even in
  the thermodynamic limit both phases can exist in coexistence.
\bibitem{note:wayscale}
In order to extrapolate correctly to the thermodynamic limit one must know how
  the finite size corrections scale with the system size. That is suppose that
  $\triangle f_N$ denotes the free energy density (given by $\triangle
  F_{BA}/N$) for a system size N, and suppose that $\triangle f$ denotes the
  corresponding quantity in the thermodynamic limit. It follows that: \[
  \triangle f_N = \triangle f + \epsilon_N \] where $\epsilon_N\sim 1/N^\alpha$
  is the finite size correction, and where $\alpha$ is some scaling exponent.
  If one knows the scaling exponent, then one may then plot a graph of
  $\triangle f_N$ versus $1/N^\alpha$ in order to obtain a straightline graph
  which yields an estimate of $\triangle f$ from the value of the intercept of
  the line with the axis.
\bibitem{note:phasereview}
For general literature on the application of computational methods to
  statistical mechanics, we refer the reader to \cite{b:frenkel},
  \cite{b:tildesley}, \cite{p:binderreview}. For an excellent introduction we
  refer the reader to \cite{p:kpnm}. For general reviews on the determination
  of phase boundaries via computer simulation, we refer the reader to
  \cite{p:brucereview}, \cite{p:acklandreview}, \cite{p:valleaureview},
  \cite{b:haile}, \cite{p:monsonreview}, \cite{p:pabloreview},
  \cite{p:allenreview}, \cite{p:shingreview}, \cite{p:panagiotopoulosreview}.
\bibitem{note:relativefreq}
Suppose that $N(\rvec)$ and $N(\rvec ')$ denote the frequencies with which the
  configurations $\rvec$ and $\rvec '$ appear in a chain of configurations
  produced by the metropolis algorithm. In the infinite limit ($t\rightarrow
  \infty$) the Metropolis method guarantees (by using Eq.
  \ref{eq:detailedbalance}) that: \[ \frac {N(\rvec)} {N(\rvec ')} \rightarrow
  \frac {\pi(\rvec)} {\pi(\rvec ')} \].
\bibitem{note:sampling}
Generally one will only have knowledge of the sampling distribution $\pi$ up to
  a normalisation constant. However this does not matter since it is only the
  ratio $\frac {\pi(\rvec ')} {\pi (\rvec)}$ which is needed in the acceptance
  probability of Eq. \ref{eq:acceptancerate}. That is the method neither
  presupposes nor provides the 'absolute probabilities' of the configurations.
\bibitem{note:mildet}
This volume has to be the same for all the particles, and should not change
  with time. If this is not the case then detailed balance will be violated
  (see for example \cite{p:mildet}).
\bibitem{note:denominator}
Strictly, even if we did know the normalisation constant a-priori, the
  employment of the metropolis method (Eq. \ref{eq:acceptancerate}) means that
  such normalisation procedures will still need to be performed in the estimate
  of such expectations (see Eq. \ref{eq:estimateexpectation}), since the method
  (Eq. \ref{eq:acceptancerate}) does not generate configurations according to
  their absolute probabilities.
\bibitem{note:otherphases}
We refer the interested reader to \cite{p:kofkecummings},
  \cite{p:panagiotopoulosreview}, \cite{p:kofkegibbs}, \cite{p:meijer1},
  \cite{p:meijer2}, \cite{p:koning}, \cite{p:chang}, \cite{p:muguruma} for
  alternative methods (which are not limited to NVT systems) for probing
  liquid-liquid, liquid-solid, liquid-vapour boundaries. For reviews we refer
  the reader to \cite{p:brucereview}, \cite{p:mezeireview},
  \cite{p:valleaureview}, \cite{p:monsonreview}, \cite{p:pabloreview},
  \cite{p:allenreview}, \cite{p:shingreview}.
\bibitem{note:waapp}
By appropriate we mean a reference system which ensures that the system does
  not undergo a phase transition on the transition to or from the reference
  state.
\bibitem{note:referencestate}
An alternative way to estimate the FEDs between the desired phase and the
  reference system is to appeal to Eq. \ref{eq:partrsestimator0}. The
  corresponding estimator then simply involves evaluating the ratio of the
  times spent in the two phases. The benefit of this latter estimator is that
  there are no systematic errors in the final estimate. The problem in practice
  is that the simulation will not naturally traverse between the desired phase
  and the reference system, and will instead have to be forced to do so. This
  may be done by appeal to an extended sampling strategy, which we describe in
  greater detail in the next section.
\bibitem{note:newsys}
Note that unlike the systematic errors alluded to earlier, these systematic
  errors are not due to the incomplete sampling of \cs , but instead arise from
  the discretisation of the integral in Eq. \ref{eq:thermody}. As a
  consequence, even if the simulation is (hypothetically) left running for an
  infinite duration of time, the systematic errors that arise, due to the
  approximation made in Eq. \ref{eq:thermodyest}, do not vanish.
\bibitem{note:absolutevsratio}
The evaluation of the FED between the two phases, as opposed to their absolute
  free energies, is preferable for a number of reasons. The first, and most
  obvious, is the fact that two separate calculations will entail more
  computational expenditure that will a single simulation. The second is that
  for the reference state technique the evaluation of the free energy
  difference, which is typically a small quantity whose magnitude is unknown
  a-priori, will be obtained by subtracting two large quantities (corresponding
  to the individual free energies). As a consequence one will need to evaluate
  the individual free energies to much greater accuracy than one would
  initially expect in order to ensure that the error of the FED is smaller than
  the actual value of the FED. A much tidier approach would be to concentrate
  on this FED in the first place.
\bibitem{note:referenceprobs}
Note that the reference state technique will also face similar problems when
  employing Eq. \ref{eq:estrba2} instead of Eq. \ref{eq:thermodyest} to
  estimate the desired FEDs. However like with the continuous path technique,
  this problem may be overcome by appeal to the appropriate extended sampling
  strategies.
\bibitem{note:magnetic}
In the case of magnetic phase transitions this problem does not exist since
  there is no structural rearrangement of the particles in the process of going
  from one phase to the other. For such systems the continuous path technique
  works well \cite{p:bergillustration}.
\bibitem{note:referenceprob2}
Unlike the continuous path technique, the reference state technique will not
  encounter such problems since the reference states are chosen so as to ensure
  that the simulation will not encounter such mixed-phase configurations.
\bibitem{note:singlepdf}
Note that since a simulation initiated in a given phase will remain in that
  phase (due to the fact that a phase is essentially a local basin of
  attraction in the configurational energy $E(\rvec)$), $\pi^c_\al$ is simply
  realised by initiating a simulation in phase \al\ and then sampling according
  to the Boltzmann distribution given in Eq. \ref{eq:boltzsamp}.
\bibitem{note:freecomplex}
The task of estimating the FEDs is not limited to the location of phase
  boundaries. For example, one may have a complex energy landscapes (as is the
  case when one deals with proteins and glassy materials) in which there exist
  several basins of attraction in the configurational energy (which in the case
  of proteins corresponds to different conformations). In this case it is one's
  desire to find the most stable of the multitude of phases that exist. In such
  a scenario it might be more desirable to evaluate the absolute value of the
  partition function associated with the different phases (via the reference
  states techniques, for example).
\bibitem{note:chooseD}
The way in which configurations of one phase are mapped onto configurations of
  the other phase does have a consequence for the efficiency of the FED
  estimators (Eq. \ref{eq:abswitchest}). We will have more to say about this in
  section \ref{sec:overlap}.
\bibitem{note:throw}
In figure \ref{pic:overlapping} (chapter \ref{chap:estsamp}) we illustrate this
  idea more generally in the context of the fast growth method, which we review
  in section \ref{sec:fg}. The figure is relevant to the current discussion if
  we take $\wba = \mba$ and $W_m = M_m$.
\bibitem{note:bothimpo}
Note that the representation \vvec\ is implicitly dependent on the reference
  configuration $\Rvec$.
\bibitem{note:choosingperfect}
In the rather artificial case of \al\ and \alp\ representing the same phases,
  it is clearly possible to construct a PM which results in a distribution as
  shown in figure \ref{bimodaldistribution}. For example suppose that the
  particles are labelled in the same way in the two identical systems, suppose
  that ${\mf T}_\al$ in Eq. \ref{eq:vtoutran} is a permutation matrix which is
  different in the two systems to the extent that there are no two common
  permutations in ${\mf T}_\al$ and ${\mf T}_\alp$, and suppose that $\Dvec =
  0$. In this case working in the \vvec\ representation essentially corresponds
  to a transformation in which the displacement of a particle of system \al\ is
  mapped onto the displacement of a different particle in system \alp\ (which
  is identical to system \al ) . Such a transformation will generally result in
  poor overlap between the two phases in \vvec\ space, which manifests itself
  as poor overlap between $P(\mba | \pi^c_A)$ and $P(\mba | \pi^c_B)$, even
  though the two systems explore the same region of $\rvec$ space. This is a
  clear illustration of the importance of choosing the correct representation.
  However the important point is that in this case it is always possible to
  construct a PS (e.g. for this particular case $\Dvec = 0$ and $S_{BA} = 1$)
  for which the overlap problem vanishes (since \mba =0 for all
  configurations). In the more general case of \alp\ and \al\ being different
  phases, it will not be possible to find a PS which overcomes this problem
  through the choice of an appropriate \Dvec\ and $S_{BA}$.
\bibitem{note:propto}
Since by virtue of the overlap identity $H(\mbai | \pia) e^{-\mbai} \propto
  P(\mbai|\pib)$.
\bibitem{note:conta}
In the sense that if one was to perform two separate simulations yielding the
  estimators $\hat{P}(\mba | \pi^c_A)$ and $\hat{P}(\mba | \pi^c_B)$, then
  there would be bins \mbai\ for which $\hat{P}(\mbai | \pi^c_A)$ is zero and
  for which $\hat{P}(\mbai | \pi^c_B)$ is non-zero.
\bibitem{note:mulweightsfree}
This follows from the fact that knowledge of the MUCA weight function
  $\eta_A(\mba)$ throughout the region spanning all of $P(\mba | \pi^c_A)$ and
  $P(\mba | \pi^c_B)$, and the regions in between them, automatically means
  that one has knowledge of the FED. To see this we note that if one knows the
  weight function over this range, so as to ensure that $P(\mba | \pi^m_A)$
  (see Eq. \ref{eq:introweights}) is flat, then one has knowledge of the value
  of $P(\mba | \pi^c_A)$ for bins \mbai\ over which $P(\mba | \pi^c_B)$ is
  non-negligible. From the overlap identity (Eq. \ref{eq:overlapidentity}) we
  see that this immediately corresponds to knowledge of the FED of the two
  phases.
\bibitem{note:mucaother}
Progress in this direction was made earlier by \cite{p:paine} and in particular
  by \cite{p:mezei}. This method was also rediscovered by \cite{p:lee}, under
  the name of "Entropic Sampling".
\bibitem{note:mucaillu}
Nice illustrations of the MUCA algorithm in action can be found in
  \cite{p:bergPRL} and \cite{p:bergillustration}.
\bibitem{note:transi}
See also \cite{p:tranmat}, \cite{p:huller}, \cite{p:huller2},
  \cite{p:fitzgerald1}, \cite{p:fitzgerald2}, \cite{p:fenwick},
  \cite{b:grahamthesis}, \cite{p:broadhistogram}, and \cite{p:kastner} for a
  more efficient method which calculates the weights based on the transition
  probabilities between macrostates.
\bibitem{note:genwl}
The transition probability based method of calculating the multicanonical
  weights (\cite{p:tranmat}, \cite{p:huller}, \cite{p:huller2},
  \cite{p:fitzgerald1}, \cite{p:fitzgerald2}, \cite{p:fenwick},
  \cite{b:grahamthesis}, \cite{p:broadhistogram}, and \cite{p:kastner}) may be
  generalised so as to work in a manner similar to the Wang-Landau algorithm.
  For example see \cite{p:shell2}. A further alternative method is given in
  \cite{p:ferkinghoff} in which several independent simulations are used to
  extract the weights via an expression reminiscent of the histogram equations
  of \cite{p:reweighting}, \cite{p:reweighting2}. More recently a completely
  different methodology which is suitable for calculating these weights had
  been presented in \cite{p:yan2}.
\bibitem{note:wlflat}
The criterion for defining flatness is highly flexible \cite{p:wanglandauprl}.
  For example, one could define flatness as the case where the differences
  between the counts of the highest bin to the mean count and the differences
  of the counts of the lowest bin to the mean count are both within 10 percent
  of the mean value. In our simulations this was the criterion that was adopted
  in deciding whether a histogram was flat or not.
\bibitem{note:launcher}
We make a point to note that in the calculation of chemical potentials of grand
  canonical systems, one estimates the FED between systems whose system sizes
  differ by a single particle. In this case careful choice of the parent phase
  is of crucial importance to the simulations. In \cite{p:kofkecummings},
  \cite{p:lukofke}, \cite{p:lusinghkofke}, \cite{p:lukofke2}, \cite{p:lukofke3}
  it is argued that in the case of insertion simulations the parent phase
  should always be the system with the fewer number of particles. This method
  will have less systematic error associated with it than doing the simulation
  in the opposite direction since the region of \cs\ associated with the system
  with more particles is essentially contained within that of the system
  composed of fewer particles. In their terminology, the FED perturbation
  calculation should always be performed in the insertion direction.
\bibitem{note:heypath}
It is in this way that a path is constructed from one phase to the other.
\bibitem{note:wham}
Eq. \ref{eq:whamcentral} is only valid when the data is uncorrelated. When the
  data is correlated, then Eq. \ref{eq:whamcentral} must be replaced by: \[
  {\tilde {Z}}_{\la_j} \est \s {k=1} {b} \{ \frac {\s {m=1} {n} g^{-1}_m
  H(\mbak | \pi_{\la_m}) e^{-\la_j \mbak}} {\s {i=1} {n} \frac {N_i g^{-1}_i}
  {{\tilde {Z}}_{\la_i}} e^{-\la_i \mbak}}\} \] where $g_i = 1 + 2 \tau_i$ and
  $\tau_i$ is the autocorrelation associated with the sampling distribution
  $\pi_{\la_i}$. In the particular case where the autocorrelation functions are
  approximately equal or in the limit of the samples being independent, the
  equation above reduces to Eq. \ref{eq:whamprob}. We stress that the set \seta
  {{\tilde {Z}}_{\la_1}, {\tilde {Z}}_{\la_2}, ..., {\tilde {Z}}_{\la_n}} is
  only estimated up to a common multiplicative constant.
\bibitem{note:canpsprob}
Given a set of data obtained via $\pi^m_{PS}$ one may obtain the canonical
  distribution simply by reweighting the data. Suppose that $H(\mbai |
  \pi^{m}_{PS})$ denotes the count for bin $\mbai$ obtained during the course
  of a sampling experiment performed with $\pi^m_{PS}$. On may then estimate
  $P(\mbai | \pips)$ via: \[ \hat{P}(\mbai | \pips) = \frac {H(\mbai |
  \pi^m_{PS}) e^{\eta_{PS}(\mbai)}} {\s {j=1} {b} H(M_{BA,j} | \pi^m_{PS})
  e^{\eta_{PS}(M_{BA,j})}} \].
\bibitem{note:whywrite}
We write it as $P(\wba | \pi^c_B)$ instead of $P(\wab | \pi^c_B)$ in order to
  bring out the similarity with the zero-equilibration cases.
\bibitem{note:whaov}
In the context of FG, when we say 'phase' we strictly mean process. Therefore
  when we say phase \al\ we are generally referring to the $\al \rightarrow
  \alp$ process.
\bibitem{note:truncationpic}
An illustration of the first nearest neighbour shell in the two structures is
  shown on pg. 58 of \cite{b:andythesis}.
\bibitem{note:modemapping}
A short investigation into the possibility of implementing permutations in the
  mapping of the effective configuration \vvec\ were made in the case of the
  FSM. However no conclusive findings were made. Looking forward, see also
  \cite{note:permutation}.
\bibitem{note:jarwork}
The issue of choosing an optimal representation has not been studied before
  except in a slightly different context in \cite{p:targetted}.
\bibitem{note:cruderep}
We have already seen (in the most simple way) in note
  \cite{note:choosingperfect} how the choice of representation affects the
  overlap.
\bibitem{note:whytaylor}
In the case of crystalline solids the truncation of the Taylor expansion (in
  powers of the displacements) of the \ce\ so as to include only the first few
  terms corresponds to the physical interpretation that one is dealing with
  'sufficiently' low temperatures, since in the limit $T\rightarrow 0$ the
  displacements $\uvec \rightarrow 0$.
\bibitem{note:permutation}
It should be noted that strictly one does not require $\avvec_A$ and $\avvec_B$
  to be identical for the harmonic excitation energies (Eq.
  \ref{eq:energy_har}) to be the same. The only requirement is that their norms
  (given by Eq. \ref{eq:einv}) are the same. More generally one may have: \[
  \avvec_B = \avec {G}_{BA} \avvec_A \mbox{ where } \avec {G}^T_{BA} . \avec
  {G}_{BA} = \mf {1} \] The additional license for improvement via this
  strategy was investigated by simulation for 3 permutations. However no
  discernible difference was seen between the choices of permutation. This
  issue was also investigated analytically by assuming that the configurations
  of a (harmonic) parent phase were mapped onto configurations of the conjugate
  phase which were anharmonic in nature. Perturbation theory shows that the
  expectation of $M_{BA}$ is unaffected by the choice of permutation under such
  conditions. The fact that the analytical analysis corresponded to a highly
  specialised scenario and the fact that only three different permutations were
  tested (via simulation) meant that our investigations on this issue were
  inconclusive, and we leave open this issue for further investigation.
\bibitem{note:meanvarspread}
For example consider figure \ref{bimodaldistribution}. For a fixed variance the
  smaller the mean, then the greater is the overlap between the two
  distributions. Likewise for fixed means, the greater the variance the greater
  the overlap is between the two phase constrained distributions.
\bibitem{note:oppositecon}
Note that in this section we illustrate the relative quantities in terms of the
  macrovariable $M_{AB}$ ($=-M_{BA}$).
\bibitem{note:prioranhar}
Prior investigations into anharmonic contributions to the free energy in the
  case of Lennard-Jones crystals can be found in \cite{p:pollock}.
\bibitem{note:implicitassumption}
We implicitly assume that as the distance of the peaks from the origin
  decrease, then any possible decrease in the spreads (given by the standard
  deviation) of the distributions are such that they decay at a slower rate
  than the corresponding separation of the two peaks. For the FSM anharmonic
  perturbation theory supports this, since it shows that at sufficiently low
  temperature the means $\omega_1$ scale as T whereas the spreads ($\propto
  \sqrt \omega_2$) scale as $T^{0.5}$. Therefore a decrease in temperature is
  accompanied by a decrease in the means of the distributions in a way which is
  faster than the associated decrease in the spreads of the two distributions.
  This ensures that the overlap between the two distributions will {\em
  increase} as the associated peak-to-peak separation decreases.
\bibitem{note:checklowtemplimit}
We verified that the distribution of the RSM in figure
  \ref{fig:comparisonofFSLSandRSLS} (a) and (b) was indeed the stationary limit
  by comparing it to the distribution obtained at a temperature several orders
  of magnitude smaller. It was also verified that the means and the variances
  of this distribution agreed with the stationary limits as predicted by Eq.
  \ref{eq:rsscumu1} and \ref{eq:rsscumu2}.
\bibitem{note:deltacentre}
The delta function has to be centred on the origin since, by construction,
  ${\cal R}^a_{BA} \rightarrow 1$ as $T\rightarrow 0$ for the FSM.
\bibitem{note:whyover}
The reason for this is that, as was mentioned in section
  \ref{sec:perturbation}, regions which contribute most significantly to the
  numerator of Eq. \ref{eq:torriereweightest} (or strictly the corresponding
  estimator for a simulation performed in phase B) will not be visited. As a
  result $<e^{-\mab}>_{\pi^c_B}$ will be underestimated, and as a consequence
  its use to estimate $\triangle F^a_{AB}$ will result in an overestimate. This
  is clearly what is observed in figure \ref{fig:anharfree}.
\bibitem{note:mulps}
The multicanonical sampling distribution was constructed so as to contain both
  $P(\mab | \pi^c_A)$ and $P(\mab | \pi^c_B)$. See section \ref{sec:umbrella}
  and section \ref{sec:phaseswitch}.
\bibitem{note:howmuca}
We recap that in the case of the PS method MUCA sampling distribution must
  ensure that the $\mab \sim 0$ regions are accessible to the simulation
  irrespective of which phase it is in. This means that one must explicitly
  construct MUCA weights so as to bias the regions of \mab\ space {\em in
  between} the two peaks (see figure \ref{bimodaldistribution}). Therefore the
  closer the peaks, the narrower the region of \mab\ space that will need to be
  reweighted. We also note that as soon as the $\mab\sim 0$ regions are
  accessible to the sampling distributions \pia\ and \pib , then extended
  sampling will not be necessary for the PS method. However on the onset of
  this happening, extended sampling will still be needed for the EP estimator
  since for this method the sampling distribution only naturally samples one
  phase, and one must therefore {\em force} the simulation into the regions
  associated with the conjugate phase.
\bibitem{note:limworkoverlap}
Or more generally between $\hat{P}(\wba | \pia)$ and $\hat{P}(\wba | \pib)$ in
  the case of arbitrary equilibration FG simulations, of which the zero
  equilibration simulations are a particular case (see section \ref{sec:fg}).
\bibitem{note:kofex}
In actual fact the N particle system is a system of N+1 particles in which one
  particle does not interact with the others.
\bibitem{note:statest}
The statistical errors in a finite time Monte Carlo simulation will depend on
  the choice of the estimator, since different estimators focus their emphasis
  on difference regions of \wba\ space. Under the phase constrained
  distributions $P(\wba | \pia)$ and $P(\wba | \pib)$ these regions are visited
  with different probabilities, and hence the statistics of the 'most
  important' regions will vary for the different estimators.
\bibitem{note:lala}
By noting that the acceptance criterion for a Monte Carlo move can be
  alternatively given by Eq. \ref{eq:fermiacceptance}, we see that the fermi
  function (FF) formula given in Eq. \ref{eq:fermiform} can, like the AR
  formula (see Eq. \ref{eq:ar}), be viewed as an estimator which involves
  sitting in the two phases and measuring the acceptance probabilities of
  phase-switch Monte Carlo moves, without actually performing.
\bibitem{note:mbacon}
We note that in section \ref{sec:fsssimulation} the discussion was presented in
  terms of the variable \mab\ (= - \mba ). However in order to preserve
  continuity of the notation in this chapter, we continue to discuss the
  results in terms of \mba .
\bibitem{note:wheremuca}
whose application to the problem of estimating FEDs in the context of
  structural phase transitions has been studied in \cite{p:LSMC},
  \cite{p:LSMCprehard}, \cite{p:LSMCpresoft}, \cite{p:solidliquid},
  \cite{p:LSMCerrington}, and in which the founding ideas were layed out in
  \cite{p:valleaucard}, \cite{p:torrievalleau}, \cite{p:tvb}, \cite{p:tvb2},
  \cite{p:torrievalleaucpl}.
\bibitem{note:welimit}
In the following we limit our discussion to the application of the MUCA
  strategy to the elementary (zero-equilibration) formulation. Generalisation
  to the arbitrary equilibration case is straightforward, see section
  \ref{sec:pfg}.
\bibitem{note:chanham}
In the case of the phase-constrained simulations the \ce\ $\ecal$ appearing in
  the acceptance rates is fixed, whereas in the case of the PS method it
  varies, depending upon which phase it is in.
\bibitem{note:wnb}
Since $\hat{P}(\mba | \pi^m_B)$ should contain the regions for which
  $w_d(\mba)$ is non-negligible and since $w_n$ is proportional to $w_d$, we
  see that $\hat{P}(\mba | \pi^m_B)$ should, like $\hat{P}(\mba | \pi^m_A)$,
  also contain the regions for which $w_n$ is non-negligible.
\bibitem{note:ch4ps}
\RBAcal\ may also be equivalently estimated from the estimator given in Eq.
  \ref{eq:psest}.
\bibitem{note:oppest}
In this case \RBAcal\ is estimated from the identity $\RBAcal =
  <e^{\mba}>_\pib$.
\bibitem{note:muloverlap}
As we have just discussed the fraction of (effective) configuration space over
  which they do not overlap tends to zero as the system size increases.
\bibitem{note:ARMH}
The MUCA-AR method, described in section \ref{sec:mul}, is a step in this
  direction. In this case one has two independent simulations which are made to
  overlap with the aid of MUCA weights.
\bibitem{note:pt}
The parallel tempering method, as it is called, has been generalised in
  \cite{p:tesi}, \cite{p:fukunishi}, \cite{p:vlugt}. The parallel tempering
  algorithm includes an additional trial Monte Carlo move in which
  configuration of adjacent replicas are periodically swapped, and accepted
  with the appropriate metropolis acceptance criterion. This acts so as to
  decorrelate the output of the simulation, and in the case of a rough energy
  landscape allows the simulation to escape from local basins of attractions.
  This swapping of configurations, which is absent from the original WHAM
  method, has been incorporated into the method in a trivial way in
  \cite{p:sugita}. Combinations of the Parallel tempering method and the ST
  method have also been constructed in \cite{p:yanpab}, \cite{p:yanpab2},
  \cite{p:fenesco}. In \cite{p:mitsu} the Parallel tempering method has also
  been shown to be an effective method of obtaining the MUCA weights. In this
  same paper it has also been shown how the MUCA and Parallel tempering methods
  can be successively incorporated into a single extended sampling strategy
  framework.
\bibitem{note:constantFG}
Some of what has been investigated here has been investigated in a different
  context in \cite{p:jarprl},\cite{p:hendrixjar}, \cite{p:jarpre},
  \cite{p:jarpre2}.
\bibitem{note:supfg}
We note that this indicates that the MH method can be viewed as a limiting case
  of the FG method. See appendix \ref{app:equivalent}.
\bibitem{note:conditionalcon}
It is important, however, to realise that this convergence is conditional on
  $\delt / \delela$ being sufficiently great. This can be better understood by
  first noting that there exists a lower bound for \delt . This lower bound
  corresponds to the case where the equilibration time \delt\ corresponds to a
  single particle perturbation. In the units that we have employed, this
  corresponds to $\delt_{min} = 1/N$, where N is the system size. Because of
  this lower bound on \delt , it turns out that the smallest value that
  \delela\ can assume, if Eq. \ref{eq:deltcon} is to hold, is: \[ \delela_{min}
  = \frac {1} {N t_W} \] \noindent Now if $t_W$ is too small, then even at the
  smallest allowed value of \delela , $\delt = 1 / N$ will not be sufficiently
  large so as to ensure that the method is free of systematic errors (since it
  will not eliminate enough of the lag). Since this value of \delela\
  corresponds to the left most point in graph \ref{fig:constantworktime}, it
  follows that all other values of \delela , subject to the constraint of Eq.
  \ref{eq:deltcon} will yield greater, if not the same, systematic errors.
\bibitem{note:discrete}
We make this assumption for the simple reason that it makes the conceptual
  transition from the classical to the quantum case easier.
\bibitem{note:newpb}
Here $H ({\sigma^s_i})$ includes both the momentum contributions and the \ce\
  contributions, and represents the total energy (or Hamiltonian \cite{b:reif})
  of the system. In the earlier formula of Eq. \ref{eq:absoluteprob}, only the
  configurational part of the total energy was included. The reason for this is
  that, as we shall see, the spatial and momentum parts of the hamiltonian $H$
  no longer decouple in a way which allows the momentum contributions to be
  integrated out (as has been done in the process of arriving at Eq.
  \ref{eq:absoluteprob}).
\bibitem{note:qfur}
That is the system cannot be in a superposition of {\em different} energy
  eigenstates. If this was the case then, because the system and reservoir may,
  together, be considered to be isolated system, the reservoir would also be in
  a superposition of states. Since the reservoir is classical (and since the
  observer is part of the reservoir!) this cannot be the case.
\bibitem{note:echno}
Briefly, the indistinguishability of particles means that if two particles are
  swapped in the wave function, then the functional form of that wavefunction
  should remain the same up to a sign. If upon the exchange (or swapping) of
  two indistinguishable particles the wavefunction remains unchanged, then the
  two particles under consideration are {\em bosons}. If the wavefunction
  changes sign upon this interchange, then the two particles are {\em
  fermions}.
\bibitem{note:discretemom}
Note that the following expression for the identity operator is only valid for
  an infinitely large system. The reason for this is that a finite system with
  periodic boundary conditions will have a discrete momentum spectrum
  associated with it. As a consequence Eq. \ref{cont_mom} is only an
  approximation, and is valid provided $\la_q \tau << L^2$, where L is the
  length of the system \cite{p:ceperleyreview}.
\bibitem{note:lowerorder}
Note the change in the leading order term in going from Eq.
  \ref{eq:separateexp} to Eq. \ref{eq:scalepart}. The reason that we go from a
  $1/P^2$ term to a $1/P^3$ is because $Z_P$ in Eq. \ref{eq:scalepart} is
  obtained by taking the trace of the product of P quantities of the form given
  in Eq. \ref{eq:separateexp}.
\bibitem{note:whypi}
The reason why simulations based on Eq. \ref{eq:prim_pot} and Eq.
  \ref{eq:partqm} are referred to as path integral methods is because these set
  of equations represent approximations of the path integral: \[ Z \equaldot
  e^{-\beta E^0} \int {{\cal {D}}} \rvec_t e^{-\beta \int dt \hcal {}
  {\rvec_t}} \] \noindent where: \[ \hcal {} {\rvec_t} = \frac {1} {4\la_q
  \beta^2} (\frac {\partial \rvec_t} {\partial t})^2 + \ecal(\rvec_t) \]
  \noindent and where the paths $\rvec_t=\rvec(t)$ have the constraint that
  $\rvec(t)=\rvec(0)$.
\bibitem{note:fpimc}
For completeness sake, we note that a radically different method that exists is
  the Fourier path integral approach of Doll and Freeman \cite{p:dollfreeman},
  \cite{p:freemandoll}. In this formulation the degrees of freedom of the
  system are parameterised by the normal modes of the path. This provides an
  alternative formulation to the one that we have discussed. A comparison of
  this method with the formulation that we have dealt with has been
  investigated in \cite{p:chakcompare} (see also \cite{p:dollcomment}). The
  differences between the two methods were not that significant, and a more in
  depth study, as commented in \cite{p:chakcomment}, needs to be done. On a
  theoretical level the two methods have been shown to be intimately related
  \cite{p:coalson} (in which it is argued that in the large timeslice/fourier
  slice limit the two methods should be more or less equivalent. It is possible
  that the differences in \cite{p:chakcompare} have more to do with the
  different estimators used than with the fact that one is employing different
  methods (see \cite{p:dollcomment}). We also note that whereas the Barker and
  Virial estimators do not show clear cut advantages in the real space
  discretisation scheme that we have employed, the virial estimator is a lot
  more efficient than the straightforward Barker estimate in the case of the
  Fourier path integral approach (see \cite{p:dollcomment}).
\bibitem{note:pimd}
In principle one should be able to simulate the classical system described by
  the hamiltonian in Eq. \ref{eq:prim_pot} via molecular dynamics. However it
  has been shown in \cite{p:hallberne} that such a method leads to
  non-ergodicity. The method can be made to be ergodic by making some
  modifications to the algorithm (see 'Path Integration via Molecular Dynamics'
  in \cite{b:julich}), though we will not concern ourselves with these issues
  any further. As pointed out in \cite{p:hallberne}, the path integral Monte
  Carlo method does not encounter such difficulties.
\bibitem{note:ceperleyterm}
We are using the terminology of \cite{p:ceperleyreview}.
\bibitem{note:nobose}
In this case phenomena such as Bose-Einstein condensation will not take place.
\bibitem{note:whysig}
The fact that the quantum effects become more significant when the gradients of
  the \ces\ are higher can be most easily seen from the Wigner-Kirkwood
  corrections to the probability distribution \cite{p:wigner},
  \cite{p:kirkwood}. This formula states that when the system is 'almost'
  classical, the probability is still given by the Gibbs distribution Eq.
  \ref{eq:absoluteprob}, but with the \ce\ $\ecal$ replaced by: \[
  \ecal(\rvec)\rightarrow \ecal(\rvec) + \frac {h^2\beta} {12} \s {i=1} {N}
  \frac {1} {m} \frac {\partial^2 \ecal(\rvec)} {{\partial \vec {r}_i}^2} -
  \frac {h^2 \beta^2} {24} \s {i=1} {N} \frac {1} {m} (\frac {\partial
  \ecal(\rvec)} {\partial \vec{r}_i})^2 \] \noindent where $\vec {r}_i$ denotes
  the position vector of particle i.
\bibitem{note:ocond}
O can either be a hermitian operator or it can be a macrovariable for which its
  expectation $<O>_P$ is obtained by taking the derivative of the P-replica
  partition function of Eq. \ref{eq:partqm} (see for example Eq.
  \ref{eq:prim_kinetic} and Eq. \ref{eq:prim_energy_estimator}).
\bibitem{note:likeclassical}
Note that the 'finite replica corrections' here are just like those that need
  to be accounted for in the classical case. See \cite{note:wayscale}.
\bibitem{note:pimcsys}
That is the accuracy when only systematic errors are taken into account. In
  reality the methods will also take different times to achieve a certain
  number of sweeps (for simulations with the same number of replicas), and
  further they will have different correlations associated with them. This
  means that for a fixed time simulation the two methods will also have
  different statistical errors associated with them.
\bibitem{note:diffdef}
We point out that our definition of the de Boer parameter corresponds to the
  square of the de Boer wavelength, as defined in \cite{p:neumannzoppi}.
\bibitem{note:handrule}
Suppose that one performs a quantum simulation with P replicas at an inverse
  temperature of $\beta$. Since the systematic errors scale as $[\beta/P]^n$
  where n=2 for the PA method and n=4 for the HOA method, it follows that if
  one wants to perform another simulation at an inverse temperature of
  $\tilde{\beta}$ and with $\tilde{P}$ replicas such that the systematic error
  associated with the estimate of the expectation of observables is of the same
  size as the original simulation, then one should have $\beta / P =
  \tilde{\beta} / \tilde{P}$. One may use this as an approximate rule of thumb
  to determine the number of replicas one needs at a lower temperature given
  the knowledge of the number of replicas needed at a higher temperature in
  order to be in the scaling regime, so as to allow determination of the
  asymptotic limit via the graphical extrapolation technique (see section
  \ref{sec:macest} and section \ref{sec:higherorder}).
\bibitem{note:ramirez}
See \cite{p:ramirez} for a nice illustration of this. We note that as the
  temperature is lowered, more replicas are needed in order for the simulation
  to converge to the correct value as predicted by theory. We also note the
  crucial feature of this graph, which is the stationary non-zero value that
  the mean total excitation energy assumes in the limit $T\rightarrow 0$, is
  only strictly obtained in the limit of the number of replicas tending to
  $\infty$. For finite replicas, the mean total energy does vanish in this
  limit \cite{b:tildesley}. However in order to see this one must perform
  simulations extremely close to the origin. If one does this, what one
  observes is that eventually the plateau of the total excitation energy turns
  downward and the curve falls to the origin. This downward fall from the
  plateau is only observed for temperatures below some threshold temperature
  $T_h$, which reduces as the number of replicas increases.
\bibitem{note:duration}
Note that since the HOA method is twice as slow as a PA simulation employing
  the same number of replicas, we see that in this case the two methods will go
  through the same number of (polymer) sweeps in a given amount of time.
\bibitem{note:meanwhat}
One may gauge the importance of the quantum effects (zero point motion) by
  looking at the deviations of the mean particle displacement from the
  classical limit (see figure \ref{pic:radi}).
\bibitem{note:extramhq}
It was found that for the MH method, the statistical error was independent of
  the number of replicas used. Therefore one may choose an arbitrarily large n
  and appropriately parallelise.
\bibitem{note:howgrad}
By gradual we mean choosing sufficiently small increments \delela\ in the
  configurational energy and sufficiently long equilibration times \delt\
  between successive work increments.
\bibitem{note:detofS}
The harmonic contribution to the FED, ${\triangle F}^h_{BA}$, which was
  inferred from \ref{eq:freetempch3} , may also be inferred from the
  determinant to $\avec {S}_{BA}$ (Eq. \ref{eq:stran2}). To find this one first
  changes to a representation in which the $\seta{{\mf e}_4, {\mf e}_5, ...,
  {\mf e}_{3N}}$ become 3N-3 column vectors. This then makes $
  \avec{\tilde{T}}_\al$ a square matrix. The subsequent derivation is then
  identical to that of Eq. \ref{eq:detS}: \begin{equation} {\det \avec
  {S}}_{BA} = \frac {{\det \avec{\tilde{T}}}_B} {{\det \avec{\tilde{T}}}_A} =
  \p {m=4} {3N} \sqrt{\frac {k^m_A} {k^m_B}} \nonumber \end{equation} Since the
  determinant is independent of this representation this result is unaffected
  on transforming back to the original representation.
\bibitem{note:alternateO}
The overlap $\tilde{O}$ can also be estimated directly from the histograms of
  the two phase constrained distributions (see the first line of Eq.
  \ref{eq:overlapfg} (a)). One may then use this to find $\RBAcal$ by starting
  off with an estimate ${\hat{\cal R}}_{BA}$ and varying this until it's
  employment in the estimator of Eq. \ref{eq:overlapfg} results in a value
  which agrees with the estimate of $\tilde{O}$ obtained directly from the
  histograms.
\end{thebibliography}

% ======================== The end! =====================================

\end{document}